\documentclass[a4paper,12pt,times,numbered,print,custommargin]{PhDThesisPSnPDF}

\usepackage[toc,nonumberlist,acronym]{glossaries}
\usepackage[utf8]{inputenc}
\usepackage{bibentry}
\usepackage{natbib}
\usepackage{gensymb}
\usepackage{enumerate}
\usepackage{enumitem} 
\usepackage{hyperref}

\usepackage{epstopdf}

\usepackage{mathtools}

\usepackage{tabularx}

\usepackage{tabularx}
\usepackage{multirow}
\usepackage{csvsimple}
\usepackage{mhchem}
\usepackage{graphicx}
\usepackage{subcaption}
\usepackage{mwe}
\usepackage{float}
\usepackage{placeins}
\usepackage{amsfonts}
\usepackage{booktabs}
\usepackage{siunitx}
\usepackage{array}

\usepackage{pdfpages}

\usepackage{tabularx}
\usepackage{multirow}
\usepackage{csvsimple}
\usepackage{mhchem}
\usepackage{graphicx}
\usepackage{subcaption}
\usepackage{mwe}
\usepackage{float}
\usepackage{placeins}
\usepackage{amsfonts}
\usepackage{booktabs}
\usepackage{siunitx}
\usepackage{array}
\usepackage{hyperref}
\usepackage{pdfpages}
\usepackage{algorithm}
\usepackage{algorithmicx}
\usepackage{algpseudocode}

\usepackage{tabularx}
\usepackage{multirow}
\usepackage{csvsimple}
\usepackage{mhchem}
\usepackage{graphicx}
\usepackage{subcaption}
\usepackage{mwe}
\usepackage{float}
\usepackage{placeins}
\usepackage{amsfonts}
\usepackage{booktabs}
\usepackage{siunitx}
\usepackage{array}
\usepackage{hyperref}
\usepackage{pdfpages}
\usepackage{algpseudocode}

\usepackage{amssymb}
\usepackage{amsthm}
\usepackage{amsmath}
\usepackage{mathtools}
 \usepackage{booktabs}
\usepackage{mhchem}
\usepackage{xcolor}
\usepackage{csvsimple,booktabs}
\usepackage{graphicx}
\usepackage{lscape}
\usepackage{algpseudocode}
\usepackage{adjustbox}
\usepackage{rotating}

\usepackage{subcaption}
\usepackage{mwe}

\usepackage{amsmath,amssymb,amsfonts}
\usepackage{graphicx}
\usepackage{textcomp}
\usepackage{xcolor}
\usepackage{csvsimple}
\usepackage{booktabs}
\usepackage{breqn}

\usepackage[utf8]{inputenc} 
\usepackage[T1]{fontenc}    
\usepackage{hyperref}       
\usepackage{url}            
\usepackage{booktabs}       
\usepackage{amsfonts}       
\usepackage{nicefrac}       
\usepackage{microtype}      
\usepackage{chemformula}

\ifsetCustomMargin
\RequirePackage[left=30mm,right=20mm,top=20mm,bottom=20mm]{geometry}
\setFancyHdr 
\fi

\ifsetCustomFont

\fi

\RequirePackage[labelsep=space,tableposition=top]{caption}

\usepackage{subcaption}

\usepackage{booktabs} 
\usepackage{multirow}

\usepackage{siunitx} 

\ifuseCustomBib
   \RequirePackage[square, sort, numbers, authoryear]{natbib} 

\fi

\usepackage{enumitem}

\makenoidxglossaries

\newglossaryentry{ANN}
{
        name=Artificial neural network,
        text=artificial neural network,
        description={A machine learning algorithm that is modelled on the brain and made up of multiple layers of neurons and weights}
}

\newglossaryentry{ABM}
{
        name=Agent-based model,
        text=agent-based model,
        description={A simulation technique made up of individual agents}
}

\newglossaryentry{SVR}
{
        name=Support vector regression,
        text=support vector regression,
        description={A machine learning algorithm used for regression which depends only on a subset of the training data}
}

\newglossaryentry{SVM}
{
	name=Support vector machine,
	text=support vector machine,
	description={A machine learning algorithm used for classification which depends only on a subset of the training data}
}

\newglossaryentry{SRMC}
{
	name=Short run marginal cost,
	text=short run marginal cost,
	description={The cost it takes to produce an additional MWh of electricity, excluding capital costs}
}

\newglossaryentry{GenCo}
{
	name=Generation company,
	text=generation company,
	description={A company which owns power plants and sells electricity to the grid}
}

\newglossaryentry{IRES}
{
	name=Intermittent renewable energy source,
	text=intermittent renewable energy source,
	description={A source of electricity from renewable energy sources which can not be dispatched}
}

\newglossaryentry{peakerplants}
{
	name=Peaker plant,
	text=peaker plant,
	description={A peaker power plant is one which which is used in times of high demand and low supply. Due to their expense they are only used when necessary and can not compete with other sources of energy during the majority of market operation}
}

\newglossaryentry{ISO}
{
	name=Independent system operator,
	text=independent system operator,
	description={An independent system operator cordinates, controls and monitors the operation of the electrical power system.}
}

\newglossaryentry{dispatched}
{
	name=Dispatched,
	text=dispatched,
	description={A power plant can be dispatched if the time and amount of electricity that can be controlled by a human operator. Examples are gas, coal and oil power plants.}
}

\newglossaryentry{digitaltwin}
{
	name=Digital twin,
	text=digital twin,
	description={A simulation of a specific real world system}
}

\newglossaryentry{montecarlo}
{
	name=Monte-Carlo,
	text=monte-carlo,
	description={A class of computational algorithms that rely on repeated random sampling to obtain results}
}

\newglossaryentry{LDC}
{
	name=Load duration curve,
	text=load duration curve,
	description={An arrangement of all electricity demand levels in descending order of magnitude}
}

\newglossaryentry{PDC}
{
	name=Price duration curve,
	text=price duration curve,
	description={An arrangement of all electricity prices in descending order of magnitude}
}

\newglossaryentry{RL}
{
	name=Reinforcement learning,
	text=reinforcement learning,
	description={A machine learning technique which learns appropriate actions through interacting with an environment}
}

\newglossaryentry{LCOE}
{
	name=Levelised cost of electricity,
	text=levelised cost of electricity,
	description={The minimum constant price at which electricity must be sold at for a power plant to break even over its lifetime}
}

\newglossaryentry{WACC}
{
	name=Weighted average cost of capital,
	text=weighted average cost of capital,
	description={The rate at which a company is expected to pay on average for its loans and in stock dividends/buybacks}
}

\newglossaryentry{CCGT}
{
	name=Combined Cycle Gas Turbine,
	text=combined cycle gas turbine,
	description={A type of gas powerplant which utilises heat as well as motion to create electricity}
}

\newglossaryentry{NPV}
{
	name=Net present value,
	text=net present value,
	description={The difference between the present value of cash inflows and the present value of cash outflows over a period of time}
}

\newglossaryentry{representativedays}
{
	name=Representative days,
	text=representative days,
	description={The difference between the present value of cash inflows and the present value of cash outflows over a period of time}
}

\newglossaryentry{GA}
{
	name=Genetic Algorithm,
	text=genetic algorithm,
	description={An optimisation algorithm which is inspired by the evolution of individuals within a population}
}

\newglossaryentry{MLP}
{
	name=Multilayer perceptron,
	text=multilayer perceptron,
	description={A class of feedforward artificial neural network (ANN)}
}

\newglossaryentry{smartmeter}
{
	name=Smart meter,
	text=smart meter,
	description={A small digital meter which records electricity consumption within a household or business premises}
}

\newglossaryentry{marketpower}
{
	name=Market power,
	text=market power,
	description={Market power refers to the ability of a firm, or group of firms, to raise and maintain prices above the level that would prevail under competition}
}

\newglossaryentry{onlinelearning}
{
	name=Online learning,
	text=online learning,
	description={Online learning is a machine learning or statistical method which data which is made available in sequential order is used to update the predictor for future data at each step}
}

\newacronym{abm}{ABM}{Agent Based Model}
\newacronym{abms}{ABMs}{Agent Based Models}

\newacronym{ann}{ANN}{Artificial Neural Network}

\newacronym{svr}{SVR}{Support Vector Regression}

\newacronym{svm}{SVM}{Support Vector Machine}

\newacronym{genco}{GenCo}{Generation Company}
\newacronym{gencos}{GenCos}{Generation Companies}

\newacronym{ires}{IRES}{Intermittent Renewable Energy Sources}

\newacronym{ldc}{LDC}{Load Duration Curve}

\newacronym{ppdc}{PPDC}{Predicted Price Duration Curve}

\newacronym{pdc}{PDC}{Price Duration Curve}

\newacronym{lcoe}{LCOE}{Levelised Cost of Electricity}

\newacronym{wacc}{WACC}{Weighted Average Cost of Capital}

\newacronym{arima}{ARIMA}{Auto Regressive Integrated Moving Average}

\newacronym{ccgt}{CCGT}{Combined Cycle Gas Turbine}

\newacronym{npv}{NPV}{Net Present Value}

\newacronym{beis}{BEIS}{UK Government Department for Business, Energy and Industrial Strategy}

\newacronym{nrmse}{NRMSE}{Normalised Root Mean Squared Error}

\newacronym{rmse}{RMSE}{Root Mean Squared Error}

\newacronym{ree}{REE}{Relative Energy Error}

\newacronym{ce}{CE}{Correlation}

\newacronym{ga}{GA}{Genetic Algorithm}

\newacronym{mase}{MASE}{Mean Absolute Squared Error}

\newacronym{mape}{MAPE}{Mean Absolute Percentage Error}

\newacronym{mae}{MAE}{Mean Absolute Error}

\newacronym{lstm}{LSTM}{Long Short Term Memory Neural Network}

\newacronym{mlp}{MLP}{Multilayer Perceptron}

\newacronym{ai}{AI}{Artificial Intelligence}

\newacronym{rbf}{RBF}{Radial Basis Function}

\newacronym{som}{SOM}{Self-Organizing Maps}

\newacronym{mdp}{MDP}{Markov Decision Process}

\newacronym{rl}{RL}{Reinforcement Learning}

\newacronym{ddpg}{DDPG}{Deep Deterministic Policy Gradient}

\newacronym{mars}{MARS}{Multivariate Adaptive Regression Spline}

\title{Modelling the transition to a low-carbon energy supply}

\author{Alexander John Michael Kell}

\dept{School of Computing}

\university{Newcastle University}
\crest{\includegraphics[width=0.2\textwidth]{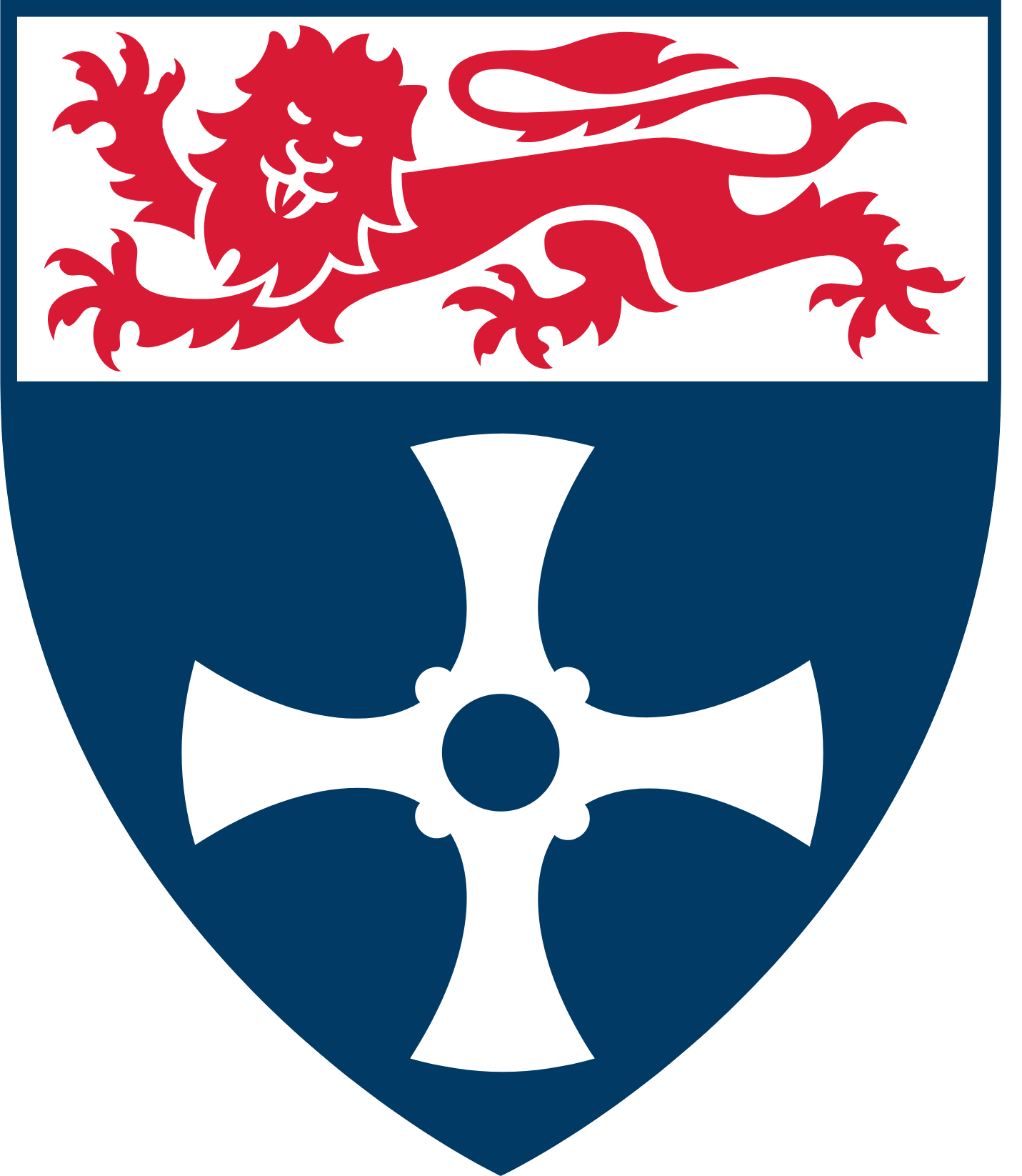}}

\degreetitle{Doctor of Philosophy}

\subject{LaTeX} \keywords{{LaTeX} {PhD Thesis} {Engineering} {University of
Cambridge}}

\ifdefineAbstract
\fi

\ifdefineChapter
\fi

\begin{document}

\frontmatter

\maketitle


\begin{dedication} 

I would like to dedicate this thesis to my family and my loving parents\dots

\end{dedication}


\begin{declaration}

I hereby declare that except where specific reference is made to the work of 
others, the contents of this dissertation are original and have not been 
submitted in whole or in part for consideration for any other degree or 
qualification in this, or any other university. This dissertation is my own 
work and contains nothing which is the outcome of work done in collaboration 
with others, except as specified in the text and Acknowledgements. This 
dissertation contains fewer than 65,000 words including appendices, 
bibliography, footnotes, tables and equations and has fewer than 150 figures.


\end{declaration}


\begin{acknowledgements}

First, I would like to express my gratitude to my supervisors Dr Matthew Forshaw and Dr Stephen McGough. Who, without their support, guidance and insight I would have been unable to develop my skills in an academic context. Through their help and encouragements I have been able to exceed my own expectations.

Secondly, I would like to thank my parents and brother for supporting me during this time. Especially my father who has proved a guiding light during the most challenging parts. For example, inspiring my work on how disruption could be modelled with computational methods and pushing me to go further.

Thirdly I would like to thank Sumiré Moncholi for putting up with me during these years and providing daily support and care.

Since joining Newcastle University I have been helped and inspired by the vast range of problems and applications tackled within the School of Computing and the School of Mathematics, Statistics and Physics. In particular Junyang Wang, George Stamatiadis, Adam Cattermole, Kathryn Garside, Alexander Brown, Michael Dunne-Willows, Ashleigh McLean, Lauren Roberts, Thomas Cooper, Shane Halloran, Jonny Law, Peter Michalák, Saleh Mohamad and Mario Parreno.

I would also like to thank all my friends outside of the department during this time, especially Thomas Smith, Clement Venard, Sam Major, Marta Fernandez, Wenijan Yang, Alessandro Boussalem, Lars Eriksson, Tom Brunt, Owen Jones, Marianne Amor, Kevin Amor and Connor Scott.

Finally, I would like to thank my colleagues at the University of Cambridge's Centre for Science and Policy, for providing a challenging and stimulating environment to undertake an internship: Nicola Buckley, Katie Cohen, Rob Doubleday, Su Ford, Kate McNeil, Lauren Milden, Jackie Ouchikh, Erica Pramauro and Laura Sayer.

\end{acknowledgements}
\begin{abstract}

A transition to a low-carbon electricity supply is crucial to limit the impacts of climate change. Reducing carbon emissions could help prevent the world from reaching a tipping point, where runaway emissions are likely. Runaway emissions could lead to extremes in weather conditions around the world - especially in problematic regions unable to cope with these conditions. 

However, the movement to a low-carbon energy supply can not happen instantaneously due to the existing fossil-fuel infrastructure and the requirement to maintain a reliable energy supply. Therefore, a low-carbon transition is required, however, the decisions various stakeholders should make over the coming decades to reduce these carbon emissions are not obvious. This is due to many long-term uncertainties, such as electricity, fuel and generation costs, human behaviour and the size of electricity demand. A well choreographed low-carbon transition is, therefore, required between all of the heterogenous actors in the system, as opposed to changing the behaviour of a single, centralised actor.

The objective of this thesis is to create a novel, open-source agent-based model to better understand the manner in which the whole electricity market reacts to different factors using state-of-the-art machine learning and artificial intelligence methods. In contrast to other works, this thesis looks at both the long-term and short-term impact that different behaviours have on the electricity market by using these state-of-the-art methods. 

Specifically, we investigate the following applications:

\begin{enumerate}
	\item Predictions are made to predict electricity demand in the short-term. We model the impact that poor predictions have on investments in electricity generators and utilisation over the long-term. We find that poor short-term predictions lead to a higher proportion of coal, gas, and nuclear power plants.
	\item We devise a long-term carbon tax for the United Kingdom using a genetic algorithm approach. We find multiple strategies that can minimise both long-term carbon emissions and electricity cost.
	\item Oligopolies can have a detrimental effect on an electricity market by raising electricity prices without an increase in benefit to users. Reinforcement learning can be used to devise intelligent bidding strategies which are based upon forecasts and predictions of other agent behaviour to maximise revenues. These behaviours can not be modelled through traditional rule-based algorithms. We use reinforcement learning to model strategic bidding into the electricity market, and find ways to limit the impact of this strategic bidding through a market cap. We find that introducing a market cap can significantly reduce the ability for oligopolies to manipulate the market.
\end{enumerate}

These studies require a number of core challenges to be addressed to ensure our agent-based model, ElecSim, is fit for purpose. These are:

\begin{enumerate}
	\item Development of the ElecSim model, where the replication of the pertinent features of the electricity market was required. For example, generation company investment behaviour, electricity market design and temporal granularity. We find that the temporal granularity of the model has a large impact on accuracy of the model, but with suitably chosen representative days calibration is possible to accurately model a time period.
	\item The complexity of a model increases with the replication of increasing market features. Therefore, optimisation of the code was required to maintain computational tractability, to allow for multiple scenario runs. This enabled us to run multiple iterations to train different machine learning techniques.
	\item Once the model has been developed, its long-term behaviour must be verified to ensure accuracy. In this work, cross-validation was used to both validate and calibrate ElecSim. We are able to accurately model a historic period observed in the real-world with this approach
	\item To ensure that the salient parameters are found, a sensitivity analysis was run. In addition, various example scenarios were generated to show the behaviour of the model. We find that the input parameters, such as the cost of capital have a disproportionate effect on the long-term electricity mix.
\end{enumerate}

The findings outlined previously demonstrate the ability for artificial intelligence, machine learning and agent-based models to perform complex analyses in an uncertain system. We find that solely focusing on the accuracy of machine learning techniques, for instance, misses out on a significant amount research potential. We instead argue, that by further developing these research themes, we are able to better understand the electricity market system of the United Kingdom.

\end{abstract}


\setcounter{tocdepth}{1}
\tableofcontents

\listoffigures	

\listoftables

	


\printnoidxglossaries 


\mainmatter


\chapter{Introduction}  
\label{chapter:intro}

\ifpdf
\graphicspath{{Chapter1/Figs/Raster/}{Chapter1/Figs/PDF/}{Chapter1/Figs/}}
\else
\graphicspath{{Chapter1/Figs/Vector/}{Chapter1/Figs/}}
\fi

\section{Motivation} 


The impacts of global warming on the earth may have profound effects on land and ocean ecosystems \cite{IPCC2018}. The release of carbon emissions into the atmosphere increases the likelihood of the most severe impacts and increases the likelihood that tipping points are reached, where runaway carbon emissions and average temperature rises are likely. 

Therefore a transition to a low-carbon energy supply is required to prevent the impacts of climate change. A low-carbon electricity supply is one which releases a lower amount of carbon dioxide over its lifetime than the current, fossil-fuel based system. 

However, such a transition is complex and contains multiple uncertainties. For instance, what carbon tax should the UK government set over the next 30 years? Are poor short-term electricity demand forecasts locking us in to higher emissions over the long-term? Can we limit the market power of generator companies? And can we rely on these models to make decisions of such importance? 

Whilst much work has been carried out investigating energy models in different electricity and energy markets, these models are not often fully validated against real-world data. This thesis seeks to validate a novel agent-based model, ElecSim, by calibrating with real-world data. Through this calibration, confidence can be gained in the underlying dynamics of the model and provide policy makers with the opportunity to better understand the system with which they are dealing with.

Secondly, much work has been undertaken to understand certain aspects of electricity markets using agent-based models and machine learning. However, this work, often, does not place these findings into a wider context. For instance, whilst a high degree of focus is placed on the ability of reinforcement learning to bid strategically within an agent-based setting, how to limit this behaviour has not been investigated to the same extent. 

Similarly, machine learning has been used to predict electricity demand at various time intervals. However, the effect that different prediction methods have on the long-term electricity market has not been explored. Finally, machine learning and simulation has the ability to optimise an entire system. However, this ability received much research attention, instead the focus has been on smaller scale changes to models. In this thesis we aim to fill this research gap by first calibrating our model,  and secondly reducing electricity price and carbon emissions from the United Kingdom's electricity mix by optimising carbon tax strategies.

\section{Research questions}

The central question of this thesis is: how can artificial intelligence (AI) and machine learning (ML) answer fundamental questions of the energy transition using an agent-based model of the electricity system?

This thesis aims to go beyond small-scale improvements to agent-based models and answering scope-limited questions by understanding first: what challenges can AI and ML tackle, and secondly: how do these methods relate back to the wider energy system?

By taking this approach, we answer multiple subquestions, which are explored below:

\begin{enumerate}

	\item \textbf{Can a simulation model an electrcity market over the long-term?} Traditional electricity market models mimic the behaviour of centralised actors with perfect foresight and information. Other models which model actors as having imperfect foresight and information lack the ability to model multiple time-steps over a long time horizon. In Chapter \ref{chapter:elecsim}, a novel open-source agent-based model called ElecSim is presented which challenges these issues. We show that it is possible to create an electricity model which can simulate multiple time-steps over a long-time horizon and generate realistic electricity mixes as model outputs.
	
	\item \textbf{Is it possible to model the variability of an electricity system?} Intermittent renewable energy can produce electricity at both maximum capacity and at zero capacity in short time intervals. It, therefore, becomes important to model these variations in power output over a long-term horizon. Otherwise, the model may overestimate the production of energy from renewables and underestimate the variability of such technologies. This is achieved in Chapter \ref{chapter:elecsim} by showing that with representative days, we are able to accurately model an entire year with a reduced computational burden. Without this additional feature, an overestimation of intermittent renewable energy and underestimation of dispatchable generation is observed.
	
	\item \textbf{Can we trust an electricity model's outputs?} Whilst long-term energy models can provide quantitative advice to experts, policymakers and stakeholders, the veracity of these models are rarely validated. The validation of long-term electricity models can highlight problems with the dynamics of the model, important components, and provide confidence in the outputs. Here, an approach is presented to provide confidence in such outputs in Chapter \ref{chapter:elecsim}. We achieve this through the use of a distributed genetic algorithm optimisation algorithm to calibrate our model. Through such a calibration we are able to observe a real-world transition of the UK's electricity market from coal to gas over a 5-year period.

	\item \textbf{Do poor short-term forecasts of electricity demand have a long-term impact?} Forecasting of electricity demand within electricity markets is critical. The settlement of markets occurs prior to the time in which the demand must be supplied. However, the long-term effect on the markets of poor forecasts has not been investigated. In Chapter \ref{chapter:demand} we investigate the long-term impact of poor short-term predictions. We find that poor short-term demand forecasts leads to increased investments in coal, gas and nuclear power, with a reduction in both onshore and offshore wind. 
	
	\item \textbf{Is it possible to use an algorithm to set carbon policy?} Setting carbon taxes has been proposed as a solution to reduce our reliance on fossil fuels. However, the impact of such carbon taxes are unknown, as are the optimal strategies from different perspectives. Such a problem can be solved using optimisation based techniques. Here, a solution to finding optimal strategies from the perspective of a benevolent government is presented in Chapter \ref{chapter:carbon}. We find that it is possible to find a variety of different carbon tax strategies to minimise both electricity price and carbon emissions, as long as a high carbon price is set (around \textsterling 200). 
	
	\item \textbf{Is it possible to limit the power of large generator companies?} It is known that oligopolies have a negative effect on markets for consumers. However, what has been explored to a lesser degree, is the proportion of capacity that \gls{GenCo} must own before they have market power. In addition, what would the effect be of a market cap on such electricity markets? Would a market cap reduce the ability for \acrfull{gencos} to inflate electricity prices artificially? Chapter \ref{chapter:reinforcement} investigates this issue. We find that if a generator company, or group of colluding generator companies own over 11\% of the total generation capacity, electricity prices start to increase. However, the impact of this market power can be limited through the setting of a market cap. In the case of the UK, a market cap of \textsterling 190/MWh suffices.
	
\end{enumerate}

Through these questions we not only answer whether AI and ML methods can be used with electricity market agent-based models, but also what is the wider impact of the behaviours on the market.

\section{Methodology}

Primarily, in this work, simulation is used as a tool to better understand and make projections for electricity markets. Specifically, in this thesis, the agent-based modelling paradigm is used. This enables us to model generator companies as individual agents, with heterogeneous strategies and characteristics. These agents have access to imperfect information and imperfect foresight. This methodology differentiates this work from the traditional centralised optimisation approach. Agent-based models are critical to model the behaviour of individual actors within an electricity system. Without this distinction, the system must be modelled as a homogenous system, which does not accurately reflect the real world. Through this approach, we hope to learn that it is possible to accurately model the UK's national electricity system with an agent-based approach.

Machine learning and statistical techniques are used to make short-term forecasts of electricity demand. We use both deep learning, offline learning and \gls{onlinelearning} to further improve our methods. Online learning is a machine learning approach which utilises new data to update model weights, and does not require the model to be completely retrained, which is the case for offline learning. In comparison, deep learning utilises neural networks with many different layers. We utilise these methods due to their data-driven approach and strong ability to forecast time-series data. Online learning is used as over the time-periods with which we are forecasting, the underlying time-series changes in structure. Online learning is able to continually use new data points to retrain the model. Deep learning, on the other hand, uses many layers to learn more complex patterns from the training set. Through taking this approach, we hope to learn that it is possible to improve predictions for energy demand data when compared to the traditional machine learning methods.

Once our simulation model is built, we are able to answer different questions using several approaches. For example, we perturb the exogenous electricity demand by the error distribution generated by the aforementioned electricity demand forecasting methods. This provides an insight into how small errors can have large impacts on long-term electricity markets in terms of both investments made and generator utilisation. This approach was taken due to its ability to mimic the behaviour of generator companies in a simple manner. We hope to learn what the impact of short-term decisions are on the long-term market.

Multi-objective genetic algorithms are used to explore carbon tax policies which will reduce both carbon emissions and average electricity price. We find that we are able to achieve both of these goals by setting a median carbon tax of ${\sim}$\textsterling200 per tonne of carbon dioxide. This methodology is chosen due to the genetic algorithm's distributed nature. We are able to run the algorithm in parallel and reduce training time significantly. From this, we hope to learn that there is an automatic method to reduce the search space for policy makers when coming up with complex policy in a high parameter space.

Finally, we explore the ability for deep reinforcement learning (DRL) to make strategic bidding decisions within a day-ahead electricity market. This work enables us to see the proportion of capacity that must be controlled to artificially inflate the electricity price in the market using \gls{marketpower}. Deep reinforcement learning was chosen due to its ability to quickly form a policy on the expected environment. The bidding environment with multiple competing agents is highly complex and difficult to solve through a rule-based approach, and so DRL is chosen to simplify the approach of forming a policy. Through this, we hope to learn the parameters which allow for the manipulation of the market, such as the size of generator companies and total capacity controlled, and how to reduce this impact of it occurs.

\section{Contributions}

The work in this thesis makes a number of key contributions:

\begin{enumerate}
	\item Development of the open-source, generalised long-term agent-based model for decentralised electricity markets, ElecSim \cite{Kell}. This can be accessed at: \url{https://github.com/alexanderkell/elecsim}. This model is parametrised to the UK electricity market, and contains the major pertinent features to model this market. This answers research question 1 by modelling an electricity market over the long-term.
	\item Validation of the aforementioned model through the use of cross-validation through five years and comparison with the established UK Government model until 2035 \cite{Kell2020}. Through this validation we are able to answer research questions 2 and 3 by modelling the inter-year variability and verify the outputs of the model.
	\item Forecasting of electricity demand using machine learning models and exploration of the impact of the prediction errors on the long-term electricity market \cite{Kell2018a}. This contribution answers research question 4 by showing that short-term errors do have a large impact on the final electricity mix.
	\item Optimisation of a carbon tax policy to reduce electricity cost and carbon emissions for the UK electricity market using a multi-objective genetic algorithm, from the perspective of a benevolent government \cite{Kell2020a}. This answers research question 5 by showing that it is possible to come up with an optimal strategy for setting a carbon price to reduce carbon emissions and electricity price.
	\item Exploration of the long-term impact of strategic bidding and collusion on decentralised electricity markets \cite{Kell2020d}. This answers research question 6 by showing that if generator companies control a large part of the market, market power occurs. However, it is possible to limit these market powers significantly by setting a market cap on electricity price.
\end{enumerate}

This work directly addresses the aim of the research. We use AI and ML to answer fundamental questions of the energy transition through the use of an agent-based model of the electricity system. With the development of a novel agent-based model, we are able to answer targeted questions on how the electricity system behaves under different pressures. We do not simply stop at validating the ability of an algorithm to perform a specific task, but rather relate this back to the wider market. 

\section{Thesis organisation and structure}

\begin{itemize}[itemindent=3em]
	\item[\textbf{Chapter \ref{chapter:intro}}] describes the motivations behind this thesis and highlights the main contributions of the research. Finally, the peer-reviewed publications produced during this PhD are presented. 
	\item[\textbf{Chapter \ref{chapter:background}}] describes the technical background material that relates to the rest of this work and investigates the different types of solutions that have been used in the current literature and differentiate this from this work. 
	\item[\textbf{Chapter \ref{chapter:elecsim}}] introduces the simulation framework developed within this work. This includes the technical details of the simulation tool, how this simulation is validated and the difficulties of validating such simulation models. Finally, a sensitivity analysis to show the impact of various variables is displayed, and some example future scenarios are produced. This chapter contains contributions 1 and 2.
	\item[\textbf{Chapter \ref{chapter:demand}}] explores the literature on electricity demand forecasting, how this can be improved with online learning, and what the long-term impact of errors are on decentralised electricity markets. This chapter contains contribution 3.
	\item[\textbf{Chapter \ref{chapter:carbon}}] demonstrates the ability for the model to come up with optimal strategies and scenarios through the use of machine learning techniques. Specifically, a carbon tax strategy between 2018 and 2035 is optimised to reduce both electricity cost and carbon emissions. This chapter contains contribution 4.
	\item[\textbf{Chapter \ref{chapter:reinforcement}}] demonstrates the ability for large or colluding generator companies to influence the price of electricity in their favour using deep reinforcement learning, as well as an approach to prevent this from occurring through the use of price caps. This chapter contains contribution 5.
	\item[\textbf{Chapter \ref{chapter:conclusion}}] summarises the conclusions of the work and motivates future directions for work in this area.
\end{itemize}

\section{Related publications}

During the course of my PhD I have authored the following peer-reviewed publications:	

\begin{itemize}
	
	\item[\textbf{\cite{Kell}}] \textbf{Kell, A., Forshaw, M., \& McGough, A. S. (2019). ElecSim : Monte-Carlo Open-Source Agent-Based Model to Inform Policy for Long-Term Electricity Planning. The Tenth ACM International Conference on Future Energy Systems (ACM e-Energy 2019), 556–565.}
	
	This work introduces the agent-based model, ElecSim. The current state-of-the-art of agent-based models is reviewed, and the technical foundations of how ElecSim works is detailed. An initial validation method of comparing the price duration curve of the model to that observed in real life is displayed. Finally, some example scenarios are presented. This work forms the basis for Chapter \ref{chapter:elecsim}.
	
	\item[\textbf{\cite{Kell2019a}}] \textbf{Kell, A., Forshaw, M., \& McGough, A. S. (2019). Modelling carbon tax in the UK electricity market using an agent-based model. E-Energy 2019 - Proceedings of the 10th ACM International Conference on Future Energy Systems, 425–427. }
	
	In this paper, further scenarios are explored by varying the carbon tax level. The effect of carbon tax on investments is demonstrated in the electricity market. This work augments the work done in Chapter \ref{chapter:elecsim}.

	\item[\textbf{\cite{Kell2020}}] \textbf{Kell, A. J. M., Forshaw, M., \& McGough, A. S. (2020). Long-Term Electricity Market Agent Based Model Validation using Genetic Algorithm based Optimisation. The Eleventh ACM International Conference on Future Energy Systems (e-Energy’20).}
	
	In this paper, further improvements are made to the ElecSim model. Through the addition of representative days, the model is validated between 2013 through 2018 by optimising for long-term predicted electricity price. The results are compared to those of the UK Government, for both a long-term and short-term validation. The results are comparable to those of the UK Government. This work further extends Chapter \ref{chapter:elecsim}.
	
	\item[\textbf{\cite{Kell2018a}}] \textbf{Kell, A., McGough, A. S., \& Forshaw, M. (2018). Segmenting residential smart meter data for short-Term load forecasting. e-Energy 2018 - Proceedings of the 9th ACM International Conference on Future Energy Systems.}
	
	In this work, various machine learning and deep learning techniques are used to predict electricity demand 30 minutes ahead using \gls{smartmeter} data. Various households are clustered using a \textit{k}-means clustering technique to further improve the accuracy. This paper forms the basis for Chapter \ref{chapter:demand}.
	
	\item[\textbf{\cite{Kell2020c}}] \textbf{Kell, A. J. M., McGough, A. S., \& Forshaw, M. (2020). The impact of online machine-learning methods on long-term investment decisions and generator utilization in electricity markets. 11th International Green and Sustainable Computing Conference, IGSC 2020.}
	
	This paper expands on the work carried out in \cite{Kell2018a}. However, instead of predicting 30 minutes ahead, electricity demand over the next day is predicted, over a 24-hour horizon. To improve results, online learning is used, which is able to update the parameters of the models as new data points become available. The results are significantly improved using this method. Finally, the errors of these predictions are taken and the long-term effects of these are shown on the electricity market using the ElecSim models, both in terms of generator utilisation and long-term investment decisions.

	\item[\textbf{\cite{Kell2020a}}] \textbf{Kell, A. J. M., McGough, A. S., \& Forshaw, M. (2020). Optimising carbon tax for decentralised electricity markets using an agent-based model. The Eleventh ACM International Conference on Future Energy Systems (e-Energy’20), 454–460.}
	
	In this paper, different carbon tax strategies are trialled using a multi-objective genetic algorithm. With the aim to minimise both the electricity price and carbon emissions. It is found that it is possible to achieve both of these goals through different carbon tax strategies. This work builds on \cite{Kell2019} by using a similar multi-objective genetic algorithm optimisation approach. 
	
		\item[\textbf{\cite{Kell2020d}}] \textbf{Kell, A. J. M., Forshaw, M., \& McGough, A. S. (2020). Exploring market power using deep reinforcement learning for intelligent bidding strategies. The 4th IEEE International Workshop on Big Data for Financial News and Data at 2020 IEEE International Conference on Big Data (IEEE BigData 2020).}
	
		This paper uses reinforcement learning to control the bidding behaviour of a single GenCo, or a group of colluding GenCos in the day-ahead market. It is found that if the other agents bid using their short run marginal costs, the GenCos which use the reinforcement learning algorithm are able to artificially inflate the market price using their market power. The work in this paper forms the work done in Chapter \ref{chapter:reinforcement}.
\end{itemize}

\subsection*{Papers not forming part of this thesis}

\begin{itemize}
	\item[\textbf{\cite{Kell2019}}] \textbf{Kell, A. J. M., Forshaw, M., \& McGough, A. S., (2019). Optimising energy and overhead for large parameter space simulations. 2019 10th International Green and Sustainable Computing Conference, IGSC 2019. }
	
	In this work, a multi-objective genetic algorithm is used to reduce both overhead and energy consumption of a cluster of computers at Newcastle University. This is achieved by varying different parameters of a reinforcement learning algorithm. The methods used in this paper influence much of the work presented in Chapters \ref{chapter:elecsim}, \ref{chapter:demand} and \ref{chapter:reinforcement}.
	
	\item[\textbf{\cite{KellA.J.M.McGoughA.S.ForshawM.MercureJ.F.Salas2020}}] \textbf{Kell, A. J. M., McGough A. S., Forshaw, M., Mercure, J. F., Salas, P. (2020). Deep Reinforcement Learning to Minimize Long-Term Carbon Emissions and Cost in the Investment of Electricity Generation. 34th NeurIPS 2020, Workshop on Tackling Climate Change with Machine Learning.}
	
	This paper modifies the FTT:Power model by using reinforcement learning as the electricity generator investment algorithm \cite{Mercure2012}. FTT:Power is a global power model which uses logistic differential equations to simulate competition between different electricity generating technologies. These logistic differential equations are replaced with the deep deterministic policy gradient reinforcement learning method. It is found find that, if the goal is to reduce both carbon emissions and electricity price, a transition to renewables occurs.
\end{itemize}

\chapter{Background and Literature Review}
\label{chapter:background}
\ifpdf
\graphicspath{{Chapter2/Figs/Raster/}{Chapter2/Figs/PDF/}{Chapter2/Figs/}}
\else
\graphicspath{{Chapter2/Figs/Vector/}{Chapter2/Figs/}}
\fi

\section*{Summary}

This chapter provides an overview of the relevant material which motivates and underpins the work carried out in this thesis and places the work in context with a literature review. Section \ref{sec:intro:elecmarkets}, introduces electricity markets and how they are regulated. In Section \ref{sec:intro:elecmarketsmodelling}, an introduction into how electricity markets are modelled is presented. We provide an introduction in Section \ref{sec:intro:simulationmodelling} to simulation and in Section \ref{sec:intro:ml} to machine learning. Finally, we conclude this chapter in Section \ref{sec:intro:conclusion}.

We also review the literature by giving an introduction to the relevant energy models and provide a systematic review on how AI has been applied to agent-based models. Section \ref{sec:litreview:energymodelling} gives an introduction to the field of energy modelling. In Section \ref{sec:model-types} we present different types of energy models, as well as a detailed view on agent-based models, the focus of this thesis. We present a table in Section \ref{sec:litreview:modelclassification}, which displays a high-level overview of the major models in the literature. Section \ref{sec:ML-abms} details a systematic review on how AI has been utilised with agent-based models in the electricity sector. We conclude this chapter in Section \ref{sec:litreview:conclusion}, where we discuss the limitations and benefits of modelling types and how AI can be integrated into agent-based models in the future.

%

\section{Background and Introduction}

\subsection{Electricity Markets}
\label{sec:intro:elecmarkets}

Electricity markets are complex. One of the principal reasons for this is the expense and difficulty of storing electricity. This is because electricity must be consumed the moment it is generated. Additionally, as electricity travels over high voltage transmission lines, electricity doesn't always follow simple or unique paths, especially when the transmission lines become congested. In addition there is power loss during transmission. Finally, electricity markets require technical overseers to ensure that the entire transmission system operates safely and reliably. 

Another aspect to consider is the fact that electricity is homogeneous. A single unit of electricity produced by a wind turbine is equivalent to a unit of electricity produced by a gas turbine. However, the functioning of different electricity producers, or generators, are not homogenous. Coal, gas and oil power plants can be \gls{dispatched} at the will of a human operator. Their ramp rates are well understood, as is the amount of fuel that is available. Where the ramp rate is defined as the increase or decrease in output per minute and is usually expressed as megawatts per minute (MW/min). \acrfull{ires}, however, such as solar, wind and tidal are dependent on the supply of solar irradiance, wind speed and the tide at any moment. Whilst these can be predicted, predictions are often wrong, and perfect knowledge is impossible. Therefore, at times where there is too much supply from \Gls{ires} generators must be curtailed if all regulator generators are already offline or can not be shut down or ramped down. In the opposite case, where there is too little supply from \acrshort{ires}, supply must be made up from other sources, such as coal, gas or hydro.

The environmental impact from different electricity generators differs significantly. Whilst gas and coal can be dispatched at a time convenient to the grid operators, they emit \ce{CO2} along with other toxic substances. In this context, dispatched refers to electricity that can be generated on demand at the request of power grid operators, according to market needs. Wind and solar do not dispatch such gases and substances, and therefore can not be controlled as easily. Storage technologies can be used to fill these gaps; however, large-scale storage depends on large pumped reservoirs that can move water to a higher position when demand for electricity is low, and supply is high. Not all geographies have access to such reservoirs, and therefore would rely on battery technology made from chemicals. However, reaching such high storage capabilities are expensive and are yet to have been done in the real world \cite{cole2019cost}. Another option is converting electricity to hydrogen. However, this technology is also expensive and uncompetitive with traditional fossil fuels such as coal, gas and oil. Currently, \gls{peakerplants} are used to fill these gaps. Peaker plants are plants which are used in times of high demand and low supply. However, these plants are expensive to operate, highly polluting and use fossil fuels \cite{lin2011energy}. It is expected that these \gls{peakerplants}s will be used increasingly due to the intermittent nature of renewable technologies \cite{lin2011energy}.

The electricity grid must match supply with demand at all times. Failure to do so results in an imbalance of supply and demand, and affects the frequency of the electricity network. Large differences between supply and demand can lead to blackouts or oversupply and may damage equipment. A number of different markets exist to regulate the supply and demand, running from within seconds, to days-ahead and bilateral contracts which settle electricity for years ahead \cite{conejo2010electricity}.

There are a number of different market mechanisms that can be used to balance the supply and demand of electricity. Largely these can be divided between ancillary services and wholesale transactions. Wholesale transactions can occur as bilateral trades or on a day-ahead market \cite{conejo2010electricity}. Bilateral trades can occur between two electricity suppliers and those that have an electricity demand \cite{conejo2010electricity}. In this case, suppliers and customers create contracts for electricity in advance. Typically, these agents must let the market operator know of their trades \cite{conejo2010electricity}. In a day-ahead market, the system price is, in principle, determined by matching offers from generators to bids from consumers at each node to develop a supply and demand equilibrium price \cite{conejo2010electricity}. 

Ancillary markets, on the other hand, provide a method to facilitate and support the continuous flow of electricity so that supply continually meets demand \cite{conejo2010electricity}. These include markets to regulate power and voltage control as well as frequency control. These markets make use of increasing supply or reducing demand at the times where this is required \cite{conejo2010electricity}. 

There exist many markets to better regulate the supply and demand at different time intervals. For example the intra-day market, capacity market. These markets ensure that electricity supply continues to meet demand as more intermittent renewable energy comes online. The intra-day market is able to dynamically meet fluctuations in supply and demand which were not matched using the day-ahead and bilateral markets. Whereas, the capacity market pays participants a per MW rate for the capacity they offer to the market.

Other mechanisms exist, such as the contracts for difference and the carbon price floor to transform the UK's electricity system. Contracts for difference pays a fixed price to generators irrespective of the market price. The carbon price floor places a minimum tax on carbon. 

An imbalanced system can also be resolved through the means of demand response (the reduction of electricity demand), interconnection with other countries, and through storage for example with vehicle to grid or by other means. In this thesis, however, we focus primarily on the day-ahead market and carbon price floor due to these markets having the largest impact on the macro-scale of the electricity market.

\subsection{Introduction to Electricity Market Modelling}
\label{sec:intro:elecmarketsmodelling}

Energy models are useful tools for insight into the functioning of electricity markets. For example, modellers and analysts can develop an intuition of how such a system works, and through modelling, they can challenge untested hypothesise. The argument that these models are used for insight and not just informational data is as old as the models themselves \cite{Huntington1982, Pfenninger2014}, and is true for models in many different disciplines \cite{Geoffrion1976}. We argue that energy models should not be taken as truth, because as previously mentioned, no model can perfectly model the real world. However, the intuition that can be learned can prove to be a valuable resource.

Energy modelling and energy policy as a distinct field began after the oil crisis in the 1970s, where long-term planning was deemed as important in the electricity field \cite{Craig2002}. Models which make use of optimisation techniques have been used since these times for diverse applications, from the global energy market to small off-grid systems. Optimisation techniques are models in which a central actor invests in the cost-optimal (cheapest) energy system over a multi-year time period.

The utilisation of energy models has lately been redirected to ensuring that there exists a security of supply, the resilience of the energy system, affordability and that there is a transition to a low-carbon supply. These models can also be used to investigate the impact that different technologies have on investments made in the future. 

However, since the traditional models have been established, various changes in the energy industry have occurred. Originally, electricity systems were built upon large-scale centralised electricity production based on fossil fuels \cite{foxon2010developing}. Since then there has been a transition towards decentralised, distributed, intermittent renewable energy sources, such as solar and wind \cite{IEA2015a}. In addition, there has been an advent in flexible demand driven by new technologies such as smart meters \cite{avancini2019energy}. This paradigm shift requires models which can work with higher temporal and spatial detail to account for fluctuations in demand, supply and distributed electricity generators.

Modelling electricity markets is a complex task. There exist many variables, actors, services and behaviours within electricity markets which make it impossible to perfectly model the system. Often simplifications must be made, where models are designed for a specific task \cite{Pfenninger2014}. Large established models exist which model every possible detail; however, with the increase in temporal and spatial resolution required, the computational tractability of these models can be negatively impacted. Many of the large models used today have existed for a long time, before the advent of the Internet \cite{Pfenninger2014}. Therefore, these models and modellers risk becoming outdated, as their models are not updated.

Energy and electricity models generally follow two approaches: bottom-up or a top-down approach \cite{Ringkjob2018}. Bottom-up models are often referred to as the engineering approach and are based on detailed technological representations of the energy system. Top-down models, on the other hand, follow an economic approach and consider the long-term changes and macroeconomic relationships \cite{Mai2013}. It is possible to combine both the technological properties and long-term changes by creating a hybrid approach \cite{Fortes2014}.

\subsection{Introduction to Simulation}
\label{sec:intro:simulationmodelling}

A computer simulation is a virtual model of a real-world system which is programmed into computer software. These models can be used to study how such a system works. One is able to change parameters in the system and make predictions as to how a system might behave. These simulations are particularly beneficial when the system one is analysing is difficult to experiment with \cite{fishman1978principles}. For example, the system exhibits a high financial costs of experimentation, or negative consequences may have large impacts \cite{mitrani1982simulation}. Additionally, for systems that operate on long timescales, such as energy markets, one may not have the ability to repeat experiments in a controlled environment.

Digital twins are a particular instance of a simulation. Digital twins have often been instantiated to a particular system, as opposed to a general system. For example, a digital twin can be made of the UK electricity market, whilst a simulation can be generalised to any decentralised market. By having a digital twin of a particular system, we are able to remove the risks associated with interacting with a system and iterate many experiments within a short time to find an optimal set of parameters. In addition, a digital twin is able to behave more like the system in question.

Whilst simulations must be built with expert knowledge, and through a thorough understanding of the system that one would like to model, machine learning is a data-driven approach. Data-driven approaches need not require an understanding of the system in which they are trying to model. Rather, they infer properties of the system entirely from data. These models are desirable in cases where a system is too complicated to have a full understanding of how the system works. These models have been shown to generate accurate results in many different disciplines \cite{Covington2016,WarrenLiao2005,Wiens2009}.

Simulations can be used to learn how a system may evolve, find optimal input parameters, learn the key features of a system and to extract data from the system. Another key use of simulations is to interact with a physical system. This list is non-exhaustive, as there exist many different uses for simulations.

\subsection{Introduction to Machine Learning}
\label{sec:intro:ml}

Machine learning (ML) is the study of computer algorithms that improve automatically with the use of data. By using training data, these algorithms are trained to make predictions or decisions without being explicitly programmed to do so.

Machine learning methods can be split into three different categories: (1) supervised learning, (2) unsupervised learning and (3) reinforcement learning. Each of these methods can be used in the following cases:

\begin{enumerate}
	\item Supervised learning is used where the data used has labelled data. Labelled data is where the true value that one is trying to predict is available in the data.
	\item Unsupervised learning is where there are no labels associated with the data. The model must, therefore infer from distinct clusters in the data where the divides in the values may lie.
	\item Reinforcement learning is concerned with how software agents must take actions in an environment in order to maximise a cumulative reward.
\end{enumerate}

In addition to the aforementioned machine learning methods, there exists an additional paradigm: optimisation methods. These methods explore a mathematical or software function to find a minimum or maximum value of an objective. These can be used to minimise an expected error, minimise total cost or maximise total return from a system, for example.

In the following sections we explore an overview of the different machine learning techniques that are applicable to the problems addressed in this thesis:

\subsubsection{Supervised Learning}

Supervised learning uses training data, which contains both the inputs and desired outputs, to predict the output associated with new inputs. A functioning supervised learning model, will therefore be able to correctly determine the output for inputs that were not part of the original training data. Supervised learning can be used for both regression and classification. Regression is where a continuous output is returned, whereas classification is where a discrete value is returned.

The model widely used supervised learning algorithms are:
\begin{itemize}
	\item Support vector machines
	\item Linear regression
	\item Logistic regression
	\item Decision trees
	\item Neural networks
\end{itemize}

Support vector machines, neural networks and decision trees can be used for both regression and classification, whereas linear regression and logistic regression are used for regression and classification respectively. Therefore, decisions must be made when choosing the appropriate algorithm for the respective task.

Once a decision has been made with respect to whether classification or regression is required, it is often the case that a number of different algorithms are trialled to see which yield the best results. 

However, these algorithms have different characteristics which should be taken into account. For instance, neural networks and support vector machines are able to learn complex patterns in data to a higher level than linear regression and so are often chosen for more complex datasets and problems. 

Secondly decision trees are interpretable, in that, one is able to graphically visualise why a specific output has been returned by the model through a series of decision criteria. This is a feature which is lacking in neural networks, which can often yield good results, but remain a black-box. 

Linear and logistic regression can be used to understand the most important variables to influence the output variables. So whilst, they are unable to model complex problems as well as neural networks or support vector machines, they can be used for better model interpretation.

In this thesis, we use a variety of different supervised learning techniques for specific problems, as previously discussed.

\subsubsection{Unsupervised Learning}

As previously mentioned, unsupervised learning learns patterns from unlabelled data. Unsupervised learning can therefore exhibit self-organisation that can capture patterns in data that were previously unknown.

Some of the most common algorithms include:
\begin{itemize}
	\item Hierarchical clustering
	\item \textit{k}-means clustering
	\item Self-organizing maps
\end{itemize}

Hierarchical clustering, similar to decision trees, can be graphically visualised to observe the decisions made by the algorithm. This can be useful, particularly when it is required to see which groups are more similar to one another, and different parent groupings.

\textit{k}-means partitions observations into \textit{k} clusters, in which each observation belongs to each cluster. \textit{k}-means provides a cluster centre, which serves as a prototype of the cluster, and provides useful information about the mean of the cluster.

A self-organising map produces a low-dimensional representation of higher dimensional data while preserving the topological structure of the data. This can make high dimension data easier to visualise. Self-organising maps are a type of neural network.

\subsubsection{Reinforcement Learning}

Reinforcement learning (RL) is an algorithm which allows intelligent agents to take actions in an environment in order to maximise a cumulative reward. Reinforcement learning does not require labelled input/output data. The basis of RL is to find a balance between exploration and exploitation.

A large amount of research has been dedicated in recent times to improving the performance of reinforcement learning algorithms \cite{Arulkumaran2017, Hunt2016a}. Various different methodologies have been tried and tested, from the use of neural networks in deep reinforcement learning to updating a lookup table.

Some of the algorithms used in the literature are:
\begin{itemize}
	\item Q-learning
	\item Deep Deterministic Policy Gradient (DDPG)
	\item Deep Q Network (DQN)
\end{itemize}

A major difference between the algorithms presented are the action and state spaces. For example, Q-learning requires both a discrete action and state space. Whereas, DDPG and DQN operate with a continuous action and state space. Q-learning updates a look-up table to map observations to actions, whereas DDPG and DQN use deep neural networks to learn a policy. 

The type of algorithm chosen can therefore be dependent on the different features required. Due to Q-learning working with a look-up table, they are more interpretable than DDPG and DQNs, as one is able to simply look at the lookup table to see which action is taken with different observations. However, due to the discrete nature of the action and state space, this methodology is less useful if one would like to have more precise actions.

It is possible to discretise the action and state space of Q-learning, but the speed and efficiency of the algorithm is greatly reduced with increasing discretised steps. Deep reinforcement learning techniques, similar to neural networks, are able to learn complex patterns within data, which can lead to better results. Care must therefore be taken when making decisions on the type of algorithm used.

\subsection*{Conclusion}
\label{sec:intro:conclusion}

In this section, we have introduced key concepts that have been used as part of this thesis: electricity markets and their modelling, simulation and digital twins, machine learning methods such as supervised learning and reinforcement learning as well as optimisation techniques. All of the mentioned methods have been used in this thesis for the purpose of increasing our understanding of electricity markets over both a long, and short time periods.

We have motivated the need for novel approaches to be used when understanding and modelling energy systems due to the fundamental changes that have occurred since the 1970s. From a system with a centralised actor and power stations which run on fossil fuels to one built on decentralised generation capacity and many heterogeneous actors. 

Additionally, we have discussed the complexity of modelling a system such as electricity markets. However, the insight that can be gained is invaluable and can provide further understanding to those who need to make decisions under large uncertainties. The ramifications of such decisions go far beyond the energy sector, and therefore any help that can be given to decision-makers is of utmost importance. Further details presented in this chapter will be presented in future chapters.

\ifpdf
\graphicspath{{Chapter3/Figs/Raster/}{Chapter3/Figs/PDF/}{Chapter3/Figs/}}
\else
\graphicspath{{Chapter3/Figs/Vector/}{Chapter3/Figs/}}
\fi

\section{Literature Review}

\subsection{Energy Modelling}
\label{sec:litreview:energymodelling}

Energy modelling is a broad field, so there have been multiple reviews that attempt to separate these models into different classifications \cite{Ringkjob2018,Savvidis2019a,Sensfub2007}. Examples of the metrics for classification are the mathematical underpinning, the underlying methodology, analytical approach or data requirements. This thesis focuses specifically on agent-based models and AI applied to electricity markets and not the wider literature of energy models. We, therefore, refer the reader to the papers presented in Table \ref{tab:litreview:reviews} for a more thorough investigation of this broader field.

\begin{table}[]
	\footnotesize
	\begin{tabular}{p{7.5cm}p{7.5cm}}
		\toprule
		Publication                                                                                                                 & Focus                                                                                                        \\ \midrule
		The gap between energy policy challenges and model capabilities \cite{Savvidis2019a}                                        & Assesses the ability of energy systems models to answer major energy policy questions.                       \\ \midrule
		Agent-based simulation of electricity markets: a literature review \cite{Sensfub2007}                                       & An overview of the work applying agent-based models to the analysis of electricity markets.                  \\ \midrule
		A review of modelling tools for energy and electricity systems with large shares of variable renewables \cite{Ringkjob2018} & An aid for modellers to choose an appropriate model which can cater for large shares of variable renewables. \\ \midrule
		Energy systems modeling for twenty-first century energy challenges \cite{Pfenninger2014}                                   & The issues of using existing models for twenty-first century challenges in energy.                           \\ \midrule
		A review of energy systems models in the UK: Prevalent usage and categorisation \cite{Hall2016}                            & Provide a classification schema for energy models.                                                           \\ \midrule
		A survey of stochastic modelling approaches for liberalised electricity markets \cite{Most2010}                             & Overview and classification of stochastic models dealing with price risks in electricity markets.            \\ \bottomrule
	\end{tabular}
	\caption{Different reviews of energy system models}
	\label{tab:litreview:reviews}
\end{table}

\subsection{Model types}
\label{sec:model-types}

Large, detailed bottom-up optimisation models have long been used for energy system modelling. These optimisation models are typically based upon a detailed description of the technical components of the energy system. 

The ultimate goal of optimisation models is to optimise a given quantity, for example, the minimisation of cost or the maximisation of welfare. In this context, welfare can be designed as the material and physical well-being of people~\cite{Keles2017}. However, these models do not state how likely each of these scenarios is to develop.

There are limitations to optimisation based models: traditional centralised optimisation models are not designed to describe a system that is out of equilibrium. Optimisation models assume perfect foresight and risk-neutral investments with no regulatory uncertainty \cite{Pfenninger2014}. The core dynamics which emerge from equilibrium remain a black-box. For example, the model assumes a target will be reached and does not provide information  when this is not the case. Reasons for this could be investment cycles which move the model away from equilibrium \cite{Chappin2017}.

Equilibrium models take an economic approach. They model the energy sector as a part of the whole economy and study how it relates to the rest of the economy \cite{Ringkjob2018}. POLES \cite{Soria2012} is a global detailed econometric model developed by the European Commission. E3MG is an econometric simulation developed by Cambridge Econometrics~\cite{Dagoumas2010}. MARKAL-MACRO is a hybrid model. Where MARKAL is bottom-up, and MACRO is top-down. 

Simulation models simulate an energy system based upon specified equations, characteristics and rules. These are often bottom-up models, and are designed with a high level of technological description \cite{Ringkjob2018}. Agent-based models are a specific case of simulation models, where actors are modelled explicitly as agents with heterogeneous strategies and behaviours.

Additionally, there exist a set of power system models which can help with decisions such as investment planning or decisions about generator dispatch. An example of a large power systems model is WASP~\cite{jenkins1974wein}.

\subsection{Agent-based models}

In this subsection, we outline current agent-based models available and motivate why the model, ElecSim, is required. Part of the literature review outlined here has been previously published in \cite{Kell}.

Electricity market liberalisation in many western democracies has changed the framework conditions \cite{Praca2003}. Centralised, monopolistic, decision making entities have given way to multiple heterogeneous agents acting for their own best interest~\cite{Most2010}. Policy options must, therefore, be used to encourage changes to attain a desired outcome. It has been proposed that these complex agents are modelled using agent-based models (ABMs) due to their non-deterministic nature \cite{Kell}. 

A number of ABM tools have emerged over the years to model electricity markets: SEPIA~\cite{Harp2000}, EMCAS~\cite{Conzelmann}, NEMSIM~\cite{Batten2006}, AMES~\cite{Sun2007}, GAPEX~\cite{Cincotti2013}, PowerACE~\cite{Rothengatter2007}, EMLab~\cite{Chappin2017} and MACSEM ~\cite{Praca2003}. Table \ref{table:litreview:abm_comparison} shows, however, that these do not suit the needs of an open source, long-term market model.

Table \ref{table:litreview:abm_comparison} contains five columns: tool name, whether the tool is open source or not, whether they model long-term investment in electricity infrastructure and the markets they model and we determine how the stochasticity of real life is modelled. 

We chose these columns to compare the models due to their importance in the application of understanding long-term electricity market scenarios. Firstly, an open-source tool enables independent users to verify the code developed for these models. It is important that this can be achieved, so that the outputs are not obfuscated by a closed system. Long-term investment allows for the endogenous propagation of an electricity market. 

We believe that a model which models both day-ahead and futures markets is of importance. This is because these markets are interlinked and correlated. If we did not take into account the day-ahead market, we would not be able to determine the economic success of the GenCos, where profitable GenCos are able to invest further, and those which are less successful invest less. The futures market enables GenCos to invest for the long-term, which determines the trajectory of a scenario. Other aspects, such as intra-day markets are important for the short-term, but less important for the long-term.

In addition, the ability to model stochasticity is of importance, due to the non-deterministic nature of the real-world. For instance, the price paid by GenCos is non-deterministic and can vary at any point in time between GenCos. In reality, a large number of features are stochastic, but we discuss those that, we believe, are most pertinent to long-term electricity markets.

There have been several recent studies using ABMs which focus on electricity markets. However, they often utilise ad-hoc tools designed for a particular application \cite{hadar2019, Kunzel2018, Saxena2019}. In our work, we develop the model ElecSim, which has been built for re-use and reproducibility. The survey \cite{Weidlich2008} cites that many of these tools do not release source code or parameters, which is a problem that ElecSim seeks to address by being open source and releasing parameters.

SEPIA \cite{Harp2000} is a discrete event ABM. SEPIA models plants as being always on. SEPIA does not model a spot market, instead focusing on bilateral contracts. As opposed to this, ElecSim has been designed with a merit-order, spot market in mind. As shown in Table \ref{table:litreview:abm_comparison}, SEPIA does not include a long-term investment mechanism. 

EMCAS ~\cite{Conzelmann} is a closed source ABM. EMCAS investigates the interactions between physical infrastructures and economic behaviour of agents. However, ElecSim focuses on the dynamics of the market, and provides a simplified, transparent model of market operation, whilst maintaining the robustness of results.

NEMSIM \cite{Grozev2005} is an ABM that represents Australia's National Electricity Market (NEM). Participants are able to grow and change over time using learning algorithms. NEMSIM is non-generalisable to other electricity markets, unlike ElecSim.

AMES ~\cite{Sun2007} is an ABM specific to the US Wholesale Power Market Platform. GAPEX \cite{Cincotti2013} is an ABM framework for modelling and simulating power exchanges. However, neither of these model the long-term dynamics for which ElecSim is designed.

PowerACE ~\cite{Rothengatter2007} is a closed source ABM of electricity markets that integrates short-term daily electricity trading and long-term investment decisions. PowerACE models the spot market, forward market and a carbon market. Similarly to ElecSim, PowerACE initialises GenCos with each of their power plants. However, as shown in Table \ref{table:litreview:abm_comparison}, unlike ElecSim, PowerACE does not consider stochasticity of price risks in electricity markets ~\cite{Most2010}.

EMLab ~\cite{Chappin2017} is an open-source ABM toolkit for the electricity market. Like PowerACE, EMLab models an endogenous carbon market; however, they both differ from ElecSim by not taking into account stochasticity in outages and operating costs. 

MACSEM \cite{Praca2003} has been used to probe the effects of market rules and conditions by testing different bidding strategies. MACSEM does not model long term investments or stochastic inputs.

As shown in Table \ref{table:litreview:abm_comparison}, none of the tools fill each of the characteristics that we require. We therefore propose ElecSim to contribute an open-source, long-term, stochastic investment model. For this work, we decided that a novel agent-based simulation was required to fill all of these categories, as the use-cases developed for our work is specific towards the requirements of our model. Whilst it is true that no other model meets each of our defined characteristics, however, it is not the case that these models are not useful. In fact, they are useful for their own specific purposes. 

\begin{table*}[]
	\centering
	\begin{adjustbox}{angle=90}
		\begin{tabular}{cccccc}
			
			\multicolumn{1}{c}{\textbf{Tool name}} & \textbf{Open Source} & \textbf{Long-Term Investment} & \textbf{Market} & \textbf{Stochastic Inputs}  \\ \midrule
			SEPIA \cite{Harp2000}  & \checkmark           & $\times$                             & \checkmark      & Demand                                            \\ 
			EMCAS \cite{Conzelmann}   & $\times$                    & \checkmark                    & \checkmark      & Outages                                           \\ 
			NEMSIM ~\cite{Batten2006}  & ?              & \checkmark                    & \checkmark      & $\times$                                                           \\ 
			AMES  ~\cite{Sun2007} & \checkmark           & $\times$                             & Day-ahead       & $\times$                                                   \\ 
			GAPEX  ~\cite{Cincotti2013} & ?              & $\times$                             & Day-ahead       & $\times$                                          \\ 
			PowerACE \cite{Rothengatter2007} & $\times$                    & \checkmark                    & \checkmark      & Outages Demand                          \\ 
			
			EMLab ~\cite{Chappin2017}  & \checkmark           & \checkmark                    & Futures         & Fuel prices                                \\ 
			MACSEM  ~\cite{Praca2003}  & ?              & $\times$                             & \checkmark      & $\times$                               \\ 
			ElecSim ~\cite{Kell}             & \checkmark           & \checkmark                    & Futures         & \checkmark                          \\ \hline
		\end{tabular}
	\end{adjustbox}
	\caption{Features of electricity market agent-based models.}
	\label{table:litreview:abm_comparison}
\end{table*}

\subsection{Energy models classification}
\label{sec:litreview:modelclassification}

In this section, we present a high-level overview of the various models that are in existence in a table format: Table \ref{tab:litreview:modelreview}. These models fall into various categories. However, for simplification we classify these as either agent-based models (ABM) or non-ABM. The time horizon details what the main purpose of the model is, is it a long-term model. Finally, the time-step column details the time resolution granularity.

\begin{table}[]
	\centering
	\footnotesize
	\begin{tabular}{@{}p{3cm}p{3cm}p{3cm}p{4cm}}
		\toprule
		\textbf{Model} & \textbf{Underlying methodology}                   & \textbf{Time horizon}            & \textbf{Time step}                                  \\ \midrule
		E3MG                                  & Non-ABM                    & 2100                             & Annually until 2030 and then each decade until 2100 \\
		LEAP                                  & Non-ABM                                   & Medium and long-term             & Annual                                              \\
		MARKAL                             & Non-ABM                               & Medium and long-term             & User-defined                                        \\
		MARKAL-MACRO                          & Non-ABM                                   & Medium and long-term             & User-defined                                        \\
		NEMS                                  & Hybrid                         & Medium (25 years)                & Yearly                                              \\
		OSeMOSYS                           & Non-ABM                                      & Medium and long-term (2010-2050) & 5-year                                              \\
		PRIMES                                & ABM                                     & Medium to long-term              & Yearly                                              \\
		POLES                                 & Non-ABM                           & Long-term (up to 2050)           & Yearly                                              \\
		TIMES                              & Non-ABM                                   & Medium and long-term             & User-chosen time-slices                             \\
		WASP                               & Non-ABM                                          & Medium and long-term             & 12 load duration curves per year                    \\
		MESSAGE                            & Non-ABM                                         & Short, medium and long-term      & User-defined (Multiple of number of years)          \\
		PLEXOS                             & Non-ABM                                    & Short-term                       & 1-minute                                            \\
		ELMOD                              & Non-ABM                                    & Short-term                       & Hourly                                              \\ 
		ElecSim                            & ABM                                        & Short, medium and long-term      & Hourly                                              \\ \bottomrule	\end{tabular}
	\caption{Model schema and presentation of various energy models \cite{Hall2016}}
	\label{tab:litreview:modelreview}
\end{table} 

\subsection{Validation}
\label{elecsim:sec:litreview}

This subsection covers the difficulties inherent in validating energy models and the approaches taken in the literature to validate these models.

\subsection{Limits of Validating Energy Models}

Beckman \textit{et al.} state that questions frequently arise as to how much faith one can put in energy model results. This is due to the fact that the performance of these models as a whole are rarely checked against historical outcomes~\cite{Beckman2011}.

Under the definition by Hodges \textit{et al.} \cite{Hodges} long-range energy forecasts are not validatable \cite{Craig2002}. Under this definition, validatable models must be observable, exhibit constancy of structure in time, exhibit constancy across variations in conditions not specified in the model and it must be possible to collect ample data \cite{Hodges}.

Whilst it is possible to collect data for energy models, the data covering important characteristics of energy markets are not always measured. Furthermore, the behaviour of the human population and innovation are neither constant nor entirely predictable. This leads to the fact that static models cannot keep pace with long-term global evolution. Assumptions made by the modeller may be challenged in the form of unpredictable events, such as the oil shock of 1973 \cite{Craig2002}.

This, however, does not mean that energy-modelling is not useful for providing advice in the present. A model may fail at predicting the long-term future because it has forecast an undesirable event, which led to a pre-emptive change in human behaviour—thus avoiding the original scenario that was predicted. This could, therefore, be viewed as a success of the model.

Schurr \textit{et al.} argued against predicting too far ahead in energy modelling due to the uncertainties involved \cite{Schurr_1961}. However, they specify that long-term energy forecasting is useful to provide basic information on energy consumption and availability, which is helpful in public debate and in guiding policymakers.

Ascher concurs with this view and states that the most significant factor in model accuracy is the time horizon of the forecast; the more distant the forecast target, the less accurate the model. This can be due to unforeseen changes in society as a whole ~\cite{gillespie_1979}.

It is for these reasons that we focus on a shorter-term (5-year) horizon window when calibrating our model for validation. This enables us to have increased confidence that the dynamics of the model work without external shocks and can provide descriptive advice to stakeholders. However, it must be noted that the UK electricity market exhibited a fundamental transition from natural gas to coal electricity generation during this period, meaning that a simple data-driven modelling approach would not work.

In addition to this short-term cross-validation, we compare our long-term projections to those of BEIS from 2018 to 2035. It is possible that our projections and those of BEIS could be wrong. However, this allows us to thoroughly test a particular scenario with different modelling approaches, and allow for the possibility to identify potential flaws in the models.

\subsection{Validation Examples}

In this Section, we explore a variety of approaches used in the literature for energy model validation.

The model OSeMOSYS \cite{Howells2011} is validated against the similar model MARKAL\slash TIMES through the use of a case study named UTOPIA. UTOPIA is a simple test energy system bundled with ANSWER \cite{Hunter2013}, a graphical user interface packaged with the MARKAL model generator \cite{Noble2004}. Hunter \textit{et al.} use the same case study to validate their model Temoa \cite{Hunter2013}. In these cases, MARKAL\slash TIMES is seen as the "gold standard". In this work, however, we argue that the ultimate gold standard should be real-world observations, as opposed to a hypothetical scenario.

The model PowerACE demonstrates that realistic prices are achieved by their modelling approach. However, they do not indicate success in modelling GenCo investment over a prolonged time period \cite{Ringler2012}.

Barazza \textit{et al}. validate their model, BRAIN-Energy, by comparing their results with a few years of historical data; however, they do not compare the simulated and observed electricity mix \cite{Barazza2020}. This reduces the confidence that one may have in the results of the produced electricity mix.

Work by Koomey \textit{et al.} expresses the importance of conducting retrospective studies to help improve models \cite{Koomey2003}. In this case, a model can be rerun using historical data in order to determine how much of the error in the original forecast resulted from structural problems in the model itself, or how much of the error was due to incorrect specification of the fundamental drivers of the forecast \cite{Koomey2003}.

A retrospective study published in 2002 by Craig \textit{et al.} focused on the ability of forecasters to accurately predict electricity demand from the 1970s \cite{Craig2002}. They found that actual energy usage in 2000 was at the very lowest end of the forecasts, with only a single exception. They found that these forecasts underestimated unmodelled shocks such as the oil crises which led to an increase in energy efficiency.

Hoffman \textit{et al.} also developed a retrospective validation of a predecessor of the current MARKAL\slash TIMES model, named Reference Energy System \cite{Hoffman_1973}, and the Brookhaven Energy System Optimization Model \cite{ERDA_48}. These were studies applied in the 70s and 80s to develop projections to the year 2000. This study found that the models had the ability to be descriptive but were not entirely accurate in terms of predictive ability. They found that emergent behaviours in response to policy had a strong impact on forecasting accuracy. The study concluded that forecasts must be expressed in highly conditioned terms \cite{Hoffman2011}.

\subsection{Modelling Conclusion}

In this section, we have introduced various electricity market models and the categories that they fall into. However, it can prove to be challenging to place models within a clear boundary, as many models fall within a continuous spectrum. We introduced the concept that traditional models may not have the ability to detail every single component of an electricity market without losing tractability.

The need for a new paradigm in which decentralised agents act within an environment was discussed. So was the need for a model with high temporal resolution to more accurately model the intermittency of renewable energy. Traditional optimisation models work in a normative, prescriptive way. However, it is not possible to describe a system which is out of equilibrium. Another limitation of the traditional optimisation models is that they assume perfect foresight, with risk-neutral investments and no regulatory uncertainty. It assumes that certain scenarios are possible, but does not highlight the way a target may not be reached.

It is for these reasons that in this thesis, we focus on agent-based models, which move away from the traditional optimisation approach, and allow for a more dynamic solution without rigid mathematical expressions.

Additionally, we found that there was a gap in the literature for an open-source agent-based model that could model long-term investments, was generalisable to many countries and modelled stochastic inputs. It is for this reason that we developed the model ElecSim.

\subsection{Machine Learning and Agent-Based Models}
\label{sec:ML-abms}

In this section, we review the literature that investigates how artificial intelligence and machine learning can be integrated into agent-based models for the electricity sector. To select the related articles to review, we conduct a systematic analysis of relevant research in the field. We limited our search to literature published in the five most recent years (2016-2021). As a result of this, we provide a comprehensive status of the applications of ML and AI in agent-based models for the electricity sector. For this purpose we used the Elsevier Scopus database. To find the articles, we used the following set of search terms to select our articles:

\begin{enumerate}
	\item Machine Learning, Artificial Intelligence, Deep Learning, Neural Networks, Decision Tree, Support Vector Machine, Clustering, Bayesian Networks, Reinforcement Learning, Genetic Algorithm, Online Learning, Linear regression.
	\item Agent-based modelling.
	\item Electricity.
\end{enumerate}

We searched using each of the keywords in each of the bullet points. For instance, the first keyword search was: Machine Learning, Agent-Based Modelling and Electricity. The second was: Artificial Intelligence, Agent-based modelling and Electricity. We selected these search terms to focus this review on agent-based models applied to the electricity sector and machine learning, which is the focus of this thesis.

These search terms resulted in 149 research articles. However, not all of these were related to our research focus. For instance, a number of electric vehicle, buildings and biological papers were returned. After a further manual review, these 149 papers were reduced to 55 papers which were specifically related to agent-based modelling, electricity, artificial intelligence and machine learning.

\begin{figure}
	\centering
	\includegraphics[width=0.6\linewidth]{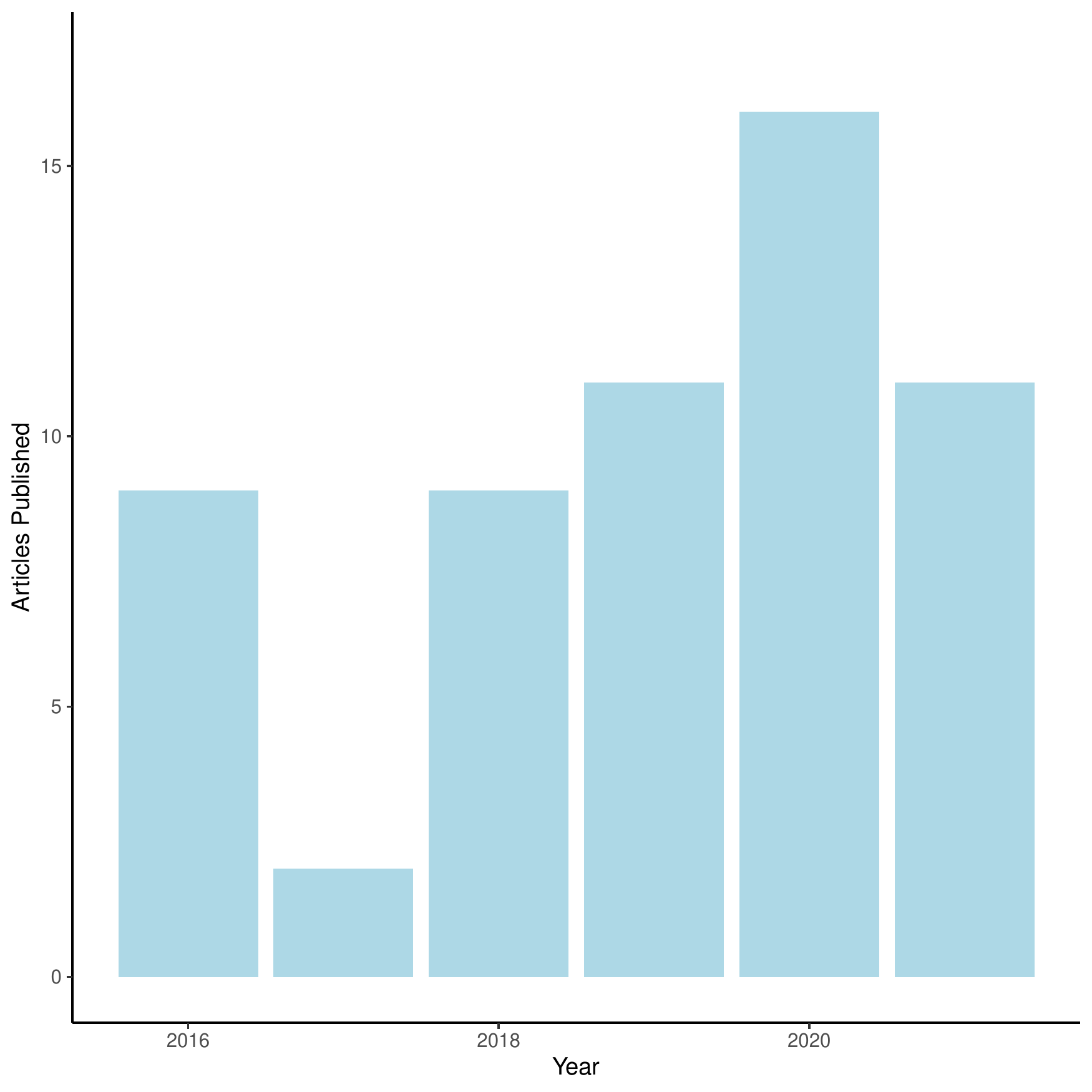}
	\caption{Number of articles published per year which work on AI in electricity focused ABMs.}
	\label{fig:articles-published}
\end{figure}

Figure \ref{fig:articles-published} shows the amount of articles published each year between 2016 and the present date. Whilst the number of articles published in this field has increased per year since 2018 to 2020, the number of papers published in 2017 was lower when compared to the other years, with a large number published in 2016. We reviewed these 55 papers systematically in the following sections.

\subsection{Market Type}

In this literature review, we make three different market type distinctions: international/national energy market, local energy market and a microgrid. The international/national energy market typically considers a country, multiple countries or the world. A local energy market is a smaller region than the international/national energy market, for instance, a city or region. Whereas a microgrid serves a discrete geographic footprint, such as a university campus, business centre or neighbourhood. Whilst there is some cross-over between a local energy market and microgrid, a microgrid can be disconnected from the traditional grid and operate autonomously. 

Tables \ref{table:RL}, \ref{table:supervised-learning} and \ref{table:unsupervised-learning}, \ref{table:optimisation} and \ref{table:game-theory} categorise  each of the market types respectively. The papers have been displayed in chronological order and categorise the market type, machine learning (ML) type used, the application in which it was used and the algorithm used. These different criteria are explored in the following subsections.

\subsection{Machine Learning Types}

Within this work, we have covered five different type of artificial intelligence paradigms. These are: supervised learning, unsupervised learning, reinforcement learning, optimisation and game theory

\begin{figure}
	\centering
	\includegraphics[width=0.85\linewidth]{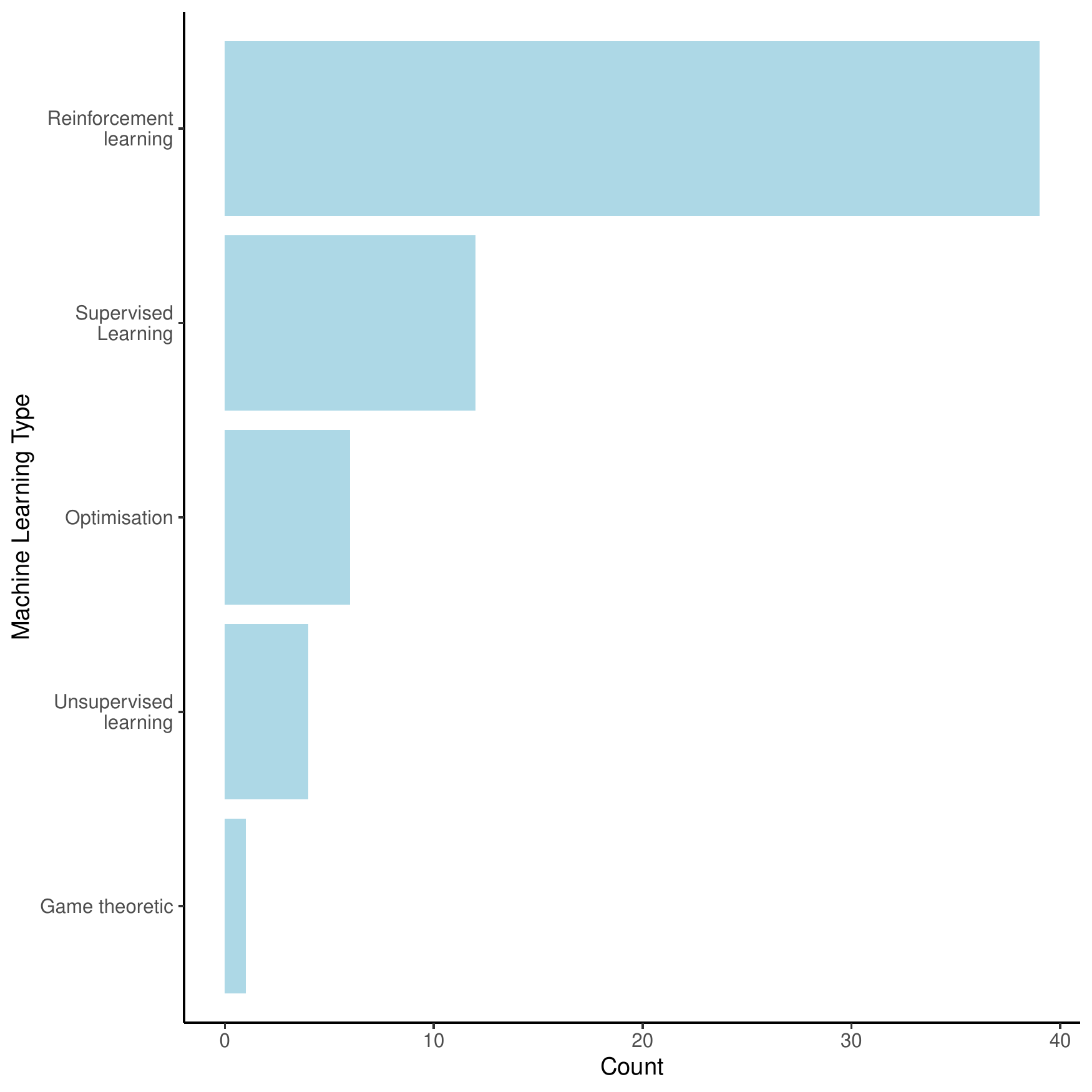}
	\caption{Number of machine learning types surveyed in papers.}
	\label{fig:ml_type}
\end{figure}

Each of these techniques have been utilised in the papers surveyed. However, a particular focus has been placed on reinforcement learning within the research community. As shown by Figure \ref{fig:ml_type}, 37 out of the 55 papers used a reinforcement learning algorithm. This greatly outweighs the other machine learning types. The second most used machine learning type was supervised learning, used by eight papers. The fact that reinforcement learning has been used so extensively within the agent-based modelling community for electricity highlights the usefulness of this technique within this field. 

Within each of the different machine learning types there exist many algorithms. The algorithms used in the papers surveyed are now presented. Within reinforcement learning the deep deterministic policy gradient (DDPG), Deterministic Policy Gradient (DPG) Deep Q-Network (DQN), Deep Q-Learning, Fitted Q-iteration (FQI), long short-term memory neural network (LSTM), Multi-Agent Deep Deterministic Policy Gradient (MADDPG), Markov Decision Process (MDP), Novel WoLF-PHC, Policy Iteration (PI), Probe and Adjust, Q-Learning, Roth-Erev, SARIMAX and Variant Roth-Erev are used.

Within supervised learning, the following algorithms were used: Artificial Neural Network (ANN), Bayesian networks, Classification trees, Extreme Machine Learning, Lasso regression,  Linear regression and Support Vector Machine (SVM) were used. Fewer algorithms were used for both unsupervised learning and optimisation. For supervised learning, the following algorithms were used: Bayesian classifier, K-Means Clustering, Naive Bayes classifier. For optimisation the following algorithms were trialled: Bi-level coordination optimisation, Genetic Algorithm, Iterative algorithm and Particle Swarm Optimisation. For the game theory method, a game theoretic algorithm was used.

\subsection{Applications}

Within this work, we classified each paper by the problem domain which they are trying to solve, or the application. The applications are: agent behaviour, bidding strategies, bilateral trading, demand forecasting, demand response, electricity grid control, expansion planning, forecasting carbon emissions, load scheduling, market investigation, microgrid management, peer to peer trading, price forecasting, risk management, scheduling of flexibility, secure demand side management and tariff design.

\begin{figure}
	\centering
	\includegraphics[width=0.95\linewidth]{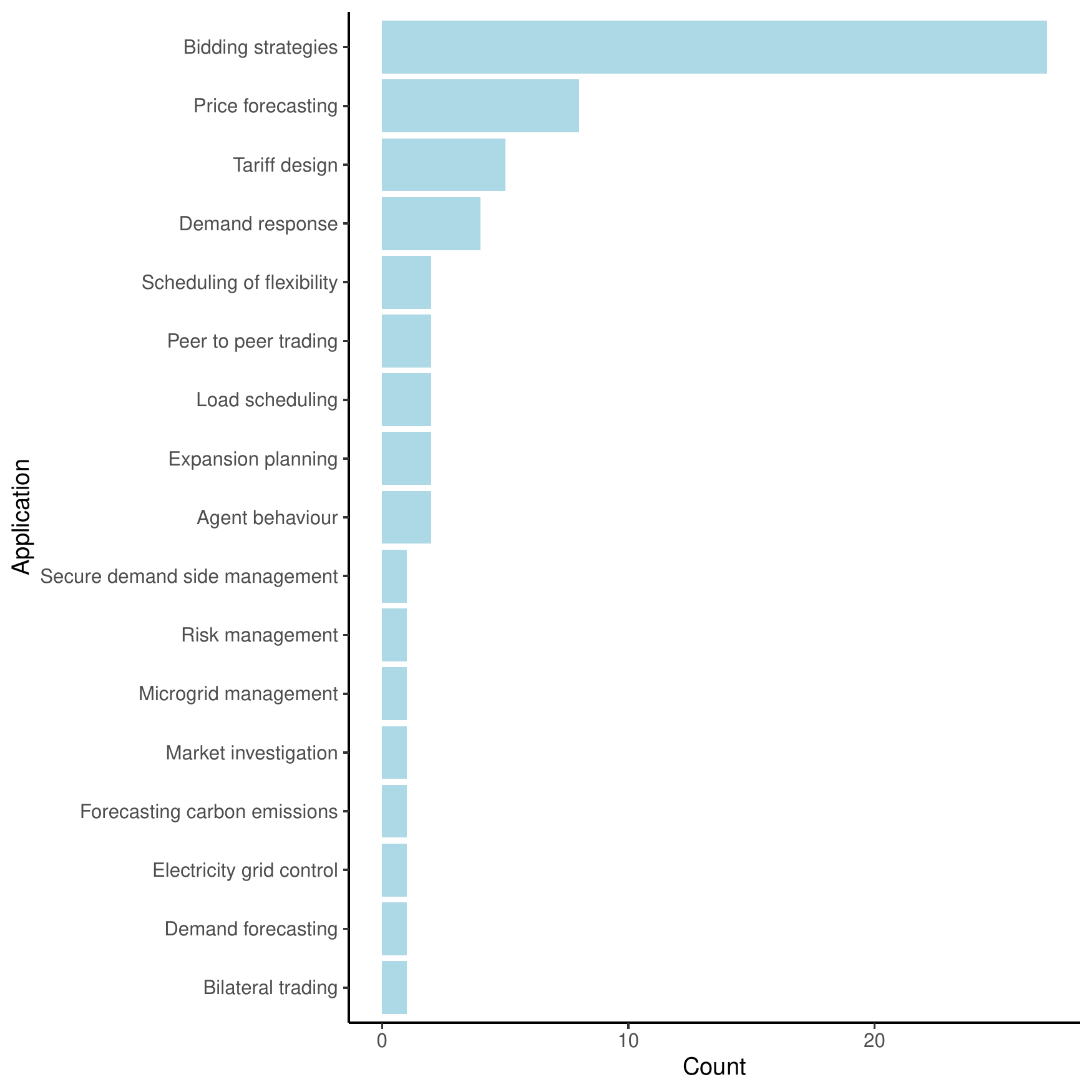}
	\caption{Application number per paper.}
	\label{fig:application}
\end{figure}

Figure \ref{fig:application} displays the number of applications used by each machine learning type. The most utilised application was bidding strategies, with price forecasting and tariff design following behind. However, the bidding strategies application was investigated 27 times, with price forecasting investigated only 8 times. This demonstrates a considerable research effort in this area.

\begin{figure}
	\centering
	\includegraphics[width=0.95\linewidth]{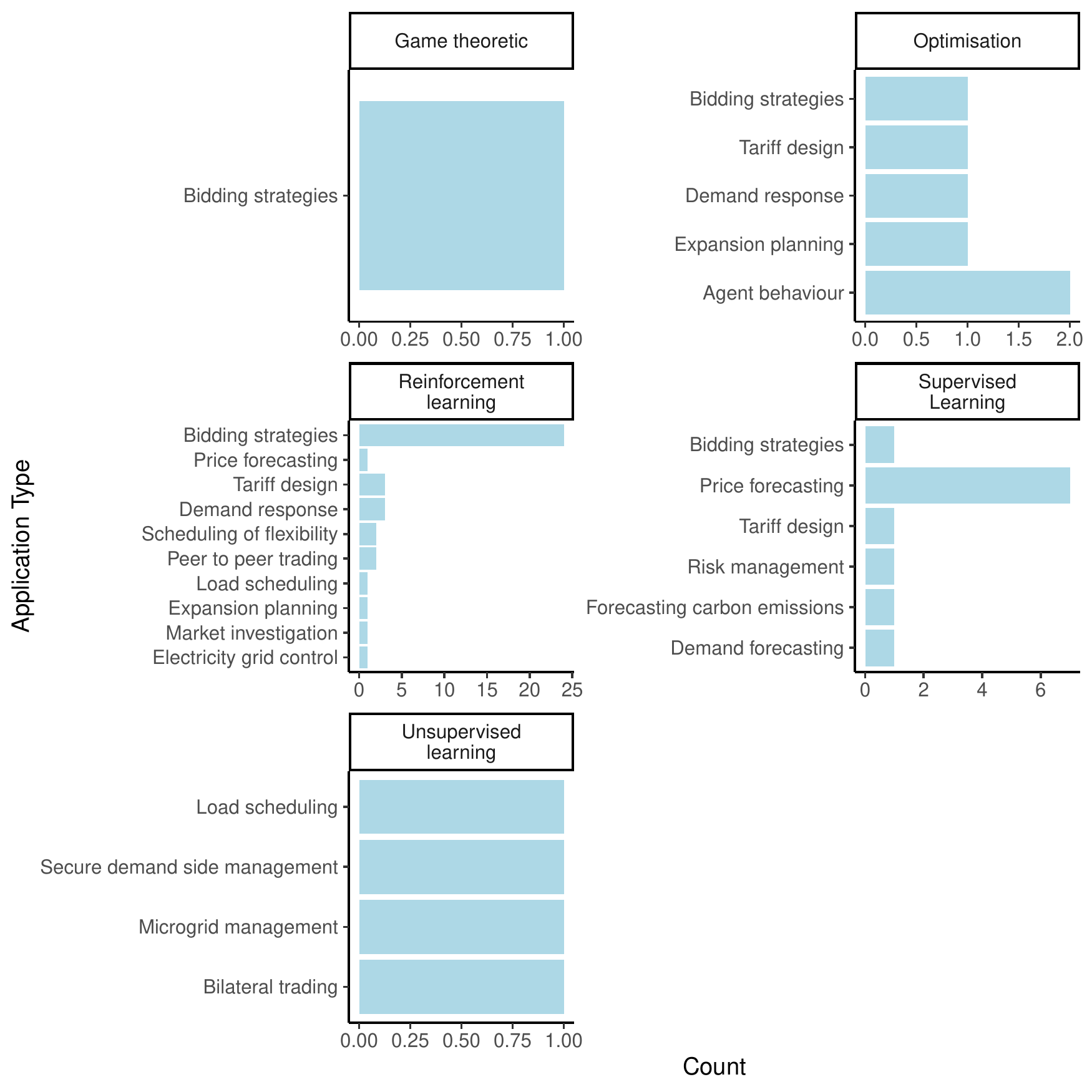}
	\caption{Applications per application type.}
	\label{fig:application-ml_type}
\end{figure}

Figure \ref{fig:application-ml_type} displays the number of applications per machine learning type area. We can see that bidding strategies is highly used within the reinforcement learning machine learning type. The reinforcement learning algorithm, however, is shown to be highly versatile, with different applications investigated, from demand response, flexibility scheduling to expansion planning. This is due to the ability for reinforcement learning to learn different policies based upon solely the reward and observations within an environment.

Within supervised learning, a large amount of research effort has been put into price forecasting. This is likely due to the strong ability of supervised learning techniques at making predictions. However, outside of classification and making predictions, supervised learning is not so versatile, when compared to reinforcement learning.

Optimisation is used for five different applications. This is because optimisation requires a problem domain to be maximised or minimised. This may not be the case for all applications.

Unsupervised learning and the game theoretic machine learning types are shown to be used less than the other machine learning types. This is because unsupervised learning is preferential when there is no labelled data. However, with labelled data, supervised learning can yield more accurate results. Within simulations it is often the case that data is available, and so supervised learning is used in preference to unsupervised learning. The game theoretic approach has been used in a single application in the papers surveyed: bidding strategies. The application of game theory is possible for the problem of bidding strategies, however, the assumptions of a Nash equilibrium and perfect information may not always exist in an electricity market.

\subsection{Reinforcement Learning}

In this section we review the papers that utilised reinforcement learning for the applications shown in Figure \ref{fig:application-ml_type}. Firstly, we cover the papers which consider the bidding strategies problem.

Liu \textit{et al.}\cite{Liu2020} establish non-cooperative and cooperative game models between thermal power companies. They show that, compared with other RL algorithms, the MADDPG algorithm is more accurate, with an increase in revenue of 5.2\%.  They show that thermal companies are more inclined to use physical retention methods to make profits in the medium and long-term power market. 

Liang \textit{et al.}\cite{Liang} use the DDPG algorithm to model the bidding strategies of GenCos. They show that the DDPG algorithm is able to converge to a Nash equilibrium even with imperfect information. They find that they are able to reflect collusion through adjusting the GenCos' patience parameter. Purushothaman \textit{et al.}\cite{Purushothaman} model the learning capabilities of power generators through the use of the Roth-Erev RL algorithm. They find that the agents are able to exhibit market power through this approach. Kiran \textit{et al.}\cite{Kiran} use a variant of the Roth-Erev algorithm to investigate the ability for a generator to bid strategically within a market. They find that agents have the ability to bid strategically, and increase their net earnings. However, the impact on the wider market and how to limit this ability was not explored in these papers.

Sousa \textit{et al.}\cite{Sousa} use an ABM to model the Iberian electricity market, with a focus on hydropower plants. They use Q-Learning to bid in the day-ahead market. Poplavskaya \textit{et al.}\cite{Poplavskaya} model the balancing services market, and investigate the effect of different market structures on price. They find that in an oligopoly, prices can deviate from the competitive benchmark by a factor of 4-5. They conclude that changing market type would not solve this issue.

Nunna \textit{et al.}\cite{Nunna} use a simulated-annealing-based Q-learning algorithm to develop bidding strategies for energy storage systems. They observe that they can effectively reinforce the balance between supply and demand in the microgrids using a mixture of local and global energy storage systems.

Lin \textit{et al.}\cite{Lin} investigate herding behaviours of electricity retailers on the monthly electricity market. Herding behaviours are where individuals act collectively as part of a group. They model herding behaviours mathematically based on the relationship network of electricity retailers which are imbedded in an ABM. They use the Roth-Erev RL to model these behaviours. They find that the herding behaviours might bring positive or negative effects to electricity retailers depending on the differing bidding strategies.

Wang \textit{et al.}\cite{Wang} investigate the bidding behaviour of all players in the electricity market. They propose a hybrid simulation model and integrate RL to bid. They find that with the hybrid simulation model, the dynamics of the entire market remain stable, the market clearing prices converge, and the market share is relatively uniform.

Machado \textit{et al.}\cite{MacHado} investigate how the energy price is affected when a government intervention is observed through the increase in number of public companies participating in auctions. They find that they are able to model the impact of public companies on the overall electricity market.

Ye \textit{et al.}\cite{Ye} propose a novel multi-agent deep RL algorithm, where they combine the DPG algorithm with LSTM for multi-agent intelligence. Their algorithm achieves a significantly higher profit than other RL methods for a GenCo.

Pinto \textit{et al.}\cite{Pinto1} investigate the ability for collaborative RL models to optimise energy contract negotiation strategies. The results show that the collaborative learning process enables players' to correctly identify which negotiation strategy to apply in each moment, context and opponent.

Feng \textit{et al.}\cite{Feng} explore the effect of a transition from a monopoly to a competitive electricity market. They simulate a day-ahead market and use RL to bid strategically. They find that they can characterise the risk characteristics and decision-making objectives of market participants.

Calabria \textit{et al.}\cite{Calabria} simulate the Brazilian power system, using 3 years worth of data which covers 98\% of the total hydro capacity of Brazil. They use RL to simulate the behaviour of virtual reservoirs in this market. They find that through the management of these virtual reservoirs, they can save water, while maintaining current efficiency and security levels.

Gaivoronskaia \textit{et al.}\cite{Gaivoronskaia} present a modification of the classical Roth-Erev algorithm to represent agents' learning in an ABM of the Russian wholesale electricity market, in a day-ahead market. Staudt \textit{et al.}\cite{Staudt} investigate the effect of the number of energy suppliers needed for a competitive market. They find that several suppliers are required to ensure a welfare optimal pricing. Where welfare is defined as a benefit to the community.

Esmaeili Aliabadi \textit{et al.}\cite{Esmaeili} study the effect of risk aversion on GenCos' bidding behaviour in an oligopolistic electricity market. They find that the change in the risk aversion level of even one GenCo can significantly impact on all GenCo bids and profits. 

Rashedi \textit{et al.}\cite{Rashedi} use Q-Learning to learn optimal bidding strategies of suppliers in a day-ahead market. They compare the competitive behaviour of players in both the multi-agent and single-agent case. Chrysanthopoulos \textit{et al.}\cite{Chrysanthopoulos} take a similar approach, by modelling the day-ahead market as a stochastic game. They use an ABM to model GenCos to maximise their profit using RL. 

Skiba \textit{et al.}\cite{Skiba} model the bidding behaviour on the day-ahead and control reserve power markets using an RL algorithm in an ABM. They investigate the resulting market prices. Tang \textit{et al.}\cite{Tang} investigate the bidding strategies of generators under three pricing mechanisms. They find they are able to achieve market equilibrium results through their novel RL approach.

Bakhshandeh \textit{et al.}\cite{Bakhshandeh} assess the ability for GenCos to withhold capacity to increase the market price using RL. The results demonstrate the emergence of capacity withholding by GenCos, which have an effect on the market price. 

Xu \textit{et al.}\cite{Xu} simulate a proactive residential demand response in a day-ahead market using RL. They use residential data in China, and test a case with 30,000 households. They find that a proactive residential demand response may yield significant benefits for both the supply and demand side.

Deng \textit{et al.}\cite{Deng} use a DDPG algorithm to also model residential demand response.  They find that the  goal of peak load-cutting and valley filling can be achieved through this method.

Shafie-Khah \textit{et al.}\cite{Shafie-Khah} develop a novel decentralised demand response model. In their model, each agent submits bids according to the consumption urgency and a set of parameters by the RL algorithm, Q-Learning. Their studies show that their decentralised model drops the electricity cost dramatically, which was nearly as optimal as a centralised approach. Najafi \textit{et al.}\cite{Najafi} also propose a decentralised demand response model. They use Q-Learning and consider small scale GenCos. Similarly to Shafie-Khah \textit{et al.}, they show the effectiveness of this technique.

Tomin \textit{et al.}\cite{Tomin} propose an RL method to interact with the electricity grid for active grid management. They demonstrate the effectiveness of this approach on a test 77-node scheme and a real 17-node network diagram of the Akademgorodok microdistrict (Irkutsk).

Huang \textit{et al.}\cite{Huang} propose a generation investment planning model for GenCos. They use a genetic algorithm and Q-learning to improve their optimisation ability, and show that the model is effective and could provide support for plant expansion planning. Manabe \textit{et al.}\cite{Manabe} also consider the generation expansion planning problem. They find through their simulation that their agents can generate higher profit using RL.

Foruzan \textit{et al.}\cite{Foruzan} use an ABM to study distributed energy management in a microgrid. They use an RL algorithm to allow generation resources, distributed storages and customers to develop optimal strategies for energy management and load scheduling. They are able to reach a Nash equilibrium, where all agents benefit through this approach.

Viehmann \textit{et al.}\cite{Viehmann} analyse the different markets: Uniform Pricing (UP) and Discriminatory Pricing. UP is where all agents pay the maximum accepted price, and DP pay their accepted bid. Through the use of RL, they find that UPs lead to higher prices in all analysed market settings.

Pinto \textit{et al.}\cite{Pinto} introduce a learning model to enable players to identify the expected prices of bilateral agreements. For this they use a Q-Learning algorithm, and they use real data from the Iberian electricity market.

Mbuwir \textit{et al.}\cite{Mbuwir} explore two model-free RL techniques: policy iteration (PI) and fitted Q-iteration (FQI) for scheduling the operation of flexibility providers - battery and heat pump in a residential microgrid. Their simulation results show that PI outperforms FQI with a 7.2\% increase in photovoltaic self-consumption in the multi-agent setting and a 3.7\% increase in the single-agent setting. 

Naseri \textit{et al.}\cite{Naseri} propose an autonomous trading agent which aims to maximise profit by developing tariffs. They find  that designing tariffs with usage-based charges and fixed periodic charges can help the agent segment the retail market resulting in lower peak demand and capacity charges.

Bose \textit{et al.}\cite{Bose} simulate a local energy market as a multi-agent simulation of 100 households. Through the use of the Roth-Erev reinforcement learning algorithm to control trading, and demand response of electricity. They are able to achieve self-sufficiency of up to 30\% with trading, and 41.4\% with trading and demand response. Kim \textit{et al.} \cite{Kim} consider a prosumer that consumes and produces electric energy with an energy storage system. They use the DQN reinforcement learning algorithm to maximise profit in peer-to-peer energy trading. Liu \textit{et al.}\cite{Liu2021} also investigate the peer-to-peer trading problem in a local energy market. They find that the community modelled are able to increase profitability through the use of DQN RL.

\begin{table}[]
\centering
\footnotesize
\begin{tabular}{@{}llp{3cm}p{3cm}p{3cm}@{}}
\toprule
Year & First Author              & Market Type                          & Application                        & Algorithm Used                                 \\ \midrule
2021 & Bose S.    \cite{Bose}          & Local energy market                  & Peer to peer trading               & Roth-Erev                                      \\
2021 & Naseri N.     \cite{Naseri}       & Local energy market                  & Tariff design                      & SARIMAX, MDP                                    \\
2021 & Tang C.   \cite{Tang}           & International/National & Tariff design                      & Novel WoLF-PHC                                 \\
2021 & Liu D.    \cite{Liu2021}           & International/National & Bidding strategies, Peer-to-Peer                 & MADDPG                                         \\
2021 & Deng C. \cite{Deng}             & International/National & Demand response                    & DDPG                                           \\
2021 & Viehmann J. \cite{Viehmann}         & International/National & Market investigation               & Q-Learning                                     \\
2020 & Tomin N.   \cite{Tomin}          & Microgrid                            & Electricity grid control           & Q-Learning                                     \\
2020 & Liang Y. \cite{Liang}            & International/National & Bidding strategies                 & DDPG                                           \\
2020 & Kim J.-G.     \cite{Kim}       & International/National & Peer to peer trading               & Deep Q-Learning                                \\
2020 & Shafie-Khah M.   \cite{Shafie-Khah}    & Local energy market                  & Demand response                    & Q-Learning                                     \\
2020 & Sousa J.C. \cite{Sousa}          & International/National & Bidding strategies                 & Q-Learning                                     \\
2020 & Poplavskaya K.   \cite{Poplavskaya}    & International/National & Bidding strategies                 & Q-Learning                                     \\
2020 & Liu Y.    \cite{Liu2020}           & Local energy market                  & Bidding strategies                 & DQN                                            \\
2020 & Nunna H.S.V.S.K.  \cite{Nunna}   & Microgrid                            & Bidding strategies                 & Q-Learning                                     \\
2020 & Purushothaman K. \cite{Purushothaman}    & International/National & Bidding strategies                 & Roth-Erev                                      \\
2020 & Mbuwir B.V.  \cite{Mbuwir}        & Microgrid                            & Scheduling of flexibility          & Policy Iteration (PI), Fitted Q-iteration (FQI) \\
2020 & Kiran P.     \cite{Kiran}        & International/National & Bidding strategies                 & Variant Roth-Erev                              \\
2019 & Lin F.   \cite{Lin}            & International/National & Bidding strategies                 & Roth-Erev                                      \\
2019 & Wang J.    \cite{Wang}          & International/National & Bidding strategies                 & Roth-Erev                                      \\
2019 & Machado M.R. \cite{MacHado}        & International/National & Bidding strategies                 & Q-Learning                                     \\
2019 & Ye Y.    \cite{Ye}            & International/National & Bidding strategies                 & DPG,LSTM                                       \\
2019 & Pinto T.    \cite{Pinto}         & International/National & Price forecasting                  & Q-Learning                                     \\
2019 & Pinto T.  \cite{Pinto1}           & International/National & Bidding strategies                 & Q-Learning                                     \\
2018 & Feng H.    \cite{Feng}          & Local energy market                  & Bidding strategies                 & Variant Roth-Erev                              \\
2018 & Foruzan E.     \cite{Foruzan}      & Microgrid                            & Load scheduling                    & Q-Learning                                     \\
2018 & Calabria F.A.   \cite{Calabria}     & International/National & Bidding strategies                 & Q-Learning                                     \\
2018 & Gaivoronskaia E.A. \cite{Gaivoronskaia}   & International/National & Bidding strategies                 & Roth-Erev                                      \\
2018 & Staudt P.       \cite{Staudt}     & Microgrid                            & Bidding strategies                 & Probe and adjust                               \\
2018 & Najafi S.    \cite{Najafi}        & Local energy market                  & Demand response, Bidding strategies & Q-Learning                                     \\
2017 & Esmaeili Aliabadi D. \cite{Esmaeili} & International/National & Bidding strategies                 & Q-Learning                                     \\
2016 & Huang X.    \cite{Huang}         & International/National & Expansion planning                 & Q-Learning                                     \\
2016 & Rashedi N.   \cite{Rashedi}        & International/National & Bidding strategies                 & Q-Learning                                     \\
2016 & Skiba L.    \cite{Skiba}         & International/National & Bidding strategies                 & Roth-Erev                                      \\
2016 & Bakhshandeh H.   \cite{Bakhshandeh}    & International/National & Bidding strategies                 & Q-Learning                                     \\
2016 & Chrysanthopoulos N. \cite{Chrysanthopoulos} & Local energy market                  & Bidding strategies                 & Q-Learning                                     \\ \bottomrule
\end{tabular}
\caption{Articles relating to reinforcement learning algorithm type.}
\label{table:RL}
\end{table}

\subsection{Supervised Learning}

In this section we review the papers that used a supervised learning approach with their agent-based models, which focus on electricity. First we consider papers which focus on price forecasting.

Fraunholz \textit{et al.}\cite{Fraunholz} use ANNs to forecast electricity price endogenously within the long-term energy model, PowerACE. They find that the ANN method outperforms the linear regression method and that this endogenous method has a significant impact on simulated electricity prices. This is of importance since these are major results for electricity market models. Pinto \textit{et al.}\cite{Pinto3} uses SVMs and ANNs for price forecasting using real data from MIBEL, the Iberian market operator. They show an ability to return promising results in a fast execution time.

Goncalves \textit{et al.}\cite{Goncalves} use multiple different methods to understand the main drivers of electricity prices. These include lasso and standard regression, and causal analysis such as Bayesian networks and classification trees. Their results are coherent and show the impact of different generators on final electricity price.

Opalinski \textit{et al.}\cite{Opalinski} propose a hybrid prediction model, where the best results from a possible large set of different short-term load forecasting models are automatically selected based on their past performance by a multi-agent system. They show an increase in prediction accuracy with their approach.

Bouziane \textit{et al.}\cite{Bouziane} forecast carbon emissions using a hybrid ANN and ABM approach from different energy sources from a city. They forecast energy production using agents and calculate the benefits of using renewable energy as an alternative way of meeting electricity demand. They find they are able to reduce emissions by 3\% per day using this approach.

Pinto \textit{et al.}\cite{Pinto4} propose an approach to addressing the adaptation of players' behaviour according to participation risk. To do this, they combine the two most commonly used approaches of forecasting: internal data analysis and sectorial data analysis. They show that their proposed approach is able to outperform most market participation strategies and reach a higher accumulated profit by adapting players' actions according to the participation risk.

Maqbool \textit{et al.}\cite{Maqbool} investigate the impact of feed-in tariffs and the installed capacity of wind power on electricity consumption from renewables. They use linear regression to understand the outputs from an agent-based model to achieve this. They find that the effect of increasing installed capacity of wind power is more significant than the effect of feed-in tariffs.

El Bourakadi \textit{et al.}\cite{Bourakadi} propose the use of an Extreme Machine Learning  (EML) algorithm to make decisions about selling/purchasing electricity from the main grid and charging and discharging batteries from an ABM. The EML algorithm predicts wind and photovoltaic power output from weather data and then makes a classification decision on whether to buy or sell power. 

Babar \textit{et al.}\cite{Babar} propose a secure demand-side management engine to preserve the efficient utilisation of energy based on a set of priorities. A model is developed to control intrusions into the smart grid using a naive Bayes classifier. Simulation is used to test the efficiency of the system, and the result reveal that the engine is less vulnerable to intrusion.

\begin{table}[]
\centering
\footnotesize
\begin{tabular}{@{}llp{3cm}lp{3cm}@{}}
\toprule
Year & First Author         & Market Type                          & Application                  & Algorithm Used                                                            \\ \midrule
2021 & Fraunholz C. \cite{Fraunholz}   & International/National & Price forecasting            & ANN                                                                       \\
2021 & Bouziane S.E.  \cite{Bouziane} & Local energy market                  & Forecasting carbon emissions & ANN                                                                       \\
2020 & Babar M.  \cite{Babar}   & International/National & Secure demand side management & Naive Bayes classifier \\
2019 & Maqbool A.S.   \cite{Maqbool} & International/National & Tariff design                & Linear regression                                                         \\
2019 & Goncalves C. \cite{Goncalves}   & International/National & Price forecasting            & Linear regression, Lasso regression, Bayesian networks, Classification trees \\
2019 & Pinto T.   \cite{Pinto4}     & International/National & Risk management              & ANN                                                                       \\
2019 & El Bourakadi D. \cite{Bourakadi} & Microgrid                            & Bidding strategies           & Extreme Machine Learning                                                  \\
2016 & Opalinski A.  \cite{Opalinski}  & International/National & Demand forecasting           & Linear regression                                                         \\
2016 & Pinto T.  \cite{Pinto3}      & International/National & Price forecasting            & ANN, SVM                                                                   \\ \bottomrule
\end{tabular}
\caption{Articles relating to supervised learning algorithm type.}
\label{table:supervised-learning}
\end{table}

\subsection{Unsupervised learning}

In this section we discuss the papers which use an unsupervised learning approach with agent-based models.

Imran \textit{et al.}\cite{Imran} develop a novel strategy for bilateral negotiations. They enable each GenCo to estimate the reservation price of its opponent using Bayesian learning. They show that the agents which use Bayesian learning gain an advantage over non-learning agents. 

Čaušević \textit{et al.}\cite{Causevic} propose a novel clustering algorithm to cluster agents into virtual cluster members for the application of load scheduling. They show that large-scale centralised energy systems can operate in a decentralised fashion when only local information is available.

Gomes \textit{et al.}\cite{Gomes} propose a management system for the operation of a microgrid by an electricity market agent. They use K-means clustering for scenario reduction and a stochastic mixed-integer linear programming problem to manage the system.

\begin{table}[]
\centering
\footnotesize
\begin{tabular}{@{}llp{3cm}lp{3cm}@{}}
\toprule
Year & First Author      & Market Type                          & Application                   & Algorithm Used         \\ \midrule
2021 & Gomes I.L.R. \cite{Gomes} & Microgrid                            & Microgrid management          & K-Means Clustering     \\
2020 & Imran K.  \cite{Imran}   & International/National & Bilateral trading             & Bayesian classifier    \\
2017 & Čaušević S. \cite{Causevic} & International/National & Load scheduling               & Novel clustering algorithm             \\ \bottomrule
\end{tabular}
\caption{Articles relating to unsupervised learning algorithm type.}
\label{table:unsupervised-learning}
\end{table}

\subsection{Optimisation}

In this subsection we review the papers which use optimisation as a basis for investigation.

Bevilacqua \textit{et al.}\cite{Bevilacqua} compare three optimisation methods to implement agent rationality in the Italian electricity market with an ABM. Through this, they observe that the moddel exhibits a very good fit to real data.

Duan \textit{et al.}\cite{Duan} propose a bi-level coordination optimisation integrated resource strategy to unify supply-side and demand side resources across China. They do this for mid-long term planning.

Gao \textit{et al.}\cite{Gao} use a genetic algorithm to determine an optimal bidding strategy.  They verify their approach by modelling a 30-bus system as an example. Meng \textit{et al.}\cite{Meng} also use a genetic algorithm optimisation approach to make dynamic pricing decisions. They model the day-ahead market and propose a two-level optimisation model. They model the price responsiveness of different customers using the optimisation algorithm, and then optimise the dynamic prices that the retailer sets to maximise its profit. They confirm the feasibility and effectiveness of their technique through simulation.

\begin{table}[]
\centering
\footnotesize
\begin{tabular}{@{}llp{3cm}lp{3cm}@{}}
\toprule
Year & First Author       & Market Type                          & Application        & Algorithm Used                                \\ \midrule
2019 & Bevilacqua S. \cite{Bevilacqua} & International/National & Agent behaviour    & Genetic Algorithm,Particle Swarm Optimisation \\
2018 & Duan W.  \cite{Duan}     & International/National & Expansion planning & Bi-level coordination optimization            \\
2018 & Gao Y. \cite{Gao}       & International/National & Bidding strategies & Genetic Algorithm                             \\
2018 & Meng F.   \cite{Meng}    & Local energy market                  & Tariff design      & Genetic Algorithm                             \\ \bottomrule
\end{tabular}
\caption{Articles relating to optimisation algorithm type.}
\label{table:optimisation}
\end{table}

\subsection{Game theory}

Filho \textit{et al.}\cite{Filho} is the only paper in this review which takes a game theoretic approach. They deal with a comparative analysis of individual strategies of generating units in auctions, using. non-cooperative game theory approach. They find that their method is best suited for second-price auctions and can be extended to more complicated networks with high precision. However, they find that it is not possible to use this methodology in some, more complex, systems. For example, bidding within a stochastic and uncertain environment, where a predictions are required for the behaviour of other actors.

\begin{table}[]
\centering
\footnotesize
\begin{tabular}{@{}llp{3cm}lp{3cm}@{}}
\toprule
Year & First Author      & Market Type                          & Application        & Algorithm Used \\ \midrule
2016 & Filho N.S.C. \cite{Filho} & International/National & Bidding strategies & Game theory    \\ \bottomrule
\end{tabular}
\caption{Articles relating to game theoretic algorithm type.}
\label{table:game-theory}
\end{table}

\subsection{AI and ABMs Conclusions}
\label{sec:litreview:conclusion}

Artificial intelligence (AI) and machine learning (ML) has been integrated with agent-based models to model the electricity sector with increasing frequency over the last years. This is due to the ability of AI to optimise agent behaviour, system parameters and add functionality to agent-based models (ABMs). This study, therefore, reviewed recent papers regarding applications of AI and ML in this space. We categorised these papers into four different criteria: market type, application, algorithm type and algorithm used. To do this, we used different search terms on Scopus and reviewed all 55 articles in the field over the past five years. It was found that the majority of papers used reinforcement learning applied to bidding strategies. However, a range of applications were investigated through a wide variety of means. This included price forecasting, demand forecasting, microgrid management and risk management. We highlight the major findings from this study in the following sections.

\subsubsection{Market Type}

Table \ref{table:market_type} displays the frequency of market types for the papers reviewed. Whilst Table \ref{table:market_type-application} shows the frequency of different applications per market type.

\begin{enumerate}
	\item 68.9\% of the papers reviewed relate to an international/national electricity market. The availability of data and the relative importance of the subject of whole system transitions in current affairs may explain why such research effort has been dedicated to this. 
	\item 32.8\% of the papers for the international/national focus on bidding strategies, as is shown by Table \ref{table:market_type-application}. This is due to the ability of reinforcement learning to make strategic decisions under uncertainty. In addition, the ability to model strategic bidding is of significance importance for international/national energy models due to the appearance of oligopolies in national energy markets. However, whilst many studies have explored on strategic bidding and a few have focused on oligopolistic bidding strategies, a study on the impact on the wider market does not exist.
	\item Price forecasting is the second largest category which is explored, and is investigated in 13.1\% of papers. This is due to the ability for supervised learning to make predictions in time-series data. It is also of importance for agents to make realistic predictions of future prices. However, there exists a gap in the literature on the long-term effects of the different accuracies of these forecasts.
	\item 18\% of papers focus on the local energy market. A significant reduction when compared to the international/national electricity market. This is because of the limited availability of publicly available data for these local energy markets. ABMs require a high amount of data to inform the behaviour of the agents and environment, and so data collection for local energy markets can be expensive and difficult to obtain. Microgrids are explored in 13.1\% of papers.  Similarly to local energy markets this is because of a smaller amount of publicly available data. It is also possible that researchers place an increasing focus on international/national electricity markets due to the availability of these models and perceived impact from a large system.
\end{enumerate}

\begin{table}[]
\centering
\footnotesize
\begin{tabular}{@{}ll@{}}
\toprule
Market Type                          & Percentage (\%) \\ \midrule
International/National & 68.9            \\
Local energy market                  & 18.0            \\
Microgrid                            & 13.1            \\ \bottomrule
\end{tabular}
\caption{Frequency of market type for papers reviewed.}
\label{table:market_type}
\end{table}

\begin{table}[]
\centering
\footnotesize
\begin{tabular}{@{}lll@{}}
\toprule
Market Type                          & Application                   & Percentage (\%) \\ \midrule
Microgrid                            & Bidding strategies            & 4.9             \\
Microgrid                            & Scheduling of flexibility     & 3.3             \\
Microgrid                            & Electricity grid control      & 1.6             \\
Microgrid                            & Load scheduling               & 1.6             \\
Microgrid                            & Microgrid management          & 1.6             \\
Local energy market                  & Bidding strategies            & 6.6             \\
Local energy market                  & Tariff design                 & 4.9             \\
Local energy market                  & Demand response               & 3.3             \\
Local energy market                  & Forecasting carbon emissions  & 1.6             \\
Local energy market                  & Peer to peer trading          & 1.6             \\
International/National & Bidding strategies            & 32.8            \\
International/National & Price forecasting             & 13.1            \\
International/National & Agent behaviour               & 3.3             \\
International/National & Expansion planning            & 3.3             \\
International/National & Tariff design                 & 3.3             \\
International/National & Bilateral trading             & 1.6             \\
International/National & Demand forecasting            & 1.6             \\
International/National & Demand response               & 1.6             \\
International/National & Load scheduling               & 1.6             \\
International/National & Market investigation          & 1.6             \\
International/National & Peer to peer trading          & 1.6             \\
International/National & Risk management               & 1.6             \\
International/National & Secure demand side management & 1.6             \\ \bottomrule
\end{tabular}
\caption{Frequency of market type and application for papers reviewed.}
\label{table:market_type-application}
\end{table}

\subsubsection{Applications}
\begin{enumerate}
	\item  A significant proportion of papers have focused on bidding strategies, with 44.3\% of papers investigating this. Bidding strategies is top for each of the market types, indicating the versatility and applicability of this technique in electricity markets.
	\item The application of control, for instance grid control, load scheduling and demand response has seen a significant amount of promising research. Particular in the application of decentralised control \cite{Najafi, Shafie-Khah, Causevic}. With the advent of distributed data collection and internet of things, it has become possible to achieve various welfare objectives through a distributed control algorithm. That is, to optimise for individuals or distributed generators from the perspective of those respective individuals and generators. Studies have shown efficiencies close to centralised algorithms. However, centralised algorithms may reduce the agency of individuals and therefore must overcome an additional barrier in uptake if the individuals do not perceive that the algorithm is working in their own best interest.
	\item Various applications have been explored within the research community. In total, 17 different applications were explored. This demonstrates the versatility of ML applied to ABMs. This, however, highlights a significant gap in the literature as the majority of applications have only been explored by one or two papers. For instance, the ability to optimise the electricity system parameters in question has not been explored to the same level of detail as forecasting or trading behaviour. 
	\item Whilst there are many studies which look at optimising certain aspects of an electricity market, there is no study which looks at optimising the whole electricity system as a whole.
\end{enumerate}

\subsubsection{Algorithm Type}

Table \ref{table:algorithm} displays the frequency of each of the algorithms used. Q-Learning is the most used algorithm, with 29\% of papers using this algorithm. Second is the Roth-Erev algorithm, which is used by 9.7\% of papers. Whilst this shows the versatility of these algorithms, further research could be placed into the use of deep reinforcement learning (DRL) to improve results. DRL uses deep neural networks to act as function approximators. DDPG, DQN and DPG are examples of DRL algorithms. 

The majority of the algorithms have only been used in a single paper, and so, there remains a significant gap in the literature to apply these algorithms to the different applications to investigate with results can be improved through other algorithms.

\begin{table}[]
\centering
\footnotesize
\begin{tabular}{@{}ll@{}}
\toprule
Algorithm Used                     & Percentage (\%) \\ \midrule
Q-Learning                         & 29.0            \\
Roth-Erev                          & 9.7             \\
ANN                                & 6.5             \\
Linear regression                  & 4.8             \\
Genetic Algorithm                  & 4.8             \\
DDPG                               & 3.2             \\
Variant Roth-Erev                  & 3.2             \\
DPG                                & 1.6             \\
DQN                                & 1.6             \\
SVM                                & 1.6             \\
SARIMAX                            & 1.6             \\
Bayesian networks                  & 1.6             \\
Bi-level coordination optimization & 1.6             \\
Probe and adjust                   & 1.6             \\
Policy Iteration (PI)              & 1.6             \\
Particle Swarm Optimisation        & 1.6             \\
Novel WoLF-PHC                     & 1.6             \\
Naive Bayes classifier             & 1.6             \\
MDP                                & 1.6             \\
MADDPG                             & 1.6             \\
Classification trees               & 1.6             \\
Lasso regression                   & 1.6             \\
LSTM                               & 1.6             \\
Bayesian classifier                & 1.6             \\
Iterative algorithm                & 1.6             \\
Clustering                         & 1.6             \\
Game theory                        & 1.6             \\
Fitted Q-iteration (FQI)           & 1.6             \\
Extreme Machine Learning           & 1.6             \\
Deep Q-Learning                    & 1.6             \\
K-Means Clustering                 & 1.6             \\ \bottomrule
\end{tabular}
\caption{Frequency of algorithms for papers reviewed.}
\label{table:algorithm}
\end{table}

\subsection{Future research direction}

Significant and increasing research interest  has been placed into agent-based models applied to the electricity sector. However, this work has shown that a lot of research has clustered around similar subjects. For instance, the top application investigated for each market type is bidding strategies, largely exploring whether it is possible to use reinforcement learning to bid into an electricity market. However, the impact of these bidding strategies on the wider market has been investigated to a lesser extent. This is a gap in the literature and something we attempt to address in Chapter \ref{chapter:reinforcement}. In this chapter, we look at the effect of strategic bidding on the wider electricity market, and how to reduce the effect of monopolies.

Another aspect that has been explored to a lesser extent is the optimisation of the electricity system parameters as a whole as applied to policy. We attempt to fill this gap in Chapter and \ref{chapter:elecsim} and \ref{chapter:carbon}. In these chapter, we minimise electricity price and carbon emissions using a genetic algorithm to optimise a carbon tax policy and also to calibrate our model, ElecSim.

Finally, a large amount of research has been placed into forecasting time series within agent-based models. However, the research typically stops here and does not consider the impact of these forecasts on the wider market. Therefore, we attempt to fill this gap in Chapter \ref{chapter:demand}. 

This chapter provides a systematic review of AI applied to agent-based models in the electricity sector. Through this analysis, various research gaps have been found, as the majority of papers focus on similar themes and algorithms. This, therefore, highlights a wide range of areas in which the research community can focus on in future work.

\chapter{ElecSim model}
\label{chapter:elecsim}

\ifpdf
\graphicspath{{Chapter3/Figs/Raster/}{Chapter3/Figs/PDF/}{Chapter3/Figs/}}
\else
\graphicspath{{Chapter3/Figs/Vector/}{Chapter3/Figs/}}
\fi

\section*{Summary}

In this chapter, we motivate and introduce the agent-based model, ElecSim. The majority of the work presented here was published in \cite{Kell} and \cite{Kell2020}. The contribution of this chapter is a new open-source framework for the long-term modelling of electricity markets. We provide curated data for the simulation, to improve realism with Monte Carlo sampling, and calibrate our model using genetic algorithm-based optimisation. We validate our model after this calibration, using observed data between 2013 and 2018, as well as compare our model to the UK Government's baseline scenario up until 2035. We also provide a sensitivity analysis of important variables. For example, these variables include the \acrfull{wacc} as well as the percentage of down payment required for a particular investment. The sensitivity analysis, as well as additional scenarios shown in Section \ref{elecsim:sec:sensitivity} and Subsection \ref{elecsim:sec:representative-days-scenarios} were previously unpublished in \cite{Kell} and \cite{Kell2020}. We show, in this work, that it is possible to validate an agent-based model which focuses on the long-term electricity market by extracting the pertinent features of the market in question.

We introduce our work in Section \ref{elecsim:sec:intro}, including why a simulation model is required to aid in a low carbon transition. In addition to the literature review on energy-based simulations presented in Chapter \ref{chapter:background}, we provide a literature review of work done to validate energy models in Section \ref{elecsim:sec:litreview}. The architecture of the model is presented in Section \ref{elecsim:sec:architecture}. The model is validated in Section \ref{elecsim:sec:validation}. Various scenarios are presented in Section \ref{elecsim:sec:scenarios}, where we vary demand until 2035. A sensitivity analysis is provided in Section \ref{elecsim:sec:sensitivity}, where we vary the weighted average cost of capital, as well as the down payment required for investments. Finally, we present the limitations of our model in Section \ref{elecsim:sec:limitations} and conclude our work in Section \ref{elecsim:sec:conclusions}.

The ElecSim model, and all related code can be accessed at: \url{https://github.com/alexanderkell/elecsim}.

\section{Introduction and Motivation}
\label{elecsim:sec:intro}

\subsection{Transition to a low-carbon energy supply}

Global carbon emissions from fossil fuels have significantly increased since 1900 \cite{boden2017global}. Fossil-fuel based electricity generation sources such as coal and natural gas currently provide 65\% of global electricity \cite{BP2018}. Low-carbon sources such as solar, wind, hydro and nuclear provide 35\% . To halt this increase in \ce{CO2} emissions, a transition of the energy system towards a renewable energy system is required. 



Due to the long construction times, operating periods and high costs of power plants, investment decisions can have long term impacts on future electricity supply \cite{Chappin2017}. Governments, therefore, have a role in ensuring that the negative externalities of emissions are priced into electricity generation. This is most likely to be achieved via a carbon tax and regulation to influence electricity market players such as \acrfull{gencos}.

Decisions made in electricity markets may have unintended consequences due to their complexity. A method to test hypothesise before they are implemented would, therefore, be useful.

To aid in such a transition, energy modelling can be used by governments, industry and agencies to explore possible scenarios under different variants of government policy, future electricity generation costs and energy demand. These energy modelling tools aim to mimic the behaviour of energy systems through different sets of equations and data sets to determine the energy interactions between different actors and the economy \cite{Machado2019}.

\subsection{ElecSim: Modelling and simulation}

Live experimentation of physical processes is not often practical. The costs of real-life experimentation can be prohibitively high, and can require a significant amount of time in order to fully ascertain the long-term trends. There is also a risk that changes can have detrimental impacts and lead to risk-averse behaviour. These factors are true for electricity markets, where decisions can have long term impacts. For instance, by investing in a coal power plant today, this will lock in the carbon emissions emitted by the coal power plant for the next 25+ years. Simulation, however, can be used for rapidly prototyping ideas. The simulation is parametrised by real-world data and phenomena. Through simulation, the user is able to assess the likelihoods of outcomes under certain scenarios and parameters \cite{Law:603360}.

Simulation is often used to increase understanding as well as to reduce risk and reduce uncertainty. Simulation allows practitioners to realise a physical system in a virtual model. In this context, a model is defined as an approximation of a system through the use of mathematical formulas and algorithms. Through simulation, it is possible to test a system where real-life experimentation would not be practical due to reasons such as prohibitively high costs, time constraints or risk of detrimental impacts. This has the dual benefit of minimising the risk of real decisions in the physical system, as well as allowing practitioners to test less risk-averse strategies.

\acrfull{abms} are a class of computational simulation models composed of autonomous, interacting agents. These can be built to model systems with many heterogeneous agents. Due to the numerous and diverse actors involved in electricity markets, \acrshort{abms} have been utilised in this field to address phenomena such as market power \cite{Ringler2016a}. 

\begin{figure}
	\centering
	\includegraphics[width=0.85\linewidth]{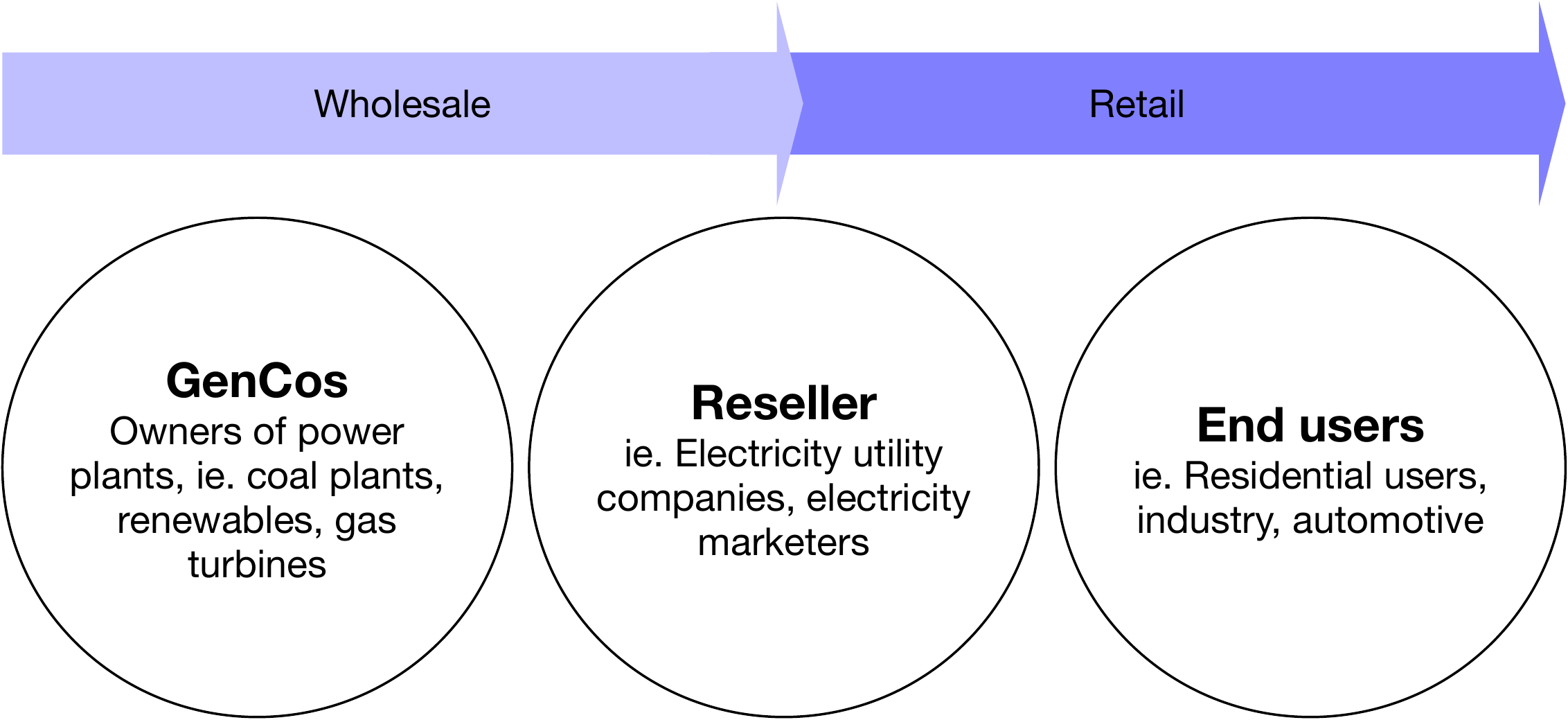}
	\caption{Overview of a wholesale electricity market.}
	\label{fig:elec_market_overview}
\end{figure}

The work presented in this chapter develops ElecSim, an open-source \acrshort{abm} that simulates GenCos in a wholesale electricity market. ElecSim models each GenCo as an independent agent. An electricity market facilitates trades between the two. Figure \ref{fig:elec_market_overview} displays an overview of a wholesale electricity market. The GenCos invest in power generation facilities, such as coal power plants, renewables or gas turbines which produce electricity. Typically, a reseller will purchase the produced electricity through bilateral contracts, or on the day-ahead market who then sell the electricity to end-users. Where end-users could be residential users, industry or the automotive sector.

GenCos make bids for each of their power plants. Their bids are based on the generators \gls{SRMC} \cite{Perloff2012}, which excludes capital and fixed costs. The electricity market accepts bids in cost order, also known as merit-order dispatch. GenCos invest in power plants based on expected profitability.	

ElecSim is designed to provide quantitative advice to policymakers, allowing them to test policy outcomes under different scenarios. A scenario is a single future which could occur under different circumstances. To achieve this, policymakers are able to modify a file to realise a scenario of their choice. For example, by changing anticipated future energy costs, carbon taxes or electricity demand. It can also be used by energy market developers who can test new electricity sources or policies, enabling the modelling of changing market conditions.

This model can be used by the following players:

\begin{itemize}
	\item {\bf Policy experts} to test policy outcomes under different scenarios and provide quantitative advice to policymakers. They can provide a simple script defining the policies they wish to use along with the parameters for these policies.
	\item {\bf Energy market developers} who can use the extensible framework to add elements such as new energy sources, policies, consumer profiles and storage types. Thus allowing ElecSim to adapt to a changing ecosystem.
\end{itemize}



\subsubsection{Current approaches}

Optimisation based solutions are the dominant approach for analysing energy policy \cite{Chappin2017}. However, the results of these models should be interpreted in a normative manner. For example, how investment and policy choices should be carried out, under certain assumptions and scenarios. The processes which emerge from an equilibrium model remain a black-box, making it difficult to fully understand the underlying dynamics of the model \cite{Chappin2017}.

In addition to this, optimisation models do not allow for endogenous behaviour to emerge from typical market movements, such as investment cycles \cite{Chappin2017, Gross2007}. By modelling these naturally occurring behaviours, policy can be designed that is robust against movements away from the optimum/equilibrium. Thus, helping policy to become more effective in the real world. 

Agent-based models differ from optimisation-based models by the fact that they are able to explore `\textit{what-if}' questions regarding how a sector could develop under different prospective policies, as opposed to determining optimal trajectories. \acrshort{abm}s are particularly pertinent in decentralised electricity markets, where a decentralised actors dictate investments made within the electricity sector. \acrshort{abm}s have the ability to closely mimic the real world by, for example, modelling irrational agents, in this case, Generation Companies (GenCos) with incomplete information in an uncertain future \cite{Ghorbani2014}.

By undertaking calculations for each day of the simulation in ElecSim, we would potentially gain accuracy. For instance, by modelling the variations within days of solar irradiance and wind speed. However, this would have the effect of slowing the simulation down significantly. Therefore, as a compromise between speed and accuracy, we use a set of representative days to model an annual time period. 

Similar to Nahmmacher \textit{et al.} we demonstrate how clustering of multiple relevant time series such as electricity demand, solar irradiance and wind speed can reduce computational time by selecting representative days ~\cite{Nahmmacher2016}. In this context, representative days are a subset of days that have been chosen due to their ability to approximate the weather and electricity demand in an entire year. Similar to Nahmacher \textit{et al.} we use a Ward hierarchical clustering algorithm  \cite{doi:10.1080/01621459.1963.10500845}, However, we also try a $k$-means clustering approach \cite{forgy65}. We chose the $k$-means clustering approach due to the previous success of this technique in clustering time series \cite{Kell2018a}.

\subsection{Validation of long-term models}

There is a desire to validate the ability of energy-models to make long-term predictions. Validation increases confidence in the outputs of a model and leads to an increase in trust amongst the public and policymakers. Energy models, however, are frequently criticised for being insufficiently validated, with the performance of models rarely checked against historical outcomes \cite{Beckman2011}.

In answer to this, we postulate that \acrshort{abm}s can provide accurate information to decision-makers in the context of electricity markets by using cross-validation of observed electricity mix compared to simulated electricity mix. We increase the temporal granularity of the work in our work in this chapter and use genetic algorithms to tune the model to observed data enabling us to perform calibration of the model. This enables us to understand the parameters required to observe certain phenomena, as well as use these fitted parameters to make inferences about the future.


We use a genetic algorithm approach to find an optimal set of price curves predicted by generation companies (GenCos) that adequately model observed investment behaviour in the real-life electricity market in the United Kingdom. Similar techniques can be employed for other countries of various sizes \cite{Kell}. 

We measure the accuracy of projections for our improved \acrshort{abm} with those of the UK Government's Department for Business, Energy and Industrial Strategy (BEIS) for the UK electricity market between 2013 and 2018. In addition to this, we compare our projections from 2018 to 2035 to those made by BEIS in 2018 \cite{DBEIS2019}. We select five years for the calibration and validation period as we believe that the transition from coal to gas in this period reflects the kind of dynamics that ElecSim must capture over the long-term. Whilst it can be argued that a five year period is too short a time to adequately model a long-term energy model, it is argued here that long-term validation is unfeasible due to the large amount of difficult to capture features over the long-term. In other words, a long-term model provides scenarios and not forecasts, where each of the scenarios could feasibly occur, but it is not feasible to predict which of these scenarios will occur due to the stochastic nature of the real-world in which we are attempting to model. 

We use numerical validation in this case to reduce the complexity required for theoretical validation. Theoretical validation requires the validation of many complex processes, and even after this process, we would not know the complex whether the interaction of these processes are realistic. We argue in this thesis, that both numerical and theoretical validation are required.

\subsection{Model outputs}
\label{elecsim:sec:results}

Through this validation process, we are able to adequately model the transitional dynamics of the electricity mix in the United Kingdom between 2013 and 2018. During this time there was an ${\sim}88\%$ drop in coal use, ${\sim}44\%$ increase in \acrfull{ccgt}, ${\sim}111\% $ increase in wind energy and increase in solar from near zero to ${\sim}1250$MW. We are therefore able to test our model in a transition of sufficient magnitude.


We show in this chapter, that agent-based models are able to mimic the behaviour of the UK electricity market under the same specific scenario conditions. Concretely, we show that under an observed carbon tax strategy, fuel price and electricity demand scenario, the model, ElecSim, closely matches the observed electricity mix between 2013 and 2018. We achieve this by determining an exogenous predicted price duration curve using a genetic algorithm to minimise the error between observed and simulated electricity mix in 2018. The predicted price curve is an arrangement of all price levels in descending order of magnitude. The predicted price duration curve achieved is similar to that of the simulated price duration curve in 2018, increasing confidence in the underlying dynamics of our model. 

In addition, we compare our projections to those of the BEIS reference scenario from 2018 to 2035~\cite{DBEIS2019}. To achieve this, we use the same genetic algorithm optimisation technique as during our validation stage, optimising for predicted price duration curves for calibration. Our model demonstrates that we are able to closely match the projections of BEIS by finding a set of realistic price duration curves which are subject to investment cycles. Our model, however, exhibits a more realistic step-change in nuclear output than that of BEIS. This is because, whilst BEIS projects a gradual increase in nuclear output, our model projects that nuclear output will grow instantaneously at a single point in time as a new nuclear power plant comes online. 

This allows us to verify the scenarios of other models, in this case, BEIS' reference scenario, by ascertaining whether the optimal parameters required to achieve such scenarios are realistic. In addition to this, we are able to use these parameters to analyse `\textit{what-if}' questions with further accuracy.

\subsection{Contributions of this Chapter}

As part of this work, we contribute a validated open-source agent-based model called ElecSim. Whilst we have used the United Kingdom as a use-case for this thesis, ElecSim is able to model decentralised markets of various sizes, and can therefore be used to model other countries. 

To improve our results as well as making the execution time tractable, we increased the temporal granularity of the model using a $k$-means clustering approach to select a subset of representative days for wind speed, solar irradiance and electricity demand. This subset of representative days enabled us to approximate an entire year and only required a fraction of the total time-steps that would be necessary to model each day of a year independently. We show that we are able to provide an accurate framework, through this addition, to allow policymakers, decision-makers and the public to explore the effects of policy on investment in electricity generators. 

We demonstrate that with a genetic algorithm approach, we are able to optimise parameters to improve the accuracy of our model. Namely, we optimise the predicted electricity price, the uncertainty of this electricity price and nuclear subsidy. We validate our model using the observed electricity mix between 2013-2018 through a process of calibration. That is, we find the predicted electricity price, uncertainty of this electricity price and nuclear subsidy which match with the stated scenario. We then use these values to project further.

A major contribution of this work is to demonstrate that it is possible for agent-based models to accurately model transitions in the UK electricity market. This was achieved by comparing our simulated electricity mix to the observed electricity mix between 2013 and 2018. In this time, a transition from coal to natural gas was observed. We demonstrate that a high temporal granularity is required to accurately model fluctuations in wind and solar irradiance for intermittent renewable energy sources.

\clearpage
\section{Architecture}
\label{elecsim:sec:architecture}

In this Section, we detail how the architecture of ElecSim has been designed. ElecSim is made up of six parts: 
\begin{itemize}
	\item \textbf{GenCos} which are made up of agents.
	\item \textbf{Demand}
	\item \textbf{Power plants}.
	\item A \textbf{power exchange}, which controls an electricity spot market.
	\item The time-steps.
	\item The \textbf{data} for parametrisation. This includes both the \textbf{configuration file} and \textbf{data sources}.
\end{itemize}

\begin{figure}
	\centering
	\includegraphics[width=0.85\linewidth]{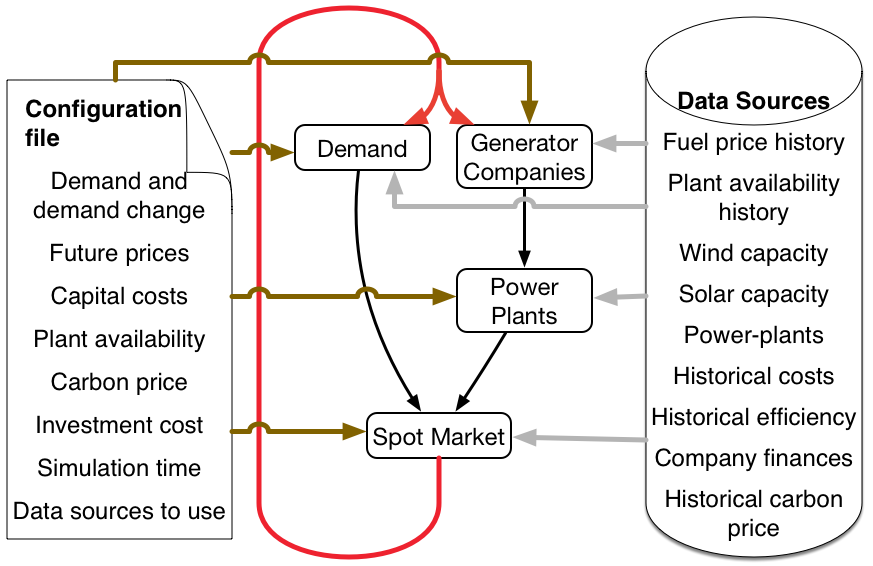}
	\caption{High level overview of the ElecSim model.}
	\label{fig:systemoverview}
\end{figure}

A schematic of ElecSim is displayed in Figure \ref{fig:systemoverview}. We have placed the items in bold which relate to each part of the figure. Next, we will describe each of the elements described in Figure \ref{fig:systemoverview} in further detail.

\subsubsection{Data parametrisation} ElecSim contains a configuration file and a collection of data sources for parametrisation. These data sources contain information such as historical fuel prices, historical plant availability, wind and solar capacity.

The configuration file allows for rapid changes to test different hypothesis and scenarios, and points to the different data sources. The configuration file enables one to change the demand growth and shape, future fuel and carbon prices, capital costs, plant availability, investment costs and simulation time.

\subsubsection{Demand Agent} 
\begin{figure}
	\centering
	\includegraphics[width=0.95\linewidth]{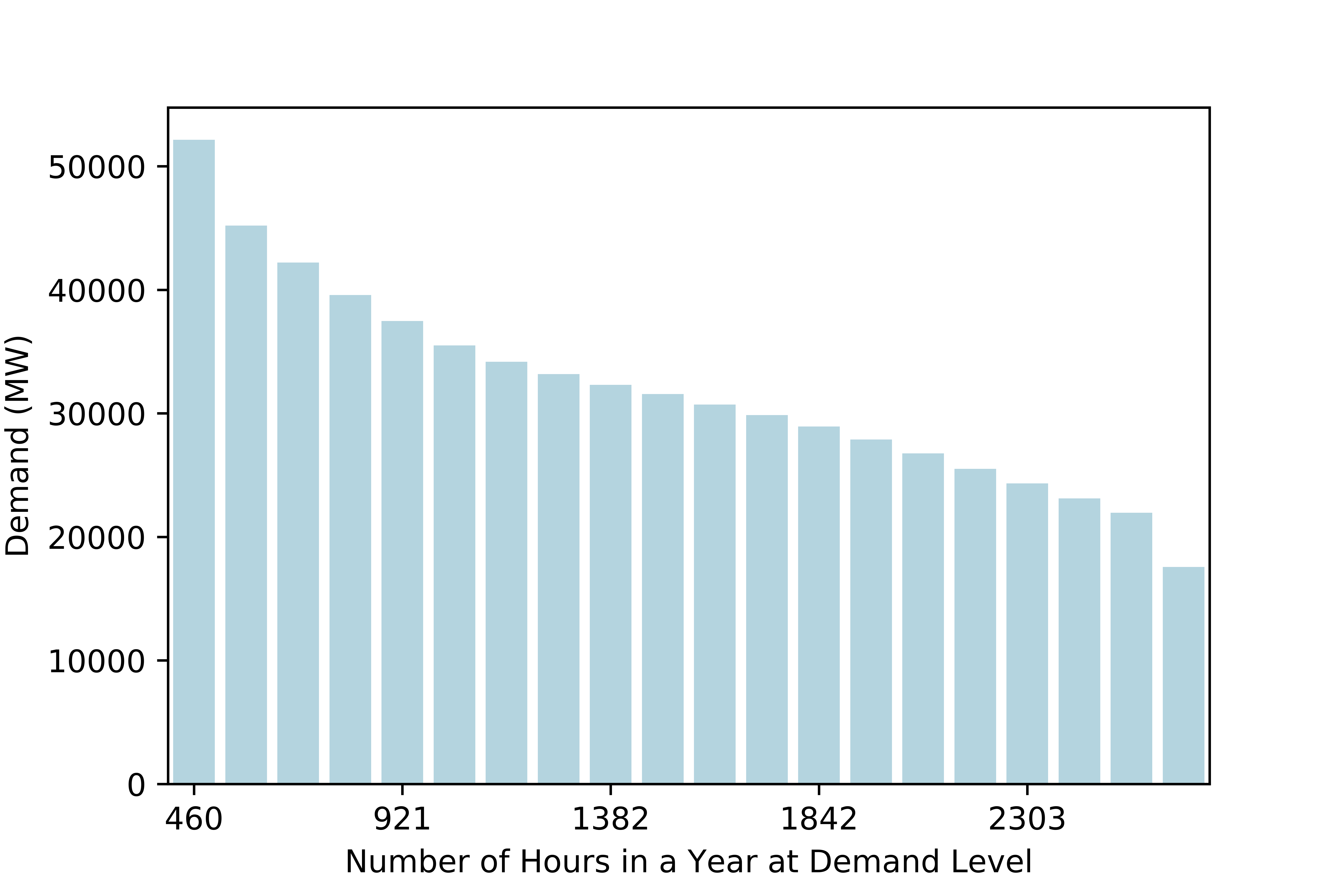}
	\caption{Example \acrfull{ldc} for a single year.}
	\label{fig:loaddurationcurve}
\end{figure}

The demand agent is a simplified representation of aggregated demand in a country. The demand is represented as a \gls{ldc}. An example load duration curve for a year is demonstrated in Figure \ref{fig:loaddurationcurve}. This Figure effectively shows the baseload level at 460 hours, and the peak load at 2400 hours and above, with the profile displayed between these points. An \acrshort{ldc} is an arrangement of all load levels in descending order of magnitude. where the lowest segment demand demonstrates baseload, and the highest segment represents peak demand. Each year, the demand agent changes each of the \acrshort{ldc} segments proportionally, by multiplying each of the levels by the factor to increase or decrease these load levels. In other words, whilst the magnitude of the LDC may change, the profile does not.

\subsubsection{Generation Company Agents} The GenCos have two main functions. Investing in power plants and making bids to sell their generation capacity. We will first focus on the buying and selling of electricity, and then cover the investment algorithm.

The power exchange (or spot market) runs every year, accepting the lowest bids until supply meets demand. Once this condition is met, the spot price or system marginal price (SMP) is paid to all generators regardless of their initial bid. Generators are motivated to bid their SRMC, to ensure that their generator is being utilised, and reduce the risk of overbidding.

\paragraph{Power Plants}\label{elecsim:ssssec:powerplantparameters} 

Power plants are made up of various parameters which dictate their behaviour within the electricity market. For example, cheaper power plants will undercut the costs of more expensive power plants, or solar will displace fossil-fuel based plants at times of high solar irradiance.

Costs form an important element of markets and investment, and publicly available data for power plant costs for individual countries can be scarce, and when available is presented in different formats. Thus, extrapolation and interpolation is required to estimate costs for power plants of differing sizes, types and years of construction.

Users are able to initialise costs relevant to their particular country by providing detailed cost parameters. They can also provide an average cost per MWh produced over the lifetime of a plant, known as the \acrfull{lcoe}. They do this within the data sources file.

The parameters used to initialise the power plants are detailed in Table \ref{table:cost_parameters}. Periods have units of years and costs in unit of currency per MW unless otherwise stated.

\begin{table}[]
	\begin{tabular}{p{5cm}p{10cm}}
		\toprule
		Parameter                                    & Definition                                                                                                                            \\ \midrule
		Efficiency ($\eta$)                          & Percentage of energy from fuel that is converted into electrical energy (\%)                                                          \\
		Operating period ($OP$)                      & The total period in which a power plant is in operation.                                                                              \\
		Pre-development period ($P_D$)               & Include the time for pre-licensing, technical and design, as well as time for regulatory, licensing and public enquiry.               \\
		Pre-development costs ($P_C$)                & Include the costs for pre-licensing, technical and design, as well as costs incurred due to regulatory, licensing and public enquiry. \\
		Construction period ($C_D$)                  & Time for development of the plant, excluding network connections.                                                                     \\
		Construction costs ($C_C$)                   & Costs incurred during the development of the plant, excluding network connections.                                                    \\
		Infrastructure costs ($I_C$)                 & The costs incurred by the developer in connecting the plant to the electricity or gas grid (\textsterling).                           \\
		Fixed operation \& maintenance costs ($F_C$) & Costs incurred in operating the plant that do not vary based on output.                                                               \\
		Variable operation \& maintenance ($V_C$)    & Costs incurred in operating the plant that depends on generator output                                                                \\
		Availability ($A$) & Percentage of the time that the power plant can run without outages and downtime.                                                     \\ 
		\bottomrule
	\end{tabular}
	\caption{Power plant parameters.}
	\label{table:cost_parameters}
\end{table}



If specific parameters are not known, the LCOE can be used for parameter estimation, through the use of linear optimisation. Constraints can be set by the user, enabling, for example, varying operation and maintenance costs per country as a fraction of LCOE. This includes future costs.

To fully parametrise power plants, availability and capacity factors are required. Availability is the percentage of time that a power plant can produce electricity. This can be reduced by forced or planned outages. We integrate historical data to model improvements in reliability over time.

The capacity factor is the actual electrical energy produced over a given time period divided by the maximum possible electrical energy it could have produced. The capacity factor can be impacted by regulatory constraints, market forces and resource availability. For example, higher capacity factors are common for photovoltaics in the summer and lower in winter. 

To model the intermittency of wind and solar power, we allow them to contribute only a certain percentage of their total capacity (nameplate capacity) for each load segment. This percentage is based upon empirical wind and solar capacity factors. In this calculation, we consider the correlation between demand and renewable resources. 

When initialised, $V_C$ is selected from a uniform distribution, with the ability for the user to set a maximum percentage increase or decrease. A uniform distribution was chosen to capture the large deviations that can occur in $V_C$, especially over a long time period. By doing this, the variance in costs between individual power plants for processes such as preventative and corrective maintenance, labour costs and skill, health and safety and chance are different per plant instant. Future work would include finding data to better inform the $V_C$ price.

Fuel price is controlled by the user; however, there is inherent volatility in fuel price. To take into account this variability, an ARIMA \cite{ARIMA} model was fit to historical gas and coal price data. The standard deviation of the residuals was used to model the variance in price that a GenCo will buy fuel in a given year. This considers differences in chance and hedging strategies.

\subsubsection{Spot market} The spot market is run by a power exchange. The power exchange effectively ranks all power plant bids on the market from lowest to highest for each demand segment, as shown by Figure \ref{fig:loaddurationcurve}. All bids are accepted until supply meets demand. This dispatch method is known as merit order dispatch. 

This method effectively ranks ensures that power plants with the lowest marginal cost are brought online first, and those with the highest marginal cost brought on last. For example, expensive peaker plants will only be used at peak times, where all cheaper electricity generation supply is being used. 

\paragraph{Investment.} Investment in power plants is made based upon a \Gls{NPV} calculation. \acrshort{npv} is a summation of the present value of a series of present and future cash flow. NPV provides a method for evaluating and comparing investments with cash flows spread over many years, making it well suited for evaluating power plants which have a long lifetime.  \acrshort{npv} is based upon the fact that current cash flow is worth more than future cash flow. This is due to the fact that money today can be invested and have a rate of return. This means that, for example, \$50,000 today is worth more than \$50,000 in 10 years time. The value in which future cash flow is worth less than present cash flow is denoted by the discount rate.

Equation \ref{architecture:eq:npv_eq} is the calculation of \acrshort{npv}, where $t$ is the year of the cash flow, $i$ is the discount rate, $N$ is total number of periods, or lifetime of power plant, and $R_t$ is the net cash flow at time $t$.

\begin{equation} \label{architecture:eq:npv_eq}
NPV(t, N) = \sum_{t=0}^{N}\frac{R_t}{(1+t)^t}.
\end{equation}
A discount rate set by a GenCo's weighted average cost of capital (WACC) is often used \cite{KincheloeStephenC1990TWAC}. WACC is the rate that a company is expected to pay on average for its stock and debt. Therefore to achieve a positive NPV, an income larger than the \acrshort{wacc} is required. However, a higher WACC is often selected to adjust for varying risk profiles, opportunity costs and rates of return \cite{KincheloeStephenC1990TWAC}. To account for these differences we sample from a Gaussian distribution, giving us sufficient variance whilst deviating from the expected price. Gaussians are often observed within real-life distributions, and therefore chosen for this work.

To calculate the NPV, future market conditions must be considered. For this, each GenCo forecasts $N$ years into the future, which we assume is representative of the lifetime of the plant. As in the real world, GenCos have imperfect information, and therefore must forecast expected demand, fuel prices, carbon price and electricity sale price. This is achieved by fitting functions to historical data. Each GenCo is different in that they will use differing historical time periods of data for forecasting.

This is a notoriously difficult task, and there are multiple different ways in which this can be achieved \cite{Tao2021}. However for this work, we forecast fuel and carbon price using two different linear regression functions. Linear regression was chosen due to the simplicity of the function, and the absence of unrealistic growth seen in other functions. Demand, however, is forecast using an exponential function, which considers compounded growth, as seen in the real-world. Linear regression is used if an exponential function is found to be sub-optimal. We define sub-optimal as the case that a solution can not be found. We use the linear regression for this, due to its characteristics of reducing unrealistic growth.

The forecasted electricity price $N$ years ahead is difficult to ascertain accurately. We therefore use two methods for forecasting these. The first is to simulate a market $N$ years ahead. The second is to optimise for the predicted \acrfull{pdc} using a genetic algorithm. We describe this optimisation in Section \ref{elecsim:sec:validation}.

For the simulated market, the forecasted data is used to simulate a market $N$ years into the future using the electricity market algorithm. We simulate a market based on the expected bids -- based on short run marginal cost (SRMC) -- that every operating power plant will make. SRMC is the cost for a power plant to produce one unit of electricity, excluding fixed and capital costs. We include the removal of plants that will be past their operating period, and the introduction of plants that are in construction or pre-development stages. This effectively runs a simulation within a simulation here. Where the GenCos must use a simulation technique to understand future market conditions.

There may be scenarios where demand is forecast to grow significantly, and limited investments have yet been made to meet that demand. The expected price would be that of lost load. Lost load is defined as the price customers would be willing to pay to avoid disruption in their electricity supply. To avoid GenCos from estimating large profits, and under the assumption that further power plant investments will be made, the lost load price is replaced with a predicted electricity price using linear regression based on prices at lower points of the demand curve. If zero segments of demand are met, then the lost load price is used to encourage investment. 

Once this data has been forecasted, the NPV can be calculated. GenCos must typically provide a certain percentage of upfront capital, with the rest coming from investors in the form of stock and shares or debt (WACC). The percentage of upfront capital can be customised by the user in the configuration file. The GenCos then invest in the power plants with the highest NPV.

\paragraph{Time-steps} Time-steps are the steps made in a simulation which simulate time. Each time-step moves the simulation forward a single iteration, and mimics time in our case. For the time-steps, two approaches were taken. For the first approach, as per Chappin \textit{et al.} \cite{Chappin2017}, we modelled the LDC of electricity demand with twenty segments. Twenty segments enabled us to capture the variation in demand throughout the year to a high degree of accuracy, whilst reducing computational complexity. However, as we show later in Section \ref{elecsim:sec:scenarios}, this led to an overestimation of the supply of \acrshort{ires}. 

For the second approach, we used representative days to model a year. Representative days in this context are a subset of days which have characteristics, that when scaled proportionally can accurately model an entire year. To select these representative days, we used a $k$-means approach. We describe this in full detail in Section \ref{elecsim:sec:representative}

\begin{landscape}
	\begin{figure*}
		\centering
		\includegraphics[width=\linewidth]{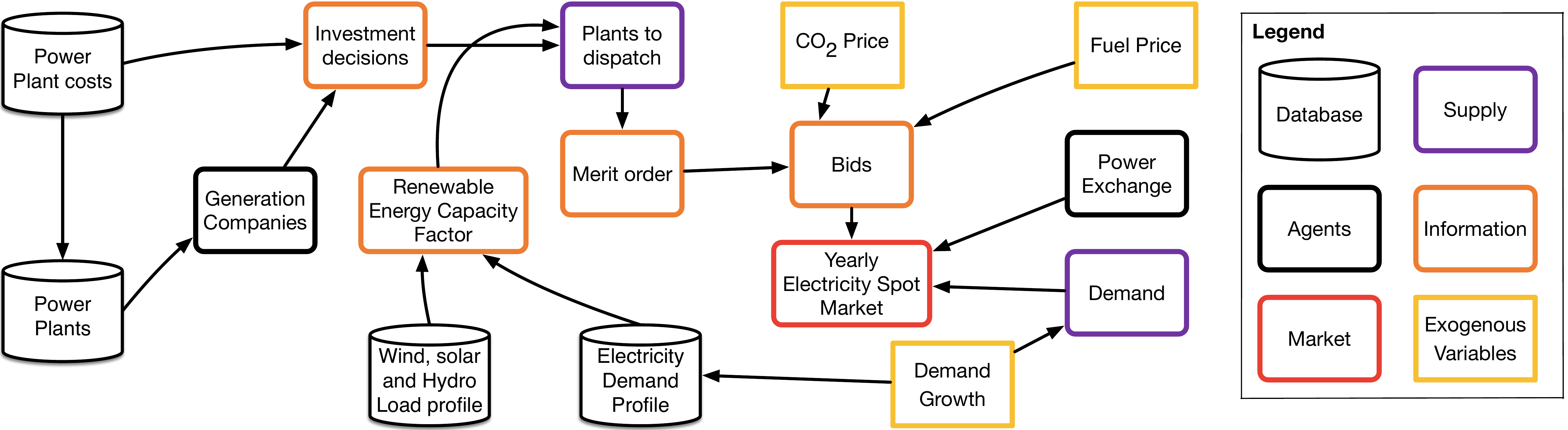}
		\caption{Detailed ElecSim simulation overview}
		\label{fig:lowlevelsystem}
	\end{figure*}
\end{landscape}

Figure \ref{fig:lowlevelsystem} demonstrates the simulation and how it co-ordinates runs. The world contains data and brings together GenCos, the Power Exchange and demand. The investment decisions are based on future demand and costs, which in turn influence bids made.

To explain the simulation, we will start from the left side of Figure \ref{fig:lowlevelsystem} and move to the right. A database of power plants is parametrised by a database of parameters, as discussed in Table \ref{table:cost_parameters}. This ensures that the plants have realistic properties. 

Generation companies (GenCos) own and invest in these power plants through the use of investment decisions. The GenCos than submit bids to dispatch these power plants. They do this using the parameters of the power plants, which for renewable energy inclues the renewable energy capacity factors. For instance, wind, solar and hydro have varying profiles throughout the day and year. 

These bids are then augmented with the \ce{CO2} price (tax) and fuel price to create the actual bid. The yearly electricity spot market then takes the demand and matches this to the submitted bids in merit order. The demand grows (or falls) each year, depending on the scenario. 

Exogenous variables include fuel and \ce{CO2} prices as well as demand growth. Once the data is initialised, the model calls on the Power Exchange to operate the yearly electricity spot market. The world also settles the accounts of the GenCos, by paying bids, and removing operating and capital costs as well as loans and dividends.

\subsection{Representative days}
\label{elecsim:sec:representative}

In this Subsection, we describe the work which allowed us to determine the granularity of time-steps. In this problem, there is a trade-off between accuracy and compute time. We visualise this trade-off and choose a cut-off point, in which we receive diminishing returns for an increase in temporal granularity, in terms of accuracy. Specifically, for this, we use representative days. Representative days, in this context, are a subset of days which, when scaled up to 365 days can adequately represent a year. 

In this work, we initialised the model to a scenario of the United Kingdom as an example. However, the fundamental dynamics of the model remain the same for other decentralised electricity markets.



Similar to findings of other authors, using a relatively low number of time-steps from the representative days leads to an overestimation of the uptake of intermittent renewable energy resources (IRES) and an underestimation of flexible technologies~\cite{Haydt2011,Ludig2011}. This is due to the fact that the full intermittent nature of renewable energy could not be accurately modelled in such a small number of time-steps. For instance, by modelling the entire year in only 20 time-steps, a reduction in the ability for ElecSim to model hourly lulls in wind speed, the absence of solar irradiance at night time and differences in electricity demand. This is critically important for a power system with high amounts of electricity generated from renewable electricity.

To address this problem, whilst maintaining a tractable execution time, we approximated a single year as a subset of proportionally weighted, representative days. This enabled us to reduce computation time whilst maintaining accuracy. Each representative day consisted of 24 equally separated time-steps, which model hours in a day. Hourly data was chosen, as this was the highest resolution of the dataset available for offshore and onshore wind and solar irradiance \cite{Pfenninger2016}. A lower resolution would allow us to model more days; however, we would lose accuracy in terms of the variability of the renewable energy sources.

Similar to Nahmmacher \textit{et al.} \cite{Nahmmacher2016} we used a clustering technique to split similar days of weather and electricity demand into separate groups. We then selected the historic day that were closest to the centre of the cluster, known as the medoid, as well as the average of the centre, known as the centroid ~\cite{Nahmmacher2016}. Similar to Nahmmacher, we used Ward's clustering algorithm and selected the centroid \cite{doi:10.1080/01621459.1963.10500845}. However, we also used the $k$-means clustering algorithm~\cite{forgy65}. This was due to the ability for the $k$-means algorithm to cluster time-series into relevant groups \cite{Kell2018a}. These days were scaled proportionally to the number of days within their respective cluster to approximate a total of 365 days. A representative day is one where solar and wind are chosen on the same day. This is because on a particular day, wind and solar are correlated. The Ward's clustering algorithm is an extension of the work published in \cite{Kell2020}.

Equation \ref{elecsim:eq:medoids_series} shows the series for a medoid or centroid, selected by the clustering algorithms:

\begin{equation}
\label{elecsim:eq:medoids_series}
P^{x,i}_{h}=\{P_1, P_2, \ldots, P_{24}\},
\end{equation}

\noindent where $P^{x,i}_{h}$ is the medoid for series $x$, where $x\in X$ refers to offshore wind capacity factor, onshore wind capacity factor, solar capacity factor and electricity demand, $h$ is the hour of the day and $i$ is the respective cluster. $\{P_1, P_2, \ldots , P_{24}\}$ refers to the capacity values at each hour of the representative day. In other words, $x$ is a set of offshore wind capacity factor, onshore wind capacity factor, solar capacity fator and electricity demand.

We then calculated the weight of each cluster. This gave us a method of assigning the relative importance of each representative day when scaling the representative days up to a year. The weight is calculated by the proportion of days in each cluster. This gives us a method of determining how many days within a year are similar to the selected medoid or centroid. The calculation for the weight of each cluster is shown by Equation \ref{elecsim:eq:cluster_weight}:

\begin{equation}
\label{elecsim:eq:cluster_weight}
w_i = \frac{n_i}{||N||},
\end{equation} 

\noindent where $w_i$ is the weight of cluster $i$, $n_i$ is the number of days in cluster $i$, and $||N||$ is the set of days that have been used for clustering.

The next step was to scale up the representative days to represent the duration curve of a full year. A duration curve is similar to the load duration curve shown in Figure \ref{fig:loaddurationcurve}, however can be defined for other processes such as wind speed and solar irradiance. We achieved this by using the weight of each cluster, $w_i$, to increase the number of hours that each capacity factor contributed in a full year. Equation \ref{elecsim:eq:scaled-medoid} details the scaling process to turn the medoid or centroid, shown in Equation \ref{elecsim:eq:medoids_series}, into a scaled day. Where $\widetilde{P}^{x,i}_{h}$ is the scaled day:

\begin{equation}
\label{elecsim:eq:scaled-medoid}
\widetilde{P}^{x,i}_{h} = P_h^{x,i}\cdot W.
\end{equation} 

\noindent where $W=\{w_1,w_2,\ldots,w_i\}$. $W$ is the set of weights, $w_1$ is the weight for the first day, $w_2$ the weight for the second day, and $w_i$ the $i^{th}$ day. Equation \ref{elecsim:eq:scaled-medoid} effectively extends the length of the day, proportional to the amount of days in the respective cluster.


Finally, each of the scaled representative days were concatenated to create a single series used for the calculations in the simulation. This concatenated number contains information of the required capacity factors and the respective duration curve. Equation \ref{elecsim:eq:total_time_series} displays the total time series of a series $x$, where each scaled medoid is concatenated to produce an approximated time series, $\widetilde{P}^x$:

\begin{equation}
\label{elecsim:eq:total_time_series}
\widetilde{P}^x=\left(\widetilde{P}^{x,1}_{h},\widetilde{P}^{x,2}_{h},\ldots, \widetilde{P}^{x,||N||}_{h}\right),
\end{equation}

\noindent the total number of hours in the approximated time series, $\widetilde{P}^x$, is equal to the number of hours in a day multiplied by the number of days in a year, which gives the total number of hours in a year ($24\times 365=8760$), as shown by Equation \ref{elecsim:eq:total_scale}:

\begin{equation}
\label{elecsim:eq:total_scale}
\sum\limits_{w\in W}\sum\limits_{t=1}^{T=24}\left(w_i t\right)=24\times 365=8760,
\end{equation}

\noindent where $w\in W$ the Set of clusters.

\subsection{Error Metrics}

To measure the validity of our approximation using representative days and also compare the optimum number of days, or clusters, we used a technique similar to Poncelet \textit{et al.} \cite{Dhaeseleer2015, Poncelet2017}. We trialled the number of clusters against three different metrics: \acrfull{ce} ($CE_{av}$), \acrfull{nrmse} and \acrfull{ree} ($REE_{av}$). 

$REE_{av}$ is the average value over all the considered time series $\widetilde{P}^x{\in} \widetilde{P}$ compared to the observed average value of the set $P^x\in P$. Where $P^x\in P$ are the observed time series and $\widetilde{P}^x{\in} \widetilde{P}$ are the scaled, approximated time series using representative days. $REE_{av}$ is shown formally by Equation \ref{eq:ree_av}:

\begin{equation}
\label{eq:ree_av}
REE_{av}=\frac
{\sum\limits_{P^x{\in} P}\left(\left|
	\frac
	{\sum\limits_{t\in T}DC_{P^x_t}-\sum\limits_{t\in T}\widetilde{DC}_{\widetilde{P}^x_t}}
	{\sum\limits_{t\in T}DC_{P^x_t}}
	\right|\right)
}
{\left|\left|P\right|\right|},
\end{equation}

\noindent where $DC_{P^x_t}$ is the duration curve for $P^x$ and $\widetilde{DC}_{\widetilde{P}^x_t}$ is the duration curve for $\widetilde{P}^x$. In this context, the duration curve can be constructed by sorting the capacity factor and electrical load data from high to low. The $x-$axis for the duration curve exhibits the proportion of time that each capacity factor represents, similar to the load duration curve shown in Figure \ref{fig:loaddurationcurve}. The approximation of the duration curve is represented in this text as $\widetilde{DC}_{\widetilde{p}^x}$. $t\in T$ refers to a specific time step of the original time series. $\widetilde{DC}$ refers to the approximated duration curve for $\widetilde{P}^x$. Note that in this text $\left|\cdot\right|$ refers to the absolute value, and $\left|\left|\cdot\right|\right|$ refers to the cardinality of a set and $\left|\left|P\right|\right|$ refers to the total number of of considered time series.

Specifically, the sum of the observed values, $P^x$, and approximated values, $\widetilde P^x$, for all of the time series are summed. The proportional difference is found, which is summed for each of the different series, $x$, and divided by the number of series, to give $REE_{av}$.



Another requirement is for the distribution of load and capacity factors for the approximated series to correspond to the observed time series. It is crucial that we can account for both high and low levels of demand and capacity factor for IRES generation. This enables us to model for times where flexible generation capacity is required.

The distribution of values can be represented by the duration curve ($DC$) of the time series. Therefore, the average normalised root-mean-square error ($NRMSE_{av}$) between each $DC$ is used as an additional metric. The $NRMSE_{av}$ is shown formally by Equation \ref{eq:nrmse_av}:

\begin{equation}
\label{eq:nrmse_av}
NRMSE_{av}=\frac
{\sum\limits_{P^x{\in} P}\left(\frac
	{\sqrt{
			\frac{1}{\left|\left|T\right|\right|}
			\cdot
			\sum\limits_{t\in T}(DC_{P^x_t}-\widetilde{DC}_{\widetilde{P}^x_t})^2}
	}
	{max(DC_{P^x})-min(DC_{P^x})}
	\right)}
{\left|\left|P\right|\right|}.
\end{equation}

Specifically, the difference between the approximated and observed duration curves for each time-step $t$ is calculated. The average value is then taken of these differences. This average value is then normalised for the respective time series $P^x$. The average of these average normalised values for each time series are then taken to provide a single metric, $NRMSE_{av}$.

The final metric used is the correlation between the different time series. This is used due to the fact that wind and solar output influences the load within a single region, solar and wind output are correlated, as well as offshore and onshore wind levels within the UK. This is referred to as the average correlation error ($CE_{av}$) and shown formally by Equation \ref{eq:ce_av}:

\begin{equation}
\label{eq:ce_av}
CE_{av}=\frac{2}{\left|\left|P\right|\right|\cdot(\left|\left|P\right|\right|-1)}\cdot
\left(
\sum\limits_{p_i\in P}\sum\limits_{p_j\in P,j>i}
\left|
corr_{p_i,p_j}-\widetilde{corr}_{p_i,p_j}
\right|
\right),
\end{equation}

\noindent where $corr_{p1,p2}$ is the Pearson correlation coefficient between two time series $p_1,p_2\in P$, shown by Equation \ref{eq:corr}. Here, $V_{p_1,t}$ represents the value of time series $p_1$ at time step t, and $\hat{V}_{p_1,t}$ refers to the mean of the values of the time series $p_1$ at time $t$:

\begin{equation}
\label{eq:corr}
corr_{p1,p2}=\frac
{\sum\limits_{t\in T}\left(\left(V_{p1,t}-\overline{V}_{p1}\right)\cdot\left(V_{p2,t}-\overline{V}_{p2}\right)\right)}
{\sqrt{
		\sum\limits_{t\in T} \left(V_{p_1,t}-\overline{V}_{p1}\right)^2\cdot\sum\limits_{t\in T}\left(V_{p2,t}-\overline{V}_{p2}\right)^2
}}.
\end{equation}

\subsubsection{Integrating higher temporal granularity}

To integrate the additional temporal granularity of the model using the representative days, extra time-steps were taken per year. This did not require a significant change to the ElecSim code, simply stepping through each hour of a representative day to gauge renewable energy capacity factors and electricity demand. This meant that we had to take an additional number of time-steps when compared to the 20 LDC case. Specifically, as we modelled each time-step as an hour, and a day contains 24 hours, we ran 24 time-steps per representative day.

The difference between using the hourly representative days and 20 segments, is that peaks and troughs can be modelled more accurately with representative days. An LDC with 20 load segments can accurately model yearly intervals where each interval has an average of 5\%, or 18.25 days (365 days $\times 5\%=18.25$ days). It is true that peaks and troughs occur for less than 18.25 days, and therefore this granularity is lost with a 20 segment LDC. This amount of granularity may have been acceptable with predictable generation technologies, but this becomes significantly more challenging when trying to model intermittent renewable electricity.

It is true that we could have modelled more than 20 segments in an LDC to have a similar number time-steps as in a set of representative days. However, this would have given us an average of a year, in regular intervals. This would not provide a diverse set of days which are significantly different enough from each other. We did not want to capture solely the averages, but also the extremes that can occur in a year, which would have been lost through a simple LDC averaging approach.

The higher temporal granularity of the model enabled us to accurately model the hourly fluctuations in solar and wind which leads to more accurate expectations of the investment opportunities of these technologies ~\cite{Haydt2011,Ludig2011}.

\begin{figure}
	\centering
	\includegraphics[width=0.7\textwidth]{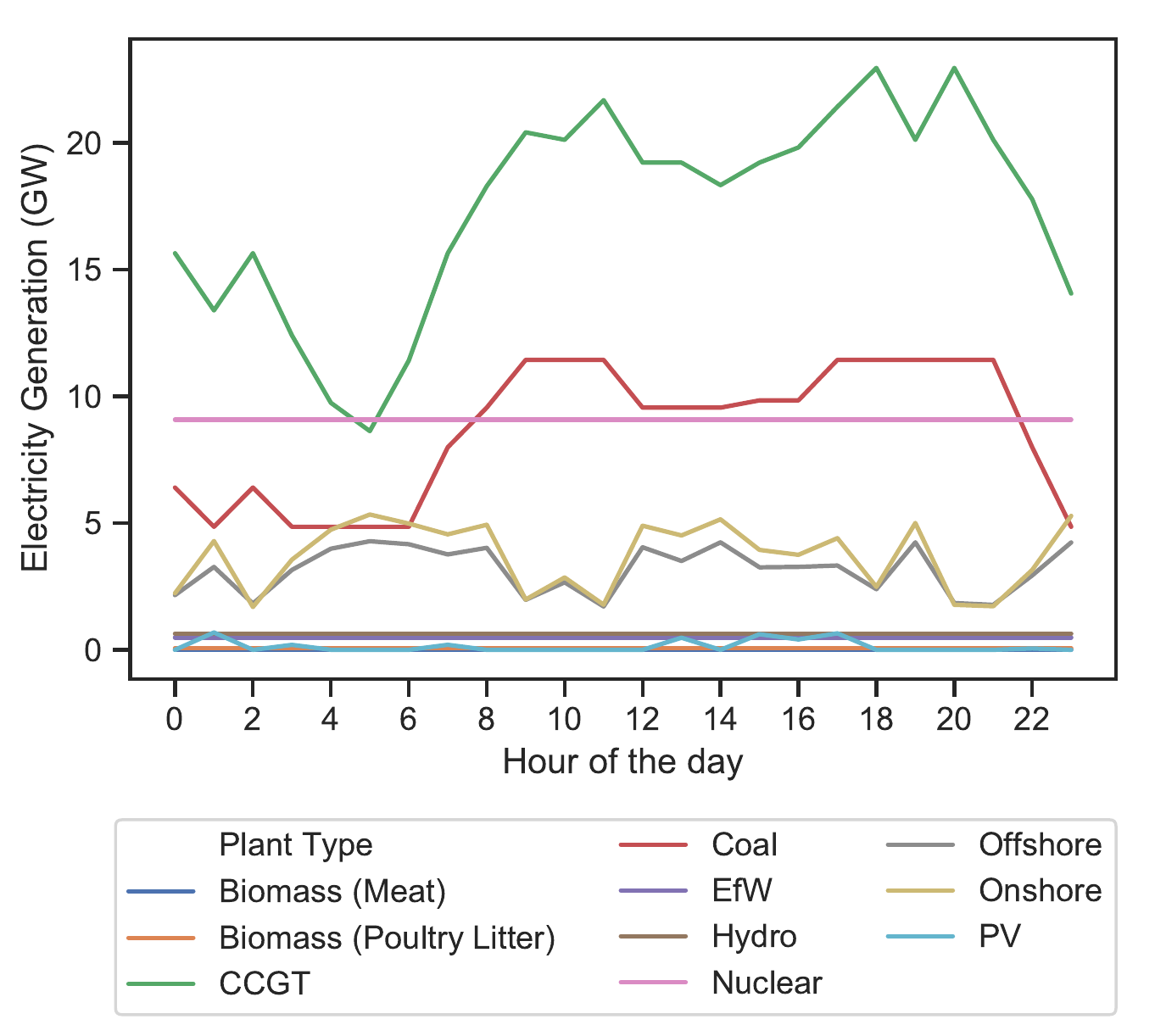}
	\caption{Visualisation of a single day of dispatched supply.}
	\label{fig:single_dispatched_day}
\end{figure}

GenCos make bids at the beginning of every time-step, and the Power Exchange matches demand with supply in merit-order dispatch using a uniform pricing market. An example of the electricity mix in a single representative day is shown in Figure \ref{fig:single_dispatched_day}. Figure \ref{fig:single_dispatched_day} displays the high utilisation of low marginal-cost generators such as nuclear, wind and photovoltaics. At hour 19, an increase in offshore wind leads to a direct decrease in CCGT. In contrast to this, a decrease in offshore and onshore between the hours of 8 and 12 lead to an increase in dispatch of coal and CCGT. One would expect this behaviour to prevent blackouts and meet demand at all times. This process has enabled us to more closely match fluctuations in IRES.

\section{Validation and performance}
\label{elecsim:sec:validation}

In this Section, we detail the validation approaches taken in our model. For this, we take two approaches. One is to compare the price duration curve of the actual vs our simulated price duration curve in 2018 using the 20 time-steps per year approach. The other is to use cross-validation between the years 2013 and 2018, using our representative days approach.

\subsection{Price Duration Curve Validation}

Validation of models is important to ascertain that the output is accurate. However, it should be noted that these long-term simulations are not predictions of the future, rather possible outcomes based upon certain assumptions. Jager posits that a certain outcome or development path, captured by empirical data, might have developed in a completely different direction due to chance. However, the processes that emerge from a model should be realistic and in keeping with expected behaviour \cite{Jager2006a}.

\begin{figure}
	\begin{center}
		\includegraphics[width=0.6\textwidth]{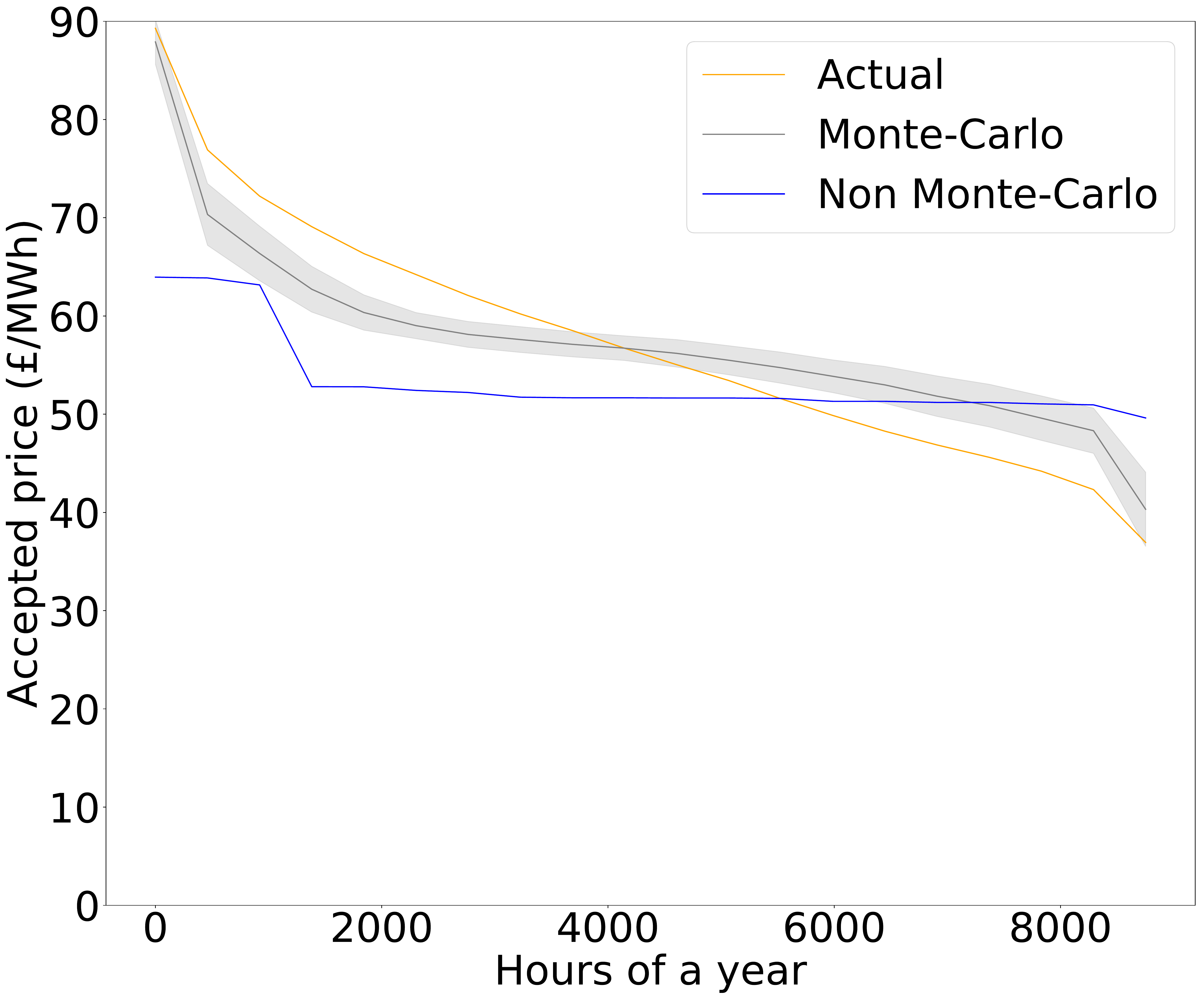}
		\caption{Price duration curve which compares real electricity prices to those paid in ElecSim (2018).}
		\label{fig:price_duration_curve}
	\end{center}
\end{figure}

We begin by comparing the price duration curve in the year 2018 for the case with 20 time-steps. Figure \ref{fig:price_duration_curve} shows the N2EX Day Ahead Auction Prices of the UK \cite{nordpool_2019}, the Monte-Carlo simulated electricity prices, and the non-Monte-Carlo electricity price throughout the year 2018. Fuel prices varying throughout a year, as does $V_C$ and WACC. WACC is sampled from a Gaussian distribution with a standard deviation of $\pm3$\%. $V_C$ is sampled from a uniform distribution between 30\% and 200\% of the mean $V_C$ price, whilst fuel price is sampled from the residuals of an ARIMA model fit on historical data. The N2EX Day Ahead Market is a day ahead market run by Nord Pool AS. Nord Pool AS runs the largest market for electrical energy in Europe, measured in volume traded and in market share \cite{nordpool_2019}.

We ran the initialisation of the model 40 times to capture the price variance. Outliers were removed as on a small number of occasions large jumps in prices at peak demand occurred which deviated from the mean. We did this, as although this does occur in real life, it occurs at a smaller fraction of the time than 5\% of the year (twenty time-steps per year). Therefore the results would be unreasonably skewed for the highest demand segment. 

Figure \ref{fig:price_duration_curve} demonstrates very little variance in the non-stochastic case. This is due to the fact that combined cycle gas turbines (CCGTs) set the spot price. These CCGTs have little variance between one another as they were calibrated using the same dataset. By adding stochasticity of fuel prices and operation and maintenance prices, a curve that more closely resembles the actual data occurs. The stochastic curve, however, does not perfectly fit the real data, which may be due to higher variance in fuel prices and historical differences in operation and maintenance costs between power plants. One method of improving this would be fitting the data used to parametrise to the curve. Figure \ref{fig:price_duration_curve} shows that up to 4,000 hours there is an overestimation. Following this, there is an underestimation. This is due to the fact that some power generators are more expensive than the average, and some are less expensive, by chance.

Table \ref{table:validation_metrics} shows performance metrics of the stochastic and non-stochastic runs versus the actual price duration curve. The stochastic implementation improves the \acrfull{mae} of the non-stochastic case by $52.5\%$.

\begin{table}[]
	\begin{tabular}{p{4cm}p{4cm}p{2.5cm}p{3cm}}
		\hline
		Metric & N2EX Day Ahead & ElecSim & Non Monte-Carlo \\ \hline
		Avg. Price (\textsterling/MWh) & 57.49 & 57.52 & 53.39 \\
		Std. dev (\textsterling/MWh) & - & 9.64 & - \\
		MAE (\textsterling/MWh) & - & 3.97 & 8.35 \\
		RMSE (\textsterling/MWh) & - & 4.41 & 10.2 \\ \hline
	\end{tabular}
	\caption{Validation performance metrics of comparing price simulated in ElecSim compared to true price values in 2018 for model with 20 time steps.}
	\label{table:validation_metrics}
\end{table}



\subsection{Validation of model with representative days}
\label{elecsim:ssec:representative_validation}

In this Section, we detail the approach taken in this work to validate our model using representative days as time-steps.

\begin{figure}
	\centering
	\includegraphics[width=1.0\textwidth]{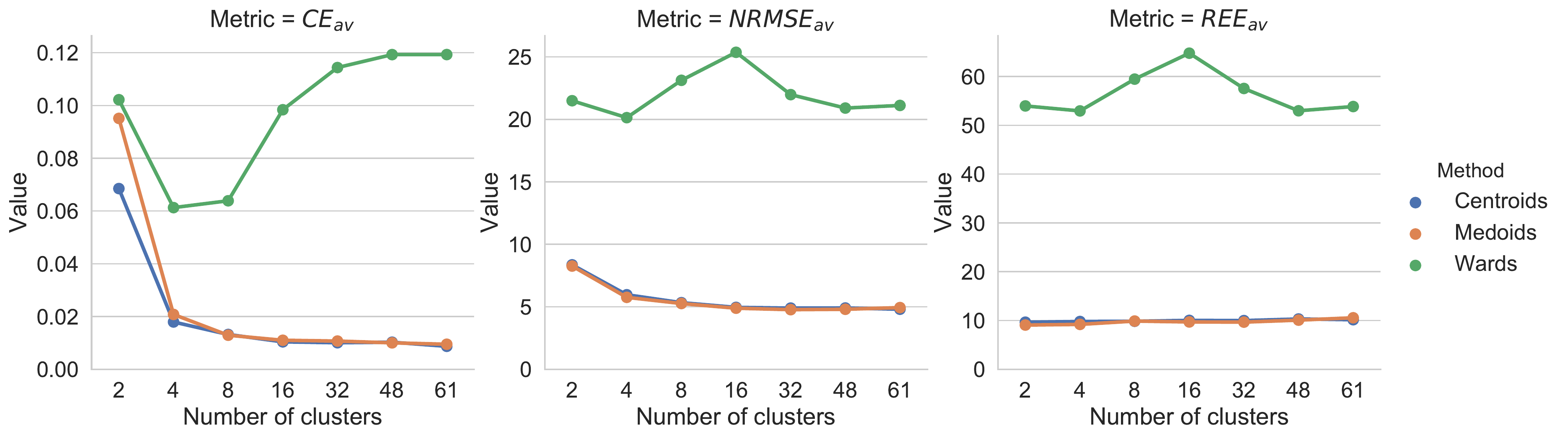}
	\caption{Number of clusters compared to error metrics for clustering methods. a) Displays the correlation metric, b) displays the normalised root mean squared error metric, c) displays the relative energy error metric.}
	\label{fig:error_metrics_vs_cluster_number}
\end{figure}

Figure \ref{fig:error_metrics_vs_cluster_number} displays the error metrics versus number of clusters, and therefore days. This is because from each cluster, the most central day of the cluster is chosen and used. Therefore, from eight clusters, eight representative days are chosen. Centroids are the days at the centre of the cluster, medoids are the average day of the cluster and Ward's is another clustering technique.

Both $CE_{av}$ and $NRMSE_{av}$ display similar behaviour for the $k$-means approach (centroids and medoids), namely the error improves significantly from a single cluster to eight clusters for both centroids and medoids. For the number of clusters greater than eight, there are diminishing returns. For $REE_{av}$, however, the error metric is best at a single cluster, and gets worse with the number of clusters. The Wards approach \cite{Salkind2013}, however, performs significantly worse for all metrics at every number of clusters.

We chose eight clusters, for centroids and medoids, as a compromise between accuracy of the three error metrics and compute time of the simulation. This is because eight was the largest number of clusters that gave us the lowest score for $CE_{av}$, $NRMSE_{av}$ and $REE_{av}$ without significantly increasing compute time. Whilst there was little significant difference between centroid and medoid, we chose to use the medoids due to the fact that the extreme high and low values would not be lost due to averaging \cite{Hilbers2019}. Therefore, a total of eight representative days were chosen.


To achieve this, we use a \acrfull{ga} to find the predicted price duration curves, which lead to the smallest error between our simulated electricity mix and the scenarios tested. The scenarios examined here are the observed electricity mix of the UK between 2013 and 2018 and the \acrfull{beis} reference scenario projected in 2018 till 2035. When projecting the BEIS reference scenario, we also optimise for nuclear subsidy and uncertainty in the price duration curves.




As mentioned in Section \ref{elecsim:sec:architecture}, GenCos make investments based upon the net present value. As shown in Equation \ref{architecture:eq:npv_eq}, an expectation of the net cash flow, $R_t$ is required. 

The net cash flow, $R_t$, is calculated by subtracting both the operational and capital costs from revenue over the expected lifetime of the prospective plant. The revenue gained by each prospective plant is the expected price they will gain per expected quantity of MWh sold over the expected lifetime of the plant. This is shown formally in Equation \ref{eq:revenue_total}:

\begin{equation}
\label{eq:revenue_total}
R_t = 
\sum\limits_{t=0}^T 
\sum\limits_{h=0}^H
\sum\limits_{m=0}^M \left(
m_{h,t}(PPDC_{h,t}
-
C_{var_{h,t}})\right)
- C_c,
\end{equation}

\noindent where $m_{h,t}$ is the expected quantity of megawatts sold in hour $h$ of year $t$. $PPDC_{h,t}$ is the \acrfull{ppdc} at year $t$ and hour $h$. $C_{var_{h,t}}$ is the variable cost of the power plant, which is dependent on expected megawatts of electricity produced, $C_c$ is the capital cost.

The predicted price duration curve ($PPDC_{h,t}$) is an expectation of future electricity prices over the lifetime of the plant. The $PPDC_{h,t}$ is a function of supply and demand. However, with renewable electricity generator costs falling  \cite{IRENA2014}, future prices are uncertain and largely dependent upon long-term scenarios of electricity generator costs, fuel prices, carbon taxes and investment decisions \cite{IRENA2014}. Due to the uncertainty of future electricity prices over the horizon of the lifetime of a power plant, we have set future electricity prices as an exogenous variable that can be set by the user in ElecSim.

To gain an understanding of expected electricity prices that lead to particular scenarios, we use a genetic algorithm optimisation approach. This enables us to understand the range of future electricity prices that lead to certain scenarios developing. In addition, it allows us to understand whether the parameters required for certain scenarios to develop are realistic. This enables us to check the assumptions of our model and the likelihood of scenarios. Further, using these optimised parameters, we are better able to further explore `\textit{what-if}' scenarios.

To verify the accuracy of the underlying dynamics of ElecSim, the model was initialised with data available in 2013 and allowed to develop until 2018. We used a genetic algorithm to find the optimum price duration curve predicted ($PPDC$) by the GenCos ten years ahead of the year of the simulation. This $PPDC$ was used to model expected rate of return of prospective generation types, as shown in Equations \ref{architecture:eq:npv_eq} and \ref{eq:revenue_total}. 

The genetic algorithm's objective was to reduce the error of simulated and observed electricity mix in the year 2018 by finding a suitable $PPDC$ used by each of the GenCos for investment evaluation.

\subsubsection{Scenario}

For this experiment, we initialised ElecSim with parameters known in 2013 for the UK. ElecSim was initialised with every power plant and respective GenCo that was in operation in 2013 using the BEIS DUKES dataset \cite{dukes_511}. The funds available to each of the GenCos was taken from publicly released official company accounts at the end of 2012 \cite{companies_house}.

To ensure that the development of the electricity market from 2013 to 2018 was representative of the actual scenario between these years, we set the exogenous variables, such as carbon and fuel prices, to those that were observed during this time period. In other words, the scenario modelled equated to the observed scenario. 

The data for the observed EU Emission Trading Scheme (ETS) price between 2013 and 2018 was taken from Ember Climate\cite{eu-ets}. Fuel prices for each of the fuels were taken from the department for business, energy and industrial strategy \cite{beis_fuel_price}. The electricity load data was modelled using data from Elexon portal ad Sheffield University \cite{gbnationalgridstatus2019}, offshore, and onshore wind and solar irradiance data was taken from Pfenninger \textit{et al.} \cite{Pfenninger2016}. There were three known significant coal plant retirements in 2016. These were removed from the simulation at the beginning of 2016.

\subsubsection{Optimisation problem}
\label{ssec:optimisation-problem}

The price duration curve was modelled linearly in the form $y=mx+c$, where $y$ is the cost of electricity, $m$ is the gradient and refers to the relative increase in price with respect to demand, $x$ is the demand of the price duration curve, and $c$ is the intercept, or the price  of electricity as demand approaches zero. A linear price duration curve was chosen due to its simplicity, small amount of parameters required for optimisation and also the similarity to the actual price duration curve.

Equation \ref{eq:problem_formulation} details the optimisation problem formally:

\begin{equation}
\label{eq:problem_formulation}
\min_{m,c} \sum\limits_{o\in O}\left(
\frac{\left|A_o-f_o(m,c)\right|}
{\left|\left|O\right|\right|}
\right),
\end{equation}

\noindent where $o\in O$ refers to the average percentage electricity mix during 2018 for wind (both offshore and onshore generation), nuclear, solar, CCGT, and coal, where $O$ refers to the set of these values. $A_o$ refers to observed electricity mix percentage for the respective generation type in 2018. $f_o(m,c)$ refers to the simulated electricity mix percentage for the respective generation type, also in 2018. The input parameters to the simulation are $m$ and $c$ from the linear $PPDC$, previously discussed, ie. $y=mx+c$. $\left|\left|O\right|\right|$ refers to the cardinality of the set.

\subsection{Long-Term Scenario Analysis}
\label{sssec:scen-analysis}

In addition to verifying the ability for ElecSim to mimic observed investment behaviour over five years, we compared ElecSim's long-term behaviour to that of the UK Government's Department for Business, Energy and Industrial Strategy (BEIS) \cite{DBEIS2019}. This scenario shows the projections of generation by technology for all power producers from 2018 to 2035 for the BEIS reference scenario. This is the same scenario as discussed in the next Section.

\subsubsection{Scenario}
\label{sssec:scenario-details}

We initialised the model to 2018 based on our previous work \cite{Kell}. The scenario for the development of fuel prices and carbon prices were matched to that of the BEIS reference scenario \cite{DBEIS2019}.

\subsubsection{Optimisation problem} The optimisation approach taken was a similar process to that discussed in Subsection \ref{ssec:optimisation-problem}, namely using a genetic algorithm to find the optimum expected price duration curve. However, instead of using a single expected price duration curve for each of the agents for the entire simulation, we used a different expected price duration curve for each year, leading to 17 different curves. This enabled us to model the non-static dynamics of the electricity market over this extended time period. The agents are dynamic, however, not heterogenous in this regard. This is due to the exponentially increasing computational time of calculating predicted price duration curves for each agent.

In addition to optimising for multiple expected price duration curves, we optimised for a nuclear subsidy, $S_n$. Further, we optimised for the uncertainty in the expected price parameters $m$ and $c$, named $\sigma_m$ and $\sigma_c$ respectively, where $\sigma$ is the standard deviation in a normal distribution. $m$ and $c$ are the parameters for the predicted price duration curve, as previously defined, of the form $y=mx+c$. Effectively, $\sigma_m$ and $\sigma_c$ represent the distribution around the price duration curve, that is sampled in a monte-carlo manner.  

This enabled us to model the different expectations of future price curves between the independent GenCos. The addition of a nuclear subsidy as a parameter is due to the likely requirement for government to provide subsidies for new nuclear \cite{Suna2016}.

A modification was made to the reward algorithm for the long-term scenario case. Rather than using the discrepancy between observed and simulated electricity mix in the final year (2018) as the reward, a summation of the error metric for each simulated year was used. This is detailed formally in Equation \ref{eq:long-term-reward}:

\begin{equation}
\label{eq:long-term-reward}
\min_{m\in M,m\in C} 
\sum\limits_{y\in Y}
\sum\limits_{o\in O}\left(
\frac{\left|A_{y_o}-f_{y_o}(m_y,c_y)\right|}
{\left|\left|O\right|\right|}
\right),
\end{equation}

\noindent where $M$ and $C$ are the sets of the 17 parameters of $m_y$ and $c_y$ respectively for each year, $y$. $y\in Y$ refers to each year between 2018 and 2035. $m_y$ and $c_y$ refer to the parameters for the predicted price duration curve, of the form $y=mx+c$ for the year $y$. $A_{y_o}$ refers to the actual electricity mix percentage for the year $y$ and generation type $o$. Finally, $f_{y_o}(m_y,c_y)$ refers to the simulated electricity mix percentage with the input parameters to the simulation of $m$ and $c$ for the year $y$.

\subsection{Genetic Algorithms}
\label{elecsim:ssec:geneticalgorithm}

Genetic Algorithms (GAs) are a type of evolutionary algorithm which can be used for optimisation. We chose the genetic algorithm for this application due to its ability to find good solutions with a limited number of simulation runs, the ability for parallel computation and its ability to find global optima. These characteristics are useful for our application, as a single simulation can take up to 36 hours. 

In this Section, we detail the genetic algorithm used in this work. Initially, a population $P_{0}$ is generated for generation 0. This population of individuals is used for the parameters to the simulation. The output of the simulations for each of the individuals are then evaluated. A subset of these individuals $C_{t+1} \subset P_{t}$ are chosen for mating. This subset is selected proportional to their fitness, with `fitter' individuals having a higher chance of reproducing to create the offspring group $C'_{t+1}$. $C'_{t+1}$ have characteristics dependent on the genetic operators: crossover and mutation. The genetic operators are an implementation decision \cite{FogelDavidB2009}. 

Once the new population has been created, the new population $P_{t+1}$ is created by merging individuals from $C'_{t+1}$ and $P_{t}$. See Algorithm \ref{genetic-algorithm} for detailed pseudocode.

We used the DEAP evolutionary computation framework to create our genetic algorithm \cite{Gagn2012}. This framework gave us sufficient flexibility when designing our genetic algorithm. Specifically, it enabled us to persist the data of each generation after every iteration to allow us to verify and analyse our results in real-time.

\begin{algorithm}[t]
	\begin{algorithmic}[1]
		\State $t=0$
		\State initialize $P_{t}$
		\State evaluate structures in $P_{t}$
		\While {termination condition not satisfied}
		\State $t=t+1$
		\State select reproduction $C_{t}$ from $P_{t-1}$
		\State recombine and mutate structures in $C_{t}$
		
		forming $C'_{t}$
		\State evaluate structures in $C'_{t}$
		\State select each individual for $P_{t}$ from $C'_{t}$ 
		
		or $P_{t-1}$
		\EndWhile
		\caption{Genetic algorithm \cite{FogelDavidB2009}}
		\label{genetic-algorithm}
	\end{algorithmic}
\end{algorithm}

\subsubsection{Parameters for Validation with Observed Data}
\label{ssec:ga_params_valid}

The parameters chosen for the problem explained in Section \ref{sssec:scen-analysis} was a population size of $120$, a crossover probability of $50\%$, a mutation probability of $20\%$ and the parameters, $m$ and $c$, as per Equation \ref{eq:problem_formulation}, were given the bounds of $[0.0, 0.004]$ and $[-30, 100]$ respectively. 

The bounds for $m$ and $c$ were calculated to ensure a positive price duration curve, with a maximum price of \textsterling300 for 50,000MW. The population size was chosen to ensure a wide range of solutions could be explored, whilst limiting compute time to ${\sim}$1 day per generation to allow for sufficient verification of the results. The crossover and mutation probabilities were chosen due to suggestions from the DEAP evolutionary computation framework \cite{Gagn2012}.

\subsubsection{Parameters for Long-Term Scenario Analysis}

The parameters chosen for the genetic algorithm for the problem discussed in Section \ref{sssec:scen-analysis} are displayed here. The population size was $127$, a crossover probability of $50\%$, a mutation probability of $20\%$. The parameters $m_y$, $c_y$ were given the bounds $[0.0, 0.003]$ and $[-30, 50]$ respectively, whilst $\sigma_m$ and $\sigma_c$ were both given the bounds of $[0, 0.001]$. Effectively, we did not allow these parameters to be larger or smaller than these bounds.

The population size was slightly increased, and the bounds reduced when compared to the parameters for Section \ref{ssec:ga_params_valid}. This was to increase the likelihood of convergence to a global optimum, which was more challenging to achieve due to the significantly higher number of parameters.

\subsection{Results}
\label{sec:results}

Here we present the results of the problem formulation of Section \ref{sssec:scen-analysis}. Specifically, we compare the ability of our model to that of BEIS in the context of a historical validation between 2013 and 2018 of the UK electricity market. We also compare our ability to generate scenarios up to 2035 with that of BEIS. 

\subsubsection{Validation with Observed Data}

\begin{figure}
	\centering
	\includegraphics[width=0.7\columnwidth]{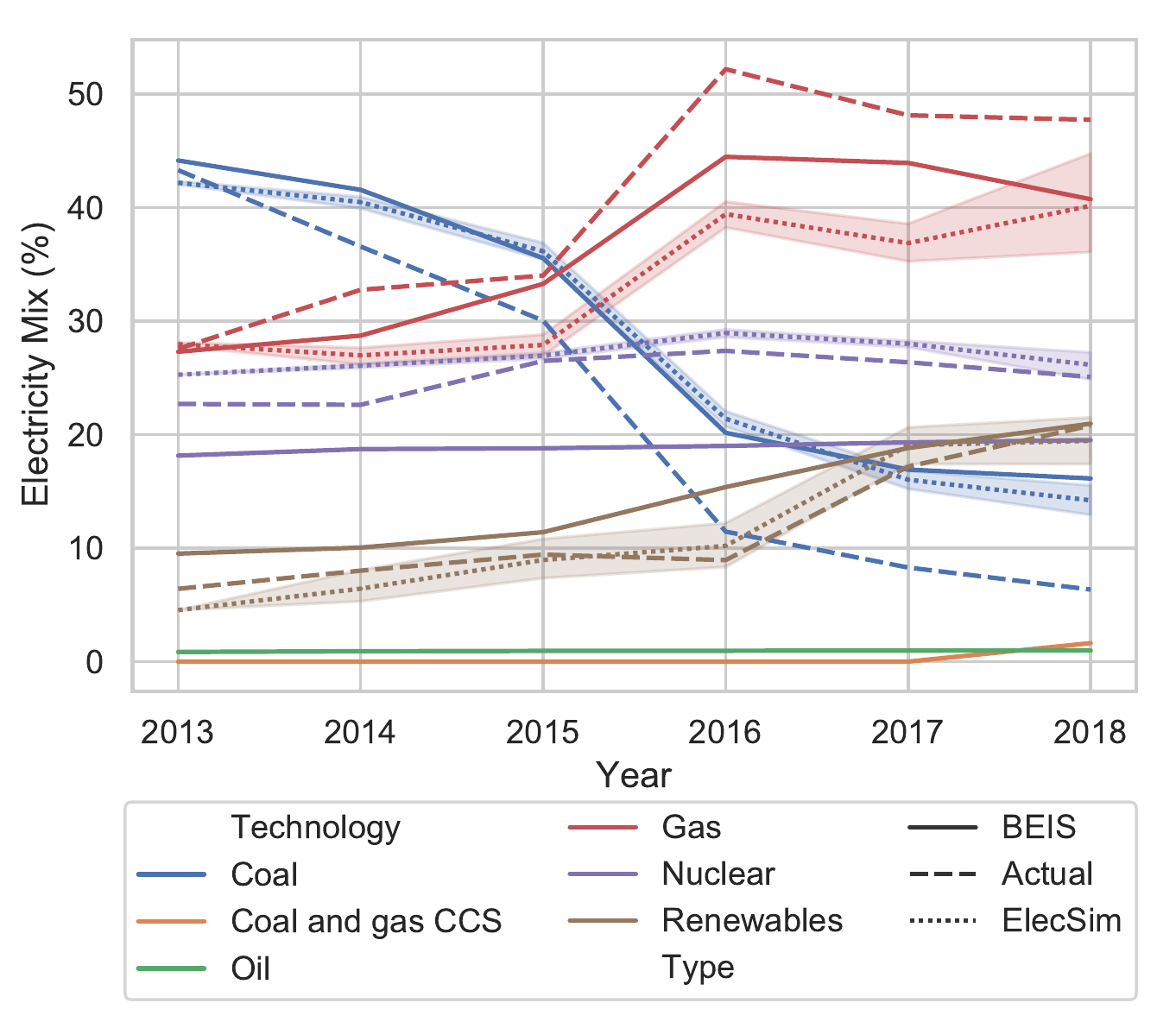}
	\caption{Comparison of actual electricity mix vs. ElecSim vs. BEIS projections and taking three coal power plants out of service. The legend here is split into two: technologies and scenario type, with the coloured lines denoting technologies, and dotted/straight lines denoting scenario.}
	\label{elecsim:fig:beis_elecsim_historic_comparison}
\end{figure}

Figure \ref{elecsim:fig:beis_elecsim_historic_comparison} displays the output of ElecSim under the validation scenario, BEIS' projections and the observed electricity mix between 2013 and 2018, as explained in Section \ref{elecsim:ssec:representative_validation}. We calibrated our data with the BEIS projection to generate these results.

The observed electricity mix changed significantly between 2013 and 2018. A continuous decrease of electricity production from coal throughout this period was observed. 2015 and 2016 saw a marked decrease of coal, which can be explained by the retirement of 3 major coal power plants. The decrease in coal between 2013 and 2016 was largely replaced by an increase in gas. After 2016, renewables play an increasingly large role in the electricity mix and displace gas.

Both ElecSim and BEIS were able to model the fundamental dynamics of this shift from coal to gas as well as the increase in renewables. Both models, however, underestimated the magnitude of the shift from coal to gas. This could be due to unmodelled behaviours such as consumer sentiment towards highly polluting coal plants, a prediction from industry that gas would become more economically attractive in the future or a reaction to The Energy Act 2013 which aimed to close a number of coal power stations over the following two decades \cite{uk_energy_act}.


ElecSim was able to closely model the increase in renewables throughout the period in question, specifically predicting a dramatic increase in 2017. This is in contrast to BEIS, who predicted that an increase in renewable energy would begin in 2016. However, both models were able to accurately predict the proportion of renewables in 2018. 

ElecSim was able to better model the observed fluctuation in nuclear power in 2016. BEIS, on the other hand, projected a more consistent nuclear energy output. This small increase in nuclear power is likely due to the decrease in coal during that year. BEIS consistently underestimated the share of nuclear power.

We display the error metrics to evaluate our models 5-year projections in Table \ref{table:metrics}. Where MAE is mean absolute squared error, MASE is \acrfull{mase} and \acrfull{rmse} is the root mean squared error.

We are able to improve the projections for all generation types when compared to the naive forecasting approach using ElecSim, as shown by the MASE. Where the naive approach is simply predicting the next time-step by using the last known time-step, in this case, the last known time-step is the electricity mix percentage for each generation type in 2013. 

\begin{table}[htb]
	\centering
	\begin{filecontents*}{error_metrics.csv}
Technology,MAE,MASE,RMSE
CCGT,9.007,0.701,10.805
Coal,8.739,0.423,10.167
Nuclear,1.69,0.694,2.002
Solar,0.624,0.419,1.019
Wind,1.406,0.361,1.498
\end{filecontents*}
\csvautobooktabular{error_metrics.csv}
	\caption{Error metrics for time series forecast from 2013 to 2018 for validation.}
	\label{table:metrics}
\end{table}

Table \ref{table:metrics_beis} displays the same set of error metrics for BEIS' projections. The technologies displayed here are slightly different to those displayed in Table \ref{table:metrics} due to the presentation of the data by BEIS. As shown in Figure \ref{elecsim:fig:beis_elecsim_historic_comparison}, we are able to predict the rise in renewables and nuclear better than that of BEIS, however, perform slightly worse for coal and gas.

\begin{table}[]
	\centering
	\begin{tabular}{@{}llll@{}}
		\toprule
		Technology       & MAE   & MASE  & RMSE  \\ \midrule
		Coal             & 6.413 & 0.311 & 7.096 \\
		Gas              & 3.99  & 0.311 & 4.884 \\
		Nuclear          & 6.192 & 2.545 & 6.406 \\
		Renewables       & 2.558 & 0.475 & 3.209 \\ \bottomrule
	\end{tabular}
	\caption{Error metrics for BEIS' scenario from 2013 to 2018.}
	\label{table:metrics_beis}
\end{table}

\begin{figure}
	\centering
	\includegraphics[width=0.65\textwidth]{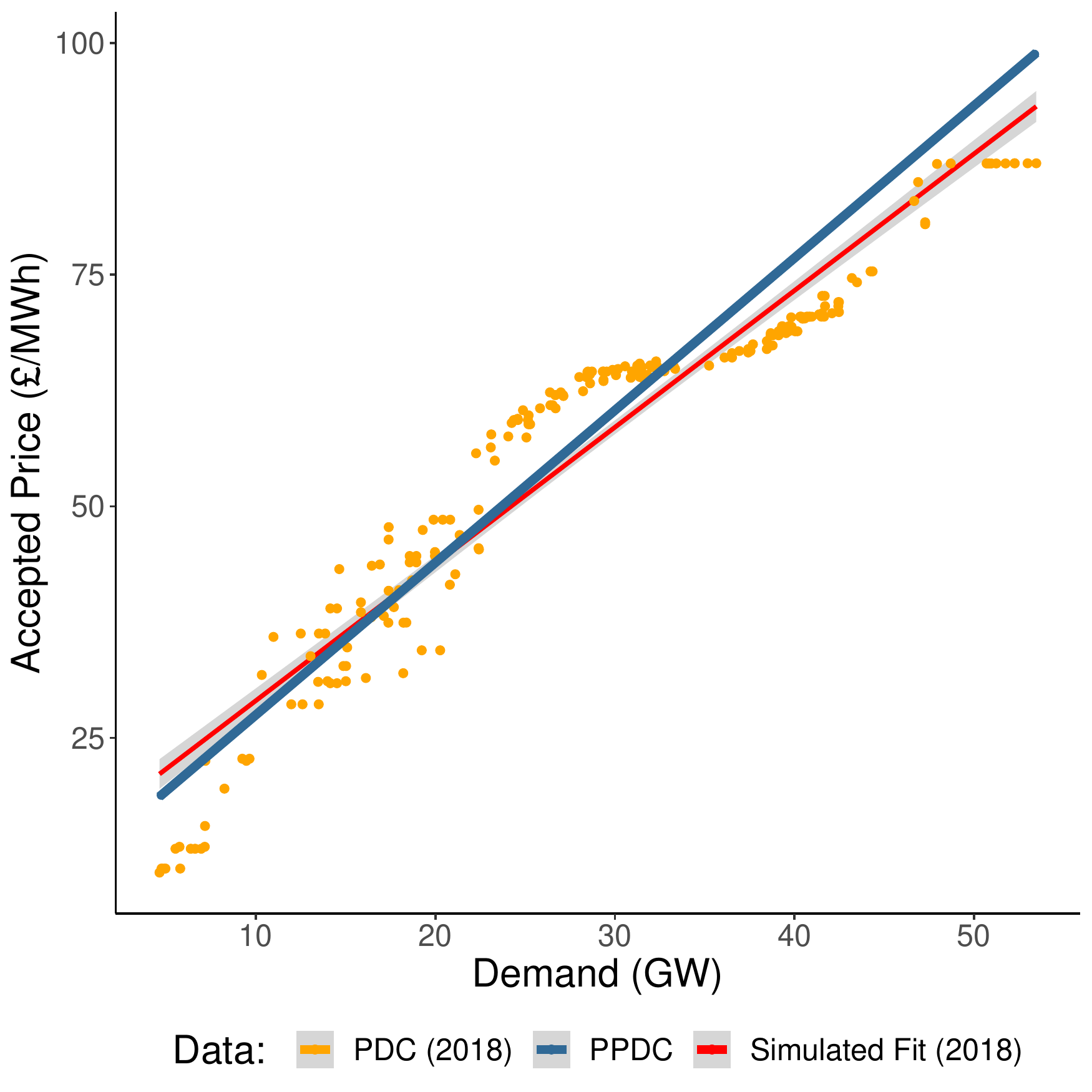}
	\caption{Predicted price duration curve for investment for most accurate run against simulated run in 2018.}
	\label{fig:best_price_curve}
\end{figure}

Figure \ref{fig:best_price_curve} displays the optimal predicted price duration curve ($PPDC$) found by the genetic algorithm. This price curve was used by the GenCos to achieve the results shown in Figure \ref{fig:forward_scenario_beis_elecsim}.

The orange points show the simulated price duration curve for the first year of the simulation (2018). The red line (Simulated Fit 2018) is a linear regression that approximates the simulated price duration curve (PDC 2018). The blue line shows the price duration curve predicted ($PPDC$) by the GenCos to be representative of the expected prices over the lifetime of the plant.


The optimal predicted price duration curve ($PPDC$) closely matches the simulated fit in 2018, shown by Figure \ref{fig:best_price_curve}. However, the $PPDC$ has a slightly higher peak price and lower baseload price. This could be due to the fact that there is a predicted increase in the number of renewables with a low SRMC. However, due to the intermittency of renewables such as solar and wind, higher peak prices are required to generate in times of low wind and solar irradiance at the earth's surface.

\begin{figure}
	\centering
	\includegraphics[width=0.65\textwidth]{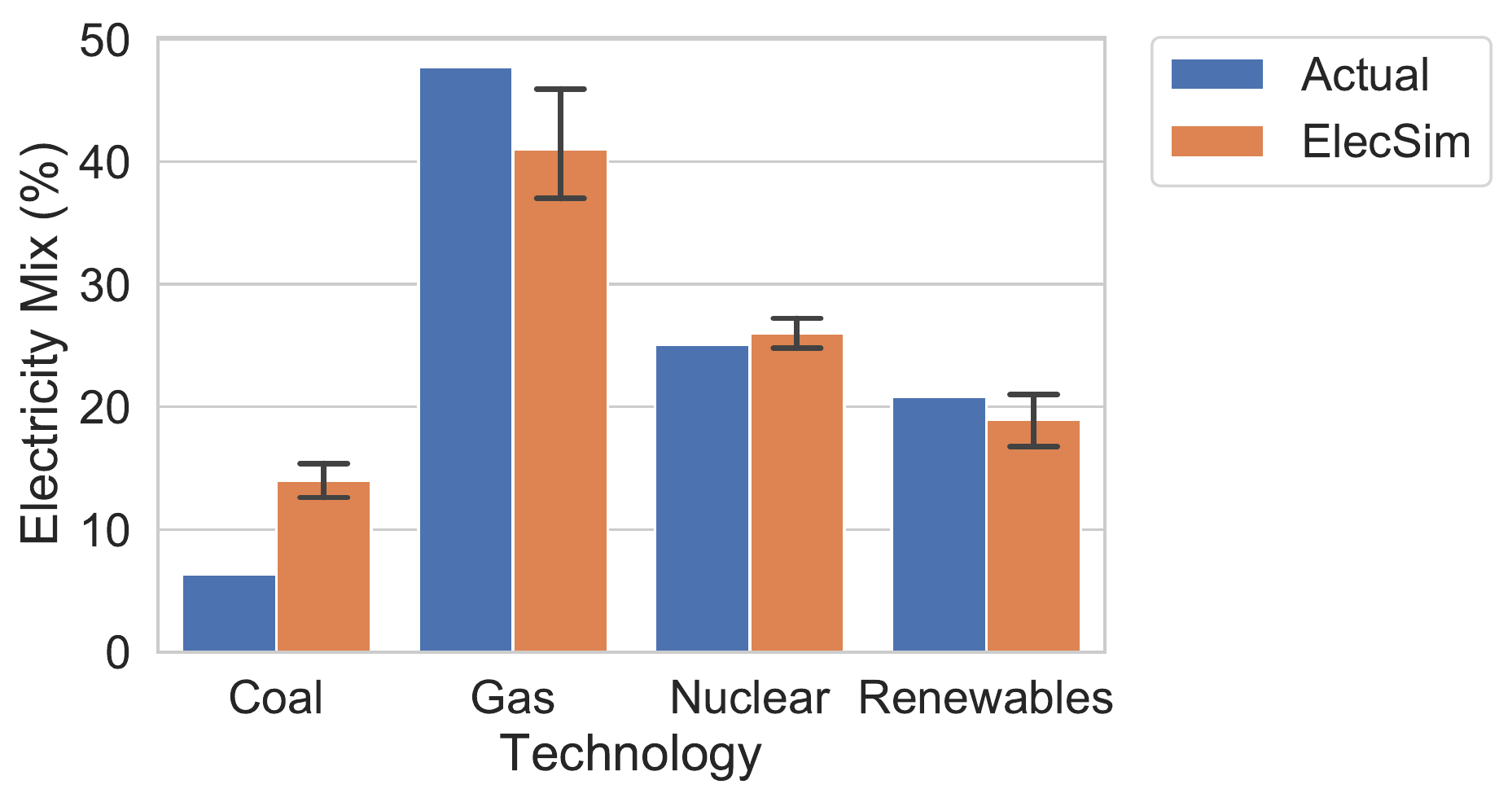}
	\caption{Electricity generation mix simulated by ElecSim from 2013 to 2018 compared to observed electricity mix in 2018.}
	\label{fig:uk_validated_results_2018}
\end{figure}

To generate Figure \ref{fig:uk_validated_results_2018}, we ran 40 scenarios with the $PPDC$ to observe the final, simulated electricity mix. The error bars are computed based on a Normal distribution 95\% confidence interval.

ElecSim was able to model the increase in renewables and stability of nuclear energy in this time. ElecSim was also able to model the transition from coal to gas, however, underestimated the magnitude of the transition. This was similar to the projections BEIS made in 2013 as previously discussed.

\subsubsection{Long-Term Scenario Analysis}

In this Section, we discuss the results of the analysis of the BEIS reference scenario explained in Section \ref{sssec:scen-analysis}. Specifically, we created a scenario that mimicked that of BEIS in ElecSim and optimised a number of parameters using a genetic algorithm to match this scenario. Through this, we are able to gain confidence in the underlying dynamics of ElecSim to simulate long-term behaviours. Further, this enables us to verify the likelihood of the scenario by analysing whether the parameters required to make such a scenario are realistic.

\begin{figure}
	\centering
	\includegraphics[width=0.60\textwidth]{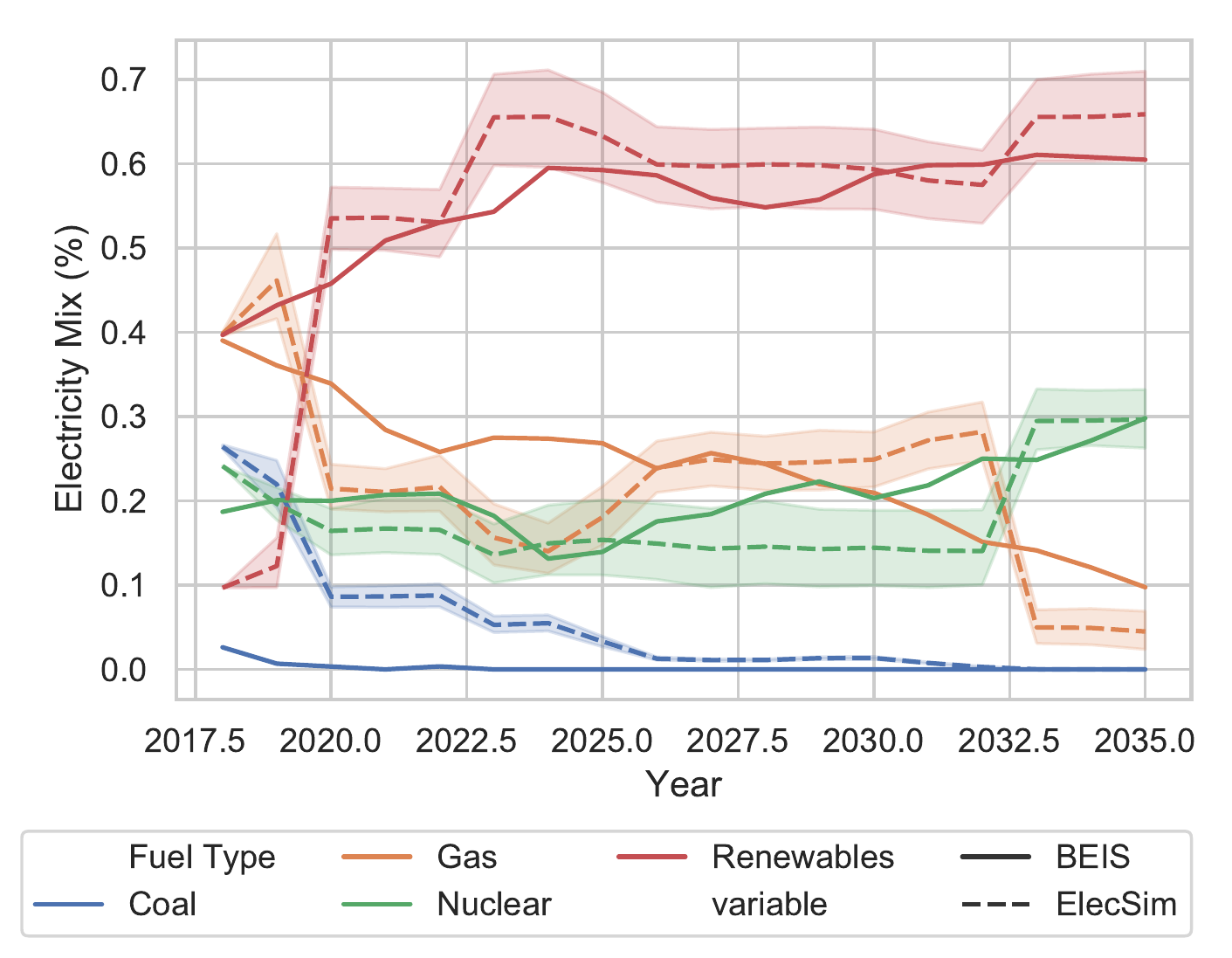}
	\caption{Comparison of ElecSim and BEIS' reference scenario from 2018 to 2035.}
	\label{fig:forward_scenario_beis_elecsim}
\end{figure}

Figure \ref{fig:forward_scenario_beis_elecsim} displays the electricity mix projected by both ElecSim and BEIS. To generate this image, we ran 60 scenarios under the optimal collection of predicted price duration curves, nuclear subsidy and uncertainty in predicted price duration curves. The optimal parameters were chosen by choosing the parameter set with the lowest mean error per electricity generation type and per year throughout the simulation, as shown by Equation \ref{eq:long-term-reward}.

\begin{figure}
	\centering
	\includegraphics[width=0.7\textwidth, keepaspectratio]{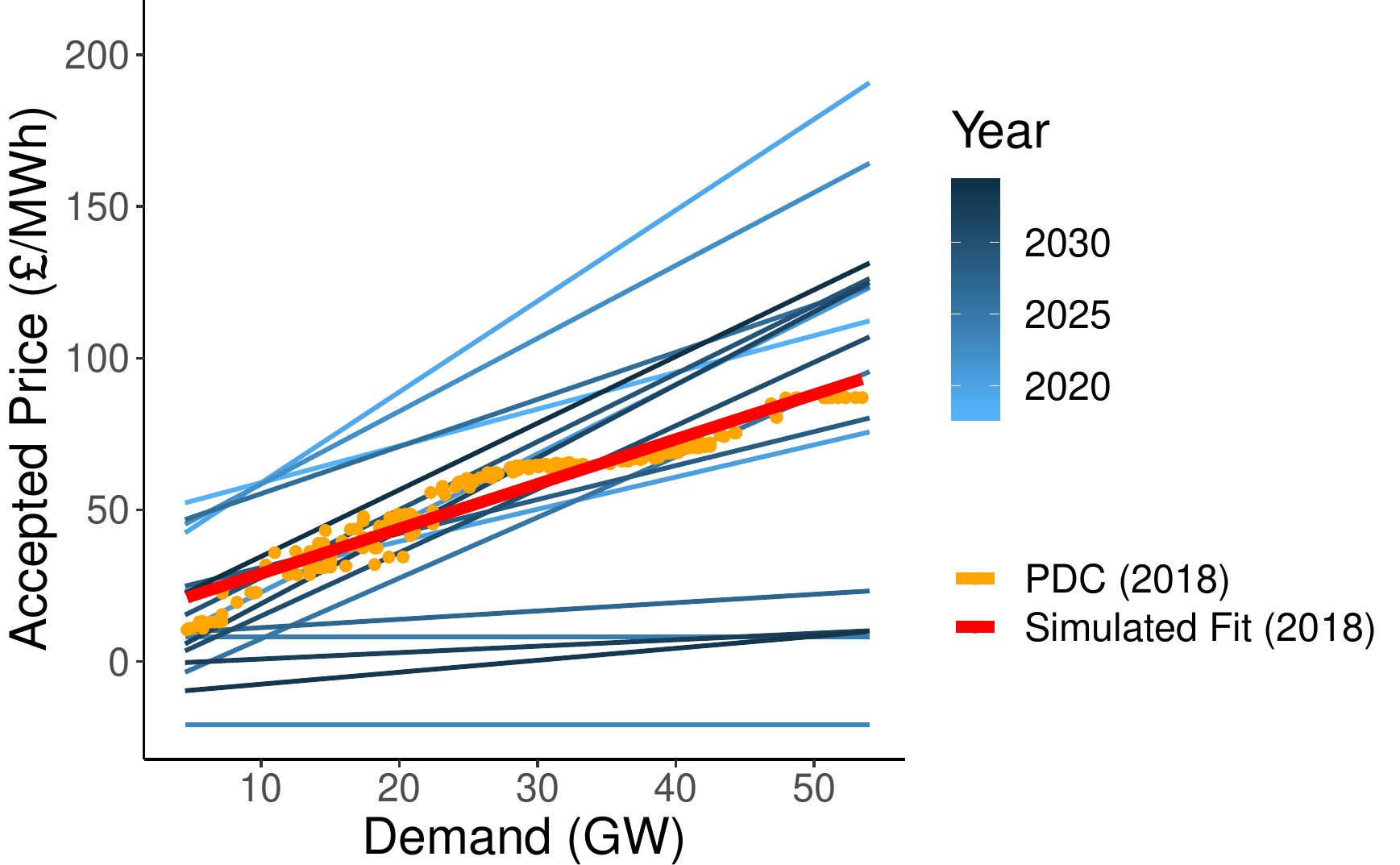}
	\caption{Comparison between optimal price duration curves and simulated price duration curve in 2018 for optimisation of ElecSim and BEIS' reference scenario from 2018 to 2035.}
	\label{fig:forward_scenario_best_pdcs}
\end{figure}

Figure \ref{fig:forward_scenario_best_pdcs} displays the optimal predicted price duration curves ($PPDC$s) per year of the simulation, shown in blue. These are compared to the price duration curve simulated in 2018, as per Figure \ref{fig:best_price_curve}. The optimal nuclear subsidy, $S_n$, was found to be ${\sim}$\textsterling $120$/MWh, the optimal $\sigma_m$ and $\sigma_c$ were found to be $0$ and ${\sim}0.0006$ respectively. In this context, optimal means the values which produced the required electricity mix scenarios.

The BEIS scenario demonstrates a progressive increase in nuclear energy from 2025 to 2035, a consistent decrease in electricity produced by natural gas, an increase in renewables and decrease to almost 0\% by 2026 of coal. ElecSim is largely able to mimic the scenario by BEIS. A large increase in renewables is projected, followed by a decrease in natural gas. A significant difference, however, is the step-change in nuclear power in 2033. This led to an almost equal reduction in natural gas during the same year. In contrast, BEIS project a continuously increasing share of nuclear. We argue that the ElecSim projection of nuclear power is more realistic than that of BEIS due to the instantaneous nature of large nuclear power plants coming on-line. This is largely due to ElecSim's discrete nature of power plants.

Figure \ref{fig:forward_scenario_best_pdcs} exhibits the price curves required to generate the scenario shown in Figure \ref{fig:forward_scenario_beis_elecsim}. The majority of the price curves are similar to the simulated price duration curve of 2018 (Simulated Fit 2018). However, there are some price curves which are significantly higher and significantly lower than the predicted price curve of 2018. These cycles in predicted price duration curves may be explained by investment cycles typically exhibited in electricity markets \cite{Gross2007}. 

In this context, investment cycles reflect a boom and bust cycle over long timescales. When electricity supply becomes tight relative to demand, prices rise to create an incentive to invest in new capacity. Price behaviour in competitive markets can lead to periods of several years of low prices (close to short-run marginal cost) \cite{white2005concentrated}. 

As plants retire or demand increases, the market becomes tighter until average prices increase to a level above the threshold for investment in new power generators. At this point, investors may race to bring new plants on-line to make the most out of the higher prices. Once adequate investments have been made, the market returns to a period of low prices and low investment until the next price spike \cite{Gross2007}.

The nuclear subsidy, $S_n$, of ${\sim}$\textsterling $120$/MWh in 2018 prices is high compared to similar subsidies, but this may reflect the difficulty of nuclear competing with renewable technology with a short-run marginal cost that tends to \textsterling $0$. However, subsidising nuclear may be an important option to gain reliable, low-carbon, base load, irrespective of the price. The low values of $\sigma_m$ and $\sigma_c$ demonstrates that the expectation of prices does not necessarily have to differ significantly between GenCos. This may be due to the fact that GenCos have access to the same market information. However, it may be true that agent's portfolios and cash reserves affect their decisions.

\clearpage
\section{Scenario Testing}
\label{elecsim:sec:scenarios}

In this section we display scenario runs of ElecSim using both 20 time-steps per year and using representative days.

\subsection{Scenarios for 20 time-steps}

In this Section, we vary the carbon tax and grow or reduce total electricity demand. This enables us to observe the effects of a carbon tax on investment. In this work, we have presented scenarios where electricity demand decreases 1\% per year, due to the recent trend in the UK.

\begin{figure}
	\centering
	\begin{subfigure}{.7\linewidth}
		\includegraphics[width=1\linewidth]{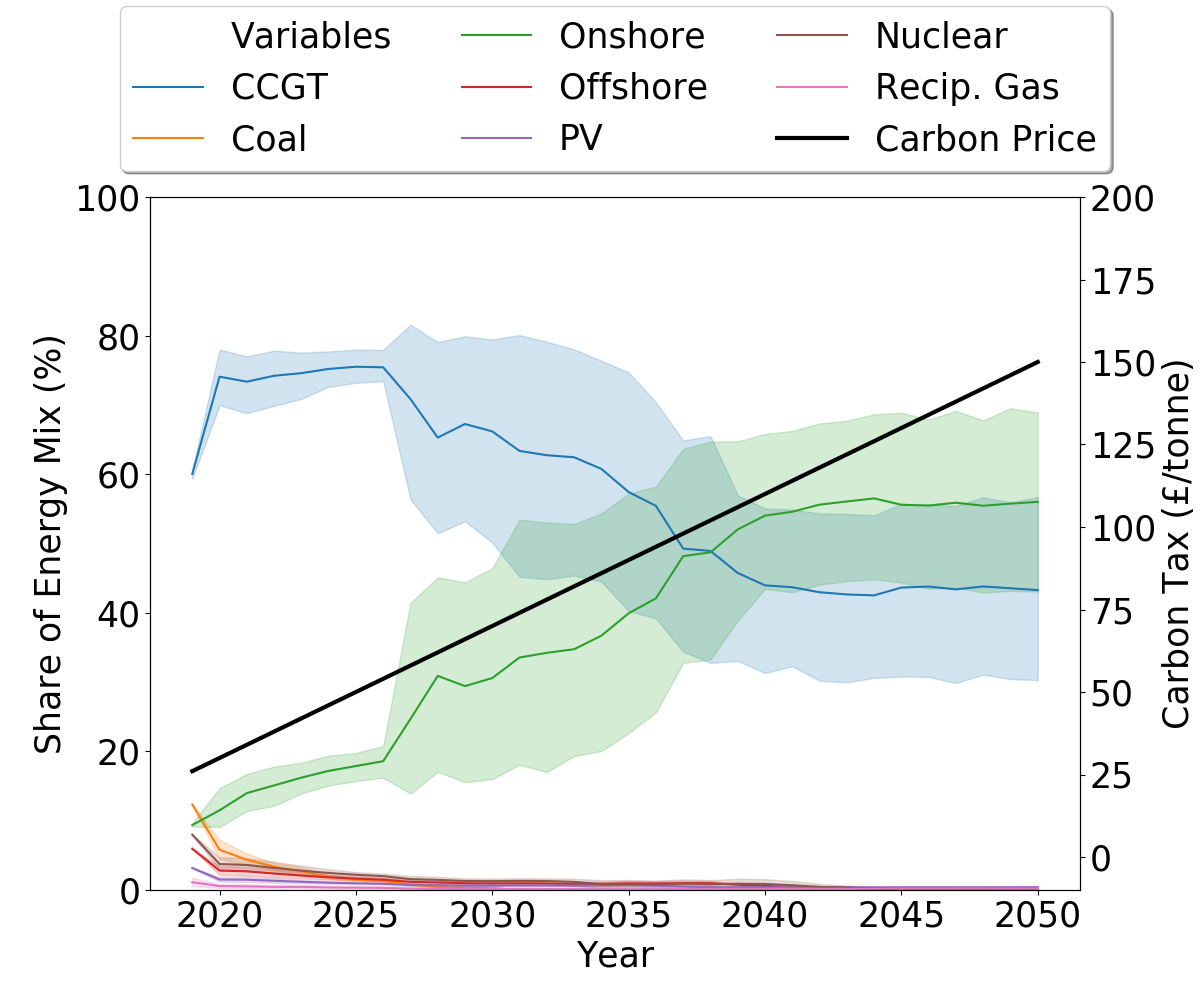}
		\caption{\textsterling26 to \textsterling150 linearly increasing carbon tax.}
		\label{fig:demand99carbon18}
	\end{subfigure}
	\begin{subfigure}{.7\linewidth}
		\includegraphics[width=1\linewidth]{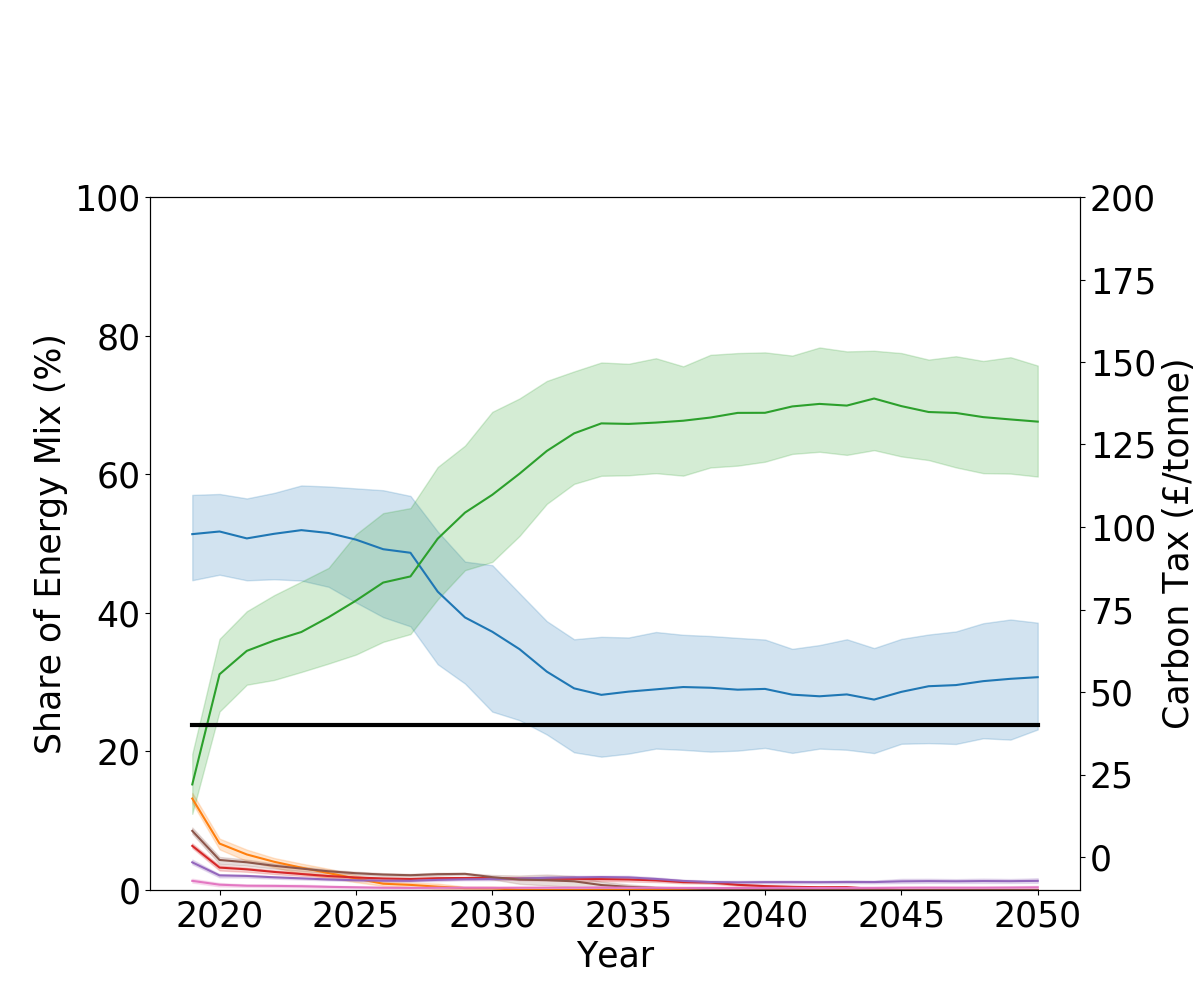}
		\caption{{\textsterling40 carbon tax}}
		\label{fig:demand99carbon40}
	\end{subfigure}
	\caption{Scenarios with varying carbon taxes and decreasing demand (-1\%/year) for 20 time steps per year.}
\end{figure}

For the first scenario run displayed, we have approximated the predictions by the UK Government, where carbon tax increases linearly from \textsterling18 to \textsterling200 by 2050 \cite{Department2016}. Figure \ref{fig:demand99carbon18} demonstrates a significant increase in gas turbines in the first few years, followed by a decrease, with onshore wind increasing.

Figure \ref{fig:demand99carbon40} displays a run with a \textsterling40 carbon tax. This run demonstrates a higher share of onshore wind than in the previous scenario. 

We experimented with the following levels of carbon tax: \textsterling10 (\$13), \textsterling20 (\$26) and \textsterling70 (\$90) with demand decreasing 1\% per year. This was chosen due to the increasing efficiency of homes, industry and technology, as this was the recent trend in the UK. We run each scenario eight times to capture the stochastic nature of the process. Via the observation of the emergent investment behaviour until 2050, an understanding of how real-life investors may behave emerges.

\begin{figure}[h]
	\centering
	\begin{subfigure}[b]{0.6\textwidth}
		\centering
		\includegraphics[width=\textwidth]{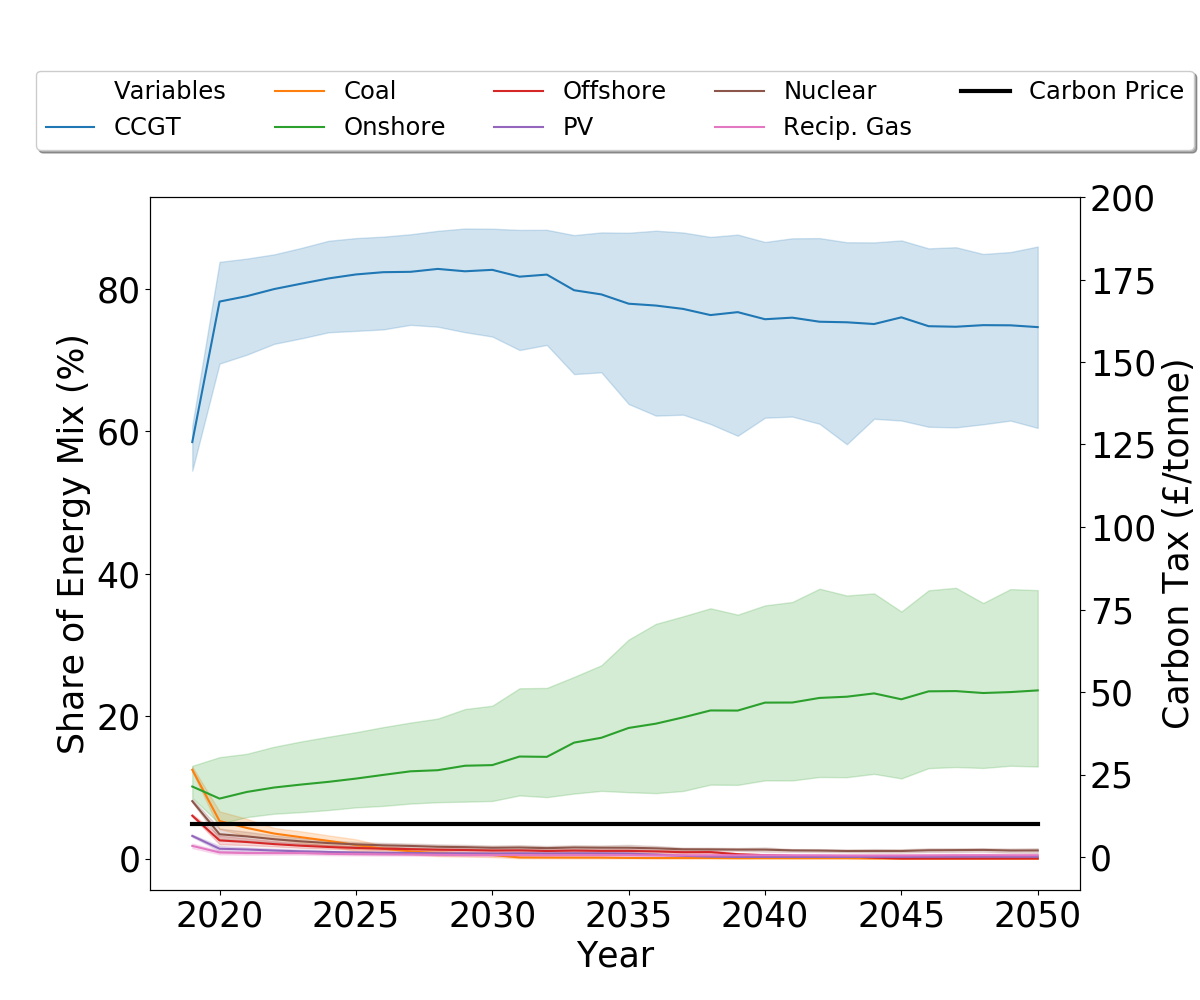}
		\caption[Network2]%
		{\small \textsterling10 carbon tax.}
		\label{fig:demand99carbon10}
	\end{subfigure}
	\hfill
	\begin{subfigure}[b]{0.6\textwidth}  
		\centering 
		\includegraphics[width=\textwidth]{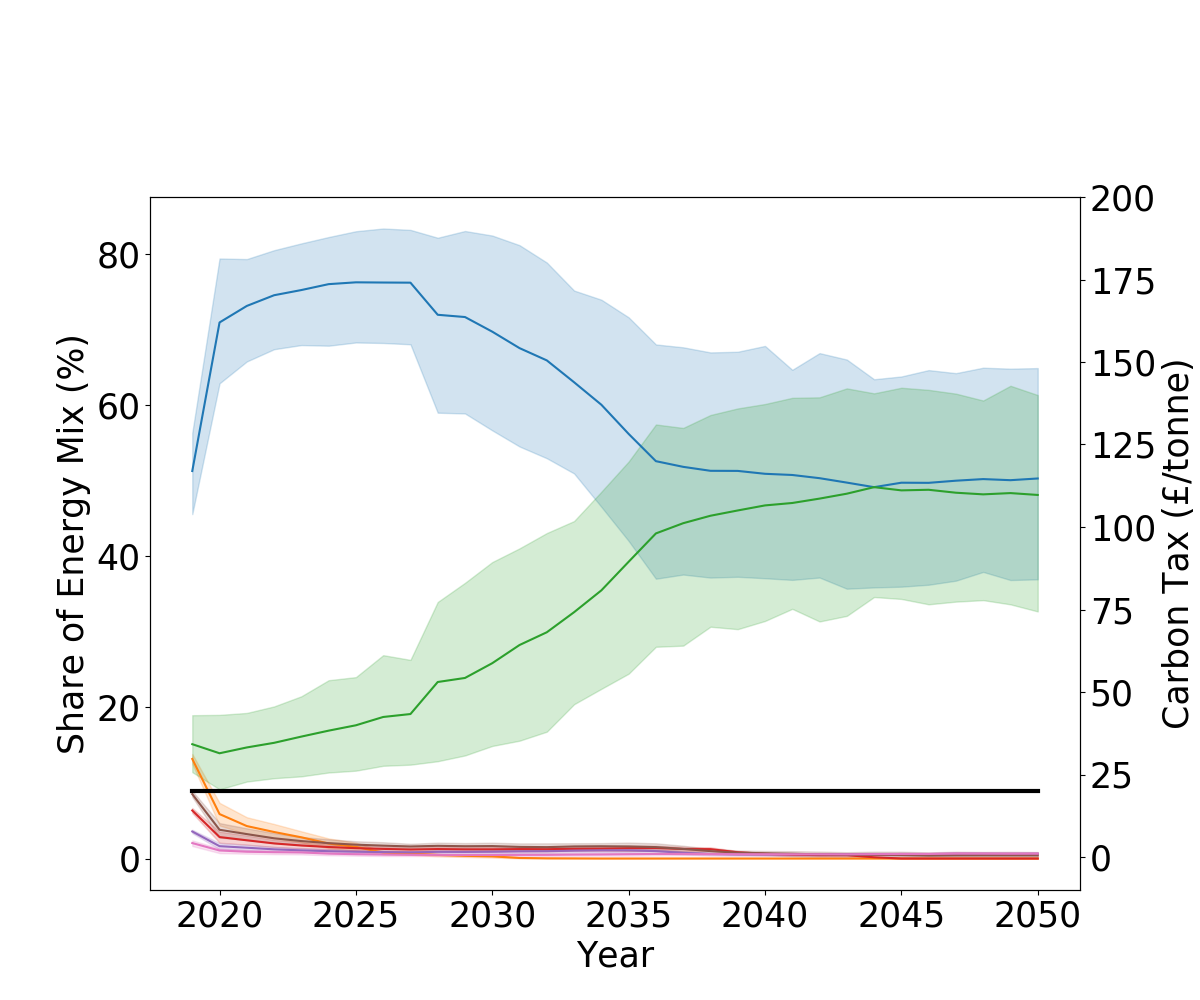}
		\caption[]%
		{\textsterling20 carbon tax.}
		\label{fig:demand99carbon20}
	\end{subfigure}
	\begin{subfigure}[b]{0.6\textwidth}
		\centering
		\includegraphics[width=\textwidth]{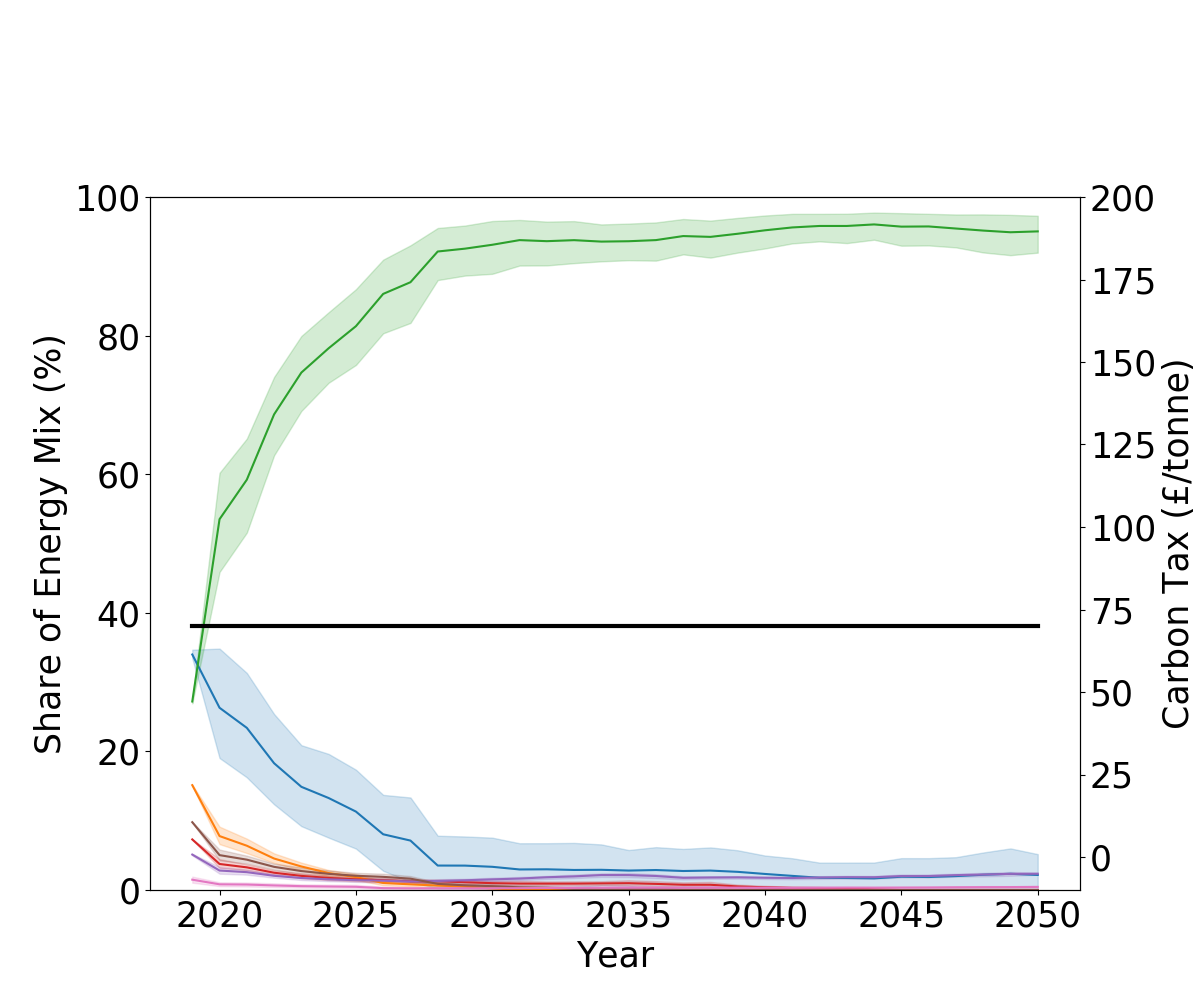}
		\caption[Network2]%
		{\small \textsterling70 carbon tax.}
		\label{fig:demand99carbon70}
	\end{subfigure}
	\caption{Scenarios from 2020 to 2050 with varying carbon tax for 20 time steps per year..}
\end{figure}

Figure \ref{fig:demand99carbon10} shows that with a carbon tax of \textsterling10, whilst renewable technology does grow, gas power plants provide the majority of supply in each year. However, at a level of \textsterling20 the increase in wind turbines is enough to match gas turbines. A carbon tax of \textsterling70, however, shows a near 100\% uptake of wind turbines. 

These results demonstrate the large impact that a carbon tax can have on the final electricity mix over the long term. Policy makers can effectively remove high proportions of carbon emitting technology lock-in by setting a carbon tax. The limitations to this work, from the perspective of policy makers, is the inability to know for certain what the final electricity mix will be as the results show a large amount of variance.

The reason that onshore wind and CCGT become dominant is due to their perceived low-cost over the lifetime of these projects when compared to the other electricity generation technologies. The capacity factor of onshore wind is lower than that of CCGT, and therefore a high investment in onshore wind is required to meet increasing shares of electricity demand. For example, if onshore wind has a capacity factor of 0.2, an installed capacity $5\times$ the total electricity demand would be required to meet 100\% of demand. This is more expensive than investing in CCGT, which has a higher capacity factor and can therefore meet more demand with a lower total installed capacity.

This trade-off between CCGT and onshore wind becomes more nuanced with different sized carbon taxes. For instance, in the £70 carbon tax case, it becomes economically efficient to install large amounts of onshore wind.

It is currently infeasible, however, for the power supply to be provided solely by wind turbines today. This overestimation, however, is due to the low time granularity of the model \cite{Collins2017}. This scenario therefore assumes perfect storage capabilities. This is because the model assumes that onshore wind's capacity factor is spread evenly across each of the LDC segments. However, this is not the case in the real world. For instance, the capacity factor of wind may be 0 at some times, and 1 at other times. This means that simply increasing the number of wind turbines will never meet electricity demand at times when there is no wind. Therefore, at these times other technologies are required, leading to a mix of different electricity generators. More detail from the representative days can capture days of zero wind and high demand.

These runs demonstrate that a consistent, but relatively low carbon tax can have a larger impact in the uptake of renewable energy than increasing carbon tax over a long time frame. We hypothesise that an early carbon tax affects the long-term dynamics of the market for many years. We, therefore, suggest early action on carbon tax to transition to a low-carbon energy supply.

%
%
%
%
%
%
%

\subsection{Discussion}

Agent-based models provide a method of simulating investor behaviour in an electricity market. We observed that an increase in carbon tax had a significant impact on investment. These findings enable policymakers to better understand the impact that their decisions may have. For a high uptake of renewable energy technology, rapid results can be seen after ten years with a carbon tax of \textsterling70 (\$90).

\subsection{Scenarios for representative days}

\begin{figure}
	\centering
	\includegraphics[width=0.7\textwidth, keepaspectratio]{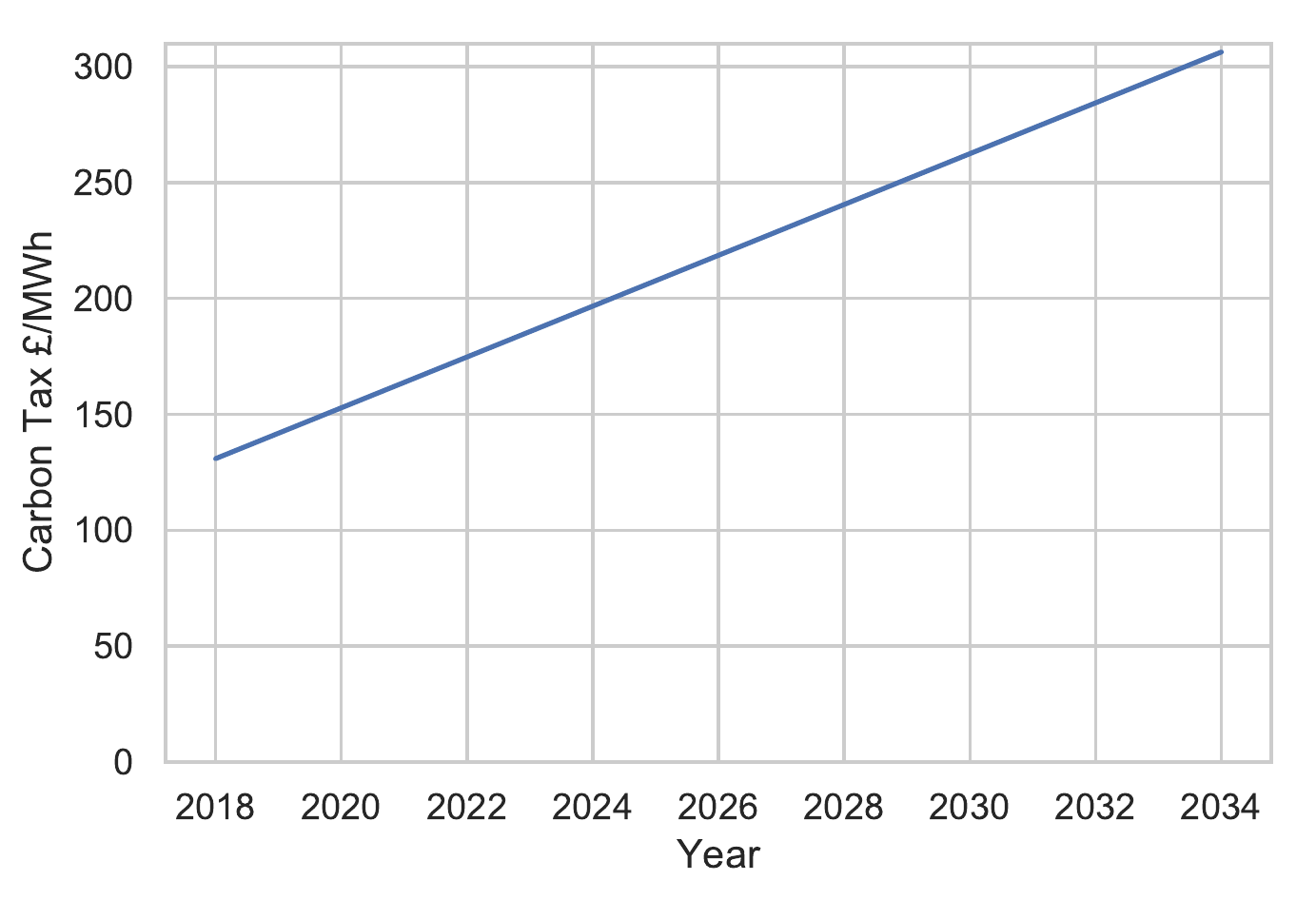}
	\caption{Optimal carbon tax strategy to reduce both electricity cost and carbon emissions for 192 time steps per year (8 days with 24 time steps per day).}
	\label{elecsim:fig:optimal_carbon_tax_strategy}
\end{figure}

In this Section, we discuss various scenarios under the model which uses representative days as time-steps. This work builds upon the work in Section \ref{elecsim:sec:validation}; we used the same predicted price duration curves as modelled on BEIS' scenario. We selected an optimal carbon tax level which would reduce both electricity price and carbon emissions, as shown later in Chapter \ref{chapter:carbon}. The optimal carbon tax strategy found in Chapter \ref{chapter:carbon} is shown by Figure \ref{elecsim:fig:optimal_carbon_tax_strategy}. Each of the scenarios were run ten times to display any variability in the results. We chose ten runs to limit both computation time and cost.

\begin{figure}
	\centering
	\begin{subfigure}{0.6\textwidth}
		\centering
		\includegraphics[width=\textwidth]{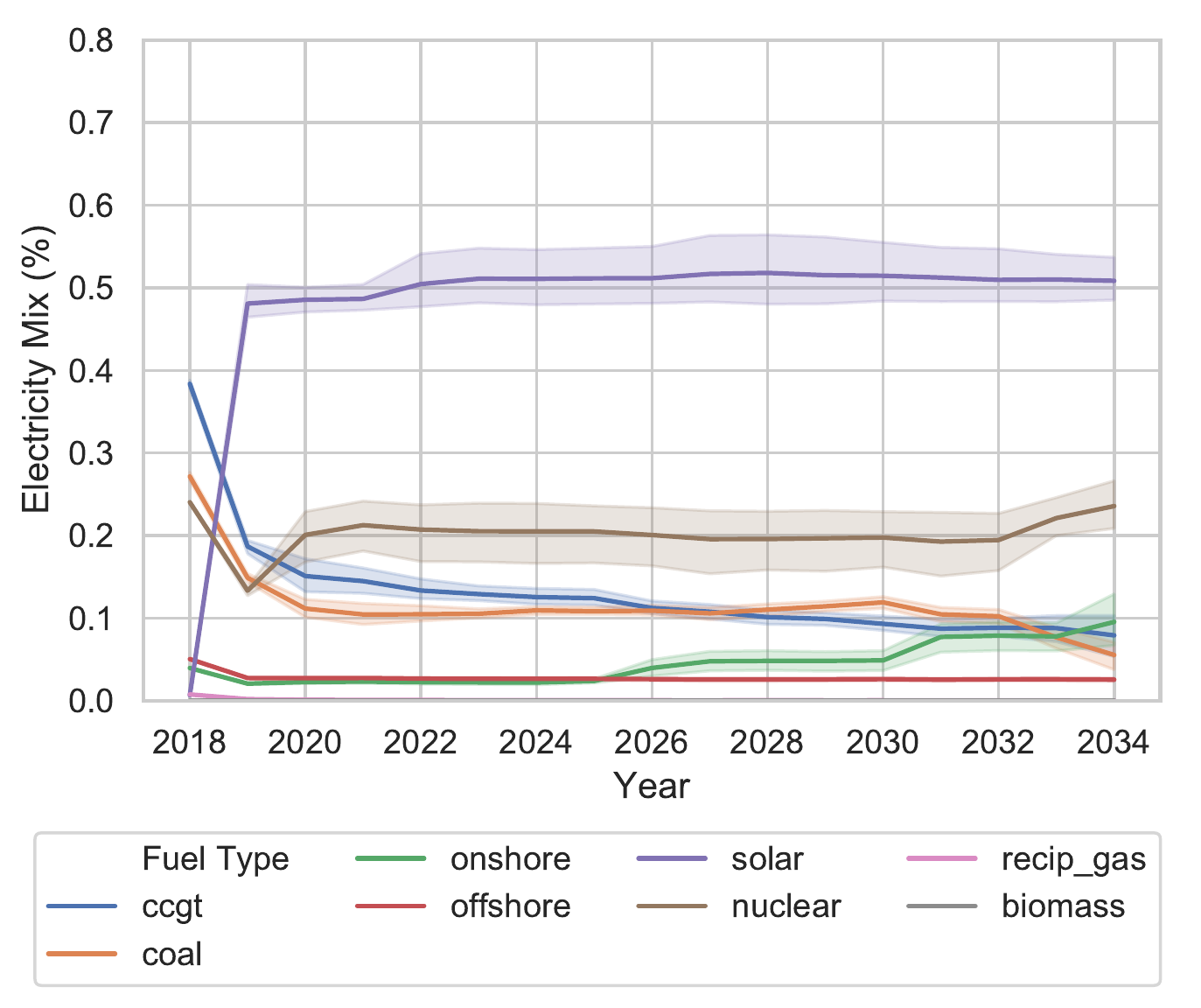}
		\caption{Demand does not increase or decrease.}
		\label{fig:demand10}
	\end{subfigure}
	\hfill
	\begin{subfigure}{0.6\textwidth}  
		\centering 
		\includegraphics[width=\textwidth]{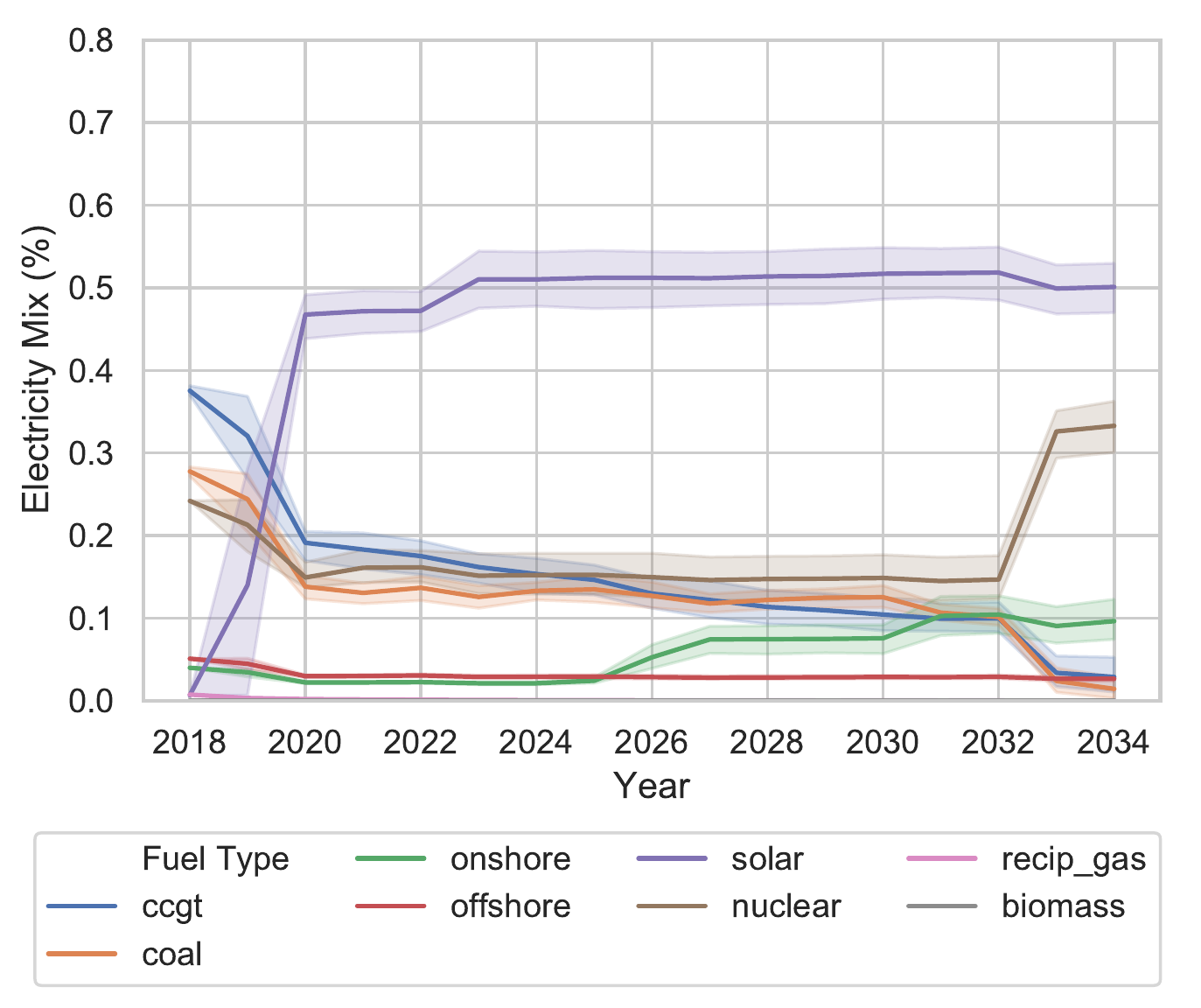}
		\caption{Demand reduces by 1\% per year.}
		\label{fig:demand099}
	\end{subfigure}
	\begin{subfigure}{0.6\textwidth}
		\centering
		\includegraphics[width=\textwidth]{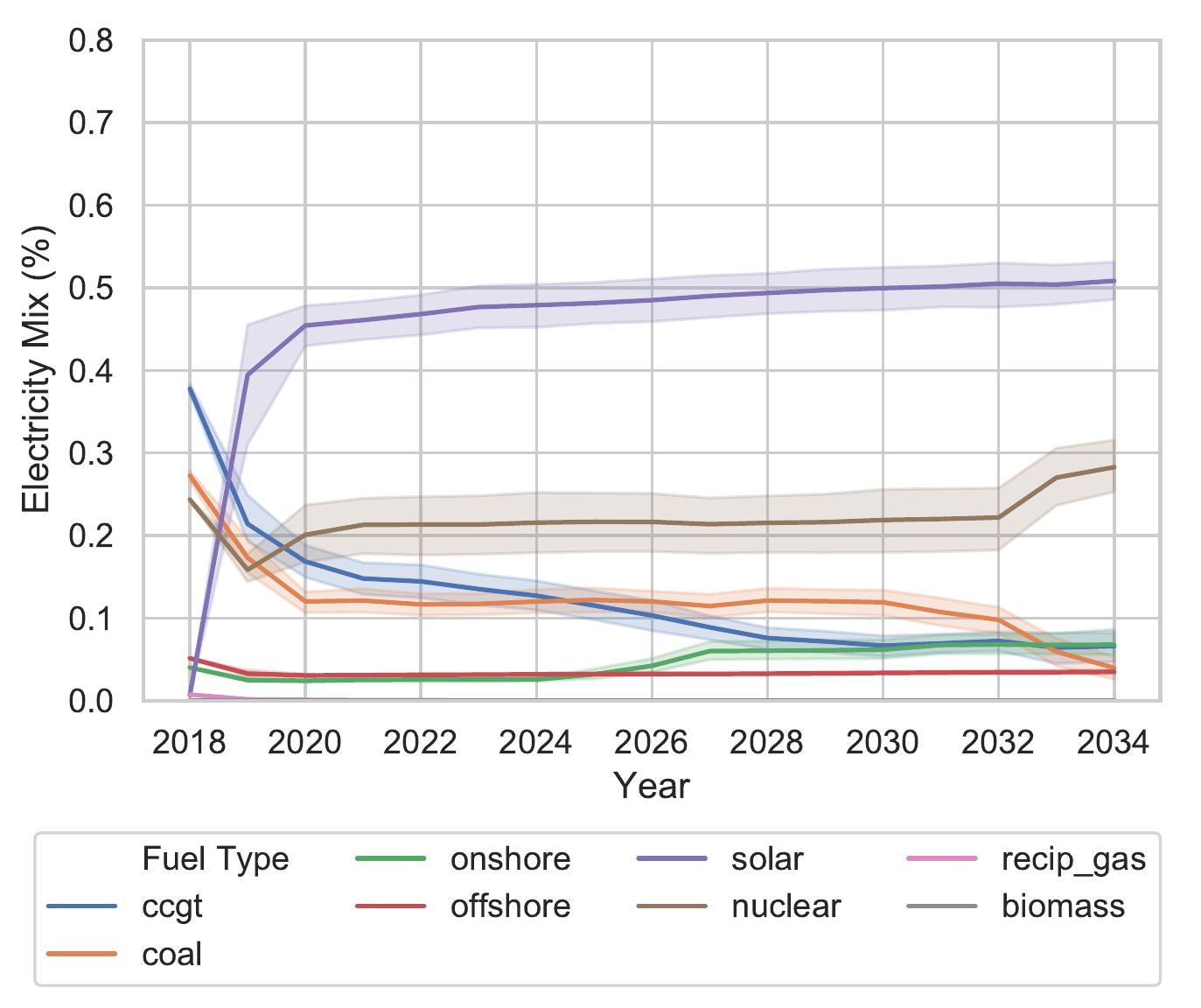}
		\caption{\small Demand reduces by 2\% per year}
		\label{fig:demand098}
	\end{subfigure}
	\caption{Scenarios from 2018 to 2035 with varying demand for 192 time steps per year (8 days with 24 time steps per day).}
	\label{elecsim:fig:increasing_demand}
\end{figure}

\begin{figure}
	\centering
	\begin{subfigure}{0.6\textwidth}
		\centering
		\includegraphics[width=\textwidth]{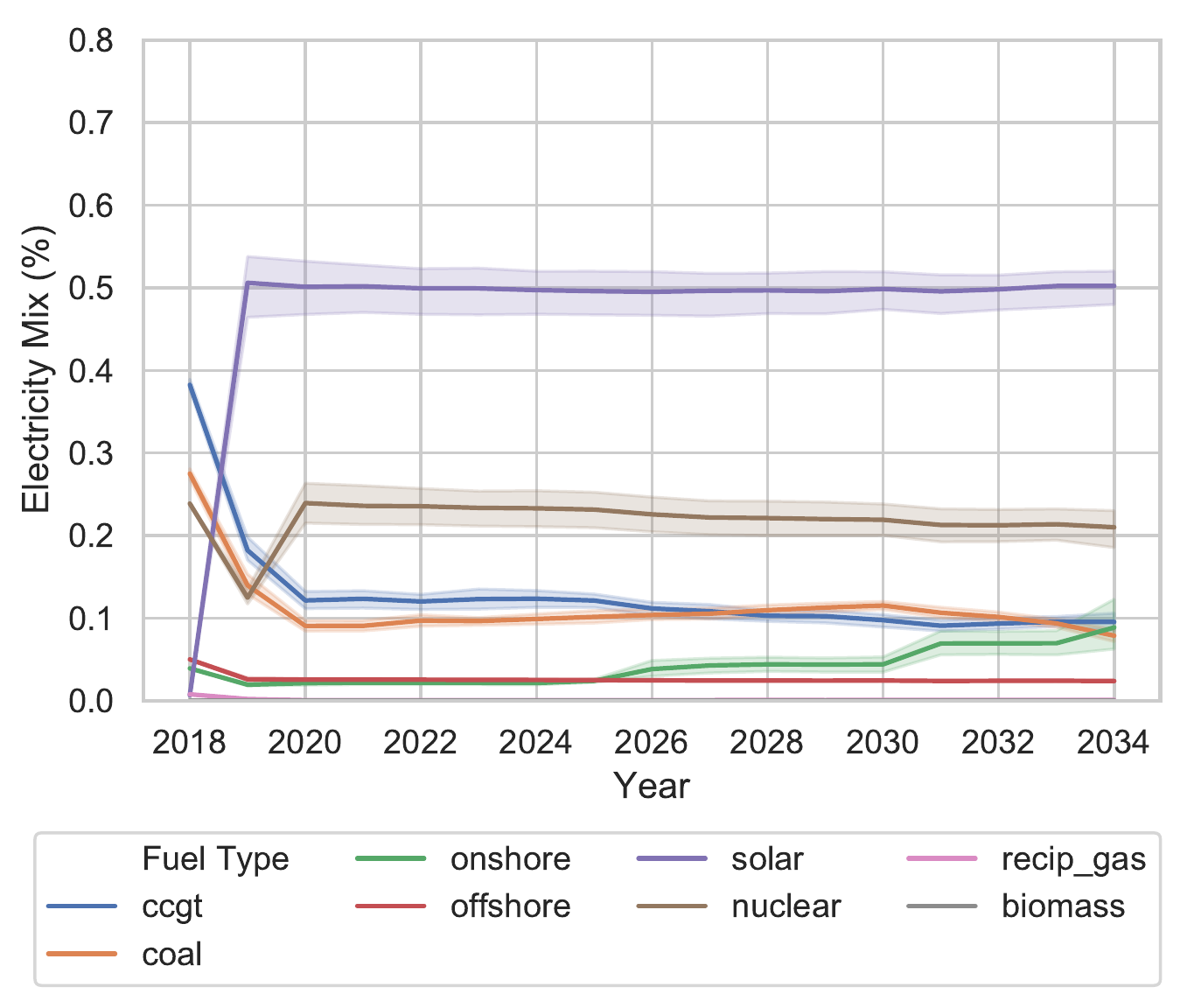}
		\caption{\small Demand increases by 1\% per year.}
		\label{fig:demand101}
	\end{subfigure}
	\hfill
	\begin{subfigure}{0.6\textwidth}  
		\centering 
		\includegraphics[width=\textwidth]{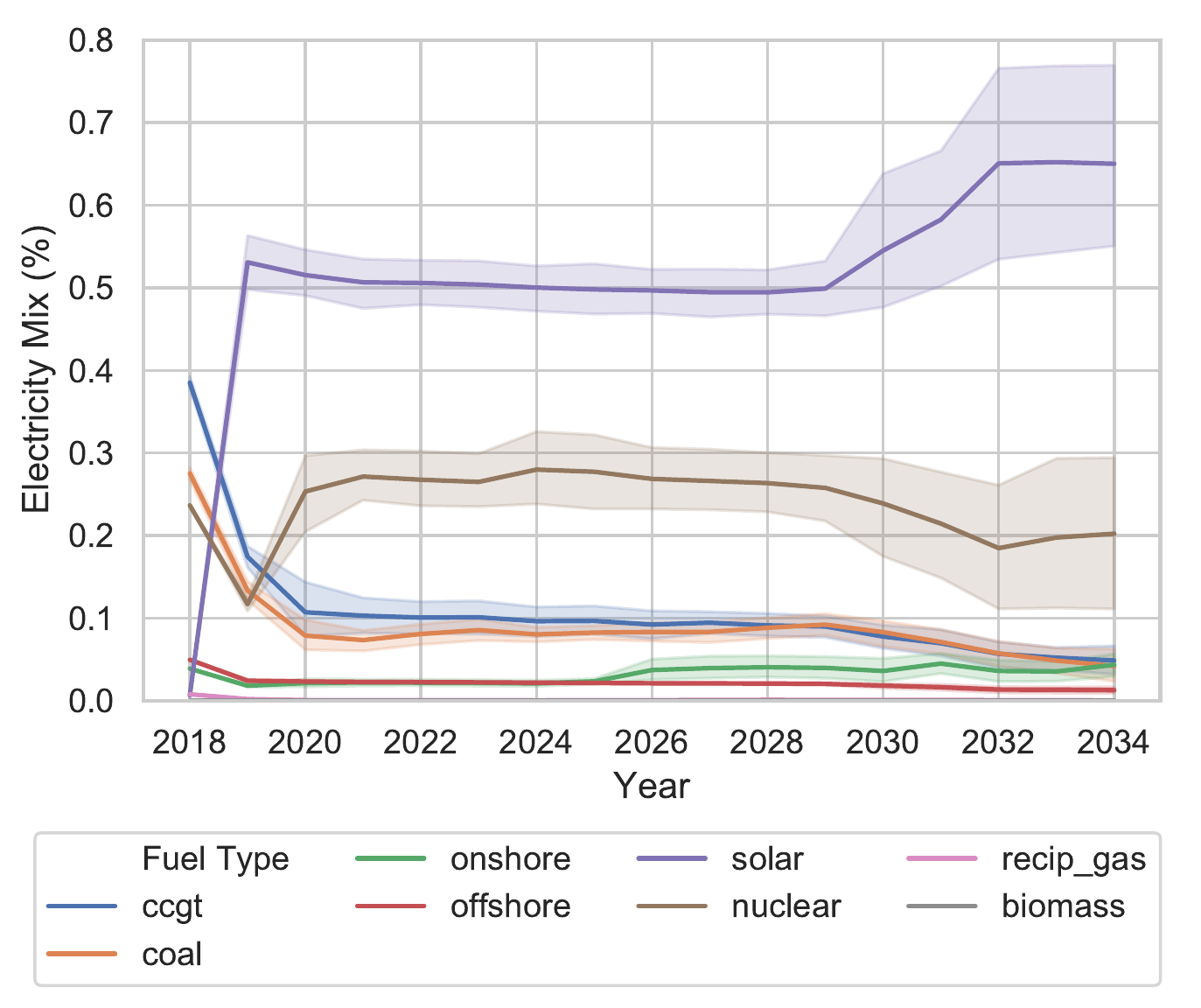}
		\caption{\small Demand increases by 2\% per year.}
		\label{fig:demand102}
	\end{subfigure}
	\begin{subfigure}{0.6\textwidth}
		\centering
		\includegraphics[width=\textwidth]{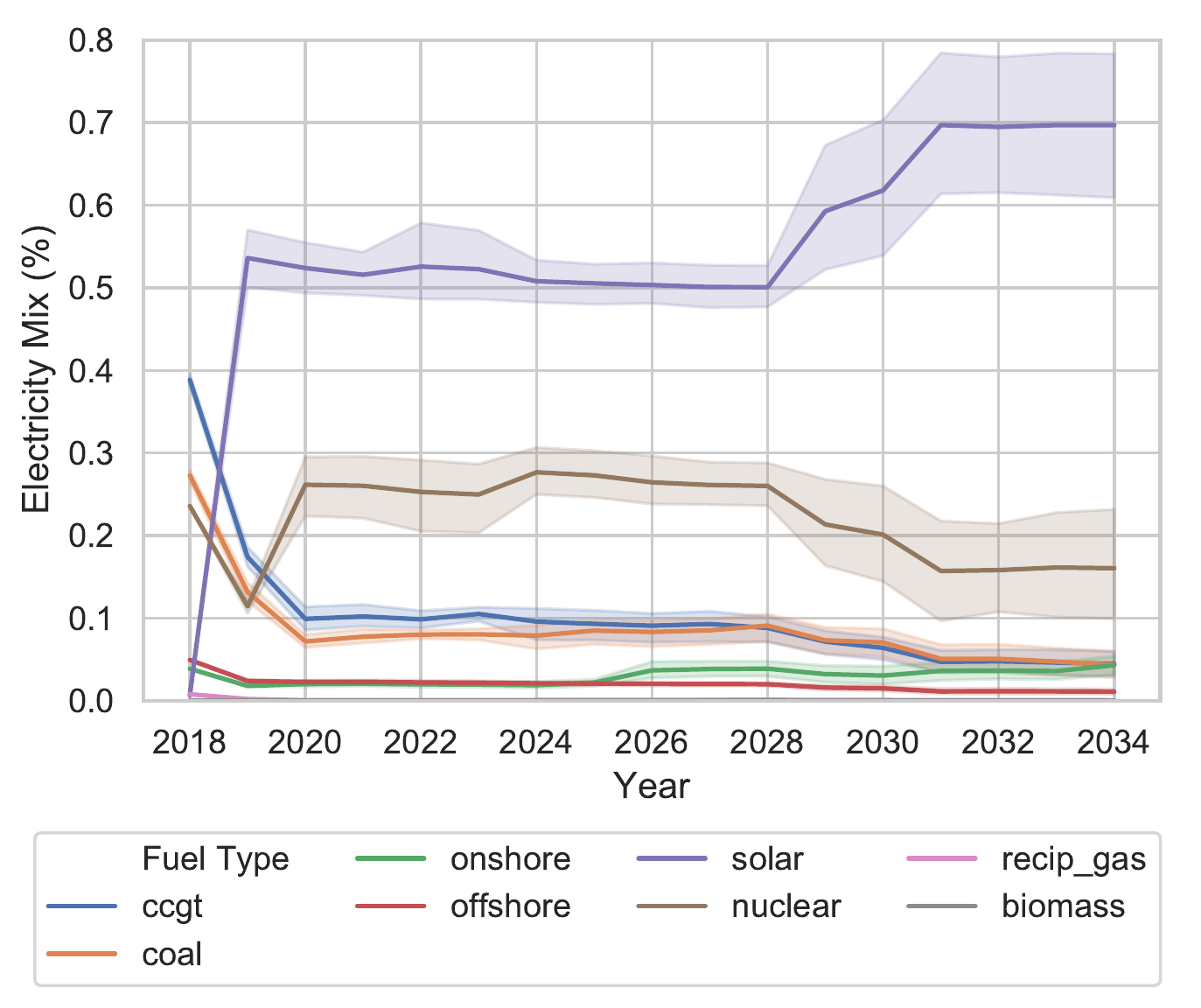}
		\caption{\small Demand increases by 2.5\% per year}
		\label{fig:demand1025}
	\end{subfigure}
	\caption{Scenarios from 2018 to 2035 with varying demand for 192 time steps per year (8 days with 24 time steps per day).}
	\label{elecsim:fig:decreasing_demand}
\end{figure}

Figures \ref{elecsim:fig:increasing_demand} and \ref{elecsim:fig:decreasing_demand} show the electricity mixes of various demand scenarios. Figure \ref{elecsim:fig:increasing_demand} displays the scenarios in which demand either stays flat, or decreases by 1\% and 2\%. For these scenarios, it can be seen that solar is the dominant electricity supply, supplying ${\sim}$50\%, with nuclear power second supplying between 20\% and 30\%. With a decreasing demand scenario of 1\% per year, as shown by Figure \ref{fig:demand099}, nuclear provides a higher proportion by the year 2034, of ${\sim}$30\%, however before the year 2033, provides a similar proportion to the other scenarios as shown by Figures \ref{fig:demand10} and \ref{fig:demand098}.

For the scenarios shown in Figure \ref{elecsim:fig:increasing_demand}, CCGT, coal and onshore provide around ${\sim}$10\% each by 2034.  Coal and CCGT, however, progress towards 0\% whereas onshore wind increases. This is to be expected due to the high carbon price, as shown by Figure \ref{elecsim:fig:optimal_carbon_tax_strategy}. Offshore does not exhibit a high amount of investment. We believe this is the case as offshore wind is more expensive than onshore wind, and in our scenario subsidies other than for nuclear are not modelled.

Figure \ref{elecsim:fig:decreasing_demand} shows scenarios where demand increases per year. Whilst electricity mix distribution is similar to the scenarios shown in Figure \ref{elecsim:fig:increasing_demand}, solar plays a significantly increased role than nuclear. This may be down to the large expense of nuclear, and the long time of deployment of this type of technology. Solar power, on the other hand, is able to be installed much more quickly to meet the high demand. This is especially true for the scenarios shown in Figures \ref{fig:demand102} and \ref{fig:demand1025} where demand rises by 2\% and 2.5\% per year respectively.

These results show the impact that different demand scenarios can have on the long-term electricity mix. It demonstrates that the electricity mix may change significantly dependent on a few factors, such as the electricity demand profile. Policy makers should therefore be conscious of the different electricity mixes that could occur, and provide support to ensure that demand is met at all times. For instance, a high nuclear subsidy may be required to meet the base-load demand, or tax credits for solar to ensure that increasing demand is met.

It is also possible for generator companies to understand which technologies will be useful in different demand scenarios. This would ensure that generator companies can make better decisions in the face of uncertainty.

The limitations, however, are that these scenarios do not place a probability to any of these scenarios. As, these results present the 2\% increase and 2\% decrease scenarios as equally likely. This may not be the case, and are dependent on more exogenous variables. This prevents policy makers and generator companies from making informed decisions about how they should adapt for relative scenarios.

\subsection{Performance}

\begin{figure}[htbp]
	\centering
	\includegraphics[width=0.6\linewidth]{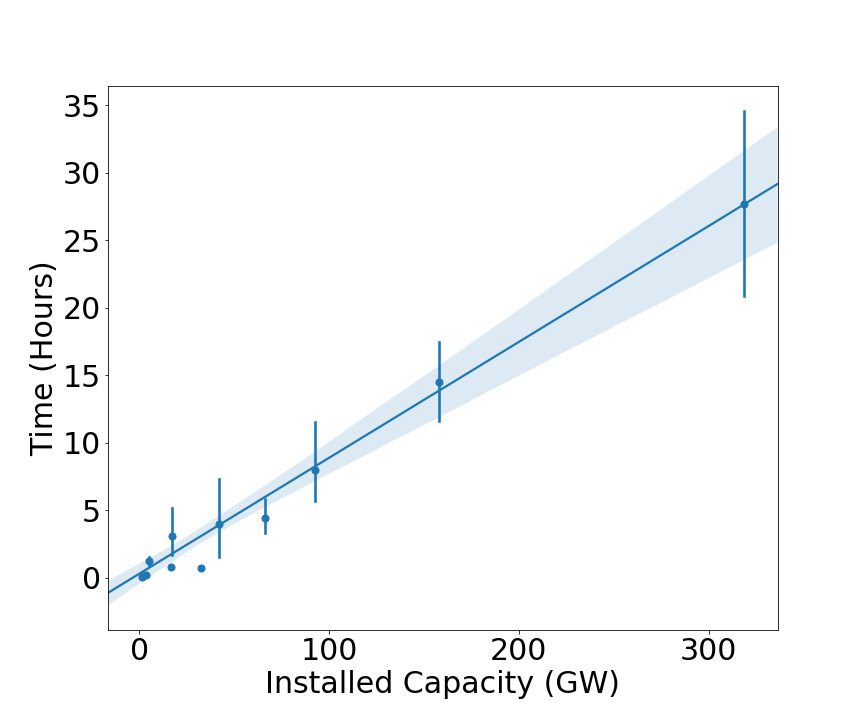}
	\caption{Run times of different sized countries for 192 time steps per year (8 days with 24 time steps per day).}
	\label{fig:timingplot}
\end{figure}

Figure \ref{fig:timingplot} shows the running time for ElecSim with varying installed capacity. We varied demand between 2GW and 320GW to see the effect of different sized countries on running time. The makeup of the electricity mix was achieved through stratified sampling of the UK electricity mix. The results show a linear time complexity.


\clearpage
\section{Sensitivity Analysis}
\label{elecsim:sec:sensitivity}

In this Section, we investigate a sensitivity analysis of ElecSim, where we vary the weighted average cost of capital and the down payment required for investment. We used the reference scenario discussed in Section \ref{elecsim:sec:scenarios}, with the optimal carbon tax to reduce both emissions and electricity price. The work presented in this section is in addition to the work published in \cite{Kell2020}.

We ran ten iterations per weighted average cost of capital and down payment element. We did this due to the monte-carlo nature of the simulation. We chose ten runs to give us sufficient variance in results, but reduce compute power, to reduce both time and cost of calculation. We show the results in the year 2035 due to this being where the largest change in electricity mix is likely to be observed over the whole horizon. Including more years in the analysis is possible, but would lead to many graphs with limited additional information.

\subsection{Results}

\begin{figure}
	\centering
	\includegraphics[width=0.9\linewidth]{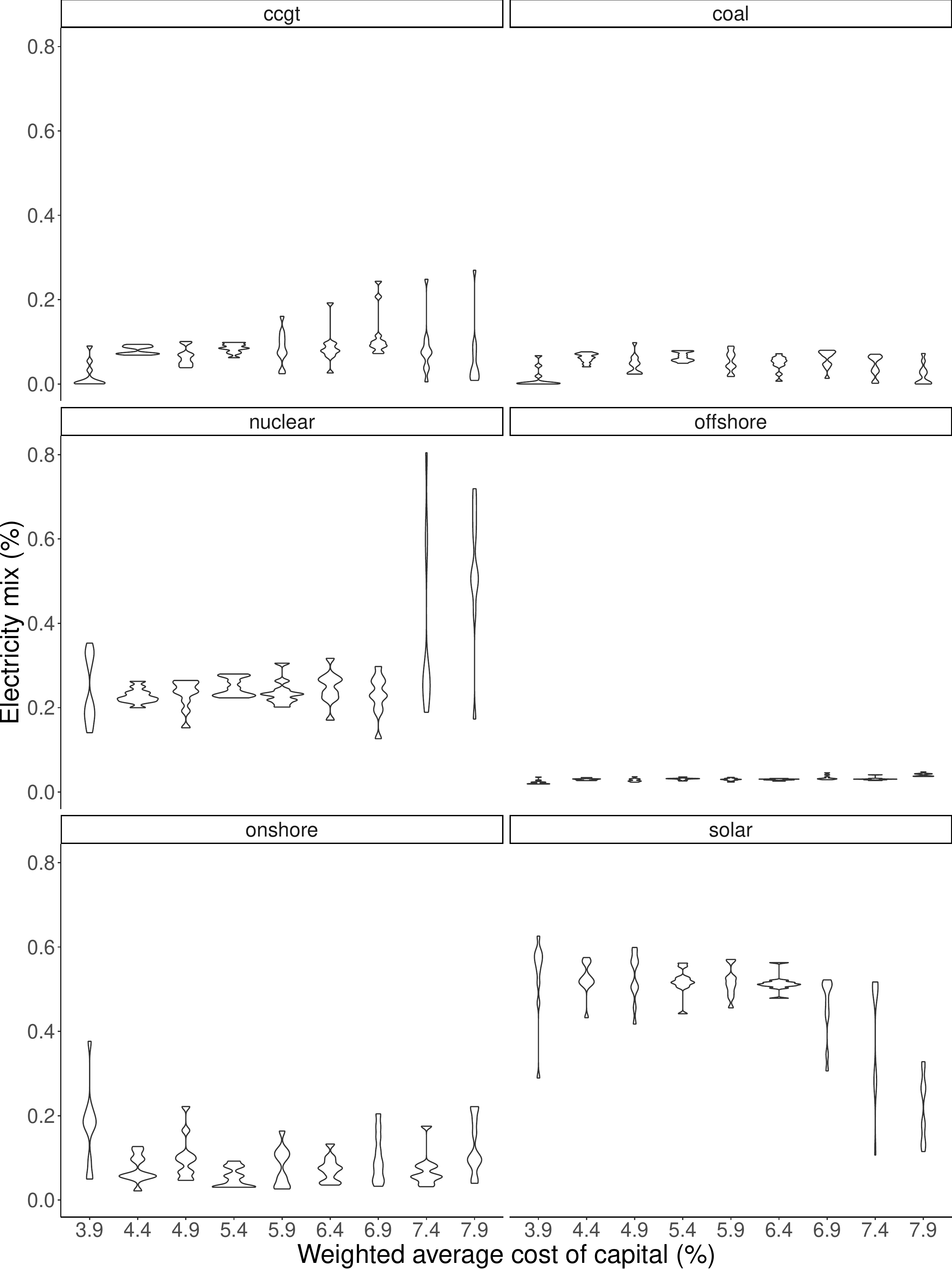}
	\caption{Sensitivity analysis where \acrfull{wacc} was varied. Results compare electricity mix in 2035.}
	\label{elecsim:fig:wacc_sensitivity}
\end{figure}

Figure \ref{elecsim:fig:wacc_sensitivity} displays the results of the sensitivity analysis for the \Gls{WACC} for non-nuclear power generators. For this, we trialled nine different \acrfull{wacc} values, where a value of 5.9\% is the reference case \cite{KincheloeStephenC1990TWAC}. 

It can be seen that the \acrshort{wacc} has an effect on the total investment in solar, nuclear and CCGT. With a \acrshort{wacc} equal to or greater than 7.4\%, nuclear increases significantly, whilst solar decreases. Nuclear has a \acrshort{wacc} of 10\%; therefore 7.4\% may be the point where nuclear becomes more competitive than solar in an environment where a low-carbon electricity supply is elicited from an optimal carbon tax.

Offshore and coal do not change significantly over different levels of \acrshort{wacc}. This may be due to the fact that CCGT and onshore are more competitive than coal and offshore, respectively, without external subsidies. 

Onshore seems to play a larger role at the lowest \acrshort{wacc}, 3.9\%. This may be due to onshore wind's high competitiveness when compared to nuclear.

\begin{figure}
	\centering
	\includegraphics[width=0.9\linewidth]{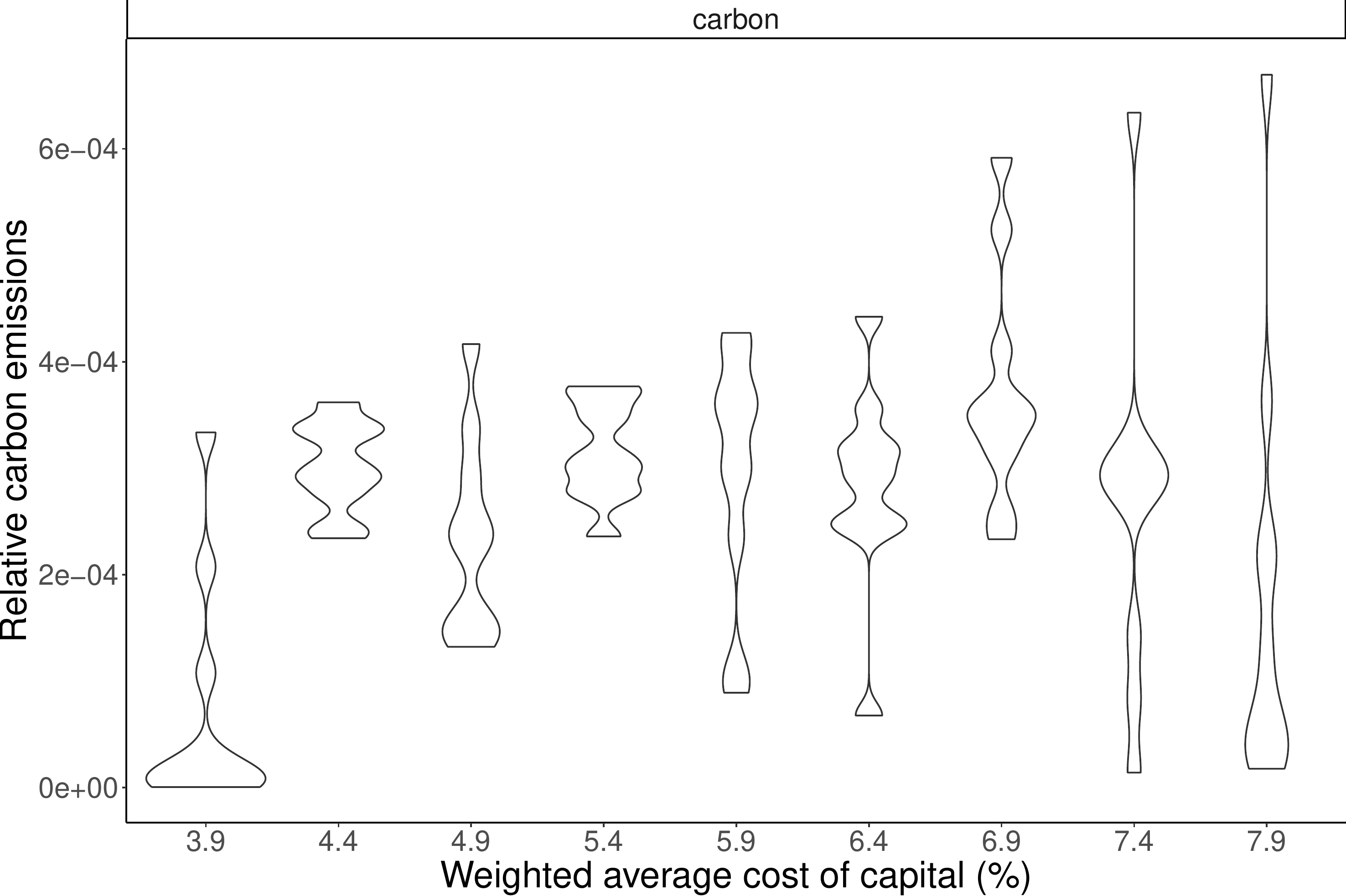}
	\caption{Sensitivity analysis where weighted average cost of capital (WACC) was varied. Results compare relative carbon emissions in 2035.}
	\label{elecsim:fig:wacc_carbon_sensitivity}
\end{figure}

Figure \ref{elecsim:fig:wacc_carbon_sensitivity} displays the relative carbon emissions in 2035. With a low \acrshort{wacc} of 3.9\%, the relative carbon emissions falls, on average, to zero. This seems to be due to the high levels of solar, onshore and nuclear.  

As \acrshort{wacc} increases, so does carbon emissions, until there is a \acrshort{wacc} of 0.5, where it reduces slightly. This is seemingly due to the higher levels of CCGT and coal, which is able to displace solar. However, it must be noted that in this scenario with the optimal carbon tax, the relative carbon emissions remains low.

\begin{figure}
	\centering
	\includegraphics[width=0.9\linewidth]{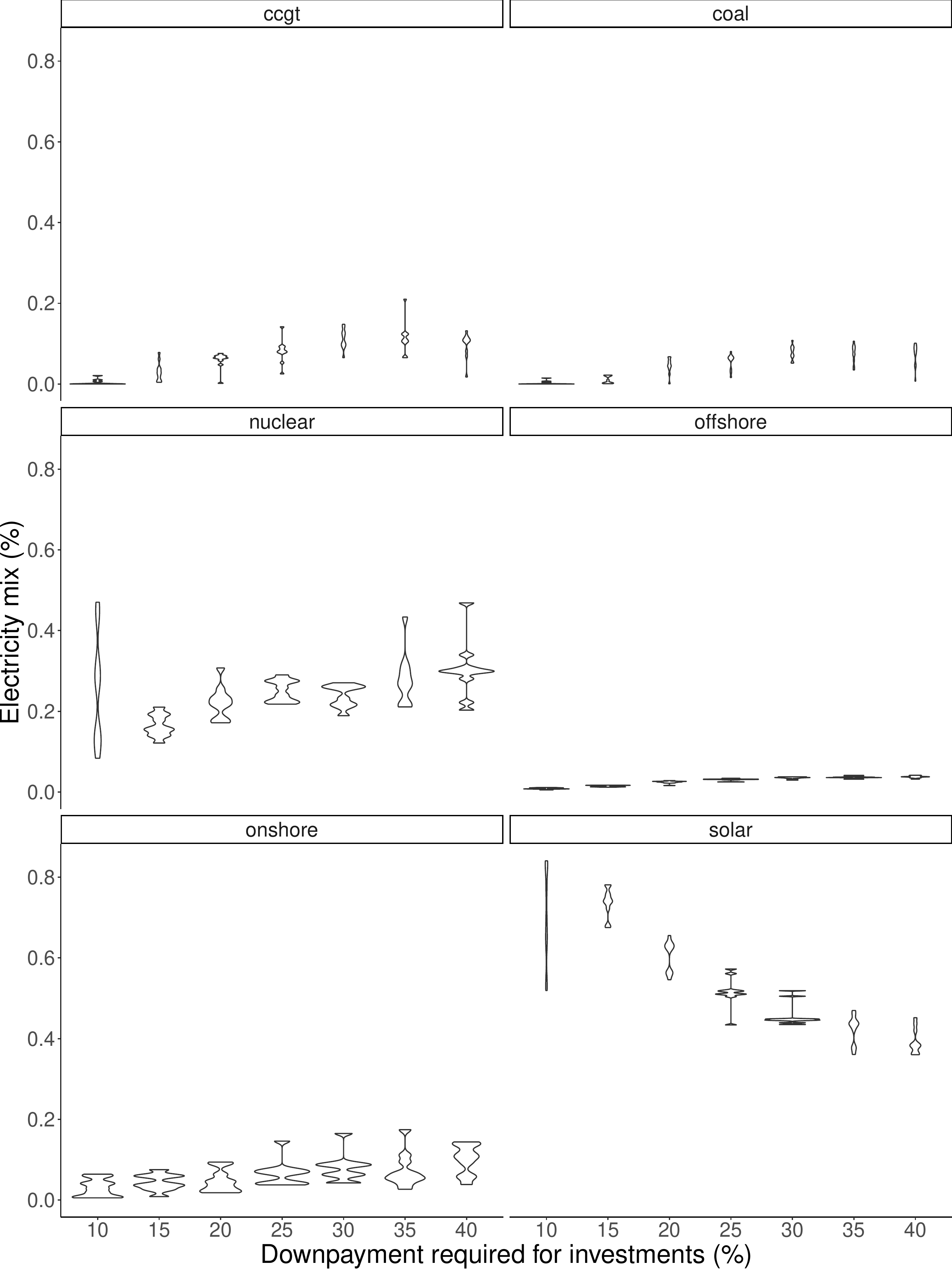}
	\caption{Sensitivity analysis where percentage of down payment was varied. Results compare electricity mix in 2035.}
	\label{elecsim:fig:downpayment_sensitivity}
\end{figure}

Figure \ref{elecsim:fig:downpayment_sensitivity} displays the sensitivity analysis results for different levels of down payment required for all investments. We varied the down payment required between the values of 10\% and 40\%.

As down payment required increases, so does nuclear and onshore, whilst solar decreases. CCGT also shows an increase with down payment required. The increase in down payment may help nuclear, due to the high costs of \acrshort{wacc}. With a higher down payment, the total costs of the project will fall when compared to other, cheaper, generators. It is likely that nuclear displaces solar in this case. CCGT increases up until a 35\% down payment required. This may be due to the increased use of onshore wind, where CCGT and coal is required to fill for times of low wind speeds.

\begin{figure}
	\centering
	\includegraphics[width=0.9\linewidth]{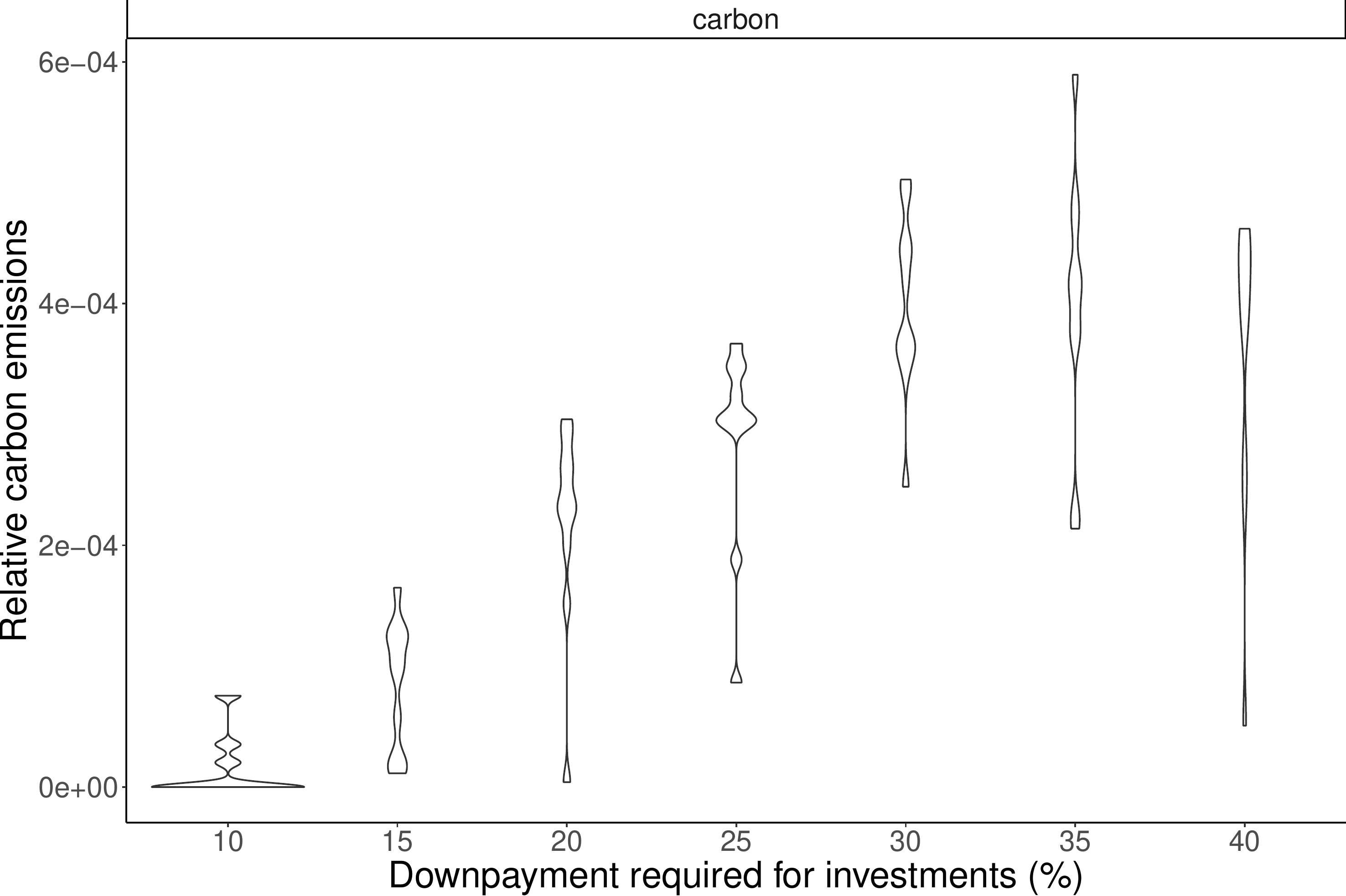}
	\caption{Sensitivity analysis where percentage of down payment was varied. Results compare relative carbon emissions in 2035.}
	\label{elecsim:fig:downpayment_carbon_sensitivity}
\end{figure}

Figure \ref{elecsim:fig:downpayment_carbon_sensitivity} displays the relative carbon emissions versus down payment required for investors. As down payment increases, so does relative carbon emissions. This is down to the increasing role that CCGT and coal play in the electricity mix and decreasing solar capacity. A down-payment of 10\% seems to have the lowest carbon emissions. This is due to the high investment in solar, and low investment in CCGT and coal. 

These results demonstrate the importance of parameters on the final electricity mix. For instance, the weighted average cost of capital can have a large impact on the long-term electricity mix. As government and policy makers have the ability to access capital at low-cost, it adds an important lever that policy makers can have to influence the final electricity mix. For instance, providing financial backing for solar, which are highly capital intensive, could significantly reduce the barriers to entry for this technology. Generator companies also have the ability to analyse the attractiveness of each technology based on their access to capital. 

However, these results do not show a sensitivity analysis on different costs. For example, capital expenditure may reduce in the future for solar. This would change the results significantly. And so, generator companies and policy makers should consider a wider range of future technology cost scenarios in addition to the sensitivity analysis presented here. 

\section{Limitations}
\label{elecsim:sec:limitations}

As with all models, ElecSim features a number of limitations. That is, ElecSim can not model the real world perfectly, and thus should be relied on only with a spectrum of other quantitative and qualitative methods. 

A number of limitations exist of ElecSim, which have been listed in this Section. This list, however, is not exhaustive, due to the possibility for unknown unknowns.

\begin{enumerate}
	\item Requirement of exogenous price prediction curve.
	\item Challenges with long-term validation
	\item National model
	\item Lack of spatial resolution
	\item Lack of detailed supply curves (such as maximum wind and solar irradiance)
	\item Comparing to other models may not be correct as all models are wrong
	\item Future may change dramatically
	\item Lack of contracts for difference modelling and other subsidies (apart from nuclear)
\end{enumerate}

\subsection{Exogenous price duration curve}

The prediction of electricity prices in the long-term future is difficult to predict with certainty. In ElecSim, we are required to predict electricity prices over the lifetime of a plant. These plants often have lifetimes of 25 years or more. Two methods, which were utilised by the GenCos, were used by ElecSim: simulating a market ten years into the future and optimising for predicted price duration curve.

The first method of simulating a market ten years into the future has two major limitations. Firstly, the investments made ten years into the future, the costs of fuel, tax all have a large effect. Predictions which may seem perfectly reasonable at the time, may be unreasonable by the time the market occurs in real life. For instance, investments in another country may significantly reduce the cost of renewables in the country where the simulation is being run. Secondly, even if the simulation ten years into the future is correct, it does not mean that it will be representative of the entire 25 years. However, to fix this problem would require many more simulations, which may not make the model more accurate.

To overcome this limitation, we removed the simulation to generate a predicted price curve and used optimisation to find an optimal predicted price duration curve to generate a scenario. We were then able to check whether the predicted price duration curve had reasonable values. Whilst this enabled us to remove the problem of simulating an uncertain market in the future, it removed the ability to rapidly and quickly trial significantly different scenarios. This is because the predicted price duration curves were optimised for a certain scenario. Whilst it remained possible to change multiple parameters, it became difficult to trial many different carbon prices, as these have a significant impact on the predicted price duration curve.

\subsection{Long-term validation}

Long-term validation, as mentioned in Section \ref{elecsim:sec:validation} is particularly difficult due to the uncertainties involved. We, therefore, do not propose that ElecSim is used as a ground truth. Instead, it should be used as a way to understand possibilities that could arise from scenarios under certain assumptions and conditions. It is only with this knowledge that the model should be used. However, we argue that, whilst limited, quantitative advice has a large role to play in understanding the future, and reducing the uncertainty through understanding complex systems, such as electricity markets. It is only through building and understanding such models that the uncertainty and complexity involved can be understood. 

Additionally, when modelling the future, many other models compare their outcomes to other models for a specific scenario. Whilst this has the benefit of checking that the underlying dynamics of the models are similar, or the same, it does not test whether the models fit correctly to real-life. It is for this reason that we used cross-validation as well as comparing our model to that of the UK Government, BEIS department.

However, a significant problem with using historical data to validate a model, is that, we know more exogenous information about what happened during the past than we do about the future. Therefore the comparison between the past and future is not equal. For instance, we knew that certain coal power plants were taken out of service during the 2013-2018 period. We do not know for certain that this will happen in the future. It is therefore not possible to equate the past and forward scenarios. This fundamental truth will always remain a problem. 

However, the modelling process is stronger by using as much available data as possible. If we did not use historic data to calibrate and validate our model, we would be losing valuable information for verifying our model, as well as finding problems within our model. For instance, we may have introduced coding errors that would not have been found without using historic data. It is for this reason that we chose this approach.

\subsection{National model}

Due to the high granularity of ElecSim on a technical and temporal level, it becomes difficult to model electricity markets on a global scale. Modelling a single run of the United Kingdom can take ${\sim}$16 hours, therefore modelling the rest of the world would take an infeasible amount of time and computing power to function. We run these on a Microsoft Azure, Dsv3-series virtual machine, which are made up of Intel Xeon Platinum 8272CL (Cascade Lake), Intel Xeon 8171M 2.1GHz (Skylake), Intel Xeon E5-2673 v4 2.3 GHz (Broadwell), or the Intel Xeon E5-2673 v3 2.4 GHz (Haswell) processors in a hyper-threaded configuration.

\subsection{Lack of spatial resolution}

Within an electricity market, a component of space becomes important. This is because electricity is not created in the locations that it is consumed. For instance, offshore wind is typically clustered on banks which are shallow enough to install wind turbines, whilst demand is typically clustered on land in residential and urban areas. In addition to this, different areas may have different demand profiles, for instance with high income earners or industry, typically consuming more than low-income areas.

By accounting for these variations in spatial resolution, one is able to optimise the location of power plants, to minimise electricity transfer losses. In addition, different distribution networks can be designed, to account for differences in demand.

\subsection{Supply curves}

As ElecSim is a national model, and the UK is a relatively small country, we did not take into account the supply-demand curve for resources such as natural gas, coal and other fossil fuels. This is due to the fact that the fossil fuel market is much bigger than solely the United Kingdom. In other words, an increase in demand in the UK would not increase the price on a global scale. 

We did not take into account limits of land usage for solar panels and wind turbines in the model either. This is due to the fact that future technologies may overcome such challenges. For example, currently, offshore wind turbines must be placed in geographies with suitable floors. However, with the advent of floating wind turbines, these limitations could be reduced. Additionally, solar panels may become more effective at being placed offshore.

\subsection{Model comparison}

Whilst it is possible to compare multiple model outputs that look at a horizon of 20+ years, it is not possible to know whether all or any of these models are correct, or incorrect. Therefore, by comparing multiple models, it may be the case that all of the models are wrong and that no correct inference can be made. This is a significant challenge of long term energy modelling. Other approaches include cross-validation, where historic data is used to compare model outputs. However, whilst all models are wrong, insights can be gained from the functioning of these models to provide additional information that would not be available without these quantitative models.

\subsection{A changing future}

Whilst it is possible that a model can closely match a scenario, it is not possible to know which scenario will occur in the future. Therefore, whilst one is able to model many different scenarios, it is reasonable to expect that none of these modelled scenarios actually occur. This means that although a model accurately models the system it is designed to replicate, none of the model outputs reflect that of real-life.

This is a significant challenge of scenario modelling, to which there is no perfect solution. Increasing the number of scenarios and placing likelihoods to each of these scenarios is one option, however, it is not perfect.

\subsection{Subsidy modelling}

In this work, we did not model subsidies such as the contracts for difference auction, which is an auction found in the UK, where generators are able to sell their electricity at a stable level, irrespective of the daily price of electricity. This is because we focused on marking ElecSim a generalisable model for different countries. To ensure this, we had to focus on global parameters and reduce complexity. We did, however, model a simple subsidy where an extra price is paid for every MWh for Nuclear.

\clearpage
\section{Conclusions}
\label{elecsim:sec:conclusions}


Liberalised electricity markets with many heterogeneous players are suited to be modelled with ABMs. ABMs incorporate imperfect information as well as heterogeneous actors. ElecSim models imperfect information through forecasting of electricity demand and future fuel and electricity prices. This leads to agents taking risk on their investments, and model market conditions more realistically.





In this chapter, we have demonstrated that it is possible to use ABMs to simulate liberalised electricity markets. Through validation, we are able to show that our model, ElecSim, is able to accurately mimic the observed, real-life scenario in the UK between 2013 and 2018. This provides confidence in the underlying dynamics of ElecSim, especially as we are able to model the fundamental transition between coal and natural gas observed between 2013 and 2018 in the UK.

In addition to this, we were able to compare our long-term scenario to that of the UK Government, Department for Business, Energy \& Industrial strategy. We show that we are able to mimic their reference scenario; however, demonstrate a more realistic increase in nuclear power. The parameters that were gained from optimisation show that the BEIS scenario is realistic, however, a high nuclear subsidy may be required.

To improve the accuracy of our model, we used eight representative days of solar irradiance, offshore and onshore wind speed and demand to approximate an entire year. The particular days were chosen using a $k$-means clustering technique, and selecting the medoids. This enabled us to accurately model the daily fluctuations of demand and renewable energy resources. 



In addition to this, a method of dealing with the non-validatable nature of electricity markets, as per the definition of Hodges \textit{et al.} is to vary input parameters over many simulations and look for general trends \cite{Hodges}. This could be achieved using ElecSim through the analysis of a reference case, and a limited set of scenarios which include the most important uncertainties in the model structure, parameters, and data, i.e. alternative scenarios which have both high plausibility and major impacts on the outcomes.

Additionally, we showed a number of scenarios, and shows that total demand has an effect on the electricity mix. An increasing demand, year-on-year, can lead to an increase in solar to accommodate this demand. However, if demand reduces, there is a higher investment in nuclear, which contributes to the electricity mix by 2035.

We ran a sensitivity analysis of \acrfull{wacc} and down payment required. We showed that these two variables have a large effect on the total electricity mix by the year 2035, which in turn affects the total carbon emissions. We, therefore, show that the input assumptions have an effect on the simulation, which must be considered when analysing model outputs.

\chapter{Electricity demand prediction}
\label{chapter:demand}
\ifpdf
\graphicspath{{Chapter3/Figs/Raster/}{Chapter3/Figs/PDF/}{Chapter3/Figs/}}
\else
\graphicspath{{Chapter3/Figs/Vector/}{Chapter3/Figs/}}
\fi

\section*{Summary}

In this chapter, we use several different machine learning and statistical methods to predict electricity demand 30 minutes ahead as well as a day ahead. By predicting electricity demand ahead, GenCos are able to better schedule their power plants to dispatch to meet this demand. The 30 minutes ahead methodology is used in the work on day-ahead work. We utilise the errors from the day ahead predictions to see what the impact of such errors are on the electricity market over the long-term using ElecSim. The work looks specifically at the difference in electricity mix with prediction error, as well as carbon emitted. Whilst much of the literature investigates the short-term benefits of using machine learning in making short-term forecasts, this work goes further in that it integrates the short-term characteristics of machine learning into the long-term model, ElecSim. This helps to identify how the long-term market can be changed through algorithms with differing accuracies. This directly relates to the aims of the thesis, which is to look beyond the short-term impact of using machine learning, but to look at the wider impacts on the long-term market. The work on 30-minute ahead forecasting was published in \cite{Kell2018a}. 

We introduce this work in Section \ref{forecast:sec:introduction}. Section \ref{forecast:sec:litreview} provides a literature review on the topic of demand forecasting. We introduce the methods used in Section \ref{forecast:sec:methods}. Sections \ref{forecast:sec:shortterm} and \ref{forecast:sec:longterm} look at 30-minute ahead predictions and day-ahead predictions respectively. Additionally, Section \ref{forecast:sec:longterm} integrates these day ahead projections into the ElecSim model. 

\section{Introduction}
\label{forecast:sec:introduction}


The need for accurate load forecasting is essential for control and planning of electricity generation in electrical grids due to the fact that supply must meet demand \cite{Lu1993}. Short-term electricity demand forecasting has become increasingly important due to the introduction of competitive energy markets. Accurate estimates of demand are required so that the correct amount of electricity is purchased on the wholesale market \cite{Dillon1991}. Electricity is unique to other commodities in that it must be either consumed the moment that it is generated or stored. The difficulties in storing electricity arise from high installation and maintenance costs, inefficiencies and low capacity \cite{Poonpun2008}. It is therefore important to match demand to supply, and thus regulate frequency. This is because, if there is a mismatch between supply and demand, frequency will either increase or decrease. Failure to accurately forecast electricity demand can lead to financial loss and/or system-wide blackouts \cite{Hines2008}. We focus on electricity demand in this work and not supply, solar or wind due to limited computing power and time. However, the conclusions found here could be extended to these timeseries.

The integration of higher proportions of intermittent renewable energy sources (IRES) in the electricity grid will mean that the forecasting of electricity demand will become increasingly important and challenging. Examples of IRES are solar panels and wind turbines, which fluctuate in terms of power output based on localised wind speed and solar irradiance. However, as supply must meet demand at all times and the fact that IRES are less predictable than dispatchable energy sources such as coal and combined-cycle gas turbines (CCGTs), this means that extra attention must be made in predicting future demand if we wish to keep, or better reduce, the current frequency of blackouts \cite{Lu1993}. A dispatchable source is one that can be turned on and off by human control and therefore, able to adjust output just in time, at a moment convenient for the grid.

Typically, peaker plants, such as reciprocal gas engines, are used to fill fluctuations in demand that had not been previously planned for. Specifically, peaker plants meet the peaks in demand where other cheaper options are at full capacity. These peaker plants are typically expensive to run and have higher greenhouse gas emissions than their non-peaker counterparts \cite{Mahmood2014}. Whilst peaker plants are also dispatchable plants; not all dispatchable plants are peaker plants. 

\begin{figure}
	\centering
	\includegraphics[width=0.85\linewidth]{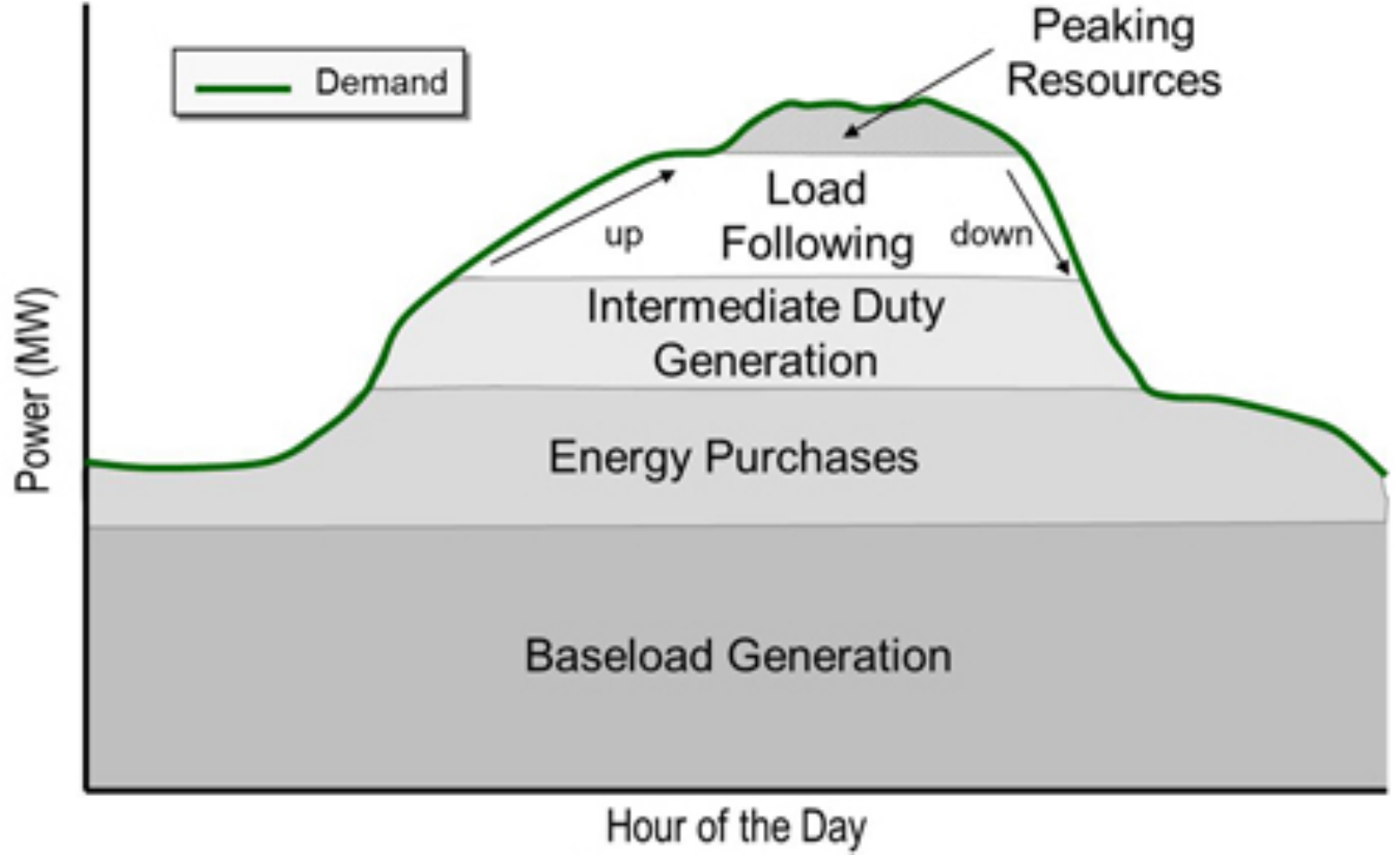}
	\caption{Types of resources for electricity generation.}
	\label{fig:peaker_plants}
\end{figure}

Figure \ref{fig:peaker_plants} shows the differences between baseload generation, peaking resources and those in between. Peaker power plants will match peak demand, as shown by the Figure, whereas baseload generation is run at all times. This includes some gas power plants and nuclear plants, which have a low short run marginal cost. It is unlikely that a nuclear power plant will be used as a peaker plant due to its inflexibility and slow ramp up and ramp down times, and low short run marginal cost. 

\begin{figure}
	\centering
	\includegraphics[width=0.85\linewidth]{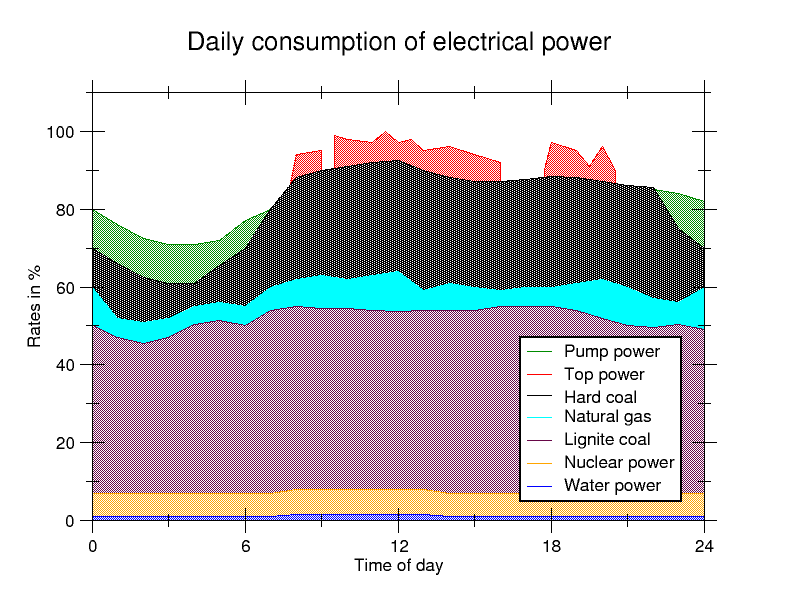}
	\caption{Types of resources for electricity generation per demand.}
	\label{fig:peaker_plants_types}
\end{figure}

Figure \ref{fig:peaker_plants_types} displays an example load profile with different types of electricity generation. As can be seen, the electricity generation types with the lowest marginal costs, such as nuclear and water power are dispatched first. Next, natural gas is dispatched which has a lower cost than coal and pump power, and finally the top power refers to the expensive peaker power plants, which are used to match peak demand only due to their high costs.

To reduce reliance on peaker plants, it is helpful to know how much electricity demand there will be in the future so that more efficient plants can be used to meet this expected demand. This is so that these more efficient plants can be brought up to speed at a time suitable to match the demand. Forecasting a day into the future is especially useful in decentralised electricity markets which have day-ahead markets. Decentralised electricity markets are ones where electricity is provided by multiple generation companies, as opposed to a centralised source, such as a government. To aid in this prediction, machine learning and statistical techniques have been used to accurately predict demand based on several different factors and data sources \cite{Kell2018a}, such as weather \cite{Hong2014}, day of the week \cite{Al-Musaylh2018} and holidays \cite{Vrablecova2017}.

The introduction of smart meters in many countries (USA, Europe, Canada and South Korea) has led to an influx of high granularity electricity consumption data that can be used for load forecasting \cite{Depuru2011a}. Smart meters are digital devices that measure the electricity consumption of individual households at regular intervals (intervals of an hour or less) and offer two-way communication between the meter and utility company. Smart meters aid customers to understand precisely how much electricity they consume at different time intervals, and enable dynamic pricing \cite{Abreu2012a}. Dynamic pricing allows utilities to charge varying prices at different times, for instance, charging a higher price when costly generation sources are used in times of peak demand, and lower prices at night time or weekends when demand is low \cite{Ito2013,Liu2016}. In this chapter, we forecast both 30-minutes ahead using smart meter data, as well as 24-hours ahead to simulate the process made for a day-ahead market.


Firstly, we explore short term load-forecasting at an interval of 30 minutes ahead and cluster similar users based on their electricity usage. A variety of different forecasting techniques were evaluated such as Random Forests \cite{ho1995random}, Long-Short Term Memory neural networks (LSTM) \cite{Hochreiter1997}, Multilayer Perceptron neural networks \cite{book:984557} and \acrfull{svr} \cite{Drucker1997}.

Random Forests are an ensemble-based learning method for classification and regression, and are made up of many decision trees. LSTM networks are recurrent neural networks which remember values over arbitrary time intervals. Multilayer Perceptrons are a popular type of neural network which are made up of a minimum of three layers and can be used to make non-linear predictions. \acrshort{svr}s are supervised learning models which analyse data used for regression analysis.

To improve forecasting results, the clustering of smart meter data was evaluated to identify similar customers, which may have a more similar, and therefore easier to predict, profile. The technique used for this was \textit{k}-means clustering. An average 24-hour electricity load profile was calculated, and the result used for clustering. The clustered sub-system is then aggregated and separate models trained on this aggregate. The yearly, weekly and daily periodicity of electricity load is accounted for by input variables into the models. Once forecasts for each cluster are made using the individual models, the results are aggregated for the final predictions. These predictions are compared to the actual results, and the accuracy measured using mean absolute percentage error (MAPE).

Whilst the work carried out for short-term forecasting does not relate directly to the long-term characteristics of ElecSim and the work presented previously, we used the methodology to inform the long-term characteristics of the electricity market. This work therefore provided the foundation for looking at the long-term using ElecSim.


Therefore, secondly, we introduce day-ahead forecasting and observe the impact errors have on the long-term dynamics of the market. Various studies have looked at predicting electricity demand at various horizons \cite{Andersen2013,Huang2003,Singh2012}. However, the impact of poor demand predictions on the long-term electricity mix has been studied to a lesser degree.

We compare several machine learning and statistical techniques to predict the energy demand for each hour over the next 24-hour horizon. We chose to predict over the next 24 hours to simulate a day-ahead market, which is often seen in decentralised electricity markets. However, our approach could be utilised for differing time horizons. In addition to this, we use our long-term agent-based model, ElecSim \cite{Kell, Kell2020}, to simulate the impact of different forecasting methods on long-term investments, power plant usage and carbon emissions for the years 2018 through 2035 in the United Kingdom. Our approach, however, is generalisable to any country through parametrisation of the ElecSim model.

As part of our work, we utilise online learning methods to improve the accuracy of our predictions. Online learning methods can learn from novel data while maintaining what was learnt from previous data. Online learning is useful for non-stationary datasets and time-series data where recalculation of a model would take a prohibitive amount of time. Offline learning methods, however, must be retrained every time new data is added. Online approaches are constantly updated and do not require significant pauses while the offline training is being re-run. By training, in an offline manner, on data that has already been used for training, the computational load and time required increases.

We trial different algorithms and train different models for different times of the year. Specifically, we train different models for the different seasons. We also split weekdays and train both weekends and holidays together. This is due to the fact that holidays and weekend exhibit similar load profiles due to the reduction in industry electricity use and an increase in domestic. This enables a model to become good at a specific subset of the data which share similar patterns, as opposed to having to generalise to all of the data. Examples of the algorithms used are linear regression, lasso regression, random forests, support vector regression, multilayer perceptron neural network, box-cox transformation linear regression and the passive aggressive model. 

We expect a-priori that online algorithms will outperform the offline approach. This is due to the fact that the demand time-series is non-stationary, and thus changes sufficiently over time. In terms of the models, we presume that the machine learning algorithms, such as neural networks, support vector regression and random forests will outperform the statistical methods such as linear regression, lasso regression and box-cox transformation regression. We expect this due to the fact that machine learning has been shown to be able to learn more complex feature representations than statistical methods \cite{Singh2012}. 

However, it should be noted, that such a-priori intuition, is no substitute for analytical evidence and can (and has) been shown to be wrong in the past, due to imperfect knowledge of the data and understanding of some of the black box models, such as neural networks.

Using online and offline methods, we take the error distributions, or residuals, and fit a variety of distributions to these residuals. We choose the distribution with the lowest sum of squared estimate of errors (SSE). SSE was chosen as the metric to ensure that both positive and negative errors were treated equally, as well as ensuring that large errors were penalised more than smaller errors. We fit over 80 different distributions, which include the Johnson Bounded distribution, the uniform distribution and the gamma distribution. The distribution that best fits the respective residuals is then used and sampled from to adjust the demand in the ElecSim model. We then observe the differences in carbon emissions, and which types of power plants were both invested in and utilised, with each of the different statistical and machine learning methods. To the best of our knowledge, this is the most comprehensive evaluation of online learning techniques to the application of day-ahead load forecasting as well as assessing the impacts of the errors that these models produce on the long-term electricity market dynamics.


We show that online learning has a significant impact on reducing the error for predicting electricity consumption a day ahead when compared to traditional offline learning techniques, such as multilayer artificial neural networks, linear regression, extra trees regression and support vector regression, which are models used in the literature \cite{Ahmad2017, Chen2004,Lu1993}. For a full list of algorithms used in this thesis see Table \ref{table:hyperparameter-tuning-offline}.

We show that the forecasting algorithm has a non-negligible impact on carbon emissions and use of coal, onshore, photovoltaics, reciprocal gas engines and CCGT. Specifically, the amount of coal, photovoltaics, and reciprocal gas used from 2018 to 2035 was proportional to the median absolute error, while both onshore and offshore wind are inversely proportional to the median absolute error.

Total investments in coal, offshore and photovoltaics are proportional to the median absolute error, while investments in CCGT, onshore and reciprocal gas engines are inversely proportional.


The contributions of this work are:

\begin{enumerate}
	\item A methodology to forecast smart-meter electricity demand using the \textit{k}-means clustering technique.
	\item The evaluation of different online and offline learning models to forecast the electricity demand profile 24 hours ahead.
	\item Evaluation of poor predictive ability on the long-term electricity market in the UK through the perturbation of demand in the ElecSim simulation.
\end{enumerate}

\section{Literature review}
\label{forecast:sec:litreview}

In this Section, we carry out a literature review on 30-minute ahead forecasting, day-ahead forecasting and online forecasting methods. In addition, we cover the literature on the impact of forecasting on electricity markets. 

\subsection{30-minute ahead forecasting}

The forecasting of aggregated and clustered electricity demand has been the focus of a considerable amount of research in recent years. The research can generally be classified into two classes, Artificial Intelligence (AI) methods \cite{Kim2000, Tiong2008,Quilumba2014} and classical time series approaches \cite{Huang2003,Nguyen2017}. We investigate both approaches in this chapter.

Singh \textit{et al.} \cite{Singh2012} produced a review of load forecasting techniques and methodologies and reported that hybrid methods, which combine two or more different techniques, are gaining traction, as well as soft computing approaches (AI) such as genetic algorithms. Our work presents a hybrid method which combines \textit{k}-means clustering with multiple different learning algorithms.

\subsubsection{Artificial Intelligence Methods}

Dillon \textit{et al.} \cite{Dillon1991} presented a neural network for short term load forecasting. Their neural network consisted of three-layers and used adaptive learning for training. They proposed the use of weather information to augment their electricity load data. They found better results with the adaptive neural network than with a linear model, or non-adaptive neural network. In contrast to Dillon our work focuses on a non-adaptive neural network and does not take into account weather information.

Chen \textit{et al.}  \cite{Chen1996} used an \Gls{ANN} to predict electricity demand of three substations in Taiwan. They integrated temperature data and reported that the best results when forecasting residential and commercial substations were during the week due to the influence of weather. In contrast to the work done by Chen \textit{et al.}, we focus on client-side prediction using smart meter data as opposed to substation data. We were, therefore, able to cluster the data based on load profile, as opposed to geographical location.

\subsubsection{Time Series Methods}

Al-Musaylh \textit{et al.}\cite{Al-Musaylh2018} proposed the use of \acrfull{svr}, an \acrfull{arima} model and a \acrfull{mars} in their short term electricity demand forecasting system. They found that for a half, and one-hour forecasting horizons, the MARS model outperformed both the ARIMA and SVR.

Taylor \cite{Taylor2008} evaluates different statistical methods including ARIMA, an adaptation of Holt-Winters' exponential smoothing \cite{Holt2004}, and an exponential smoothing method which focuses on the evolution of the intra-day cycle. He found that the double seasonal adaptation of the Holt-Winters' exponential smoothing method was the most accurate method for short lead times between 10 and 30 minutes. 

In contrast to Taylor, Fard \textit{et al.} \cite{Fard2014} proposed a novel hybrid forecasting method based on both artificial intelligence and classical time series approaches. They utilised the wavelet transform, ARIMA and ANNs for short term load forecasting. The ARIMA model is created by finding the appropriate order using the Akaike information criterion \cite{Akaike1974}. The ARIMA model models the linear component of the load time series, and the residuals contain the non-linear components. These residuals are then decomposed by the discrete wavelet transform into its sub-frequencies. ANNs are then applied to these sub-frequencies, and the outputs of both the \acrfull{ann} and ARIMA models are summed to make the final prediction. They found that this hybrid technique outperformed traditional methods. Our work does not integrate artificial intelligence and classical time series techniques, due to time constraints.

\subsubsection{Clustering}

Multiple techniques have been proposed for the clustering of electricity load data prior to forecasting. Both Shu and Luonan \cite{Shu2006}, and Nagi \textit{et al}\cite{Tiong2008} propose a hybrid approach in which self-organising maps are used to cluster the data, and Support Vector Regression is used to make predictions. This technique proved robust for different data types, and was able to tackle the non-stationarity of the data. Shu showed that this hybrid approach outperformed a single SVR technique, whilst Nagi showed superior results to a traditional ANN system. In contrast to both Nagi \textit{et al.} and Shu and Luonan our work utilises \textit{k}-means as the clustering algorithm. 

Quilumba \textit{et al.}\cite{Fard2014} also apply machine learning techniques to individual households' electricity consumption by aggregation. To achieve this aggregation, they use \textit{k}-means clustering to aggregate the households to improve their forecasting ability. The authors also use a neural network-based model for forecasting, and show that the number of optimum clusters for forecasting is dependent on the data, with three clusters optimal for the Con Edison Launches Smart Grid Pilot Program in Queens \cite{queens_dataset}, and four for the Commission for Energy Regulation (CER) dataset \cite{CER_DATASET}.

Wijaya \textit{et al.}\cite{Wijaya2010} demonstrated that implementing clusters improved load-forecasting accuracy up to a certain level. Whilst, a study by Ili\'c \textit{et al.} \cite{Ilic2013} showed that increasing the number of clusters did not improve accuracy.

Humeau \textit{et al.}\cite{Humeau2013} compare MLPs, SVRs and linear regression at predicting smart meter data. They aggregate different households and observe which models work the best at each aggregate level. They find that linear regression outperforms both MLP and SVR when forecasting individual households. However, after aggregating over 32 households, SVR outperforms linear regression.

\subsection{Online learning}

Whilst multiple papers have looked at demand-side forecasting, as shown by the review paper by Singh \textit{et al.}\cite{Singh2012}, to the best of our knowledge, the impact of online learning has been discussed with less frequency. In addition to this, our research models the impact of the performance of different algorithms on investments made, electricity sources dispatched and carbon emissions over a 17 year period. To model this, we use the model ElecSim. In our work, we trial a set of different algorithms to our problem. Due to time and compute constraints, we do not trial the additional techniques discussed in this literature review within this thesis. 

Johansson \textit{et al}.\cite{Johansson2017} apply online machine learning algorithms for heat demand forecasting. They find that their demand predictions display robust behaviour within acceptable error margins. Finding that artificial neural networks (ANNs) provide the best forecasting ability of the standard algorithms and can handle data outside of the training set. Johansson \textit{et al.}, however, do not look at the long-term effects of different algorithms on their application.

Baram \textit{et al}.\cite{Baram2003} combine an ensemble of active learners by developing an active-learning master algorithm. To achieve this, they propose a simple maximum entropy criterion that provides effective estimates in realistic settings. Their active-learning master algorithm is empirically shown to, in some cases, outperform the best algorithm in the ensemble on a range of classification problems.

Schmitt \textit{et al}.\cite{Schmitt2008} also extends on existing algorithms through an extension of the FLORA algorithm \cite{Widmer1996}. The FLORA algorithm generates a rule-based model, which has the ability to make binary decisions. Their FLORA-MC enhances the FLORA algorithm for multi-classification and numerical input values. They use this algorithm for an ambient computing application. Ambient computing is where computing and communication merges into everyday life. They find that their model outperforms traditional offline learners by orders of magnitude.

Similarly to our work, Pindoriya \textit{et al}.\cite{Pindoriya2008} trial several different machine learning methods such as adaptive wavelet neural network (AWNN). They find that AWNN has good prediction properties when compared to other forecasting techniques such as wavelet-ARIMA, multilayer perceptron (MLP) and \acrfull{rbf} neural networks as well as the fuzzy neural network (FNN).

Goncalves Da Silva \textit{et al}.\cite{GoncalvesDaSilva2014} show the effect of prediction accuracy on local electricity markets. To this end, they compare forecasting of groups of consumers in comparison to single individuals. They trial the use of the Seasonal-Naïve and Holt-Winters algorithms and look at the effect that the errors have on trading in an intra-day electricity market of consumers and prosumers. They found that with a photovoltaic penetration of 50\%, over 10\% of the total generation capacity was uncapitalised and roughly 10, 25 and 28\% of the total traded volume were unnecessary buys, demand imbalances and unnecessary sells respectively. This represents energy that the participant has no control. Uncapitalised generation capacity is where a participant could have produced energy; however, it was not sold on the market. Additionally, due to forecast errors, the participant might have sold less than it should have. Our work, however, focuses on a national electricity market, as opposed to a local market.

\section{Methods}
\label{forecast:sec:methods}

In this section, we explore the principles behind the methods used in this chapter.


%
%
%
%

\subsection{Error Metrics}

\subsubsection{Mean Absolute Percentage Error}

The \acrfull{mape} is a measure of prediction accuracy which is used in this thesis. It can be defined as follows:

\begin{equation}
MAPE=\frac{1}{n}\sum_{i=1}^n\left|\frac{y_i-\hat{y}_i}{y_i}\right|\times 100\%,
\end{equation}

\noindent where $y_i$ is the actual value, $\hat{y}_i$ is the forecast value and $n$ is the number of points forecast \cite{Li2016}.

\subsubsection{Root Mean Squared Error}

The root mean squared error (RMSE) is a measure between the values predicted by a model and the observed values. The RMSE is the sample standard deviation of the differences between the predicted and observed values.

The RMSE is defined as follows:
\begin{equation}
RMSE = \sqrt[]{\frac{\sum_{t=1}^n(\hat{y}_i-y_i)^2)}{n}},
\end{equation}

\noindent where $\hat{y}_i$ are the predicted values, $y_i$ are the observed values, and $n$ is the number of observations.

\subsubsection{Mean Absolute Scaled Error}

The mean absolute scaled error (MASE) is a measure of the accuracy of forecasts \cite{Hyndman2006}. It is defined as the mean absolute error of the forecast values, divided by the mean absolute error of the in-sample one-step naive forecast. MASE can be scaled across different scales, has symmetry for both positive and negative errors, is interpretable and has predictable behaviour for a value of zero. 

MASE can be defined as follows:

\begin{equation}
MASE = mean\left(\frac{|e_j|}{\frac{1}{T-1}\sum_{t=2}^T|Y_t-Y_{t-1}|}\right)=\frac{\frac{1}{J}\sum_j|e_j|}{\frac{1}{T-1}\sum_{t=2}^{T}|Y_t-Y_{t-1}|},
\end{equation}

\noindent where $e_j$ is the forecast error for a given period, $J$ is the number of forecasts. Where $e_j$ is defined as the actual value ($Y_j$) minus the forecast value ($F_j$) for that period. The denominator is the mean absolute error of the one-step naive forecast method on the training set. This naive forecast is the actual value from the prior period, or $F_t=Y_{t-1}$. $T$ is the total number of forecasts.


\subsection{Machine learning}

Machine learning is a methodology for finding and describing structural patterns in data \cite{Witten2011}. Offline learning models are trained with the data available at a single point in time. With non-stationary data -- where underlying distributions change -- the model must be retrained at periodic intervals, determined by how quickly the model goes out of step with the true data. With online learning, the model is able to retrain every time a new data point becomes available, without having to retrain the entire model. This makes these models good for time-series data which exhibit moderate to significant non-stationary properties, such as electricity demand profiles.

\subsection{Online learning}


Examples of online learning algorithms are Passive Aggressive (PA) Regressor \cite{Gzik2014}, Linear Regression, Box-Cox Regressor \cite{Box1964}, K-Neighbors Regressor \cite{forgy65} and Multilayer perceptron regressor \cite{Hinton1989}. For our work, we trial the stated algorithms, in addition to a host of offline learning techniques. The offline techniques trialled were Lasso regression \cite{Tibshirani1996a}, ridge regression \cite{GeladiPaul1994Mrac},  Elastic Net \cite{Geostatistics2010}, Least Angle Regression \cite{Fike1988}, Extra Trees Regressor \cite{Fike1988}, Random Forest Regressor \cite{Breiman2001}, AdaBoost Regressor \cite{Freund1997}, Gradient Boosting Regressor \cite{316} and Support vector regression \cite{Cortes1995}. We chose the boosting and random forest techniques due to our previous successes of these algorithms when applied to electricity demand forecasting \cite{Kell2018a}. We trialled the additional algorithms due to the availability of these algorithms using the scikit-learn package and online learning package, Creme \cite{creme, scikit-learn}. 


\subsection{Linear regression models}

Linear regression is an approach to modelling linear relationships between a dependent variable and one or more independent variables. Linear regressions can be used for both online and offline learning. In this thesis, we use them for both online and offline learning. Linear regression models are often fitted using the least-squares approach. The least-squares approach minimises the sum of the squares of the residuals. 

Other methods for fitting linear regressions are by minimising a penalised version of the least-squares cost function, such as in ridge and lasso regression \cite{GeladiPaul1994Mrac,Tibshirani1996a}. Ridge regression is a useful approach for mitigating the problem of multicollinearity in linear regression. Multicollinearity is where one predictor variable can be linearly predicted from the others with a high degree of accuracy. This phenomenon often occurs in models with a large number of parameters. 

In ridge regression, the OLS loss function is augmented so that we not only minimize the sum of squared residuals but also penalized the size of parameter estimates, in order to shrink them towards zero:
\begin{equation}
L_{ridge}(\hat{\beta})=\sum^n_{i=1}(y_i-x'_i\hat{\beta})^2+\lambda\sum^m_{j=1}\hat{\beta^2_j}=||Y-X\hat{\beta}||^2+\lambda||\hat{\beta}||^2,
\end{equation}

\noindent we solve for $\hat{\beta}$, where $\hat{\beta}$ are the OLS parameter estimates. $\hat{\beta}$  is a vector. $n$ is the number of observations of the response variable, $Y$, Where $Y=\{y_0, y_1, \ldots,y_n\}$. With a linear combination of $m$ predictor variables, $X$, where $x=\{x_0,x_1,\ldots,x_n\}$. $\lambda$ is the regularisation penalty which can be chosen through cross-validation, or the value that minimises the cross-validated sum of squared residuals, for instance.

Lasso is a linear regression technique which performs both variable selection and regularisation. It is a type of regression that uses shrinkage. Shrinkage is where data values are shrunk towards a central point, such as the mean. The lasso model encourages models with fewer parameters. This enables the selection of models with fewer parameters, or automate the process of variable selection.

Under lasso the loss is defined as:

\begin{equation}
L_{lasso}(\hat{\beta})=\sum^n_{i=1}(y_i-x'_i\hat{\beta})^2+\lambda\sum^m_{j=1}|\hat{\beta}_j|.
\end{equation}

The only difference between lasso and ridge regression is the penalty term. Elastic net is a regularization regression that linearly combines the penalties of the lasso and ridge methods. Specifically, Elastic Net aims to minimize the following loss function:
\begin{equation}
L_{enet}(\hat{\beta})=\frac{\sum^n_{i=1}(y_i-x'_i\hat{\beta})^2}{2n}+\lambda(\frac{1-\alpha}{2}\sum^m_{j=1}\hat{\beta}^2_j+\alpha\sum^m_{j=1}|\hat{\beta_j}|),
\end{equation}
where $\alpha$ is the mixing parameter between ridge ($\alpha=0$) and lasso ($\alpha=1$). The two parameters $\lambda$ and $\alpha$ can be tuned. Least Angle Regression (LARS) provides a means of producing an estimate of which variables to include in a linear regression, as well as their coefficients.

%

\subsection{Decision tree-based algorithms}

The decision tree is a model which goes from observations to output using simple decision rules inferred from data features \cite{Quinlan}. To build a decision tree, recursive binary splitting is used on the training data. Recursive binary splitting is a greedy top-down algorithm used to minimise the residual sum of squares. The RSS, in the case of a partitioned feature space with $M$ partitions, is given by:

\begin{equation}
RSS=\sum^M_{m=1}\sum_{i\in R_m}(y-\hat{y}_{R_m})^2.
\end{equation}

\noindent Where $y$ is the value to be predicted, $\hat{y}$ is the predicted value for partition $R_m$.

Beginning at the top of the tree, a split is done into two branches. This split is carried out multiple times and the split is chosen that minimises the current RSS. To obtain the best sequence of subtrees cost complexity, pruning is used as a function of $\alpha$. $\alpha$ is a tuning parameter that balances the depth of the tree and the fit to the training data. This parameter is used to see how deep the tree should be. This parameter can be tuned using cross-validation. 

The AdaBoost training process selects only the features of a model known to improve the predictive power of the model \cite{Freund1997}. By doing this, the dimensionality of the model is reduced and can improve compute time. This can be used in conjunction with multiple different models. In this thesis, we utilise the decision tree-based algorithm with AdaBoost.

Random Forests are an ensemble learning method for classification and regression \cite{Breiman2001}. Ensemble learning methods use multiple learning algorithms to obtain better predictive performance. They work by constructing multiple decision trees at training time, and outputting the predicted value that is the mode of the predictions of the individual trees.

To ensure that the individual decision trees within a Random Forest are not correlated, bagging is used to sample from the data. Bagging is the process of randomly sampling with replacement of the training set and fitting the trees. This has the benefit of reducing the variance of the model without increasing the bias. 

Random Forests differ in one way from this bagging procedure. Namely, using a modified tree learning algorithm that selects, at each candidate split in the learning process, a random subset of the features, known as feature bagging. Feature bagging is undertaken due to the fact that some predictors with a high predictive ability may be selected many times by the individual trees, leading to a highly correlated Random Forest.

ExtraTrees adds one further step of randomisation \cite{Fike1988}. ExtraTrees stands for extremely randomised trees. There are two main differences between ExtraTrees and Random Forests. Namely, each tree is trained using the whole learning sample (and not a bootstrap sample), and the top-down splitting in the tree learner is randomised. That is, instead of computing an optimal cut-point for each feature, a random cut-point is selected from a uniform distribution. The split that yields the highest score is then chosen to split the node.

\subsection{Gradient Boosting}


Gradient boosting is a machine learning technique which is used for both regression and classification problems \cite{316}. Typically, decision trees are used to produce a prediction model in the form of an ensemble of weak prediction models. The model is built in a stage-wise fashion. These model types are generalisable by allowing for an optimisation of an arbitrary differentiable loss function.

The outline of the algorithm is defined here. As discussed, in many supervised learning problems, one has an output variable $y$ and a vector of input variables $x$. These two variables are described via a joint probability distribution $P(x,y)$. The goal of supervised learning is to use a training set $\{(x_1,y_1),\ldots,(x_n,y_n))\}$ of known $x$ and $y$ values to find an approximation of $F(x)$, $\hat{F}(x)$. $\hat{F}(x)$ is found by minimizing the expected value of some specified loss function $L(y,F(x))$:

\begin{equation}
\hat{F}=\underset{F}{\arg\min}\mathbb{E}_{x,y}[L(y,F(x))].
\end{equation}

The gradient boosting method assumes a real-valued $y$ and seeks an approximation $\hat{F}(x)$, where $\hat{F}(x)$ is a weighted sum of function $h_i(x)$ from some class $H$. Where $H$ are called weak leaners:

\begin{equation}
\hat{F}(x)=\sum_{i=1}^{M}\gamma_i h_i(x)+const.
\end{equation}

This method tries to find an approximation $\hat{F}(x)$ that minimizes the average value of the loss function on the training set. It does so, by starting with a constant function model. $F_0(x)$ and incrementally expands it in a greedy fashion:

\begin{equation}
F_0(x)=\underset{\gamma}{\arg\min}\sum_{i=1}^{n}L(y_i,\gamma),
\end{equation}

\begin{equation}
F_m(x)=F_{m-1}(x)+\underset{h_m\in\mathcal{H}}{\arg\min}\left[\sum_{i=1}^{n}L(y_i,F_{m-1}(x_i)+h_m(x_i))\right],
\end{equation}

\noindent where $h_m\in H$ is a weak learner function. Due to the fact that choosing the best function $h$ is computationally infeasible, an approximation must be sought. This approximation is found using functional gradient descent to this minimization problem. If we consider the continuous case, i.e. where $H$ is the set of arbitrary differentiable functions on $\mathbb{R}$, we would update the model in accordance with the following equations:

\begin{equation}
F_m(x)=F_{m-1}(x)-\gamma_m\sum_{i=1}^{n}\nabla_{F_{m-1}}L(y_i,F_{m-1}(x_i)),
\end{equation}

\begin{equation}
\gamma_m=\underset{\gamma}{\arg\min}\sum_{i=1}^{n}L(y_i,F_{m-1}(x_i)-\gamma\nabla_{F_{m-1}}L(y_i,F_{m-1}(x_i))),
\end{equation}

\noindent where the derivatives are taken with respect to the function $F_i$ for $i \in \{1,\ldots,m\}$ and $\gamma_m$ is the step length.

\subsection{Support vector regression}

Support vector regression is an algorithm which finds a hyperplane and decision boundary to map an input domain to an output \cite{Cortes1995}. The hyperplane is chosen by minimising the error within a certain tolerance.

Suppose we have the training set: $(x_1,y_1), \ldots,(x_i,y_i),\ldots,(x_n,y_n)$, where $x_i$ is the input, and $y_i$ is the output value of $x_i$. Support Vector Regression solves an optimization problem \cite{Chen2004, Shu2006}, under given parameters $C>0$ and $\varepsilon >0$, the form of support vector regression is \cite{Drucker1997}: 

\begin{equation}
\min_{\omega,b,\xi,\xi^{*}}\frac{1}{2}\omega^T\omega+C\sum_{i=1}^{n}(\xi_i+\xi_i^*),
\end{equation}

\noindent subject to:
\begin{align}
\begin{multlined}
\label{svr:constrains}
y_i-(\omega^T\phi(x_i)+b)\leq\varepsilon+\xi_i^{*},\\
(\omega^T\phi(x_i)+b)-y_i\leq\varepsilon+\xi_i,\\
\text{with }\xi_i,\xi^*_i\geq0,i=1,\ldots,n,
\end{multlined}
\end{align}

\noindent $x_i$ is mapped to a higher dimensional space using the function $\phi$. The $\varepsilon$-insensitive tube $(\omega^T\phi(x_i)+b)-y_i\leq\varepsilon$ is a range in which errors are permitted. $\xi_i$ and $\xi^*_i$ are slack variables which allow errors for data points which fall outside of $\varepsilon$. This enables the optimisation to take into account the fact that data does not always fall within the $\varepsilon$ range \cite{Smola2004}.

The constant $C>0$ determines the trade-off between the flatness of the support vector function. $\omega$ is the model fit by the SVR. The parameters which control regression quality are the cost of error $C$, the width of the tube $\varepsilon$, and the mapping function $\phi$ \cite{Chen2004, Shu2006}.

\subsection{\textit{k}-Neighbours Regressor}

\textit{k}-Neighbours regression is a non-parametric method used for regression \cite{forgy65}. One such \textit{k}-NN algorithm uses a weighted average of the \textit{k} nearest neighbours, weighted by the inverse of their distance.

The algorithm works as follows:
\begin{enumerate}
	\item Compute the Euclidean or Mahalanobis distance from the test example to the labelled examples.
	\item Order the labelled examples by increasing distance.
	\item Find a heuristically optimal number \textit{k} of nearest neighbours, based on cross-validation.
	\item Calculate an inverse distance weighted average with the \textit{k}-nearest multivariate neighbours.
\end{enumerate}

\subsection{Neural Networks}

\begin{figure}
	\centering
	\includegraphics[width=0.4\textwidth]{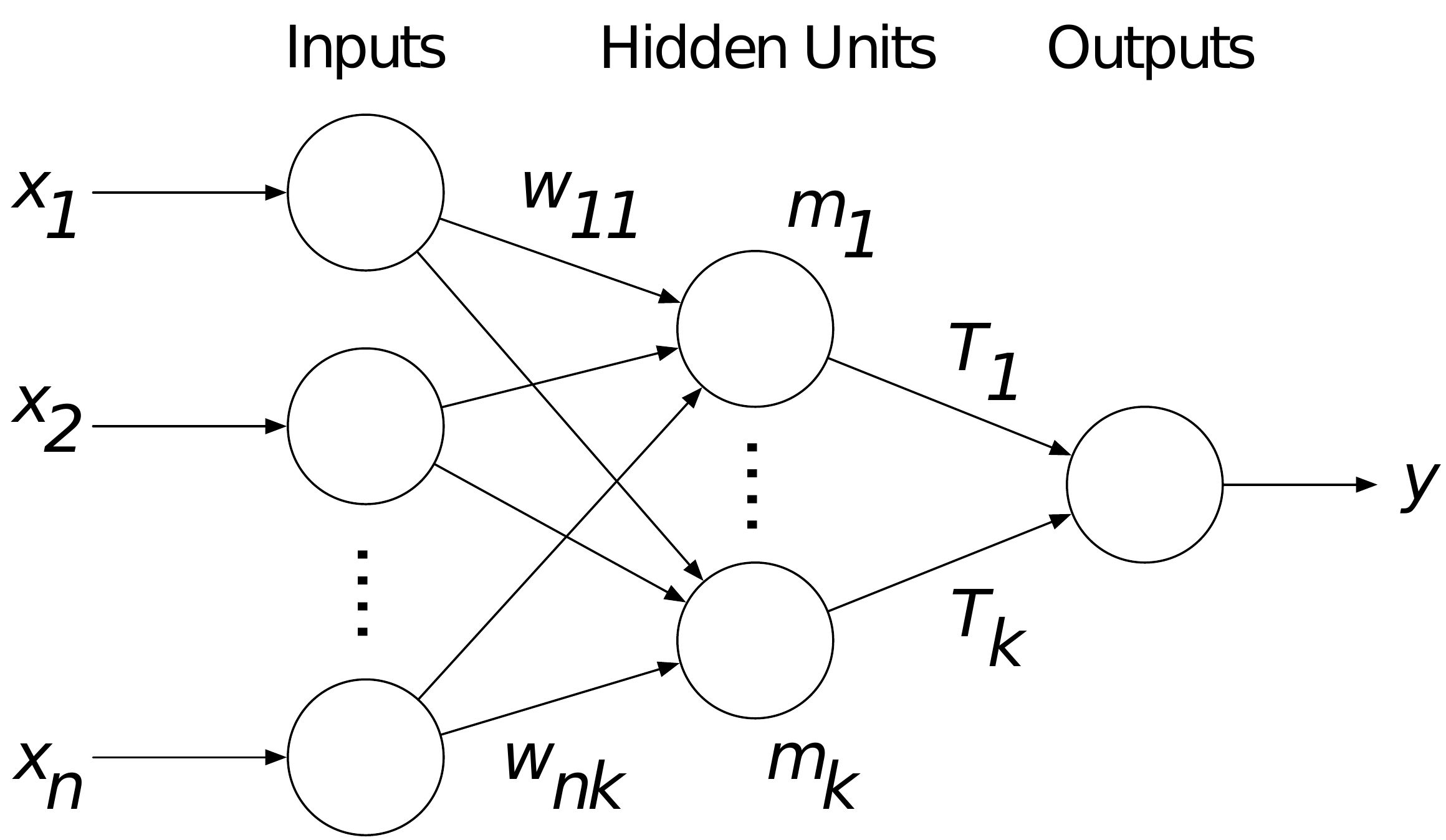}
	\caption{A three-layer feed forward neural network.}
	\label{fig:mlp}
\end{figure}

A neural network can be used in both offline and online cases. In this thesis, we used them for both online and offline.

Artificial Neural Networks are a model which can model non-linear relationships between input and output data \cite{Akaike1974}. A popular neural network is a feed-forward multilayer perceptron. Fig. \ref{fig:mlp} shows a three-layer feed-forward neural network with a single output unit, \textit{k} hidden units, $n$ input units. $w_{ij}$ is the connection weight from the $i^{th}$ input unit to the $j^{th}$ hidden unit,  and $T_j$ is the connecting weight from the $j^{th}$ hidden unit to the output unit \cite{Pao2007}. These weights transform the input variables in the first layer to the output variable in the final layer using the training data. 


For a univariate time series forecasting problem, suppose we have N observations $y_1, y_2, \ldots, y_N$ in the training set, and $m$ observations in the test set, $y_{N+1}, y_{N+2}, \ldots, y_{N+m}$, and we are required to predict \textit{m} periods ahead \cite{Pao2007}. 

The training patterns are as follows:
\begin{align}
y_{p+m} & =f_{W}(y_p, y_{p-1},\ldots,y_1)\\
y_{p+m+1} & =f_{W}(y_{p+1}, y_{p},\ldots,y_2)\\
&\vdotswithin  \notag \\
y_{N} & =f_{W}(y_{N-m},y_{N-m-1},\ldots,y_{N-m-p+1})
\end{align}

\noindent where $f_{W}(\cdot)$ represents the MLP network and $W$ are the weights. For brevity we omit $W$. The training patterns use previous time-series points, for example, $y_p, y_{p-1},\ldots,y_1$ as the time series is univariate. Where $p$ is the total number of points in the univariate time-series. That is, we only have the time series in which we can draw inferences from. In addition, these time series points are correlated, and therefore provide information that can be used to predict the next time point.

The $m$ testing patterns are:

\begin{align}
y_{N+1} & =f_{W}(y_{N+1-m}, y_{N-m},\ldots,y_{N-m-p+2}),\\
y_{N+2} & =f_{W}(y_{N+2-m}, y_{N-m+1},\ldots,y_{N-m-p+3}),\\
&\vdotswithin  \notag \\
y_{N+m} & =f_{W}(y_{N},y_{N-1},\ldots,y_{N-p+1}).
\end{align}

The training objective is to minimise the overall predictive mean sum of squared estimate of errors (SSE) by adjusting the connection weights. For this network structure the SSE can be written as:
\begin{equation}
SSE = \sum_{i=p+m}^N(y_i-\hat{y}_i),
\end{equation}

\noindent where $\hat{y}_i$ is the prediction from the network. The number of input nodes corresponds to the number of lagged observations. Having too few or too many input nodes can affect the predictive ability of the neural network \cite{Pao2007}.

It is also possible to vary the hyperparameter or the number of input units. Typically, various different configurations of units are trialled, with the best configuration being used in production. The weights $W$ in $f_W$ are trained using a process called backpropagation \cite{rumelhart1986learning}, which uses labelled data and gradient descent to update and optimise the weights.

\subsection{Long-Short Term Neural Networks}

\acrfull{lstm} \cite{Hochreiter1997} is an artificial recurrent neural network (RNN) architecture, which is used in the field of deep learning. They are able to process entire sequences of time-series data. A common LSTM unit is composed of a cell, an input gate, an output gate and a forget gate. The cell is able to remember values over arbitrary time intervals and the three gates regulate the flow of information into and out of the cell. 

LSTMs are well suited to time-series data, due to the presence of lags between important events, with an unknown duration. LSTMs have an advantage over traditional RNNs by solving the vanishing gradient problem. This is where the gradients which are back-propagated can tend to zero or infinity. This is due to the computations involved in the process, which use finite-precision numbers. LSTMs partially solve this problem by allowing gradients to flow unchanged in the back-propagation process.

\subsection{Online Algorithms}

In this section we discuss the algorithms which were used exclusively for online learning in this thesis.

\subsection{Box-Cox regressor}

In this subsection, we discuss the Box-Cox regressor. Ordinary least squares is a method for estimating the unknown parameters in a linear regression model. It estimates these unknown parameters by the principle of least squares. Specifically, it minimises the sum of the squares of the differences between the observed variables and those predicted by the linear function.

The ordinary least squares regression assumes a normal distribution of residuals. However, when this is not the case, the Box-Cox Regression may be useful \cite{Box1964}. It transforms the dependent variable using the Box-Cox Transformation function and employs maximum likelihood estimation to determine the optimal level of the power parameter lambda. The Box-Cox Regression requires that no dependent variable has any negative values.

Variable selection and ordinary least squares output dialogues are identical to that of linear regression. 

The Box-Cox regression will transform the dependent variable as follows:

\begin{equation}
y^{(\lambda)} = \frac{y^{\lambda}-1}{\lambda}\:\text{if} \:\lambda\neq0,
\end{equation}
\begin{equation}
y^{(\lambda)} = Ln(y)\; \text{if}\: \lambda=0.
\end{equation}

\noindent Where $\lambda$ is the power parameter, and the data vectors are $y_i=(y_1,\ldots,y_n)$. The optimal value of ($\lambda$) is determined by maximising the following log-likelihood function:

\begin{equation}
L^{(\lambda)}=-\frac{n}{2}Ln(\hat{\sigma}^2_{(\lambda)}+(\lambda - 1)\sum_{i=1}^nLn(y_i),
\end{equation}

\noindent where $\hat{\sigma}^2_{(\lambda)}$ is the estimate of the least squares variance using the transformed y variable. 

\subsection{Passive-Aggressive regressor}

The goal of the Passive-Aggressive (PA) algorithm is to change itself as little as possible to correct for any mistakes and low-confidence predictions it encounters \cite{Gzik2014}. Specifically, with each example PA solves the following optimisation \cite{Ma2009}:

\begin{align}
\boldsymbol{w}_{t+1}\leftarrow \text{argmin} \frac{1}{2}\left|\left|{\boldsymbol{w}_t-\boldsymbol{w}}\right|\right|^2 \\
s.t. \; \; y_i(\boldsymbol{w}\cdot \boldsymbol{x}_t)\geq1.
\end{align}

\noindent Where $x_t$ is the input data and $y_i$ the output data, and $w_t$ are the weights for the passive aggressive algorithm. Updates occur when the inner product does not exceed a fixed confidence margin - i.e., $y_i(\boldsymbol{w}\cdot \boldsymbol{x}_t)\geq1$. The closed-form update for all examples is as follows:
\begin{equation}
\boldsymbol{w}_{t+1}\leftarrow \boldsymbol{w}_{t} + \alpha_t y_t \boldsymbol{x}_t,
\end{equation}

\noindent where 

\begin{equation}
\alpha_t=max\left\{\frac{1-y_t(\boldsymbol{w}_t\cdot\boldsymbol{x}_t)}{\left|\left|\boldsymbol{x}_t\right|\right|^2},0\right\}. 	
\end{equation}

\noindent $a_t$ is derived from a derivation process which uses the Lagrange multiplier. For full details of the derivation see \cite{Gzik2014}.

\section{Short-term demand forecasting}
\label{forecast:sec:shortterm}

In this section we present the work undertaken for short-term demand forecasting. We forecast 30-minutes ahead using smart meter data. 

\subsection{Methodology}
\subsubsection{The Data}

Smart meter data obtained from the Irish Social Science Data Archive (ISSDA) on the 28th of September 2017 was used in this study \cite{cer_2012}. The Commission for Energy Regulation released a public dataset of anonymised smart meter data from the "\textit{Electricity Smart Metering Customer Behaviour Trials}" \cite{cer_2012}. This dataset is made up of over 5,000 Irish homes and businesses and is sampled at 30-minute intervals.

The data was recorded between the 14th July 2009 and 31st December 2010, providing 17 months worth of data. For the purposes of cross-validation, the data was split into two partitions, the training set and the testing set. The training set made up the first 11 months of data and was used to parametrise the models, whereas the test set is made up of the remaining six months of data. This split was chosen to balance the amount of training data with the test data and to give the models a chance to learn the periodicity inherent in a twelve month period of electricity load. The test set was used for evaluation of the models proposed. Due to the long training times for these algorithms, we worked with a sub-sample of 709 individual Irish homes from the whole dataset. However, we believe that our results would hold over the full dataset.

\begin{figure}
	\centering
	\includegraphics[width=0.95\textwidth]{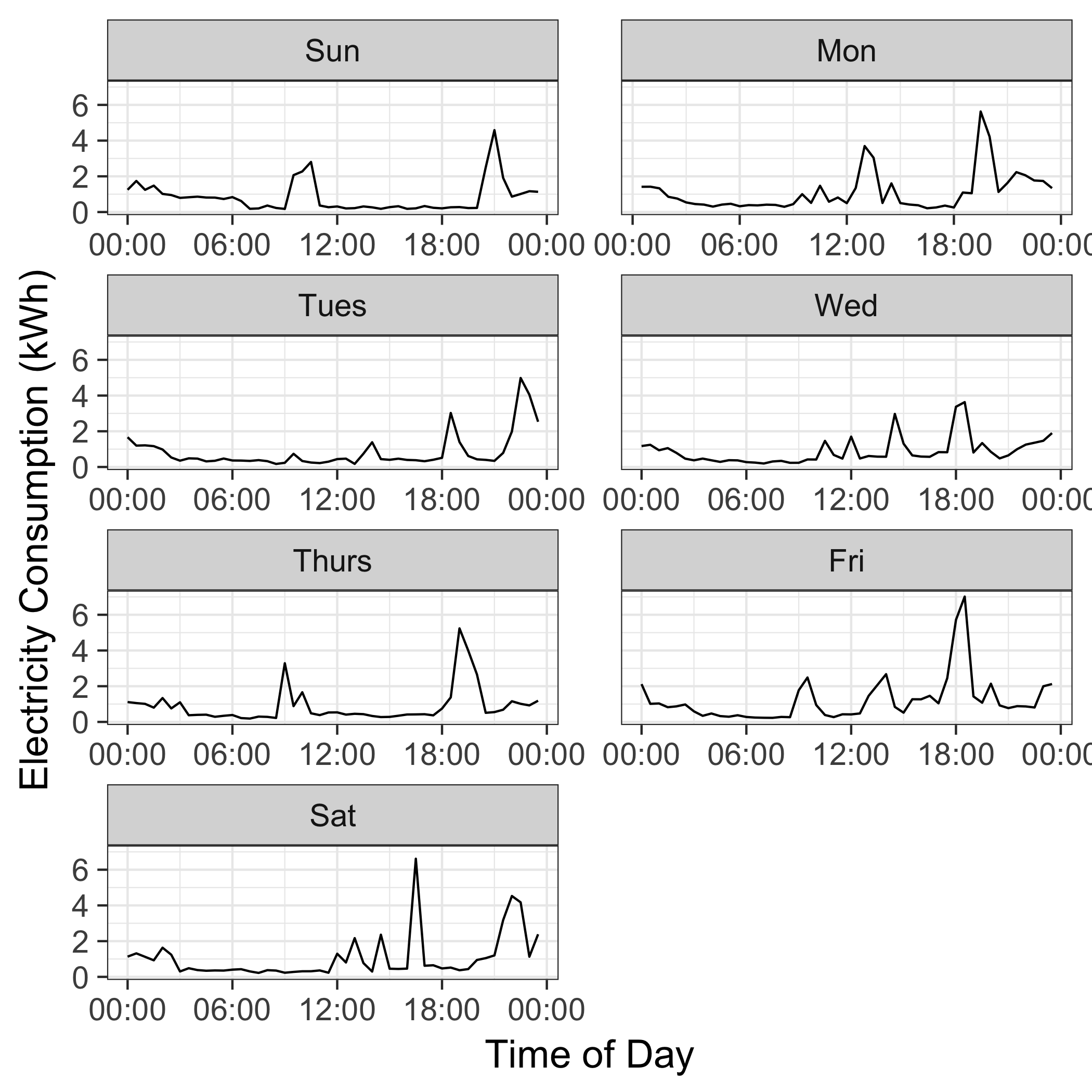}
	\caption{Daily smart meter data load profiles of a single customer over a week between 20th July 2009 and 27th July 2009. }
	\label{fig:single_user}
\end{figure}

Figure \ref{fig:single_user} demonstrates the electricity consumption profile of a single week for a single user. Whilst it can be seen that electricity usage changes significantly between days, a pattern of behaviour is exhibited. There is a large peak displayed each day in the evening, as well as a peak earlier during the day. It can, therefore, be assumed that this customer has some form of habitual behavioural pattern. 

\begin{figure}
	\centering
	\includegraphics[width=0.95\textwidth]{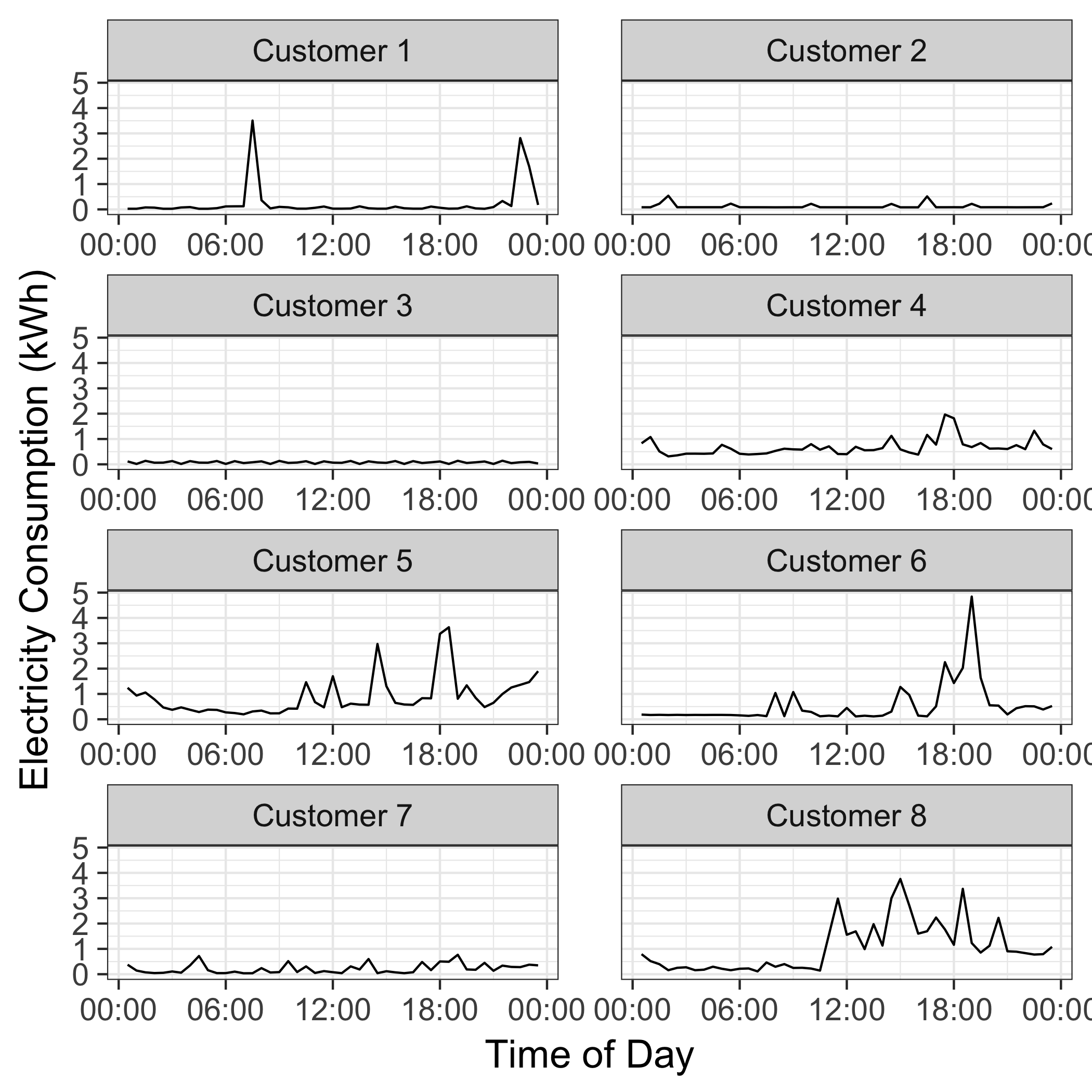}
	\caption{Daily smart meter data load profiles of different customers over a single day on the 22nd June 2009.}
	\label{fig:multiple_users}
\end{figure}

Figure \ref{fig:multiple_users} shows eight different residential customer load profiles on the 22nd June 2009. It can be seen that the daily load profile changes between each customer. The consumers use varying quantities of electricity and at different times.

These figures display that electricity consumption changes per person per day. To capture this variability between customer types, these customers are clustered and then aggregated. Each of the different aggregated electricity consumptions should provide a less stochastic load profile, and therefore increase the accuracy of the models.

\subsubsection{Clustering}

We propose that clustering similar customer load profiles and aggregating each cluster's electricity consumption improves the accuracy of the models. 

\begin{figure}
	\centering
	\includegraphics[width=0.8\textwidth]{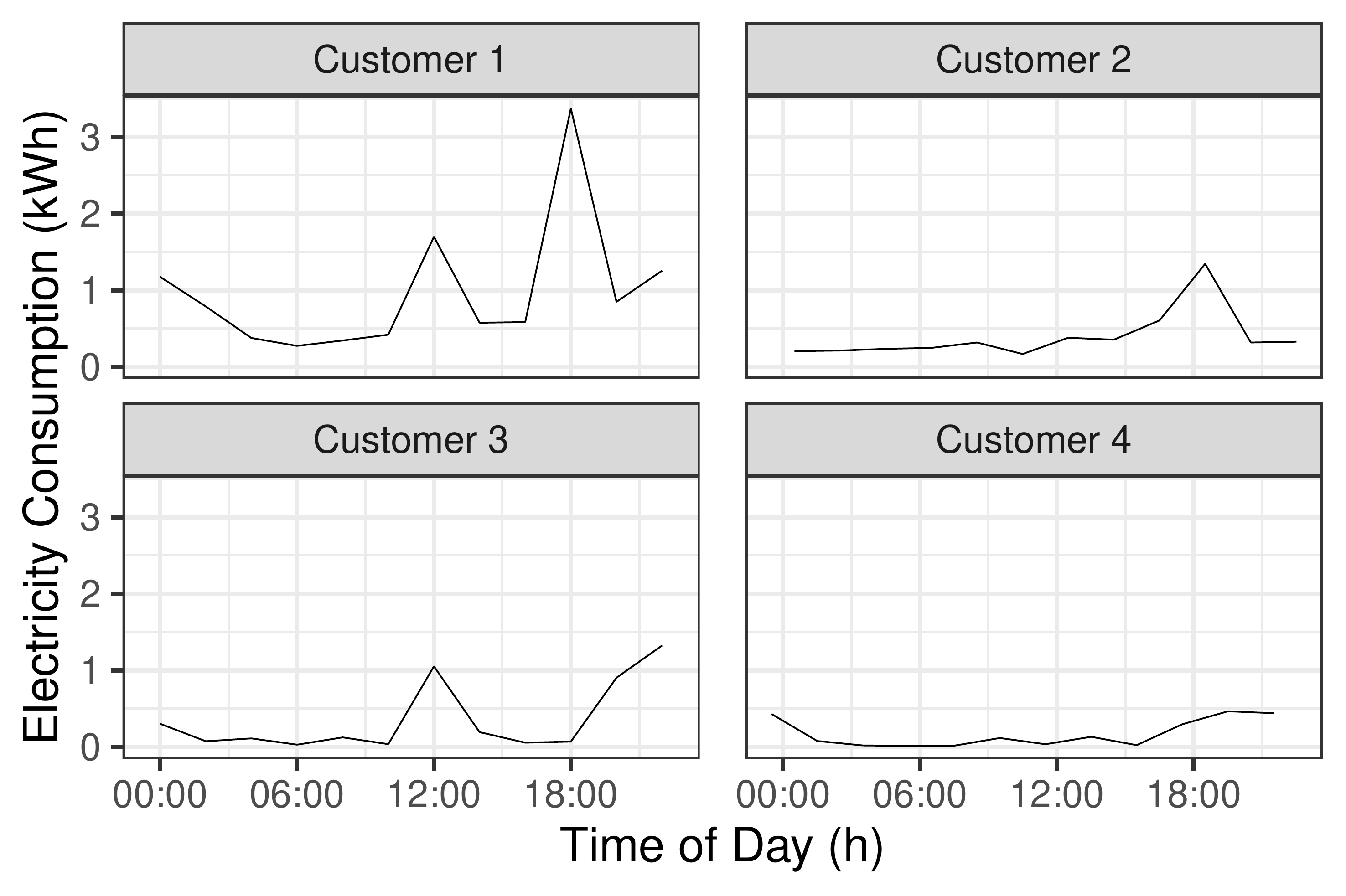}
	\caption{Figure showing similar smart meter data load profiles for four different customers on the 22nd July 2009.}
	\label{fig:similar_customers}
\end{figure}

Figure \ref{fig:similar_customers} displays four different customers with similar load profiles. Each of the users display a strong peak in electricity consumption during the evening and less consumption during the day. These customers may potentially be clustered together by the \textit{k}-means clustering algorithm.

To cluster the load profiles, different options were considered. Hierarchical clustering using metrics such as Euclidean and wavelet distance metrics were evaluated \cite{BIMJ:BIMJ4710240520}, as was \textit{k}-means \cite{forgy65}.\textit{ K}-means was demonstrated to be the most robust and best-performing clustering algorithm, and thus was chosen for use in this thesis.

To select the optimum number of clusters (\textit{k}) cross-validation was explored. This allowed us to compare the results of each of the models and select \textit{k} with the highest MAPE accuracy.

The cross-validation method proposed, worked by trying a different number of clusters per model, and testing for the resulting MAPE. The optimum number of clusters with a low MAPE is then chosen. In this thesis we varied \textit{k} between 1 and 7; this range was chosen due to the fact that the error did not vary greatly past seven clusters. We fit multiple models per cluster and predicted six months of electricity consumption.

With \textit{k}-means clustering, it is possible that with the same initialisation number of clusters, different clusters are formed. This is due to the algorithm converging at a local minima. To overcome local minima, the \textit{k}-means algorithm is run multiple times, and the partition with the smallest squared error is chosen \cite{Jain2010}. In our case, the \textit{k}-means clustering algorithm is run 1,000 times to reduce the chance of finding a local minima. 

The clustering technique utilised in our work was a scaled input approach. The daily load profile was averaged for each customer based on each day of the training data. The data was then scaled so that households of different sizes, but with similar usage profiles were clustered together. This data, which is made up of a \textit{m-by-n} matrix, where \textit{m} is equal to the total number of meters and \textit{n} is equal to 48 (two readings for each hour in the day).

To find the optimum number of clusters it is recommended that the user selects a value of $k$ that is high enough that distinct average load profiles are displayed, however, not so high that well-clustered customers are split. By doing this, the stochasticity of the load profiles in each of the clusters will be reduced, and thus lead to the best results.

\subsubsection{Aggregating Demand}

Once each customer is assigned to their respective cluster using \textit{k}-means clustering, the total electricity consumed per cluster is aggregated. This is achieved by summing the electricity consumed at each time interval per cluster. This creates a partial system load. A different model is trained on each of the different partial system loads, and the resultant forecasts are aggregated to generate the total system load forecast. The total system load forecast is then used to evaluate the accuracy of each of the different models using MAPE. 

Random Forests, Support Vector Regression, Multilayer Perceptron neural networks and Long-Short Term Memory neural networks were evaluated, and a comparison between the different models were made. 

These models were chosen due to their ability to model multivariate non-linear relationships. They are data-driven methods and therefore suited to this type of problem.

\subsubsection{Feature Selection}

Each component of the training data is known as a feature. Features encode information from the data that may be useful in predicting electricity consumption. 

\subsubsection{Calendar Attributes}

Due to the daily, weekly and annual periodicity of the electricity consumption daily calendar attributes may be useful to model the problem. The calendar attributes included are as follows:

\begin{itemize}
	\item Hour of day
	\item Day of the month
	\item Day of the week
	\item Month
	\item Public holidays
\end{itemize}

These attributes enable the daily, weekly and annual periodicity to be taken into account by the model. We chose these due to the daily, monthly and weekly variations for electricity demand. It is also observed that electricity demand is changed on a public holiday.

It is noted that electricity consumption changes on a public holiday such as Christmas or New Year's Eve. For example, a hotel may consume more electricity and an office space less. It is therefore proposed that public holidays in Ireland are input into the model as features. 

For testing purposes, two sets of models for Random Forests, Multilayer Perceptrons and Support Vector Regression were fit. One set omitted these calendar attributes whilst the other didn't. This is done to evaluate the importance of periodicity in electricity consumption prediction.

\subsubsection{Time Series Data}

As well as the calendar attributes, it is important to consider the historical load demand. This allows the time-series element to be modelled.  

To do this, a lagged input of the previous 3 hours, the equivalent three hours from the previous day, and the equivalent 3 hours from the previous week were used. For example, to predict the electricity consumed on the 21st December 2010 at 12:00 pm the electricity between 9:00 pm and 11:30 pm on the 21st of December are used as inputs, as are the times between 9:00 pm and 12:00 pm on the 20th and 14th of December.

Long-Short Term Memory neural networks remember values over arbitrary time intervals. They can remember short-term memory over a long period of time; for this reason, five lagged inputs of the previous two and a half hours were used as features to the Long-Short Term Memory network.

\subsubsection{Data Representation}

Once useful information is selected, we must encode the data for input into the models. To encode the day of the week, seven binaries are utilised in the form of one-hot encoding. Six of the binaries are for Monday through to Saturday. When all six binaries are equal to zero Sunday is encoded. A single binary for public holidays is included. Eleven binaries are used for month of the year, with the first eleven representing January to November, with December represented by all zeros in the calendar binaries. The current hour and date are input using a numerical attribute. The lagged data inputs, such as previous hour's electricity usage are also inputted using a numerical attribute for each entry, totalling 20 attributes (six half-hourly entries for each 3 hour period multiplied by three days plus 2 entries for the time to be predicted on the previous day and week). Table \ref{tab:feature} displays these features.

\begin{table}
	\begin{tabular}{p{3cm}p{3cm}p{8cm}}
		\toprule
		Input & Variable      & Detail description \\
		\midrule
		1     & Hour          & Single numeric input representing hour of the day                                                                                              \\
		2     & Day of month  & Single numeric input representing day of the month                                                                                             \\
		3-9   & Day of week   & Six binary digits representing calendar information regarding day of the week                                                                                            \\
		10-21 & Month         & Eleven binary digits representing calendar information regarding month                                                                                         \\
		22-42 & Lagged inputs & Twenty numeric inputs representing lagged inputs of previous 3 hours, previous 3 hours of previous day including hour to be predicted, and previous 3 hours of previous week including hour to be predicted \\
		43    & Holiday       & One binary digit representing whether the day was a public holiday  \\     \bottomrule                                                           
	\end{tabular}
	\caption{Input smart meter data features for 30-minute ahead forecasting.}
	\label{tab:feature}
\end{table}

\subsection{Overfitting}

Overfitting can be a significant problem within the machine learning domain. Overfitting is where a model is fit too closely to the training data. Therefore, whilst the model is able to predict the original dataset well, it is unable to predict data it has not seen before to the same degree of accuracy. 

To avoid this problem, we used both a training and a test set. The training set is used to fit the data, however, to see the performance of the model, we tested the model using previously unseen data, which is called the test set. This therefore allowed us to verify that the models trained were not overfitting solely to the training data.

\subsection{Experiments}

This section explores the different methods used to select the model parameters, and the tests to evaluate our models. Once the parameters were chosen the models were trained on the different clusters of residential customers. Each model was run five times to explore the variance of the results.  

\subsubsection{Support Vector Regression}

To implement a Support Vector Regression model, a variety of parameters must be chosen. These parameters influence the performance of the model. The parameters were evaluated using cross-validation. To do this, the data was split 75\% into training data, and the remaining 25\% into test data. This split was chosen to balance the trade-off between having enough training data so that the model can accurately learn the underlying form of the data, but also to have enough data to validate each model.

To choose the optimum support vector machine kernel cross-validation was also used. Again, with 75\% acting as the training data and 25\% as the test. The kernels compared were polynomial, radial basis function (RBF) and the linear kernel \cite{Chang2010, theodoridis2009pattern}. These were chosen due to their popularity, support and relative speed of computation.

The parameter values selected are shown in Table \ref{forecasting:tab:kernel}. These parameter values were chosen from the results of cross-validation for each of the different kernels.  From the cross-validation, the linear kernel was found to be the best performing. For this reason, the linear kernel was utilised for prediction of electricity consumption in this thesis.

\begin{table}
	\centering
	\begin{tabular}{ccl}
		\toprule
		Kernel Type& Kernel Parameters & RMSE\\
		\midrule
		Linear & No values & 0.02103\\
		RBF & C=2, $\gamma=0.016$ & 0.0245\\
		Polynomial & C=2, $d=2, r=2$ & 0.0315 \\
		\bottomrule
	\end{tabular}
	\caption{Prediction accuracy based on type of kernel for support vector regression for 30-minute ahead forecasting.}
	\label{forecasting:tab:kernel}
\end{table}

\subsubsection{Random Forest}

\begin{figure}
	\centering
	\includegraphics[width=0.6\textwidth]{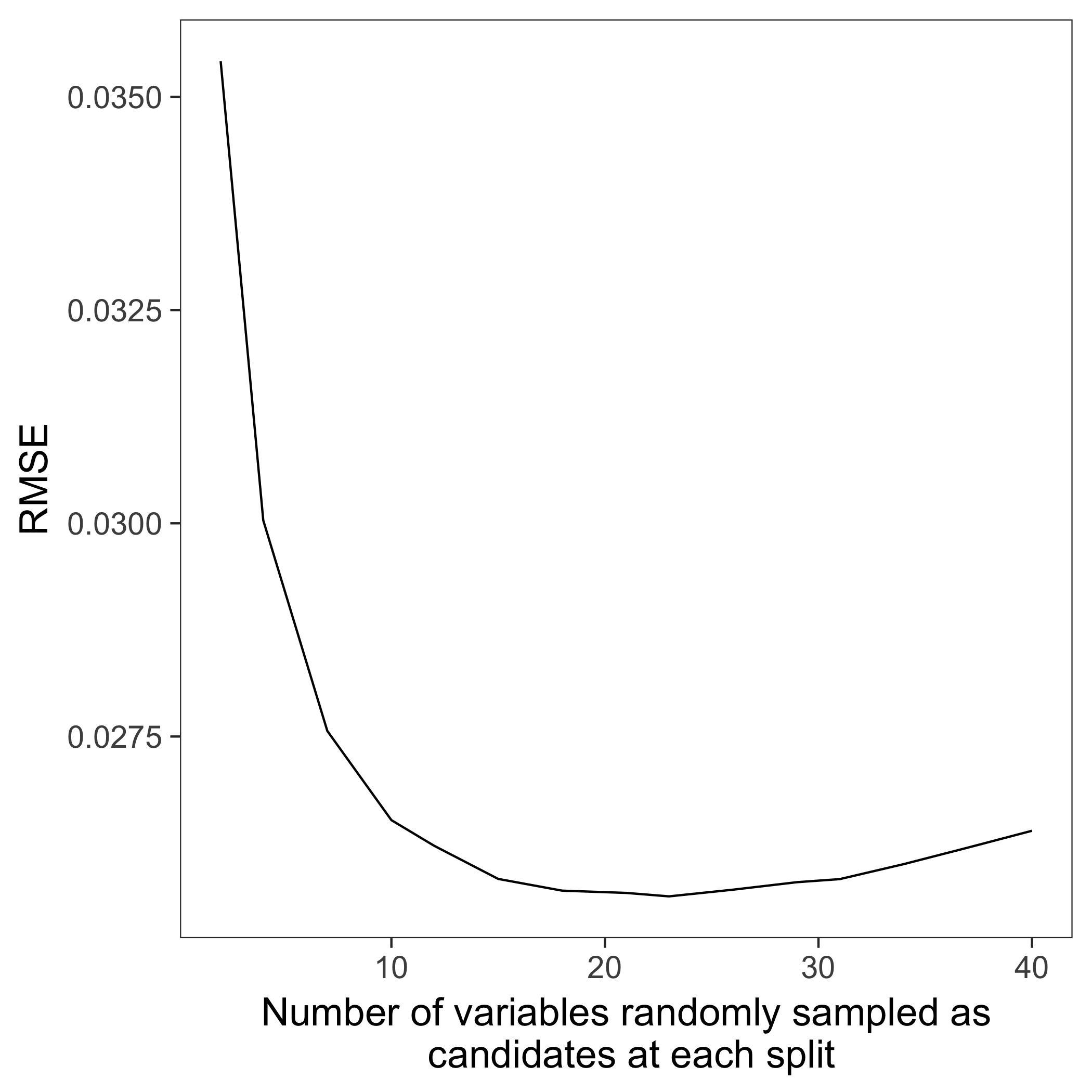}
	\caption{RMSE vs Number of variables randomly sampled as candidates at each split in the Random Forest model.}
	\label{fig:rf_param_tune}
\end{figure}

To initialise the Random Forest algorithm with the number of variables randomly sampled as candidates at each split, cross-validation was used. Once again, 75\% of the data was used for training and the remaining 25\% for testing due to the trade-off between training and testing.

Figure \ref{fig:rf_param_tune} shows the results of tuning the parameter of the number of variables randomly sampled as candidates at each split. The optimum number was found to be 23. Either side of this value the RMSE increases. Therefore the value 23 was selected to be the number of variables randomly sampled as candidates at each split in the Random Forest model. It is proposed that the value 23 was found to be optimum due to the 20 lagged inputs, as this data is crucial for the Random Forest to learn the underlying nature of electricity load.

\subsubsection{Neural Networks}

A feed-forward Multilayer Perceptron is a common neural network architecture used for the prediction of time-series data, which has comparable, and in some cases better, results than statistical models \cite{Hill1994}. 

The first step when designing a Multilayer Perceptron neural network is to design the architecture. For this case, the number of input neurons is set to 41 (see Table \ref{tab:feature}). Once an input for each neuron is entered, the output layer must be designed. Due to the fact that we are forecasting only one-time step ahead (30 minutes ahead) one output neuron is required.

The next step is to design the architecture of the hidden layers. To accomplish this, cross-validation is utilised as per the previous models. A maximum of 3 hidden layers were tested, and the results analysed. A similar method to Fan \textit{et al.}  \cite{Fan2009} was evaluated to choose the number of neurons and hidden layers, a technique known as the Levenberg-Marquardt technique \cite{more1978levenberg}. The Levenberg-Marquardt is a technique suitable for training medium-sized Artificial Neural Networks with a low mean-squared error. The fundamental rule is to select the minimum number of neurons in the hidden layer so as to capture the complexity of the model, but not too many as to introduce over-fitting, which results in a loss in the generalisation of the algorithm.

The method begins by choosing a small number of neurons and gradually increasing the number each time the model is trained, and the forecast error obtained. The forecast error is monitored until an optimum value is found, to which no further improvement is noted. Once the optimum number of neurons in the layer is obtained, an additional layer is added, and the same technique is used.

Using this technique, an optimal architecture with three layers is obtained. The first layer contained two neurons, the second contained five, and the third contained four.

\subsubsection{LSTM}

To initialise the LSTM, cross-validation was used to select the number of stacked layers and memory units. We trialled the number of stacks from the following list: $[1,2,3,4]$ and the number of memory units from the following list: $[5,6,7,8,9,10,20,30,40,50,75,100]$.

The optimum number of layers was found to be 2, with a total of 50 memory units. Different combinations of layers and memory units displayed worse results.

\subsection{Results}

\begin{figure}
	\centering
	\includegraphics[width=0.6\textwidth]{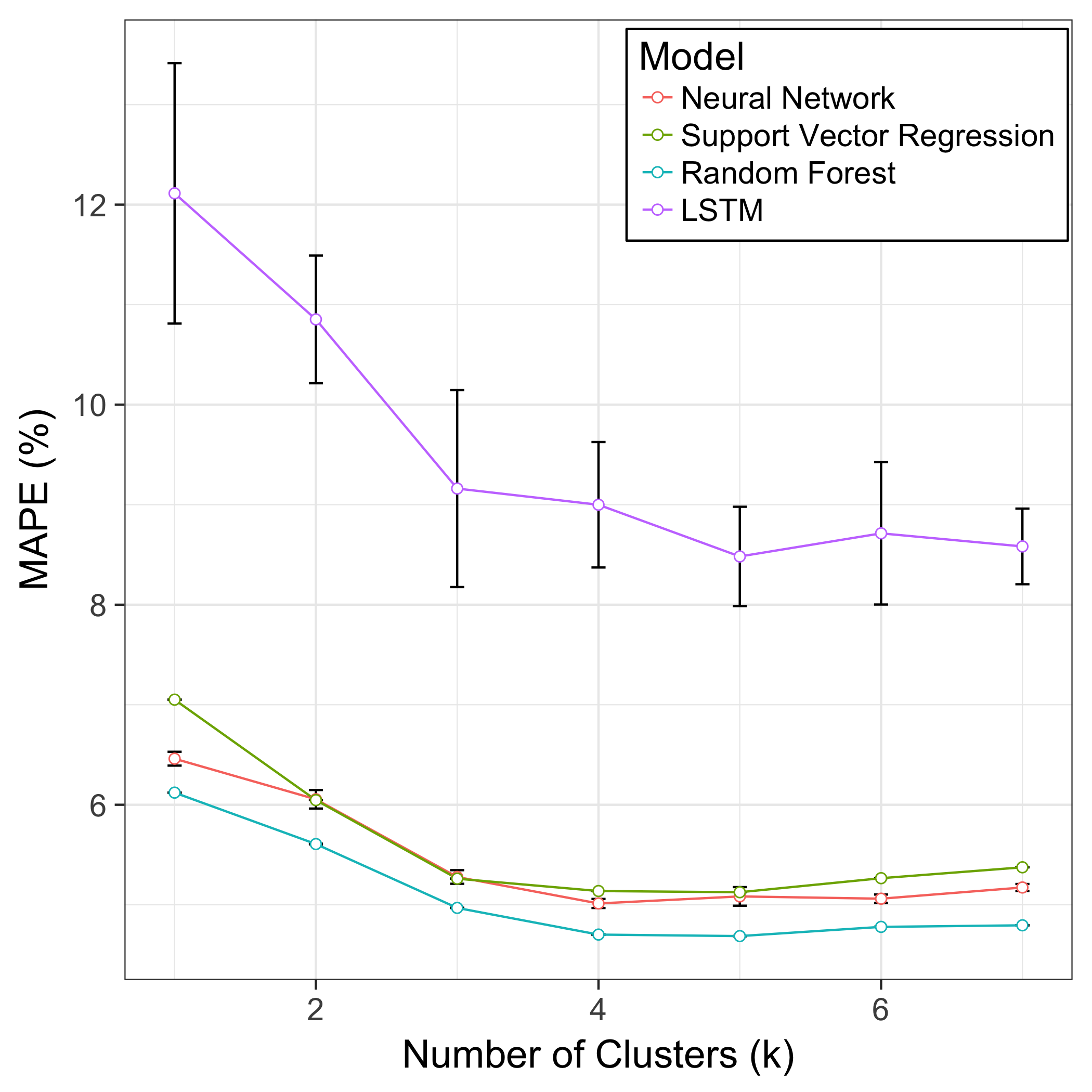}[H]
	\caption{Comparison of accuracy of models 30-minute ahead forecasting electricity with varying number of clusters.}
	\label{fig:results}
\end{figure}

\begin{figure}
	\centering
	\includegraphics[width=0.6\textwidth]{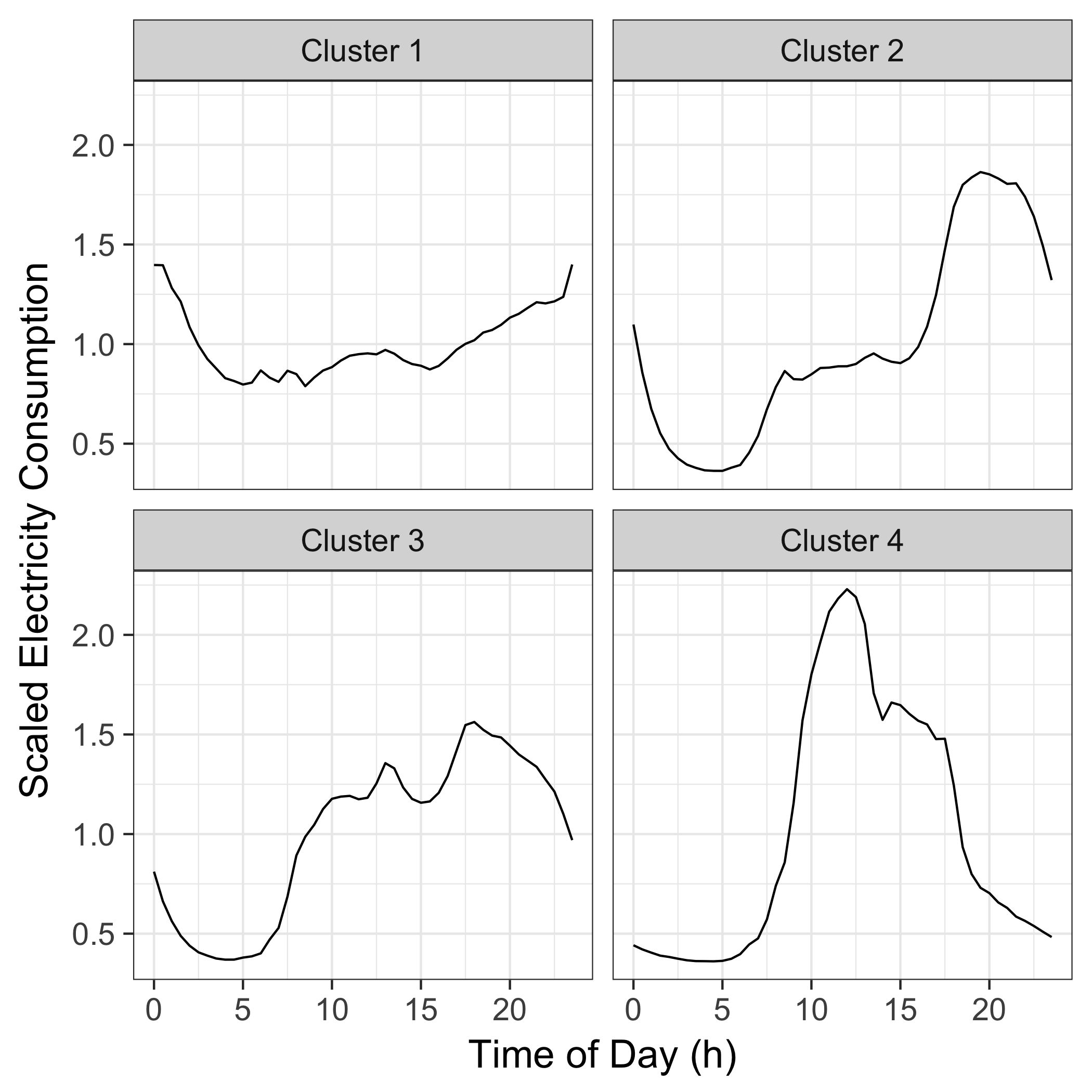}
	\caption{Average load profile for each cluster of the smart meter data.}
	\label{fig:clustercentre}
\end{figure}

To test the accuracy of the trained model, the data was split into a training and test set. The data between the 14th of July 2009 and the 15th of June 2010 was used as the training data, whilst the data between the 15th of June 2010 and 31st of December 2010 was used for testing purposes. The test set is separate from the training set and not used during training. We did not, however, consider seasonal issues for this work.

28 independent forecasting models are constructed for each of the Random Forests, Support Vector Regression, LSTMs and Multilayer Perceptron neural networks for each of the groups with \textit{k} varying from 1 to 7. This was done to determine the optimal number of clusters.  Each of the 28 models are trained independently, five times each so that the standard deviation results of MAPE for each cluster could be displayed. We evaluated the MAPE of the overall prediction. 

\begin{figure}
	\centering
	\includegraphics[width=0.6\textwidth]{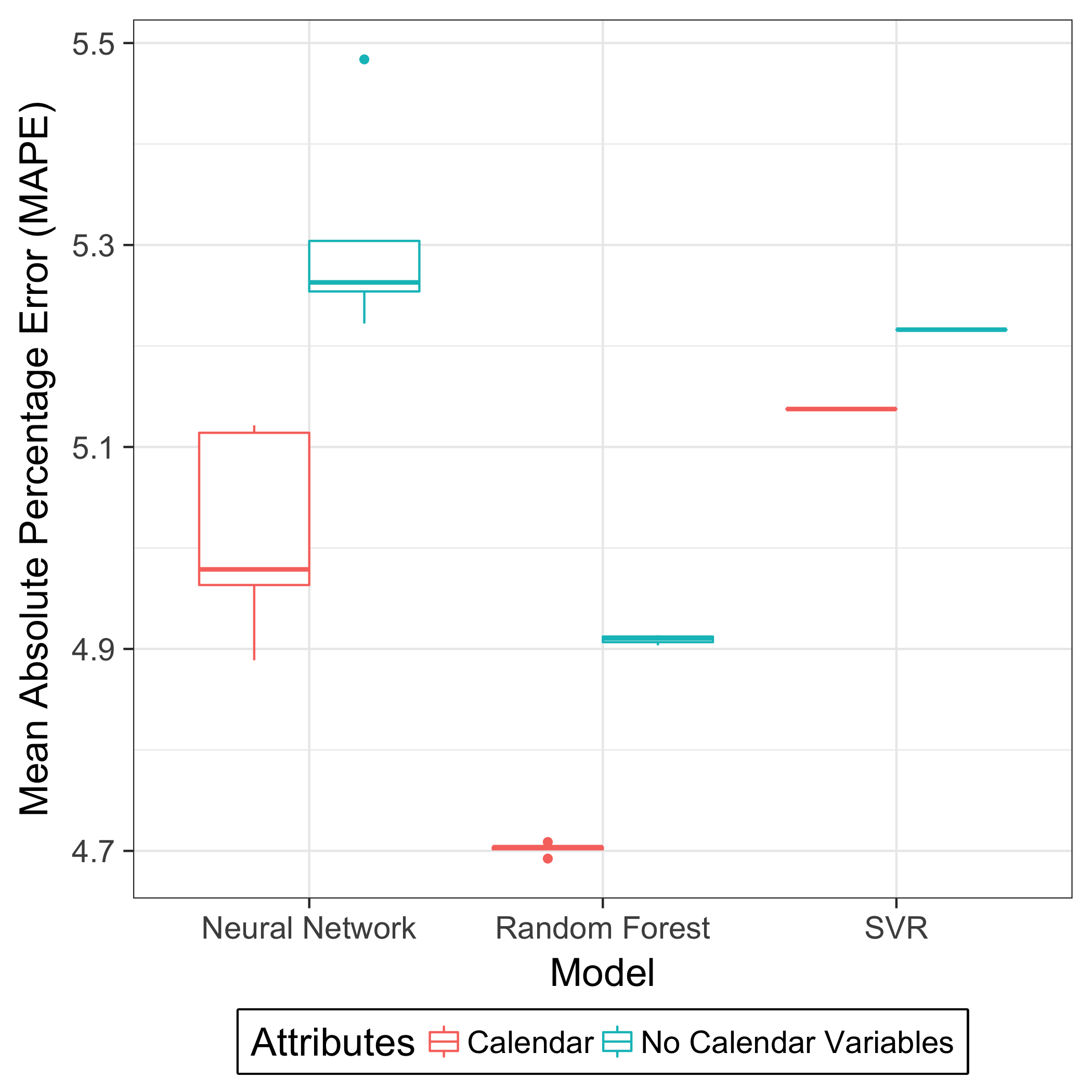}
	\caption{Comparison of accuracy of models with or without calendar attributes for 30-minute ahead forecasting.}
	\label{fig:calendar_attr}
\end{figure}

Figure \ref{fig:calendar_attr} demonstrates the impact of using calendar attributes such as month, day of the month, and day of the week on prediction accuracy. The results show an increase in prediction accuracy of 6\% for neural networks, 4\% for Random Forests and 1\% for support vector regression when taking into account these variables. It is proposed that the ability for the models to take into account the yearly, monthly and weekly seasonal behaviour improves the results.

Figure \ref{fig:results} displays the accuracy of the models trained at different numbers of clusters (\textit{k}). The results demonstrate that introducing clusters to group similar customers improve results in all cases. The optimum value for \textit{k} for Random Forests, Support Vector Regression and neural networks was shown to be four for our dataset. After this, the accuracy diminishes slightly. The error bars shown in Figure \ref{fig:results} show a small variance in MAPE in \acrshort{svr}s, ANNs and Random Forests. However, the MAPE of the LSTMs seem to vary by up to 11\% in the five models run.

Figure \ref{fig:clustercentre} shows the average load profiles of different clusters when $k=4$. It is proposed that the optimum number of clusters is four due to the distinct load profiles that can be seen in Figure \ref{fig:clustercentre}. The four different distinct patterns seen are high night time use in cluster 1, a typical residential load profile is shown in cluster 2, a spread of usage in cluster 3, and high daytime usage in cluster 4. At $k=3$ these distinct patterns are not adequately clustered, and at $k=5$ one of the distinct clusters is split, leading to an increase in stochasticity.

It is true that the optimum number of clusters will vary for different datasets. Whilst residential smart meter datasets may be similar, it is entirely possible that different geographies display different usage characteristics based on factors such as culture, temperature and economic reasons. It is therefore important to choose an optimal number of clusters for each dataset.

The results demonstrate that SVR, Random Forests and the Multilayer Perceptrons have a similar overall accuracy. The LSTM shows a similar pattern in increasing accuracy with a number of clusters. However, the Random Forest seems to outperform each of the models at every point. This may be due, in part, to the internal operation of the Random Forest which undertakes its own cross-validation using out-of-bag samples and only having a few tuning parameters. This implies that neural networks and support vector regressions require more tuning.

It has been shown that neural networks, SVR and Random Forests all perform within an adequate range of predicting electricity consumption. Whilst LSTMs perform poorly. This may be due to the features given to the LSTM, which only had previous two and a half hours of data as input. 

However, it is well known that the best machine learning technique for predicting energy consumption cannot be chosen \textit{a priori}. Therefore it is necessary to compare different techniques to find the best solution to a particular regression problem \cite{Ahmad2017}.

For this work, the training time was tested by timing how long the models would take to fit to one cluster (single model trained on the training set). The Support Vector Regression took much less time than all of the other methods, whereas the LSTM took the longest. The Artificial Neural Network required 9 minutes and 5 seconds to run. The Support Vector Regression model required 3 minutes and 32 seconds to run. The Random Forest, on the same data, required 9 minutes and 44 seconds to run, whilst the LSTM took 12 minutes 55 seconds.

\section{Day-ahead forecasting}
\label{forecast:sec:longterm}

In this section, we expand on the work undertaken in Section \ref{forecast:sec:shortterm} by utilising further time-series prediction algorithms, including online machine learning methods. We take the error distributions and perturb the exogenous electricity demand of ElecSim, and observe the long-term impacts of poor error forecasts on the UK electricity market. It should be noted that this could be applied to any decentralised electricity market.

\subsection{Methods}
\label{sec:methods}


\subsubsection{Data preparation}

Similarly to our previous work in Chapter \ref{chapter:elecsim} \cite{Kell2018a}, we selected a number of calendar attributes and demand data from the GB National Grid Status dataset provided by the electricity market settlement company Elexon, and the University of Sheffield \cite{gbnationalgridstatus2019}. This dataset contained data between the years 2011-2018 for the United Kingdom. The calendar attributes used as predictors to the models were hour, month, day of the week, day of the month and year. These attributes allow us to account for the periodicity of the data within each day, month and year.

It is also the case that electricity demand on a public holiday which falls on a weekday is dissimilar to load behaviours of ordinary weekdays \cite{Kim2000}. We, therefore, marked each holiday day to allow the model to account for this.

As demand data is highly correlated with historical demand, we lagged the input demand data. In this context, the lagged data is where we provide data of previous time steps at the input. For example, for predicting t+1, we use $n$ inputs: {t, t-1, t-2, $\ldots$, t-n}. This enabled us to take into account correlations on previous days, weeks and the previous month. Specifically, we used the previous 28 hours before the time step to be predicted for the previous 1st, 2nd, 7th and 30th day. We chose this as we believe that the previous two days were the most relevant to the day to be predicted, as well as the weekday of the previous week and the previous month. We chose the previous 28 hours to account for a full day, plus an additional 4 hours to account for the previous day's correlation with the day to be predicted. We could have increased the number of days provided to the algorithm, as well as marking holiday days. However, due to time and computational constraints, we used our previously described intuition for lagged data selection. The number of lagged inputs to trial increases exponentially with each additional day added, therefore making the problem intractable when also trialling such a high number of algorithms and hyperparameters. 

In addition to this, we marked each of the days with their respective seven seasons. These seasons were defined by the National Grid Short Term Operating Reserve (STOR) Market Information Report \cite{ESO2019}. These differ from the traditional four seasons by splitting autumn into two further seasons, and winter into three seasons. Finally, to predict a full 24-hours ahead, we used 24 different models, 1 for each hour of the day.

The data is standardised and normalised using min-max scaling between -1 and 1 before training and predicting with the model. This is due to the fact that the inputs such as day of the week, hour of day are significantly smaller than that of demand. Therefore, the demand will influence the result more due to its larger value; however, this does not necessarily mean that demand has greater predictive power.

\subsubsection{Algorithm Tuning}

To find the optimum hyperparameters, cross-validation is used. As this time-series data was correlated in the time-domain, we took the first six years of data (2011-2017) for training and tested on the remaining year of data (2017-2018).

Each machine learning algorithm has a different set of parameters to tune. To tune the parameters in this thesis, we used a grid search method. Grid search is a brute force approach that trials each combination of parameters at our choosing; however, for our search space, this was small enough to make other approaches not worth the additional effort.

Tables \ref{table:hyperparameter-tuning-offline} and \ref{table:hyperparameter-tuning-online} display each of the models and respective parameters that were used in the grid search. Table \ref{table:hyperparameter-tuning-offline} shows the offline machine learning methods, whereas Table \ref{table:hyperparameter-tuning-online} displays the online machine learning methods. Each of the parameters within the columns ``Values'' are trialled with every other parameter.

Whilst there is room to increase the total number of parameters, due to the exponential nature of grid-search, we chose a smaller subset of hyperparameters, and a larger number of regressor types. Specifically, with neural networks, there is a possibility to extend the number of layers as well as the number of neurons, to use a technique called deep learning. Deep learning is a class of neural networks that use multiple layers to extract higher levels of features from the input. 

\begin{sidewaystable*}[ph!]
	\centering
	\begin{tabular}{@{}lllllll@{}}
		\toprule
		\textbf{Regressor Type} & \textbf{Parameters} & \textbf{Values}   & \textbf{Parameters} & \textbf{Values} & \textbf{Parameters} & \textbf{Values}       \\ \midrule
		Linear                  & N/A                 & N/A               &                     &                 &                     &                       \\
		Lasso                   & N/A                 & N/A               &                     &                 &                     &                       \\
		Elastic Net             & N/A                 & N/A               &                     &                 &                     &                       \\
		Least-Angle             & N/A                 & N/A               &                     &                 &                     &                       \\
		Extra Trees             & \# Estimators       & {[}16, 32{]}      &                     &                 &                     &                       \\
		Random Forest           & \# Estimators       & {[}16, 32{]}      &                     &                 &                     &                       \\
		AdaBoost                & \# Estimators       & {[}16, 32{]}      &                     &                 &                     &                       \\
		Gradient Boosting       & \# Estimators       & {[}16, 32{]}      & learning rate       & {[}0.8, 1.0{]}  &                     &                       \\
		Support Vector          & Kernel              & {[}linear, rbf{]} & C                   & {[}1, 10{]}     & Gamma               & {[}0.001, 0.0001{]}   \\
		Multilayer Perceptron   & Activation function & {[}tanh, relu{]}  & hidden layer sizes  & {[}1, 50{]}     & Alpha               & {[}0.00005, 0.0005{]} \\
		K-Neighbours            & \# Neighbours       & {[}5, 20, 50{]}   &                     &                 &                     &                       \\ \bottomrule
	\end{tabular}%
	\caption{Hyperparameters for offline machine learning regression algorithms for day-ahead forecasting.}
	\label{table:hyperparameter-tuning-offline}
	
	
	\qquad
	\qquad
	\qquad
	\qquad
	\qquad
	\qquad
	\qquad
	\qquad
	\qquad
	
	\centering
	\begin{tabular}{@{}llp{2.5cm}lllp{1.6cm}@{}}
		\toprule
		\textbf{Regressor Type} & \textbf{Parameters} & \textbf{Values}                                  & \textbf{Parameters} & \textbf{Values}   & \textbf{Parameters} & \textbf{Values}        \\ \midrule
		Linear                  & N/A                 & N/A                                              &                     &                   &                     &                        \\
		Box-Cox                 & Power               & {[}0.1, 0.05, 0.01{]}                            &                     &                   &                     &                        \\
		Multilayer Perceptron   & Hidden layer sizes  & {[}(10, 50, 100), (10),  (20), (50), (10, 50){]} & 
		&                   &                     &                        \\ 
		Passive Aggressive      & C                   & {[}0.1, 1, 2{]}                                  & Fit intercept?      & {[}True, False{]} & Max iterations      & {[}1, 10, 100, 1000{]} \\
		\bottomrule
	\end{tabular}%
	\caption{Hyperparameters for online machine learning regression algorithms for day-ahead forecasting.}
	\label{table:hyperparameter-tuning-online}
\end{sidewaystable*}%

\subsubsection{Prediction Residuals in ElecSim}

Each of the previously mentioned models trialled will have a certain degree of error. Prediction residuals are the difference between the estimated and actual values. We collect the prediction residuals to form a distribution for each of the models. We then trial 80 different closed-form distributions to see which of the distributions best fits the residuals from each of the models. These 80 distributions were chosen due to their implementation in scikit-learn \cite{scikit-learn}.

Once each of the prediction residual distributions are fit with a sensible closed-form distribution, we sample from this new distribution and perturb the demand for the electricity market at each time step within ElecSim.

By perturbing the market by the residuals, we can observe what the effects are of incorrect predictions of demand in an electricity market using the long-term electricity market model, ElecSim. We are able to understand the differences that prediction residuals have on long-term investment decisions as well as generators utilised.

\subsection{Results}
\label{sec:results}

In this section, we detail the accuracy of the algorithms and statistical models to predict 24 hours ahead for the day-ahead market. In addition to this, we display the impact of the errors on electricity generation investment and electricity mix from the years 2018 to 2035 using the agent-based model ElecSim.

\subsubsection{Offline Machine Learning}

To generate these results, we use a training set to train the data, and a test set to see how well each algorithm performs on the testing data. That is, how well the algorithm can predict data it is yet to see. In our case, the training data was from 2011 to 2017, and the testing data was from 2017 to 2018.

\begin{figure}
	\centering
	\includegraphics[width=0.65\columnwidth]{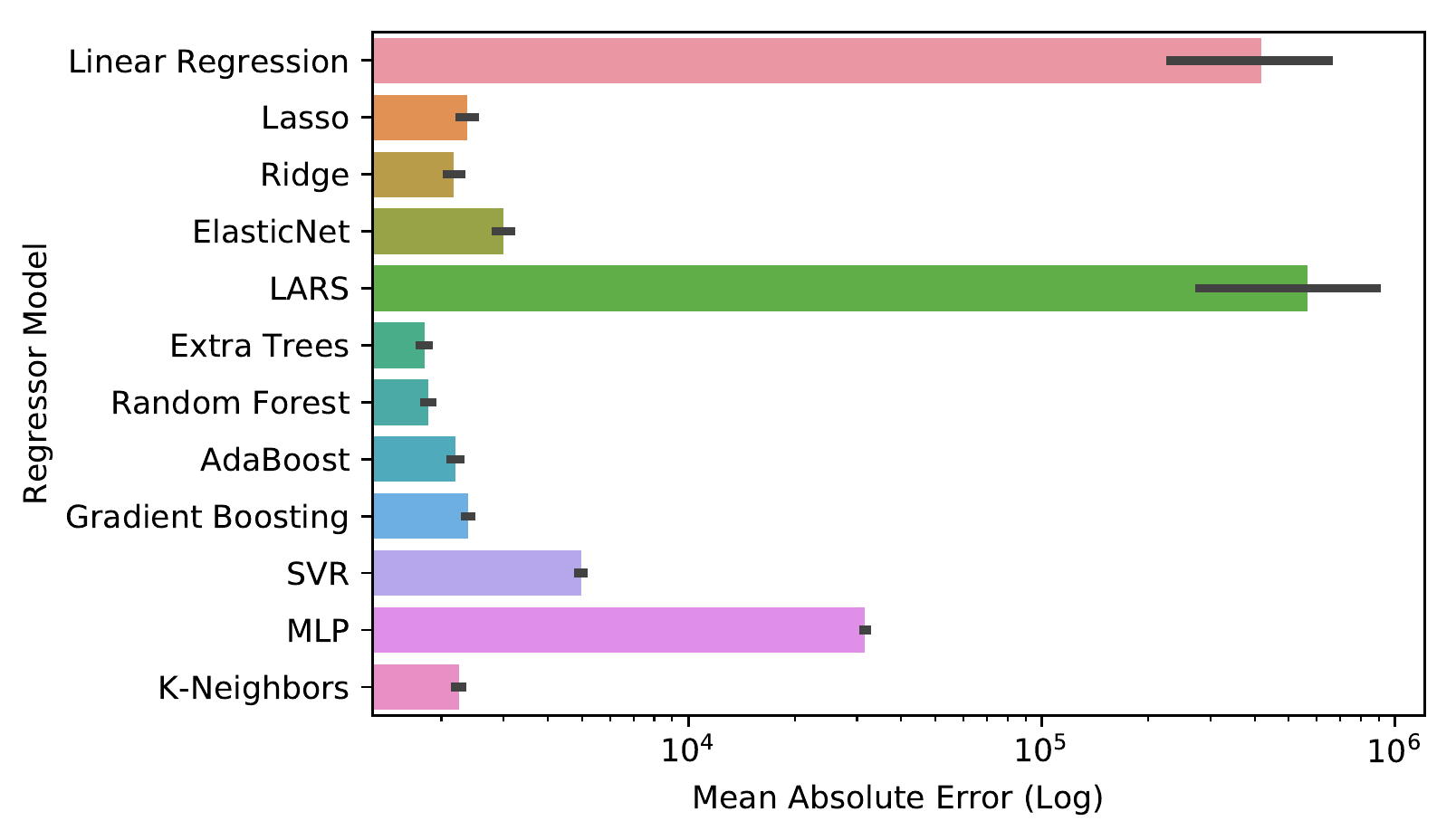}
	\caption{Offline models mean absolute error comparison, with 95\% confidence interval for 5 runs of each model. Forecasting day-ahead.}
	\label{fig:beis_elecsim_historic_comparison}
\end{figure}

Figure \ref{fig:beis_elecsim_historic_comparison} displays the mean absolute error of each of the offline statistical and machine learning models on a log scale. It can be seen that the different models have varying degrees of success. The least accurate models were linear regression, the \acrfull{mlp} model and the Least Angle Regression (LARS). These all have mean absolute errors over 10,000MWh. This error would be prohibitively high in practice; the max tendered national grid reserve is 6,000MWh, while the average tendered national grid reserve is 2,000MWh \cite{ESO2019}.

A number of models perform well, with a low mean absolute error. These include the lasso, gradient Boosting Regressor and K-neighbours regressor. The best model, similar to the work presented in the 30-minute ahead forecasting  \cite{Kell2018a}, was the decision tree-based model, Extra Trees Regressor, with a mean absolute error of $1,604$MWh. This level is well within the average national grid reserve of 2,000MWh.

\begin{figure}
	\centering
	\includegraphics[width=0.65\columnwidth]{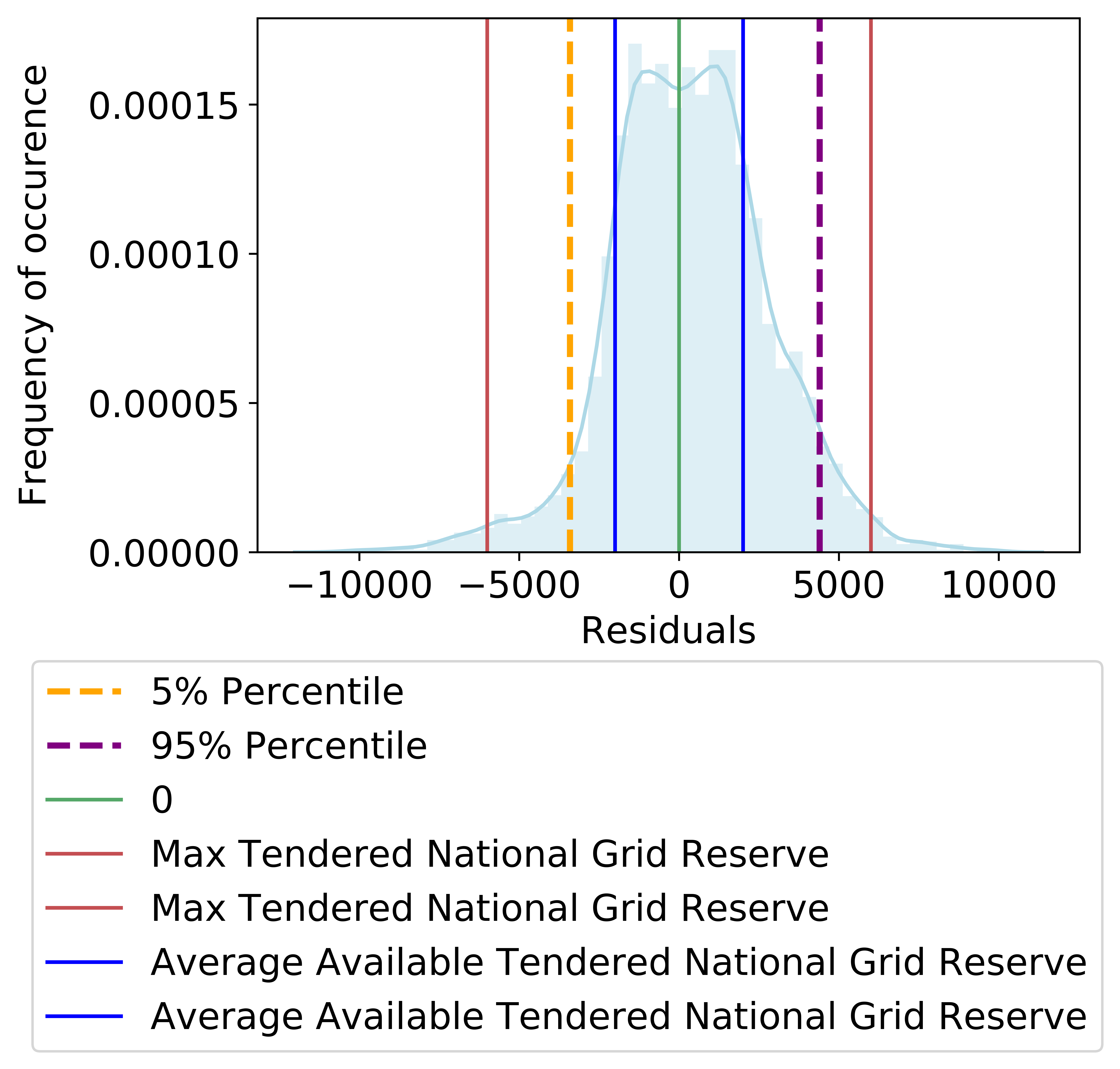}
	\caption{Best offline machine learning algorithm (Extra Trees Regressor) distribution for day-ahead forecasting.}
	\label{fig:best_offline_learning_day_distribution}
\end{figure}

Figure \ref{fig:best_offline_learning_day_distribution} displays the distribution of the best offline machine result (Extra Trees Regressor). It can be seen that the max tendered national grid reserve falls well above the 5\% and 95\% percentiles. However, there are occasions where the errors are greater than the maximum tendered national grid reserve. In addition, the majority of the time, the model's predictions fall within the average available tendered national grid reserve.

%
%

\begin{figure}[H]
	\centering
	\includegraphics[width=0.65\columnwidth]{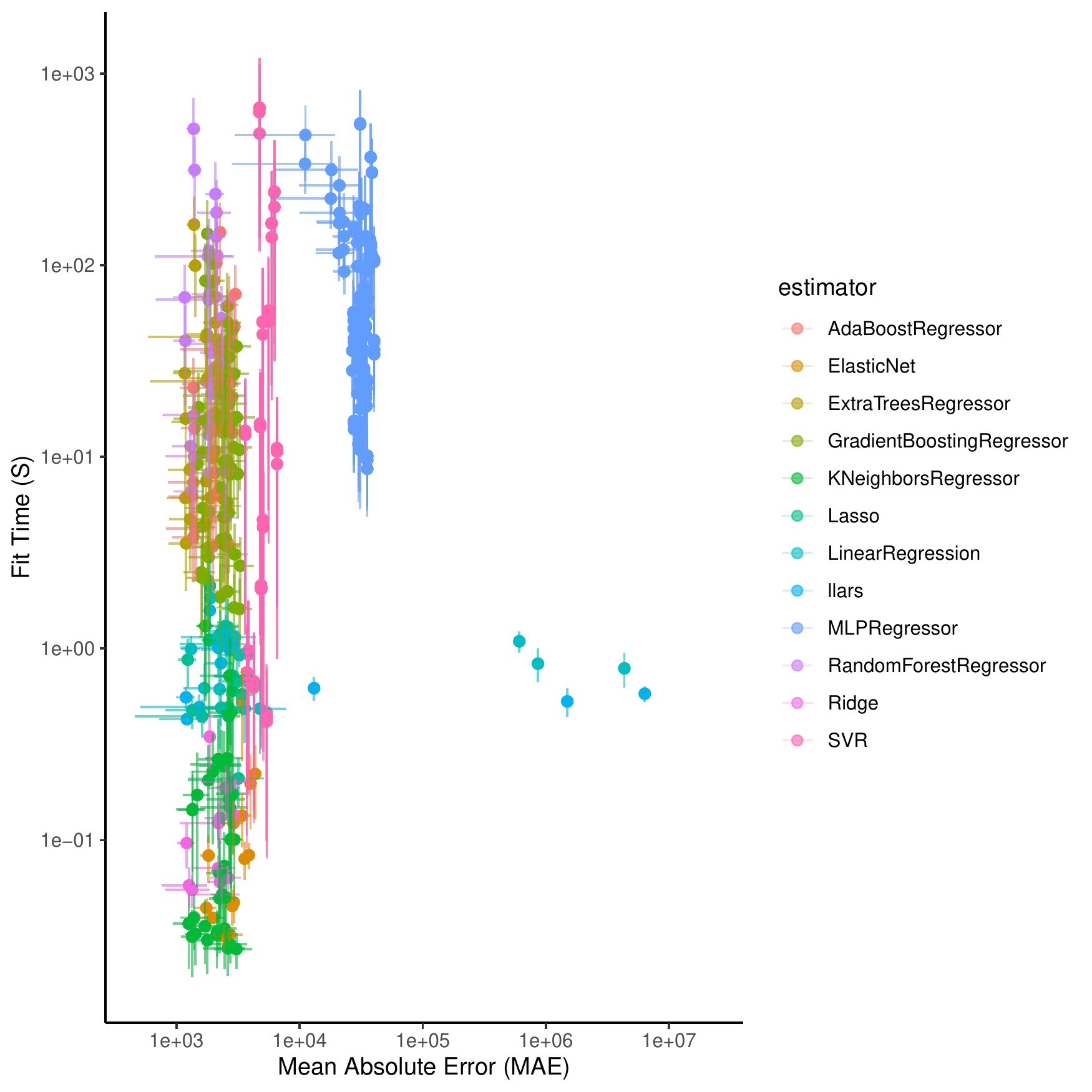}
	\caption{Time taken to train the offline models for day-ahead forecasting versus mean absolute error. Error bars display standard deviation between points.}
	\label{fig:offline_fit_time_vs_mae}
\end{figure}

\begin{figure}
	\centering
	\includegraphics[width=0.65\columnwidth]{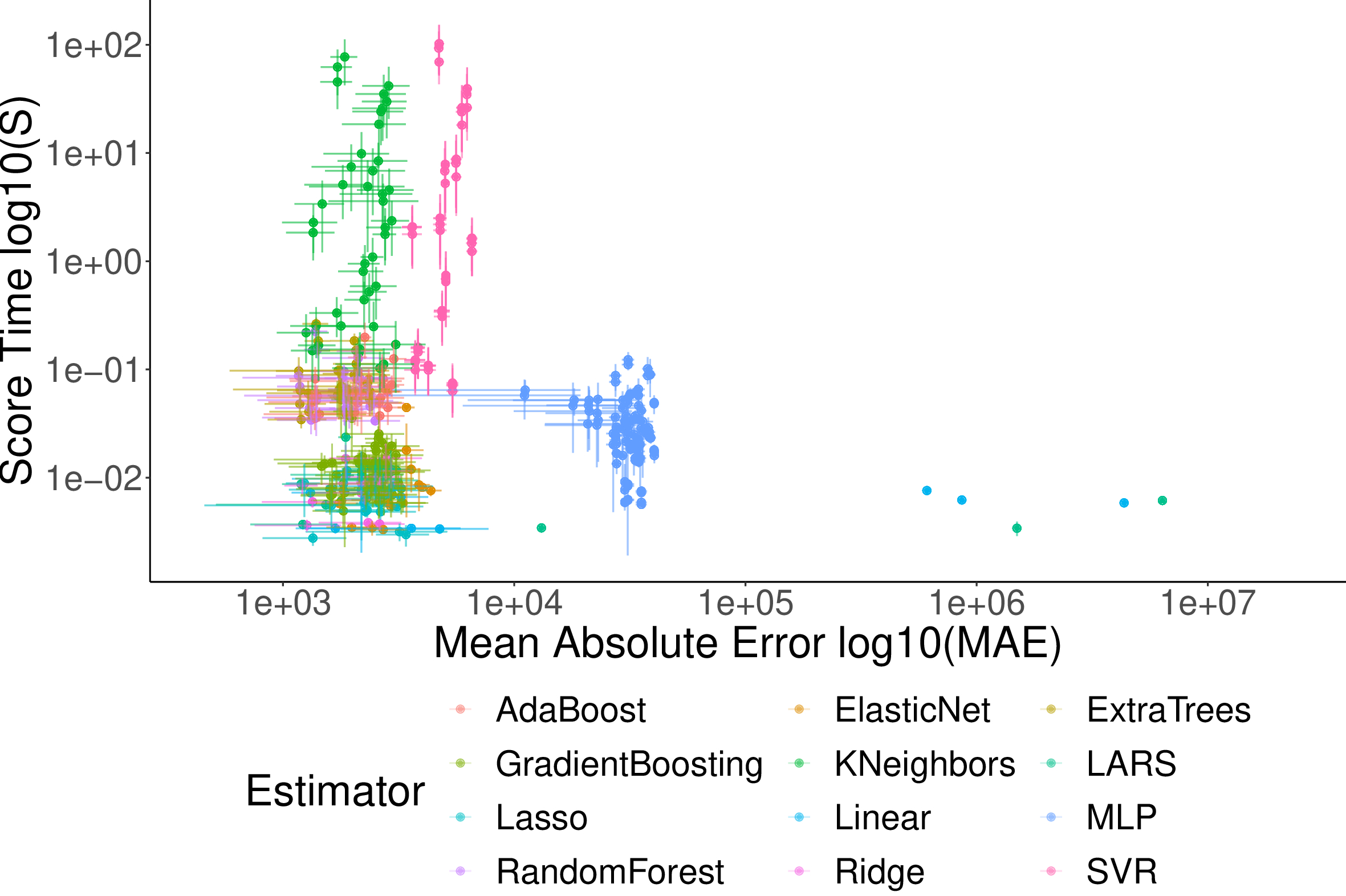}
	\caption{Time taken to score the offline models versus mean absolute error. Error bars display standard deviation between points.}
	\label{fig:offline_score_time_vs_mae}
\end{figure}

Figures \ref{fig:offline_fit_time_vs_mae} and \ref{fig:offline_score_time_vs_mae} display the time taken to train the model and time taken to sample from the model versus the absolute error respectively for the offline algorithms. Multiple fits are trialled for each parameter type for each model. The error bars indicate the results of multiple cross-validations.

It can be seen from Figure \ref{fig:offline_fit_time_vs_mae} that the time to fit varies significantly between algorithms and parameter choices. The multilayer perceptron consistently takes a long time to fit, when compared to the other algorithms and performs relatively poorly in terms of MAE. There are many models, such as the random forest regressor, and extra trees regressors which perform well, however, take a long time to fit, especially when compared to the K-Nearest neighbours.

For a small deterioration in MAE it is possible to decrease the time it takes to train the model significantly. For example, by using the K-Nearest neighbours or support vector regression (SVR).


The scoring time, displayed in Figure \ref{fig:offline_score_time_vs_mae}, also displays a large variation between model types. For instance, the MLP regressor takes a shorter time to sample predictions when compared to the K-Neighbors algorithm and support vector regression. It is possible to have a cluster of algorithms with low sample times and low mean absolute errors. However, often a trade-off is required, with a fast prediction time requiring a longer training time and vice-versa.

\subsubsection{Online Machine Learning}

\begin{figure}
	\centering
	\includegraphics[width=0.65\columnwidth]{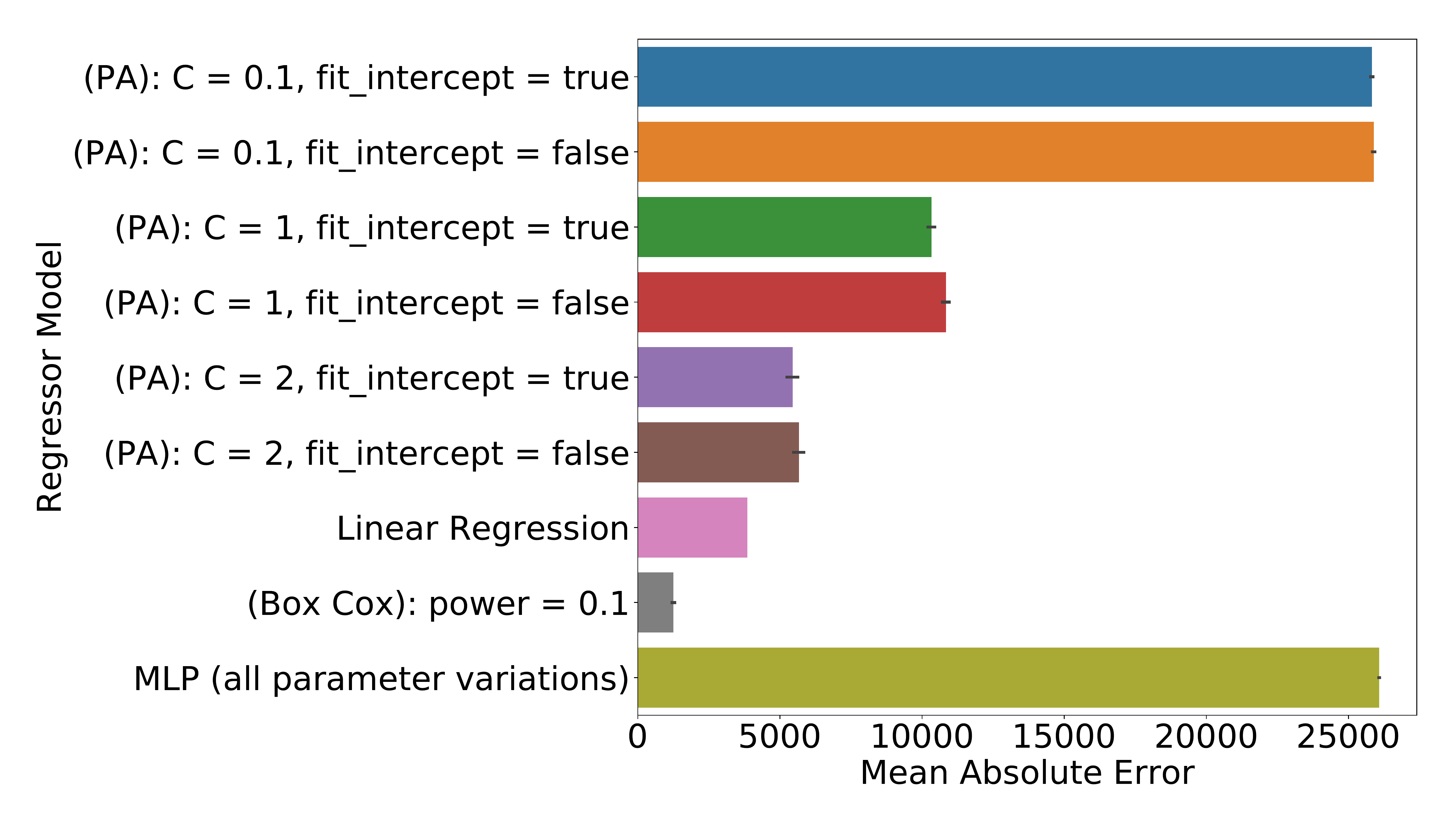}
	\caption{Comparison of mean absolute errors (MAE) for different online regressor models for day-ahead forecasting. MLP results for all parameters are shown in a single barchart due to the very similar MAEs for the differing hyperparameters.}
	\label{fig:online_model_mae_barplot}
\end{figure}

To see if we can improve on the predictions, we utilise an online machine learning approach. If we are successful, we should be able to reduce the national grid reserves, reducing cost and emissions.

Figure \ref{fig:online_model_mae_barplot} displays the comparison of mean absolute errors for the different trialled online regressor models. To produce this graph, we performed various hyperparameter trials. Where the hyperparameters had the same results, we removed them. For the multilayer perceptron (MLP), we aggregated all hyperparameters, due to the similar nature of the predictions. This is due to the fact that no additional information would be gained from presenting this work, as the results are the same or very similar.

It can be seen that the best performing model was the Box-Cox regressor, with an MAE of 1,100. This is an improvement of over 30\% on the best offline model. The other models perform less well. However, it can be seen that the linear regression model improves significantly for the online case when compared to the offline case. The passive aggressive (PA) model improve significantly with the varying parameters, and the MLP performs poorly in all cases.

\begin{figure}
	\centering
	\includegraphics[width=0.6\columnwidth]{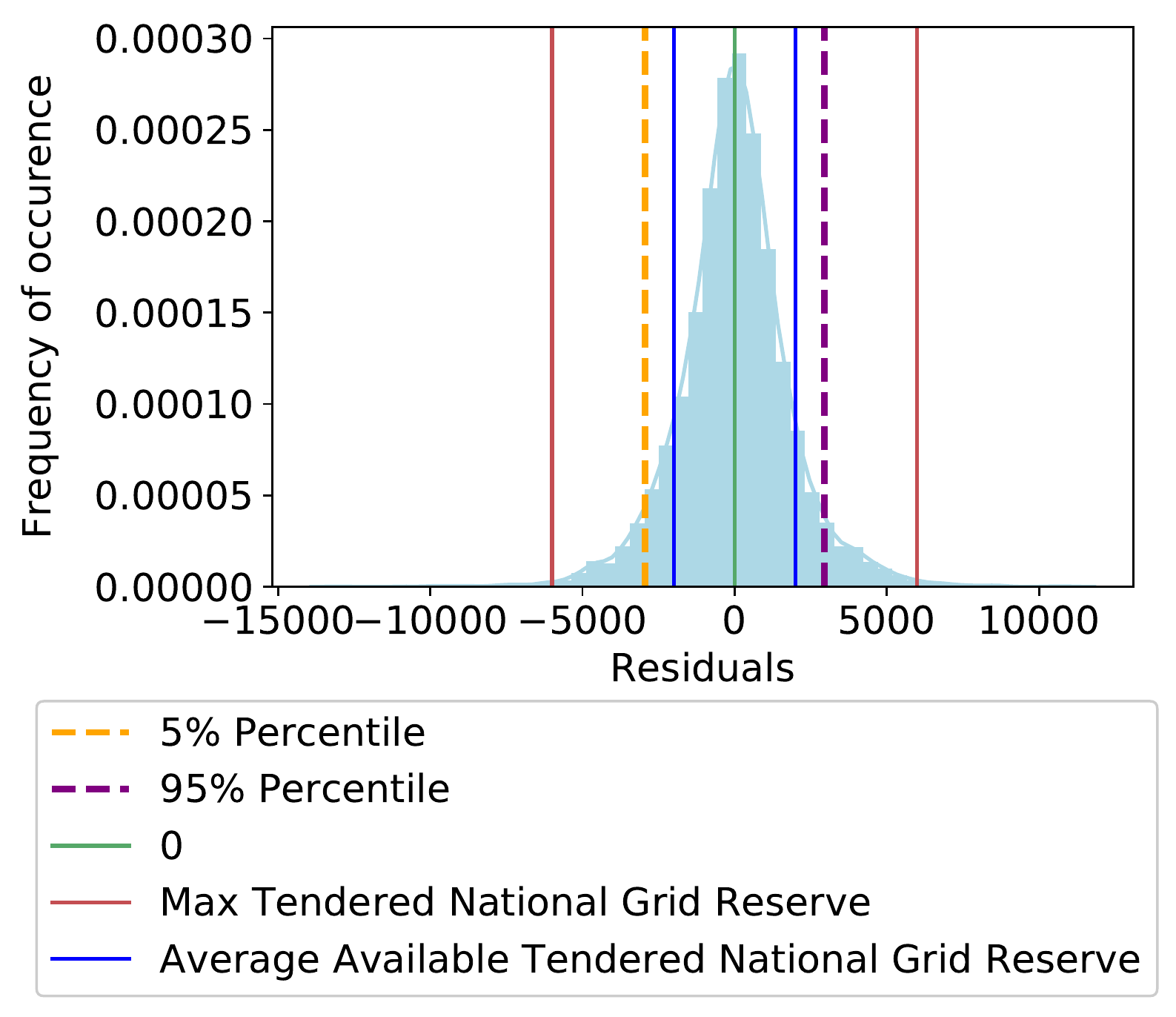}
	\caption{Best online model (Box-Cox Regressor) distribution for day-ahead forecasting.}
	\label{fig:best_online_learning_day_distribution}
\end{figure}

Figure \ref{fig:best_online_learning_day_distribution} displays the best online model. We can see a significant improvement over the best online model distribution, shown in Figure \ref{fig:best_offline_learning_day_distribution}. We remain within the max tendered national grid reserve for 98.9\% of the time, and the average available tendered national grid reserve is close to the 5\% and 95\% percentiles.

\begin{figure}
	\centering
	\includegraphics[width=0.6\columnwidth]{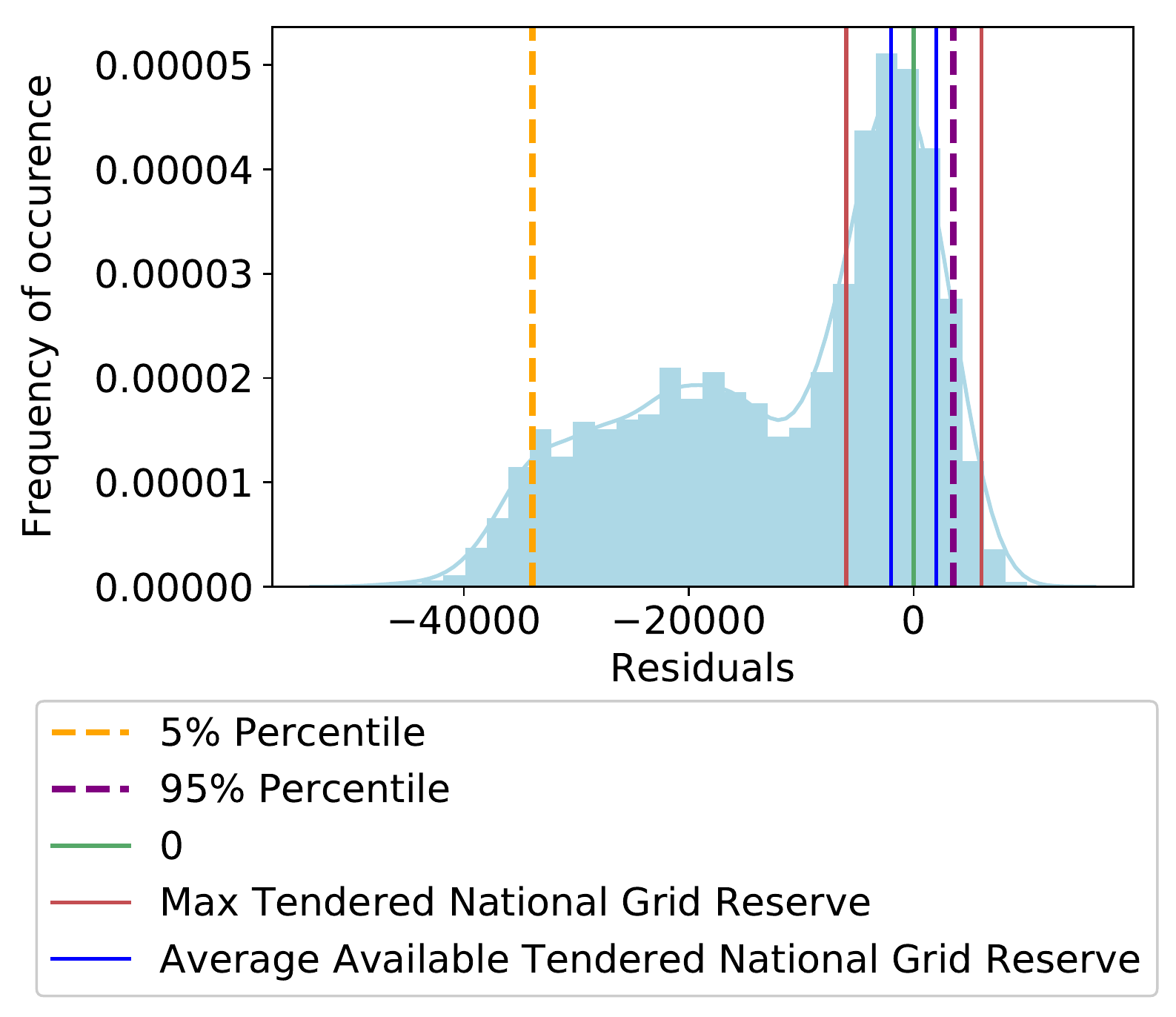}
	\caption{Online machine learning algorithm distribution. (Passive Aggressive Regressor (C=0.1, fit intercept = true, maximum iterations = 1000, shuffle = false, tolerance = 0.001), chosen as it was the worst result for the passive aggressive model for day-ahead forecasting.}
	\label{fig:bad_online_learning_day_distribution}
\end{figure}

Figure \ref{fig:bad_online_learning_day_distribution} displays the residuals for a model with poor predictive ability, the passive aggressive regressor. It displays a large period of time of prediction errors at -20,000MWh, and often falls outside of the national grid reserve. These results demonstrate the importance of trying a multitude of different models and parameters to improve prediction accuracy.

\begin{figure}
	\centering
	\includegraphics[width=0.6\columnwidth]{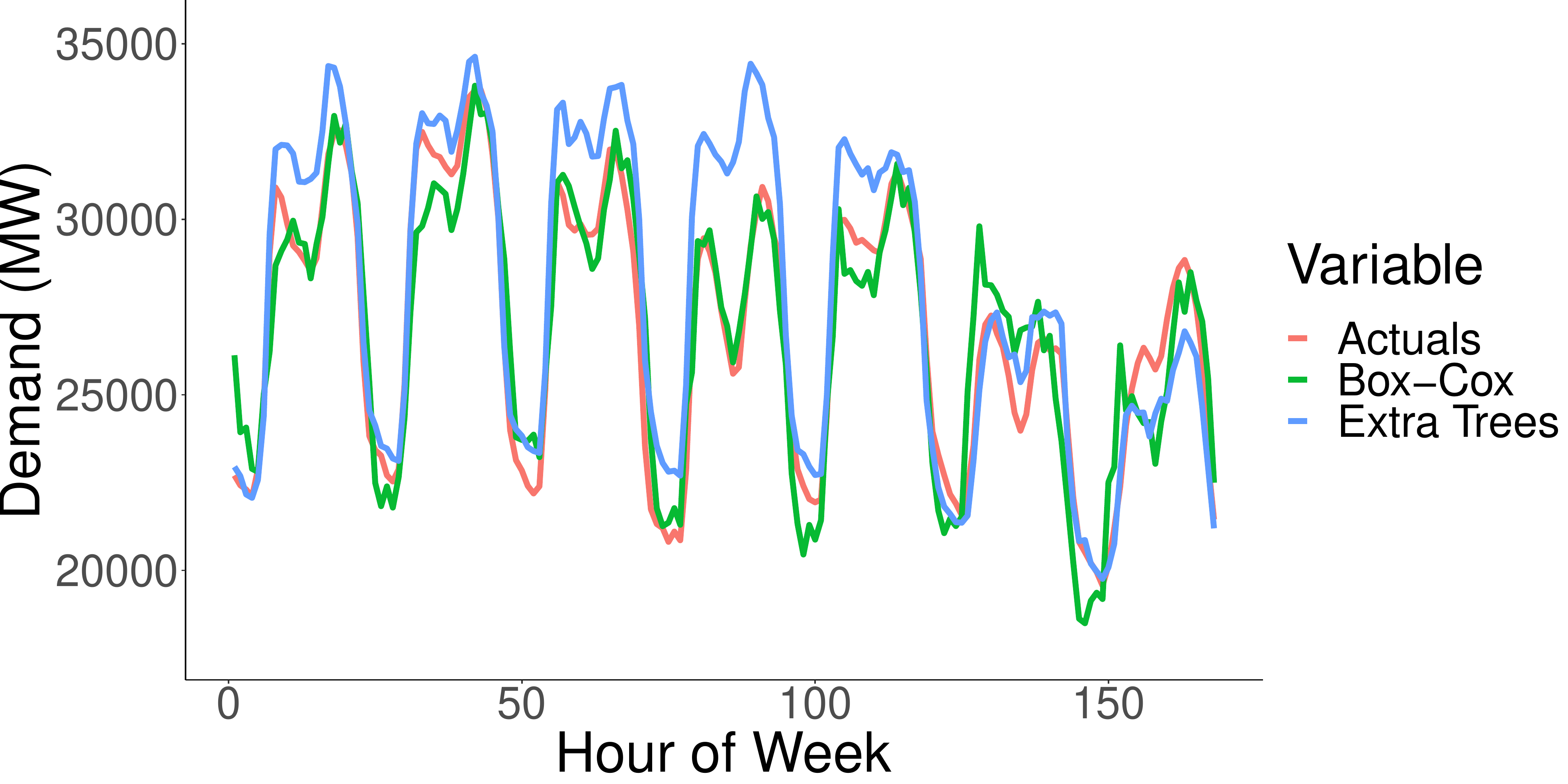}
	\caption{Best offline model compared to the best online model over a one week period for day-ahead forecasting.}
	\label{fig:both_actual_predicted}
\end{figure}

Figure \ref{fig:both_actual_predicted} displays a comparison between the actual electricity consumption compared to the predictions. It can be seen that the Box-Cox model better predicts the actual electricity demand in most cases when compared to the best offline model, the Extra Trees regressor. The Extra Trees regressor often overestimates the demand, particularly during weekdays. Whilst the Box-Cox regressor more closely matches the actual results. During the weekend (between the hours of 120 and 168), the Extra Trees regressor performs better, particularly on the Saturday (between hours of 144 and 168).

\begin{figure}
	\centering
	\includegraphics[width=0.75\columnwidth]{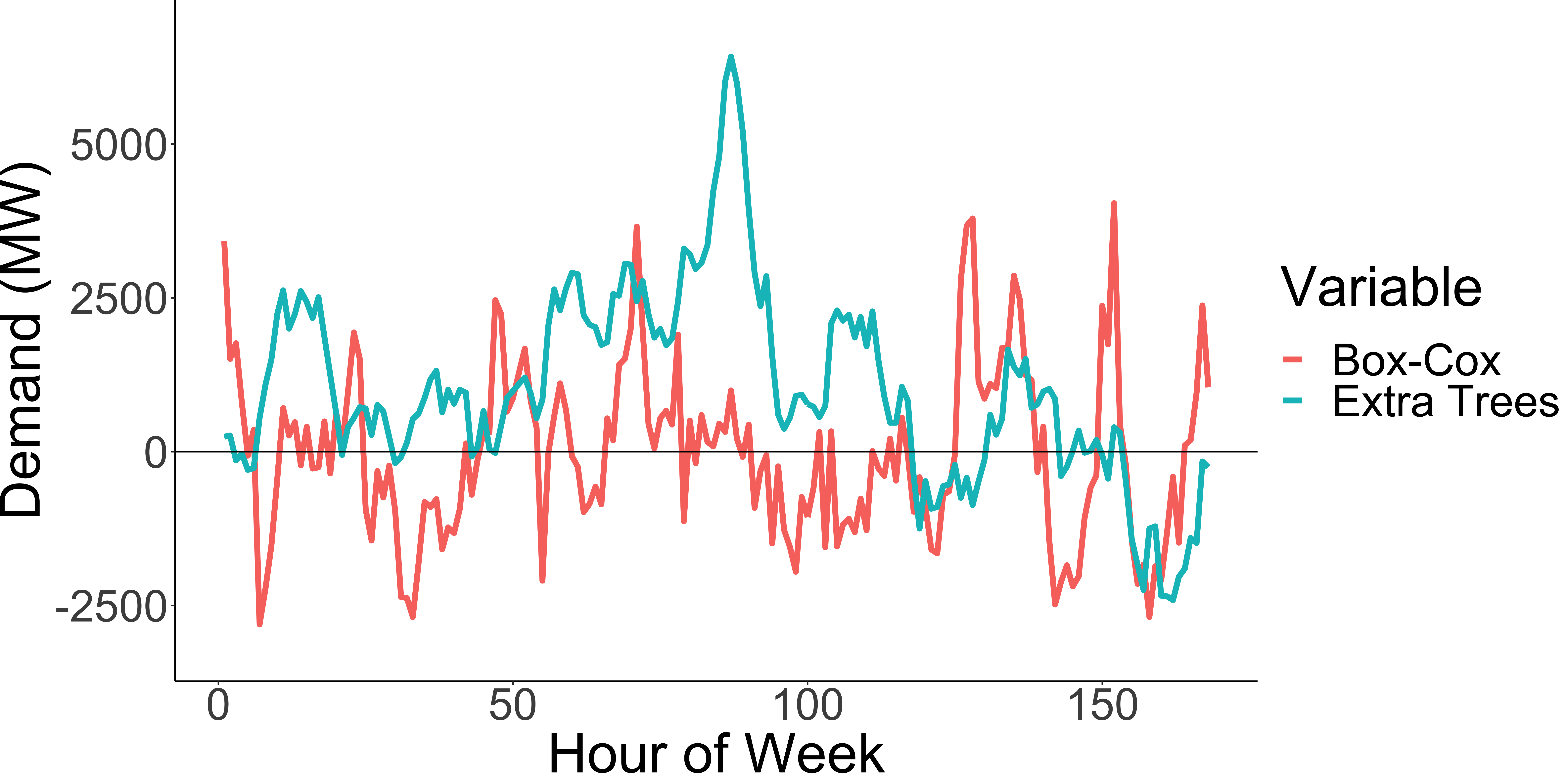}
	\caption{Residuals of best offline model compared to the best online model over a one week period for day-ahead forecasting.}
	\label{fig:both_pred_actuals_plot_difference}
\end{figure}

Figure \ref{fig:both_pred_actuals_plot_difference} shows the differences between predicted values and actuals. Where the horizontal line $y=0$ would be perfect predictions. This figure shows that the Extra Trees over predicts during the week. The Box-Cox however, struggles with the weekends. 

\begin{figure}
	\centering
	\includegraphics[width=0.6\columnwidth]{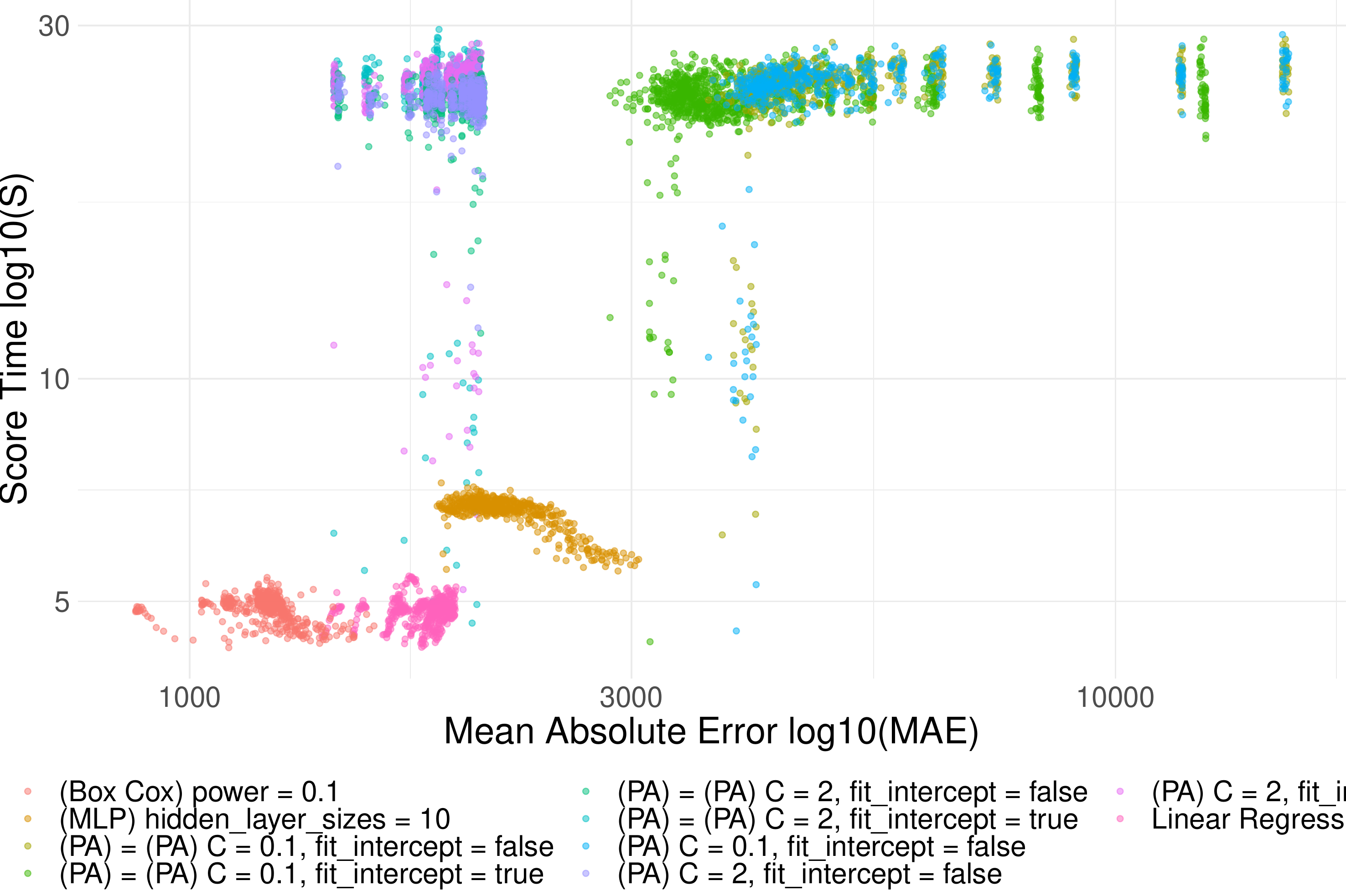}
	\caption{Time taken to test the online models versus mean absolute error for day-ahead forecasting.}
	\label{fig:online_test_vs_mae}
\end{figure}

\begin{figure}
	\centering
	\includegraphics[width=0.6\columnwidth]{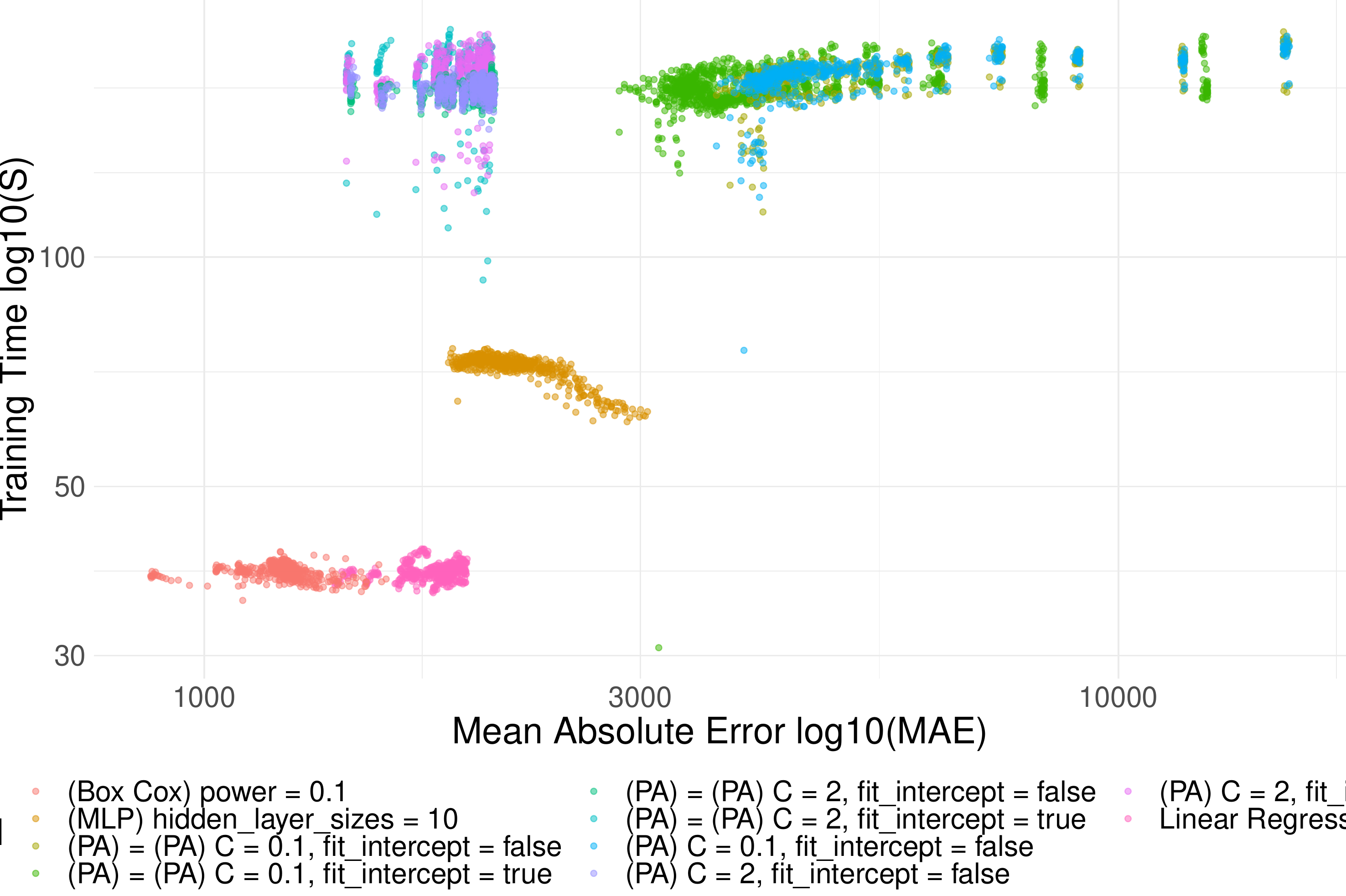}
	\caption{Time taken to train the online models versus mean absolute error for day-ahead forecasting.}
	\label{fig:online_train_vs_mae}
\end{figure}

Figures \ref{fig:online_test_vs_mae} and \ref{fig:online_train_vs_mae} display the mean absolute error versus test and training time respectively. In these graphs, a selection of models and parameter combinations are chosen. Clear clusters can be seen between different types of models and parameter types, with the passive aggressive (PA) model performing the slowest for both training and testing. Different parameter combinations show different results in terms of mean absolute error.

The best performing model is the Box-Cox model, which is also the fastest to both train and test. The linear regression, which performs worse in terms of predictive performance, is as quick to train and test as the Box-Cox model. Additionally, the multilayer perceptron (MLP) is relatively quick to train and test when compared to the PA models. 

It is noted that when compared to the offline models, the training time is a good indicator to the testing time. In other words, models that are fast to train are also fast to test and vice-versa. This also tells us that these algorithms are highly suited to this problem domain, as the training and testing time are much shorter than the available time to solve these problems.

\subsection{Scenario Comparison}

In this section we explore the effect of these residuals on investments made and the electricity generation mix.  To generate these graphs, we perturbed the exogenous demand in ElecSim by sampling from the best-fitting distributions for the respective residuals of each of the online methods. We did this for all of the online learning algorithms displayed in Figure \ref{fig:online_model_mae_barplot}. We let the simulation run for 17 years from 2018 to 2035. 

Running this simulation enabled us to see the effect on carbon emissions on the electricity grid over a long time period. For instance, does underestimating electricity demand mean that peaker power plants, such as reciprocal gas engines, are over utilised when other, less polluting power plants could be used?

\subsubsection{Mean Contributed Energy Generation}

\begin{figure*}
	\centering
	\begin{subfigure}{0.3\textwidth}
		\includegraphics[width=\columnwidth]{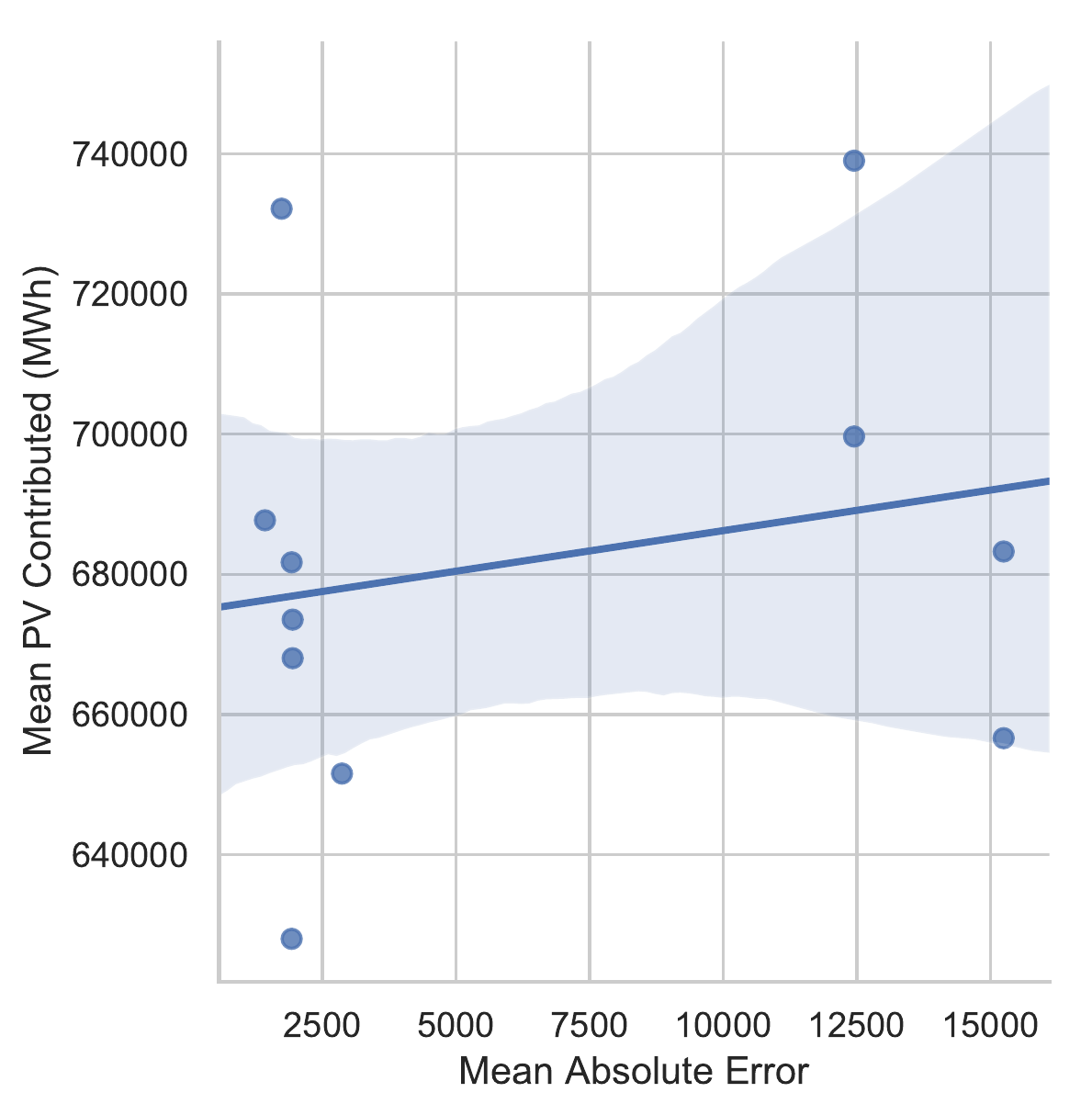}
		\caption{Photovoltaic output}
		\label{fig:contributed_PV_mean_output}
	\end{subfigure}
	\hfil
	\begin{subfigure}{0.3\textwidth}  
		\includegraphics[width=\columnwidth]{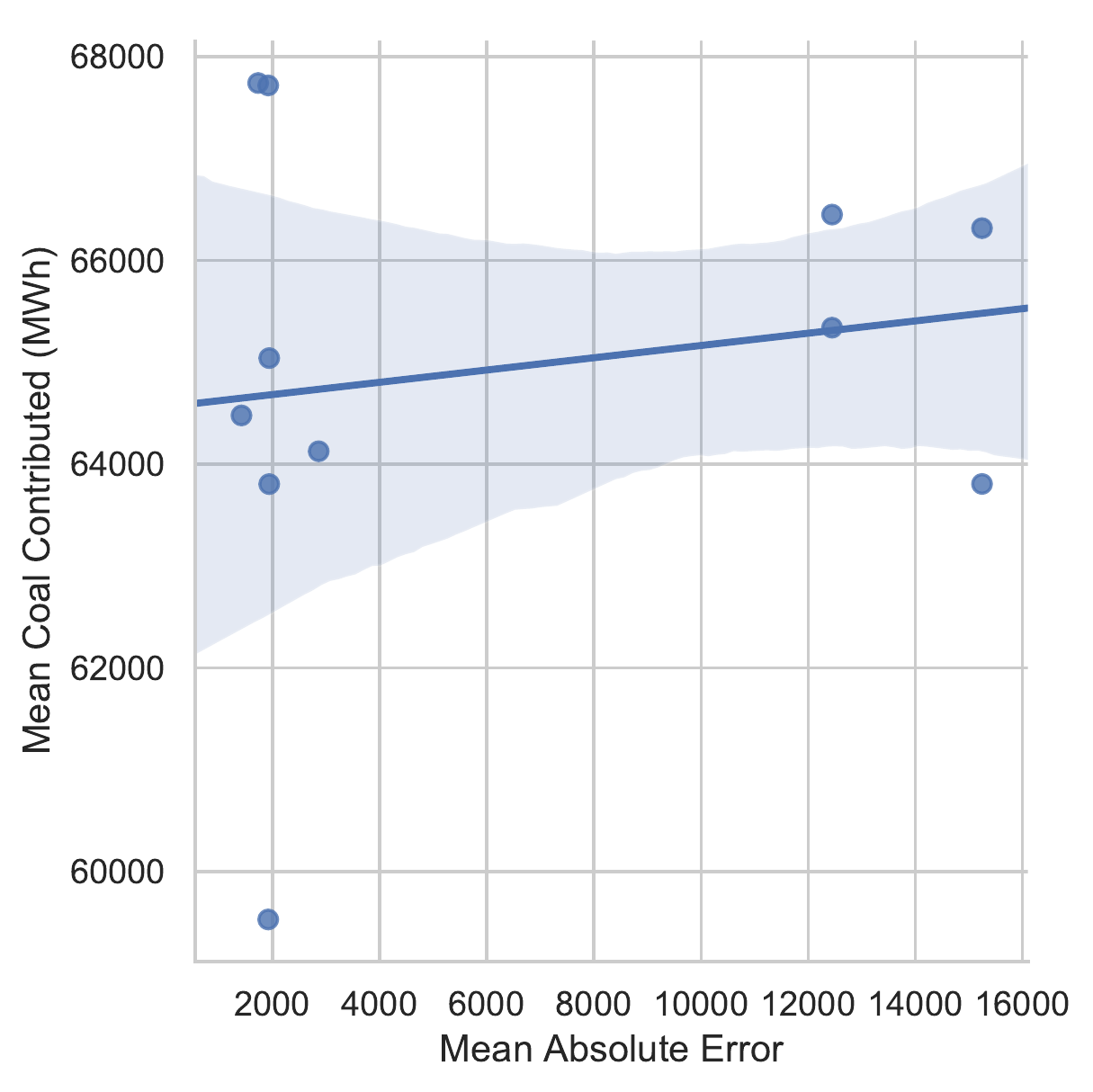}
		\caption{Coal output}
		\label{fig:contributed_Coal_mean_output}
	\end{subfigure}
	\hfil
	\begin{subfigure}{0.3\textwidth}   
		\includegraphics[width=\columnwidth]{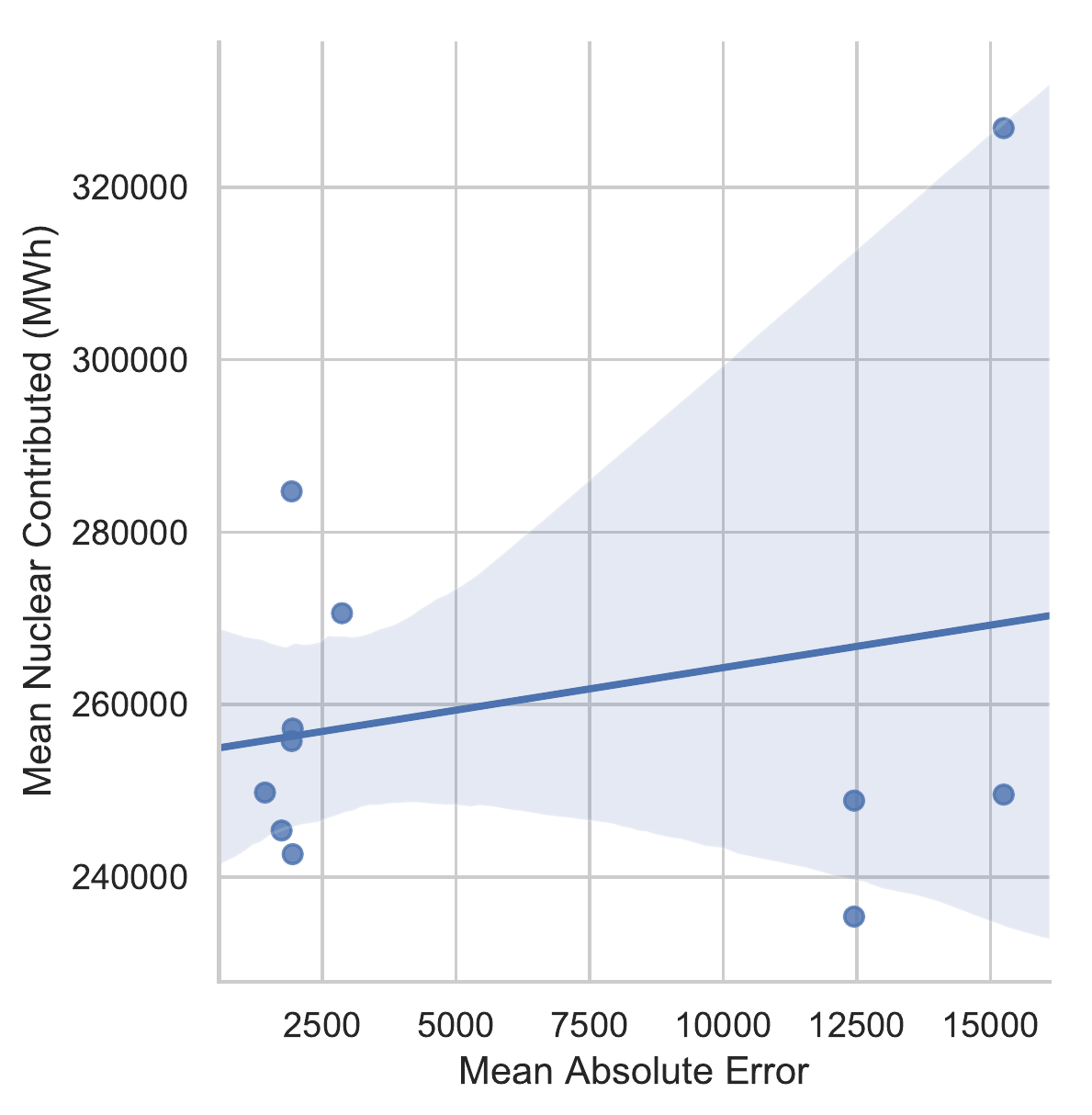}
		\caption{Nuclear output}
		\label{fig:contributed_Nuclear_mean_output}
	\end{subfigure}
	\medskip
	\begin{subfigure}{0.3\textwidth}   
		\includegraphics[width=\columnwidth]{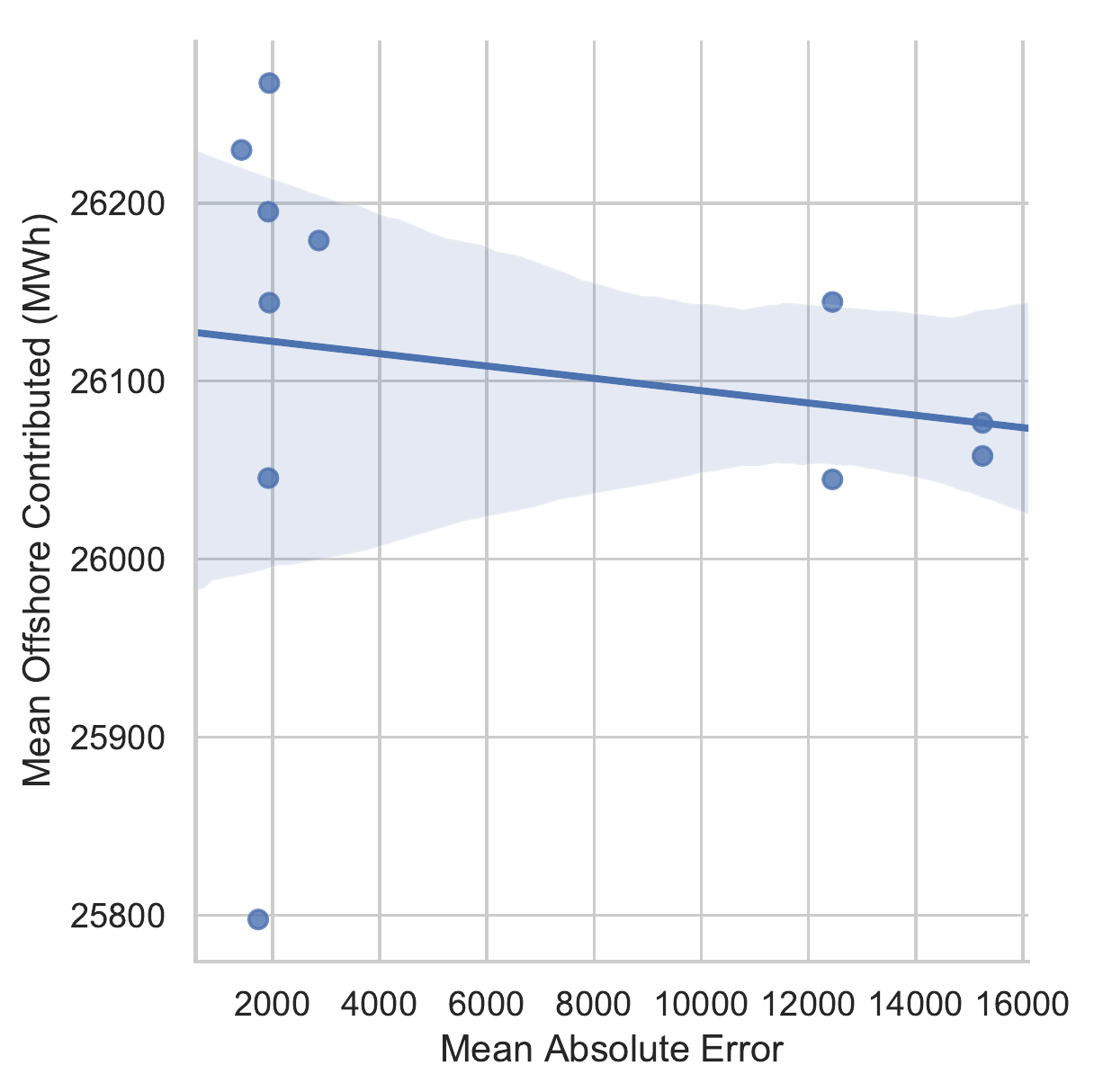}
		\caption{Offshore output}
		\label{fig:contributed_Offshore_mean_output}
	\end{subfigure}
	\hfil
	\begin{subfigure}{0.3\textwidth}   
		\includegraphics[width=1\columnwidth]{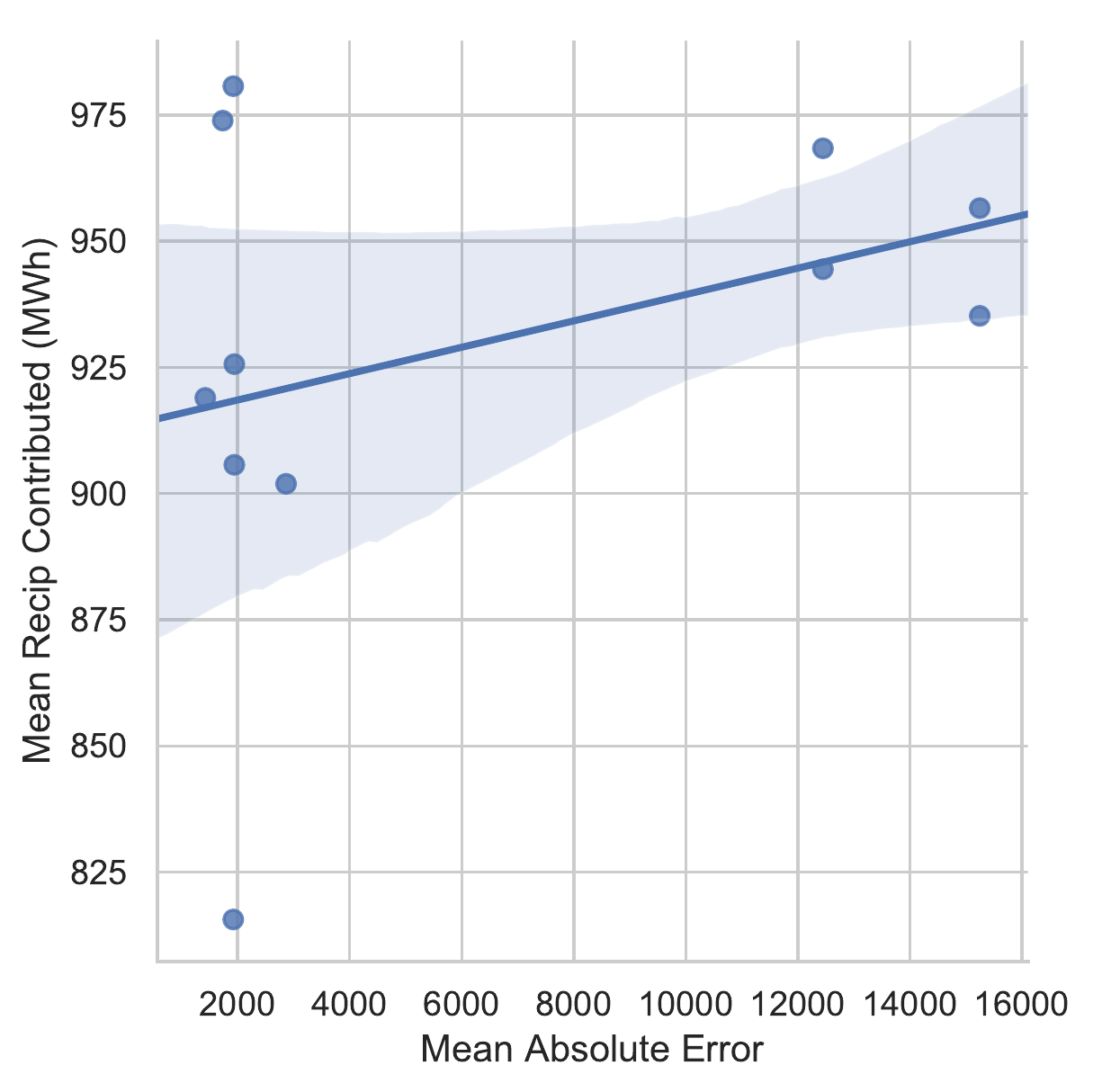}
		\caption{Reciprocal gas engine output}
		\label{fig:contributed_Recip_gas_mean_output}
	\end{subfigure}
	\label{fig:pv_coal_nuclear_offshore_outputs}
	\caption{Mean outputs of various technologies vs. mean absolute error from 2018 to 2035 in ElecSim.}
\end{figure*}

In this section we display the mean electricity mix contributed by different electricity sources over the years 2018 to 2035. 

Figure \ref{fig:contributed_PV_mean_output} displays the mean photovoltaic (PV) contributed between 2018 and 2035 vs. mean absolute error of the various online regressor models displayed in Figure \ref{fig:online_model_mae_barplot}. A positive correlation can be seen with PV contributed and mean absolute error. This is similar for coal and nuclear output, shown in Figures \ref{fig:contributed_Coal_mean_output} and \ref{fig:contributed_Nuclear_mean_output} respectively. However, as shown by Figure \ref{fig:contributed_Offshore_mean_output}, offshore wind reduces with mean absolute error. Figure \ref{fig:contributed_Recip_gas_mean_output} displays the mean reciprocal gas engine output vs mean absolute error between the same time period. Output for the reciprocal gas engine also increases with mean absolute error.

The reciprocal gas engine was expected to increase with times of high error. This is because, traditionally, reciprocal gas engines are peaker power plants. Peaker power plants provide power at times of peak demand, which cannot be covered by other plants due to them being at their maximum capacity level or out of service. It may also be the case, that with higher proportions of intermittent technologies, there is a larger need for these peaker power plants to fill in for times where there is a deficit in wind speed and solar irradiance.

It is hypothesised that coal and nuclear output increase to cover the predicted increased demands of the service. As these generation types are dispatchable, meaning operators can choose when they generate electricity, they are more likely to be used in times of higher predicted demand.

Photovoltaics may be used more with higher errors due to the times at which the errors were greatest. For example, during the day, where demand is higher, as is solar irradiance.

\subsubsection{Total Energy Generation}

In this Section, we detail the difference in total technologies invested in over the time period between 2018 to 2035, as predicted by ElecSim.

\begin{figure*}
	\centering
	\begin{subfigure}{0.3\textwidth}
		\includegraphics[width=\columnwidth]{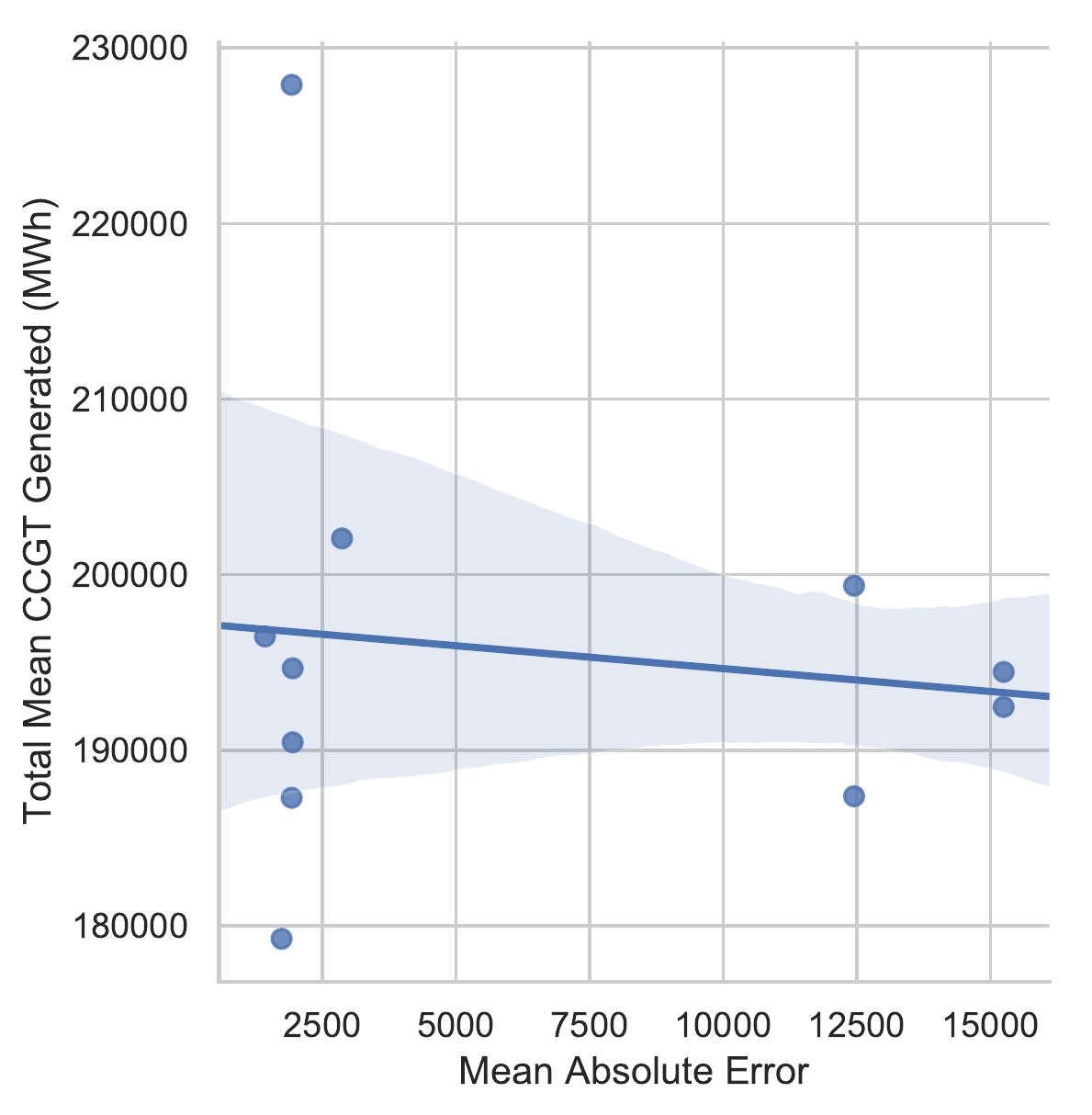}
		\caption{Total CCGT}
		\label{fig:total_CCGT_mean_output}
	\end{subfigure}
	\hfil
	\begin{subfigure}{0.3\textwidth}  
		\includegraphics[width=\columnwidth]{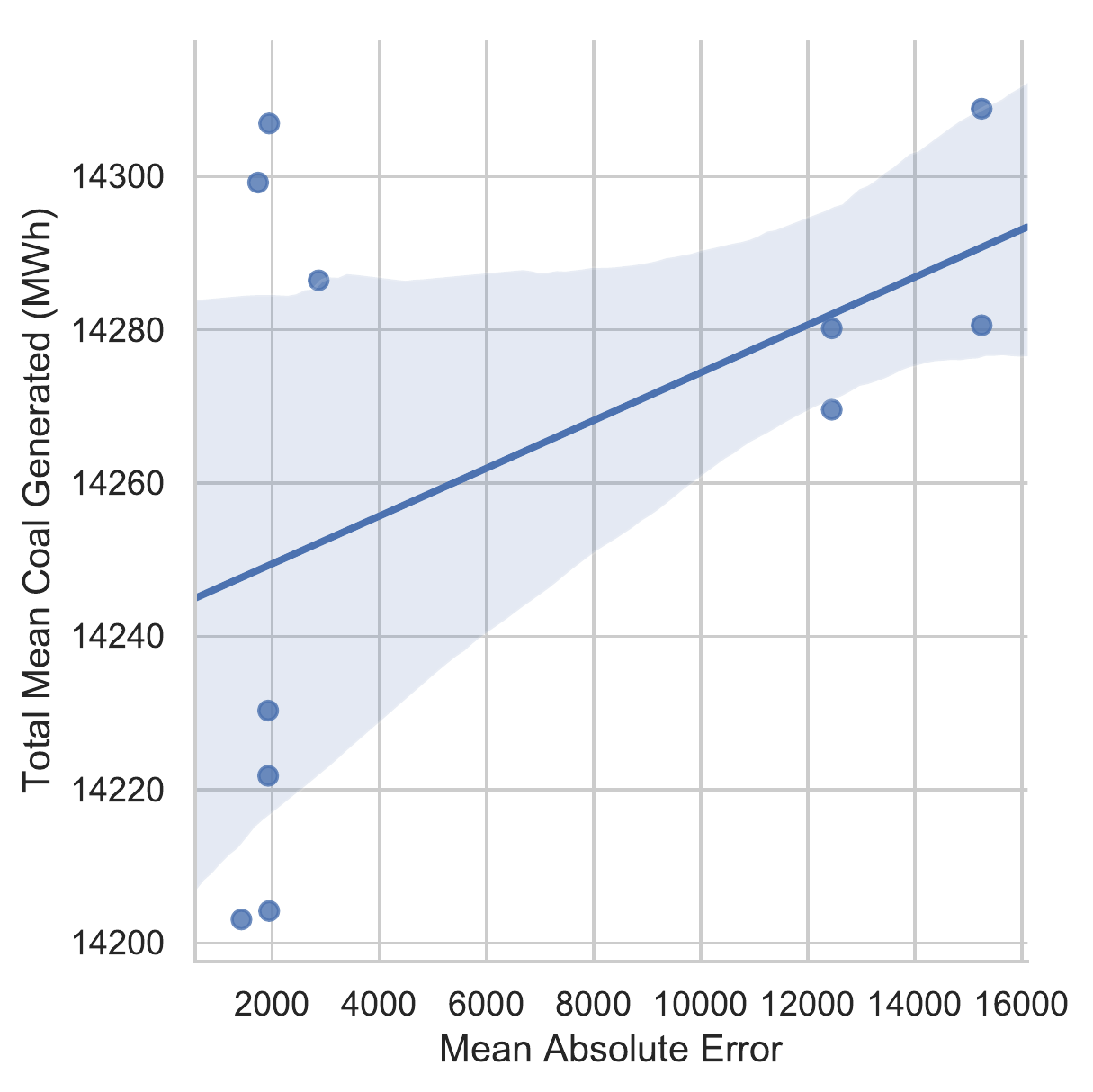}
		\caption{Total Coal}
		\label{fig:total_Coal_mean_output}
	\end{subfigure}
	\hfil
	\begin{subfigure}{0.3\textwidth}   
		\includegraphics[width=\columnwidth]{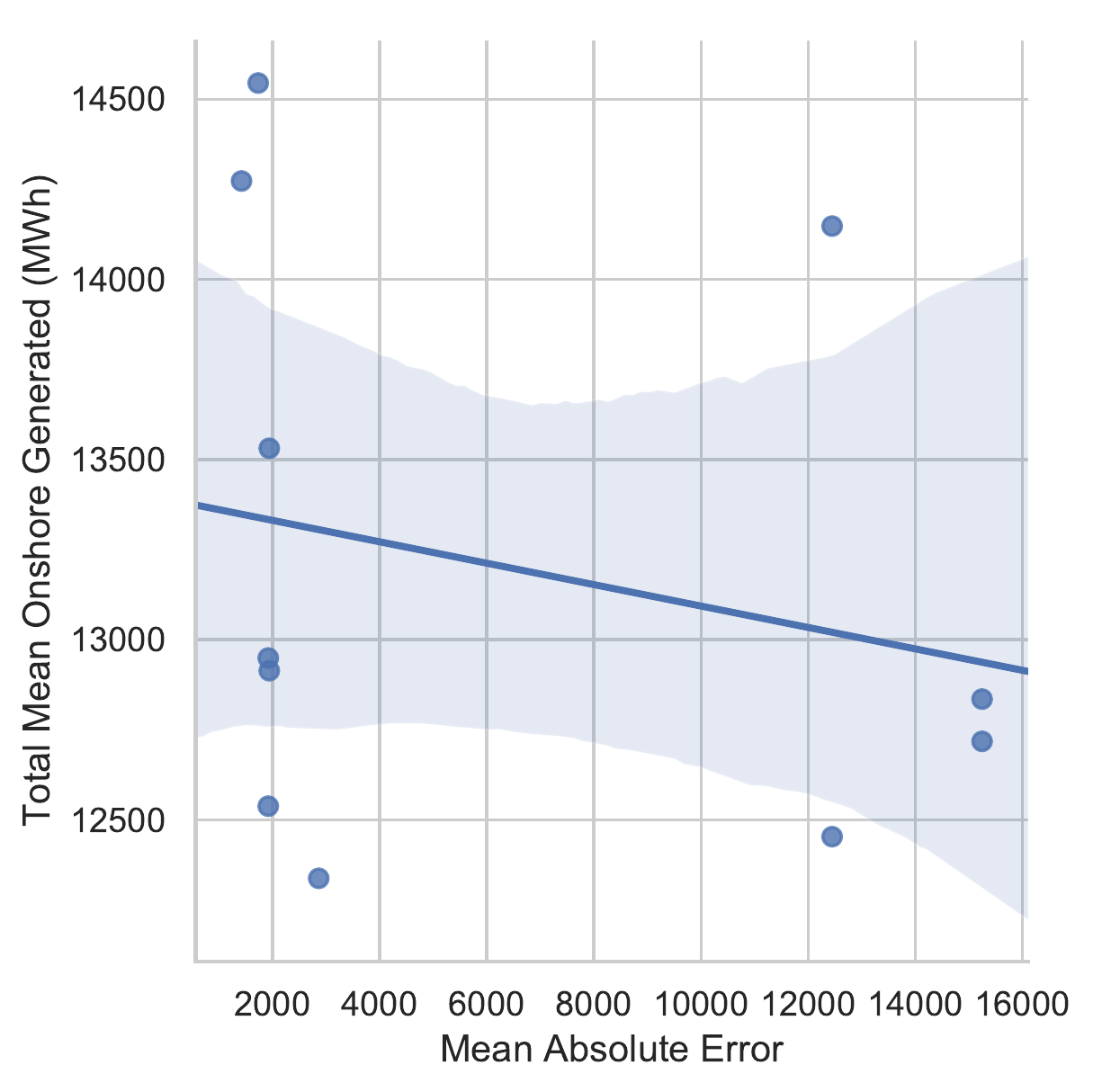}
		\caption{Total Onshore}
		\label{fig:total_Onshore_mean_output}
	\end{subfigure}
	\medskip
	\begin{subfigure}{0.3\textwidth}   
		\includegraphics[width=\columnwidth]{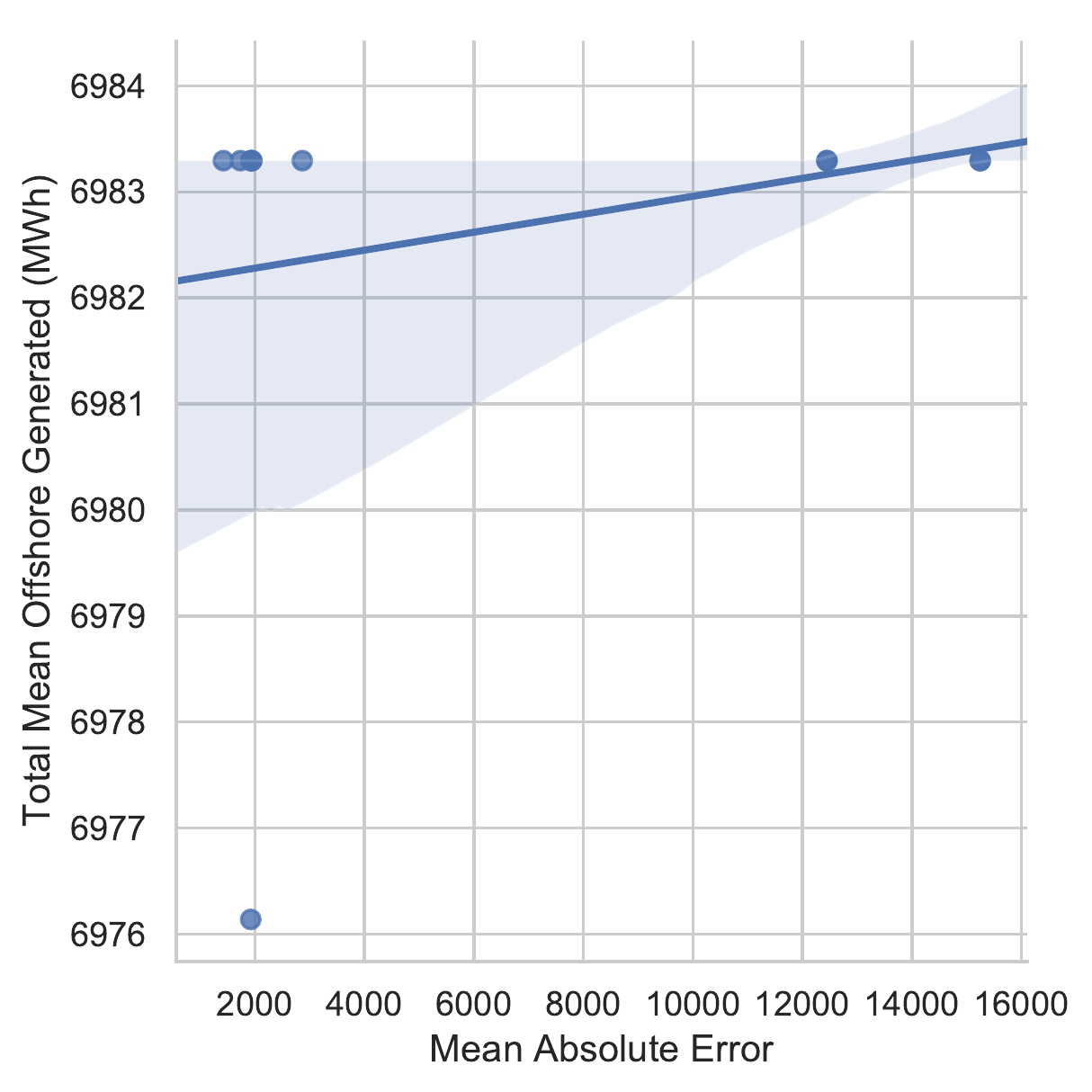}
		\caption{Total Offshore}
		\label{fig:total_Offshore_mean_output}
	\end{subfigure}
	\hfil
	\begin{subfigure}{0.3\textwidth}
		\includegraphics[width=\columnwidth]{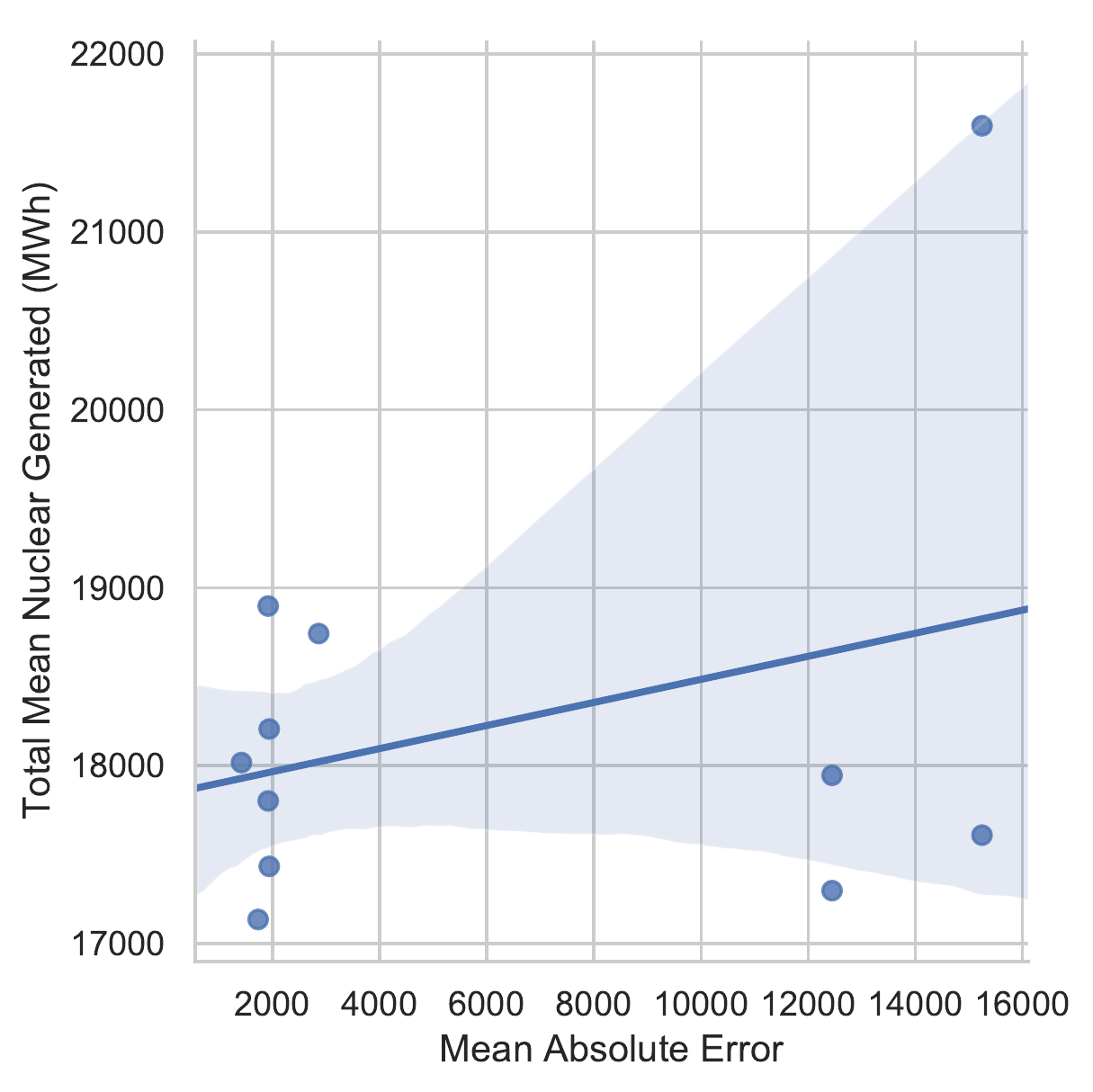}
		\caption{Total nuclear}
		\label{fig:total_nuclear_mean_output}
	\end{subfigure}
	\hfil
	\begin{subfigure}{0.3\textwidth}  
		\includegraphics[width=\columnwidth]{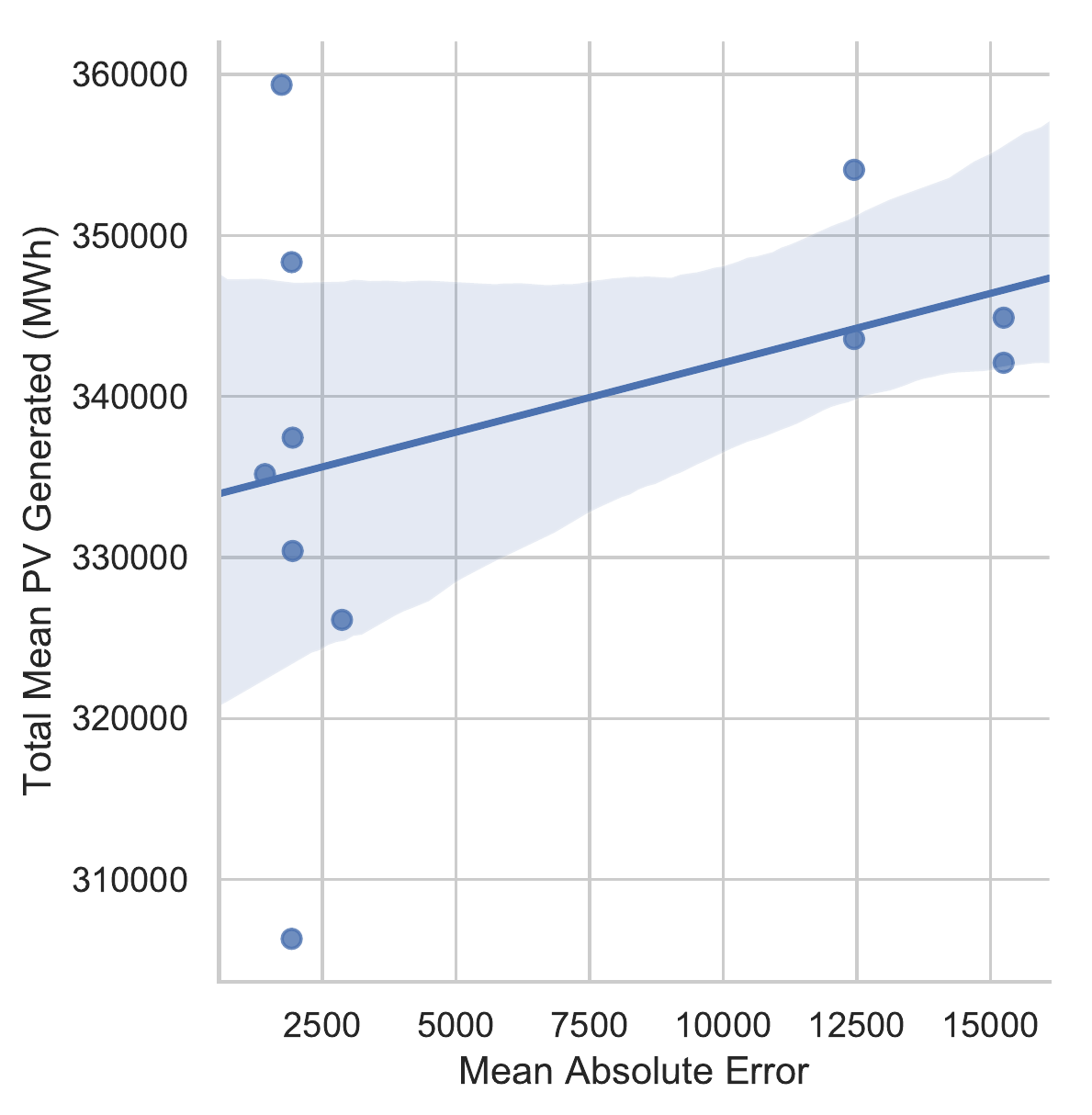}
		\caption{Total photovoltaics.}
		\label{fig:total_PV_mean_output}
	\end{subfigure}
	\label{fig:ccgt_coal_onshore_offshore_totals}
	\caption{Total technologies invested in vs. mean absolute error from 2018 to 2035 in ElecSim}
\end{figure*}

\begin{figure}
	\centering
	\begin{subfigure}{0.33\textwidth}   
		\includegraphics[width=\columnwidth]{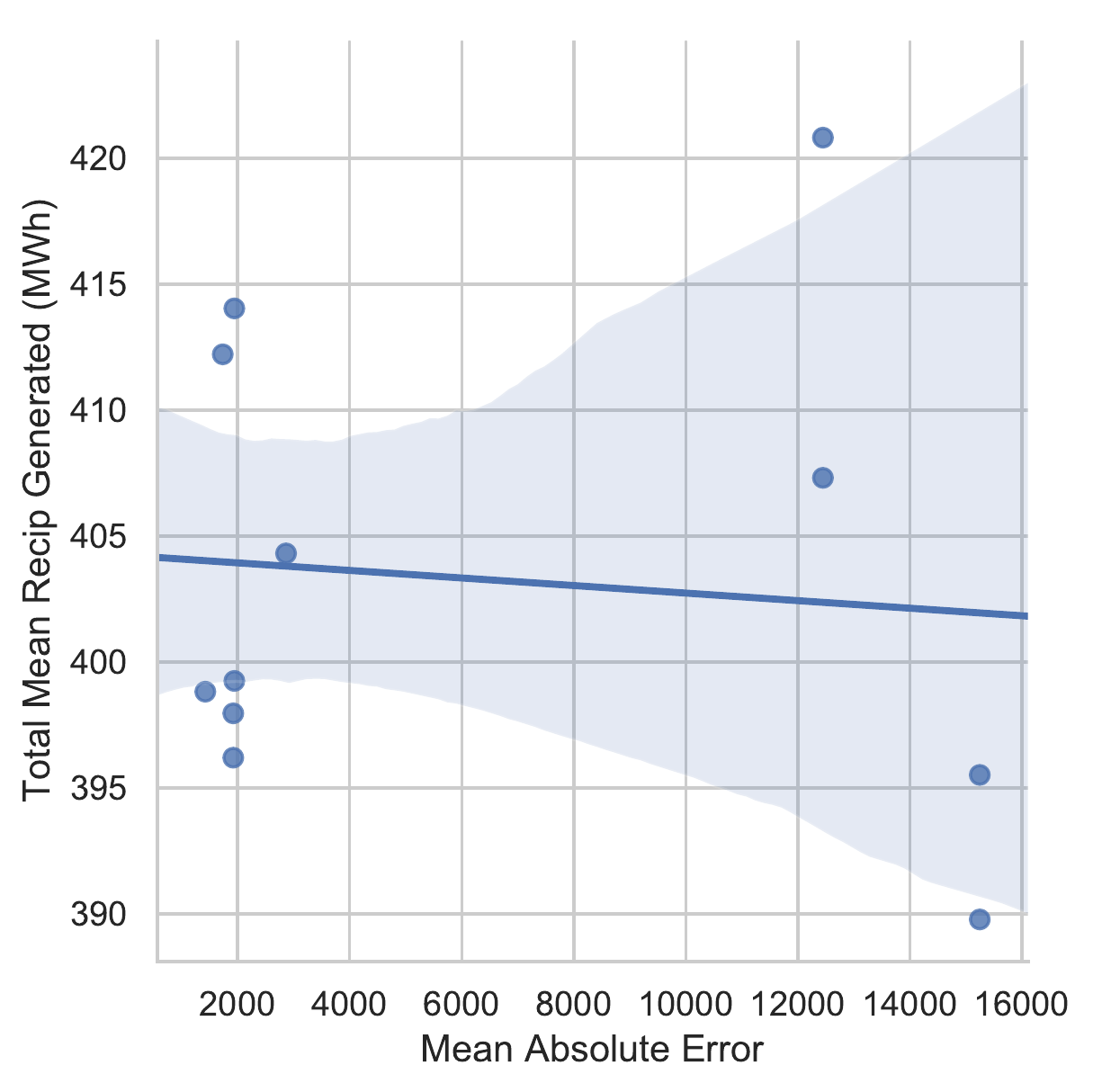}
		\caption{Total Reciprocal gas engine}
		\label{fig:total_Recip_gas_mean_output}
	\end{subfigure}
	\hfil
	\begin{subfigure}{0.33\textwidth}   
		\includegraphics[width=\columnwidth]{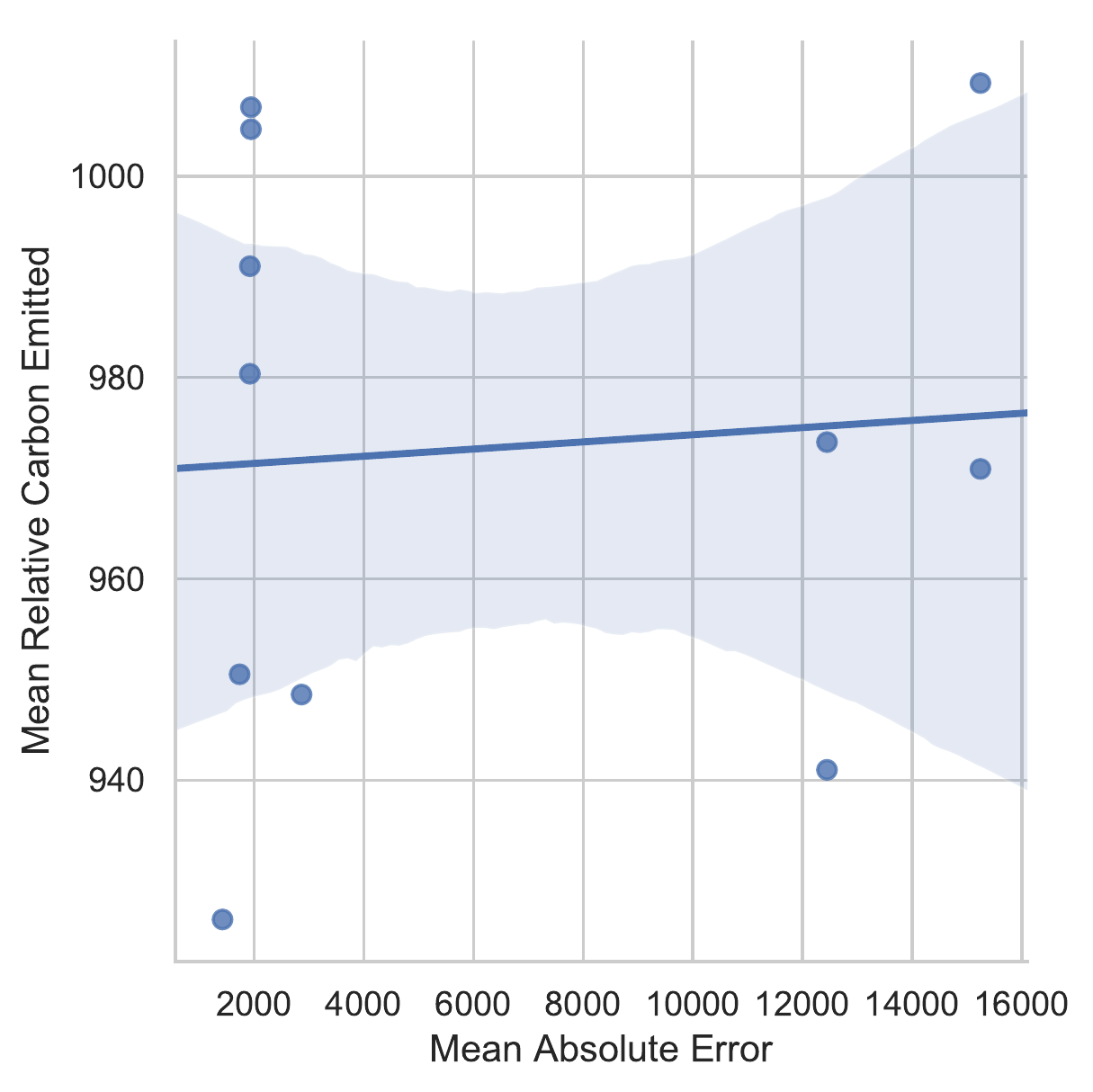}
		\caption{Mean carbon emitted}
		\label{fig:Carbon_emitted_mean_output}
	\end{subfigure}
	\label{fig:nuclear_pv_carbon_totals}
	\caption{a) Investments in reciprocal gas engine technologies vs. mean absolute error from 2018 to 2035 in ElecSim and d) mean carbon emissions between 2018 and 2035.}
\end{figure}

CCGT, onshore, and reciprocal gas engines are invested in less over the time period, as shown by Figures \ref{fig:total_CCGT_mean_output}, \ref{fig:total_Offshore_mean_output}, \ref{fig:total_Recip_gas_mean_output} respectively. While coal, offshore, nuclear and photovoltaics all exhibit increasing investments.

It is hypothesised that coal and nuclear increase in investment due to their dispatchable nature. While onshore, non-dispatchable by nature, become a less attractive investment when compared to the other technologies.

CCGT and reciprocal gas engines may have decreased in capacity over this time, due to the increase in coal. This could be because of the large consistent errors in prediction accuracy that meant that reciprocal gas engines were perceived to be less valuable.

Figure \ref{fig:Carbon_emitted_mean_output} shows an increase in relative mean carbon emitted with mean absolute error of the predictions residuals. The reason for an increase in relative carbon emitted could be due to the increased output of utility of the reciprocal gas engine, coal, and decrease in offshore output. Reciprocal gas engines are peaker plants and, along with coal, can be dispatched. By being dispatched, the errors in predictions of demand can be filled. It is therefore recommended that by improving the demand prediction algorithms, significant gains can be made in reducing carbon emissions.

\subsection{Discussion}
\label{sec:discussion}


From our results, it can be seen that different algorithms yield differing prediction accuracies. Online models can result in a decrease in 30\% of prediction error on the best offline models. We calculated this by comparing the MAE for Extra Trees to the MAE for the Box-Cox regressor. We, therefore, recommend the use of online machine learning for predicting electricity demand in a day-ahead market.

Similar to our assumptions, the online learning algorithms were able to outperform the offline models. This is due to the non-stationary nature of the data. An online method is able to use the most up-to-date knowledge of the complex system of energy demand. For instance, a certain year may have a particularly warm winter when compared to previous years, reducing the amount of electricity used for heating.

However, contrary to our assumptions, the online linear regression techniques outperformed the online machine learning techniques. This may be due to their simpler nature and ability to learn from a smaller subset of new data as opposed to relying on a large historic subset. For the offline models, the best performing algorithms were the decision tree approaches such as extra trees and random forests. This is a similar outcome to our previous work, which showed that the best performing method for demand forecasting were random forests \cite{Kell2018a}. Contrary to our assumptions, however, the lasso and ridge regression outperformed the machine learning techniques support vector regression and multilayer perceptron. This may be due to the ability of feature selection by lasso and ridge regression, which only uses the most important features.

To the best of our knowledge, more work has been done using offline learning to predict electricity demand. This may be due to the additional complexity of running online algorithms, and a smaller number of available models to run in an online fashion.

In terms of computing power, finding the optimal input parameters, hyperparameters and models to use can be a large undertaking. This is due to the exponential growth of the number of choices that can be made. This can be an issue where accuracy is of importance, especially when the data changes over time, meaning it may be necessary to retest previous results. However, due to the financial and sustainability implications, we believe the trade-off between compute time and accuracy is balanced towards compute time. There are also large implications if the model were to break at a certain point in time. We, therefore, recommend the reliance on multiple well-performing models, as opposed to solely the best performing model at any one time. That is, qualitatively compare and contrast models. However, a quantitative approach could be undertaken here. For example, taking the mean or median of decisions of various outputs of models. 

In addition, the time it takes to make a prediction using the algorithm is much smaller than the time you have to make the decision. That is, where an algorithm can be run in seconds, we have hours to submit bids onto the day ahead market. This therefore gives the user the ability to use these predictions in a real setting.

For training time and prediction time, there is often a trade-off between training and predicting. For instance, the k-nearest neighbours is fast to train, but slow to sample from. Therefore stakeholders must make a decision based upon accuracy, speed of training and sampling. 

Additionally, the impact on the broader electricity market has been shown to be significant. Principally, the investment behaviours of generation companies change as well as the dispatched electricity mix. The relative mean carbon emitted over this time period increases, due to an increase in the utilisation of coal and reciprocal gas engines, at the expense of offshore wind.

This work shows the importance for generator companies to predict electricity demand with high accuracy. Without this work, it is possible to have a sub-optimal electricity mix over the long-term from the perspective of cost minimisation. The generator companies can use these results to improve their predictions, and policy makers can use these results to ensure that a framework is in place to provide companies with high accuracy forecasts.

The limitations of this work, from the perspective of a generator company or policy maker, is that it does not provide a fool-proof way of improving accuracy of predictions with online learning. Rather, a methodology to follow to improve results. In addition, we do not model the impact of weather for solar and onshore renewable energy sources. This may have a long-term significant impact on the electricity mix. Particularly with climate change effects, which may change weather patterns.

\section{Conclusion}
\label{sec:conclusion}

The availability of high granularity data produced by the smart grid enables network operators to gain greater insights into their customer behaviour and electricity usage. This enables them to improve customer experience, utility operations and power management. In the first piece of work, we demonstrated that implementing the \textit{k}-means clustering algorithm to group similar customers improved the accuracy of every one of the different models tested. Distinct models were trained for each of the clusters and the individual forecasts aggregated for the total aggregated forecast. It was found that Random Forests outperformed the other models at all levels of clustering and that the optimum number of clusters was 4. Whilst the dataset used focused on residential data it is expected that applying a similar clustering technique on commercial properties would have a similar effect.

Next, we evaluated 16 different machine learning and statistical models to predict electricity demand in the UK for the day-ahead market. Specifically, we used both online and offline algorithms to predict electricity demand 24 hours ahead. We compared the ability for the offline models: lasso regression, random forests, support vector regression, for both online and offline learning: linear regression, multilayer perceptron and for just online learning: the Box-Cox transformation and the passive aggressive regressor, amongst others. The Box-Cox, as well as the passive aggressive regressors, were used as online learning algorithms, the multilayer perceptron and linear regression were used as both, whereas the rest were used as offline learning algorithms.

We measured the errors and compared these to each model as well as the national grid reserve. We found that through the use of an online learning approach, we were able to significantly reduce error by 30\% on the best offline algorithm.  We were also able to reduce our errors to significantly below the national grid's mean and maximum tendered reserve, thus significantly reducing the chances of blackouts.

In addition to this, we took these errors, or residuals, and perturbed the electricity demand within the market of the agent-based model ElecSim. This enabled us to see the impact of different error distributions on the long-term electricity market, both in terms of investment and in terms of the electricity mix.

We observed that with an increase in prediction errors, we get a higher proportion of electricity generated by coal, offshore, nuclear, reciprocal gas engines and photovoltaics. This could be due to the fact that more peaker and dispatchable plants are required to fill in for unexpected demand. In addition, a higher proportion of intermittent renewable energy sources leads to a higher use of peaker power plants to fill in the gaps of intermittency of wind and solar irradiance. However, by reducing the mean absolute error, we are able to significantly reduce the amount of reciprocal gas engines and coal usage.

In future work, we will look into the features that best aid in the forecasting of electricity consumption, try a wider variety of models in an ensemble manner and try different clustering techniques such as \acrfull{som} to obtain better accuracy measures. We will also compare different prediction error measures. 

We would also like to trial a different selection of algorithms and statistical models and trial different inputs to the models, for instance, by providing the model with two months worth of historical data as dependent variables. Additionally, we would like to see the impact of predicting wind speed and solar irradiance to see how these impact the overall investment patterns and electricity mix.

\chapter{Carbon optimization}
\label{chapter:carbon}
\ifpdf
\graphicspath{{Chapter3/Figs/Raster/}{Chapter3/Figs/PDF/}{Chapter3/Figs/}}
\else
\graphicspath{{Chapter3/Figs/Vector/}{Chapter3/Figs/}}
\fi

\section*{Summary}


Carbon taxes have been shown to be an efficient way to aid in a transition to a low-carbon electricity grid. In this work, we demonstrate how to find optimal carbon tax policies through a genetic algorithm approach, using the ElecSim model. To achieve this, we use the NSGA-II genetic algorithm \cite{Valkanas2014} to minimise average electricity price and relative carbon intensity of the electricity mix. We demonstrate that it is possible to find a range of carbon taxes to suit differing objectives. 

In previous chapters we investigated the building of a long-term agent-based model for electricity markets in the UK, and investigated the impact of short-term predictions on long-term investments. However, a policy maker may have specific objectives within the electricity market that they are overseeing. For example, two key objectives in the current climate are a reduction in carbon emissions, as well as a reduction in electricity prices. A common tool for this are carbon taxes. 

However, a policy maker is unable to trial all carbon taxes and know the interactions between carbon price, carbon emissions and electricity price. Therefore, in this chapter we propose a machine learning method that carries out this process automatically. This goes beyond the manual work of setting carbon taxes exhibited in Chapter \ref{chapter:elecsim}. It fits within this thesis by exploring the ability for machine learning to have an impact on the wider electricity market over the long-term.

Our results show that we are able to minimise electricity cost to below \textsterling10/MWh as well as carbon intensity to zero in every case. In terms of the optimal carbon tax strategy, we found that an increasing strategy between 2020 and 2035 was preferable. Each of the Pareto-front optimal tax strategies are at least above \textsterling81/tCO2 for every year. The mean carbon tax strategy was \textsterling240/tCO2. Whilst this work was undertaken for the UK, it could be applied elsewhere.

This chapter is structured as follows: we introduce our work in Section \ref{carbonoptim:sec:intro}. Section \ref{carbonoptim:sec:litreview} covers examples of optimisations using genetic algorithms and different carbon strategies. Section \ref{carbonoptim:sec:optimisation} details the optimization techniques applied. We present our results in Section \ref{carbonoptim:sec:results}, and conclude in Section \ref{carbonoptim:sec:conclusion}. The work in this chapter has been published in \cite{Kell2020a}.

\section{Introduction}
\label{carbonoptim:sec:intro}

In this work, we use the electricity market agent-based model ElecSim to find an optimum carbon tax policy \cite{Kell}. Specifically, we use a genetic algorithm to find a carbon tax policy to reduce both average electricity price and the relative carbon density by 2035 for the UK electricity market. We compare this optimal strategy to the carbon tax policy of the UK British government.

Carbon taxes have been shown to quickly lower emissions and lower the costs to the public \cite{Wittneben2009}. Carbon taxes are able to send clear price signals, as opposed to a cap-and-trade scheme, such as the EU Emissions Trading System (ETS), which has been shown to be unstable~\cite{Wittneben2009}.

In this work, we use the reference scenario projected by the UK Government's Department for Business \& Industrial Strategy (BEIS) with model parameters calibrated by Kell \textit{et al.} \cite{DBEIS2019,Kell2020}. This reference scenario projects energy and emissions until 2035. We undertake various carbon tax policy interventions to see how we can reduce relative carbon density whilst at the same time, reduce the average electricity price.

We did not model the EU ETS within ElecSim due to the significant additional complexity required to model it in a simulation. However, the carbon tax strategy that we took allows for policy makers to understand, in principle, the range of carbon tax price that are required. This information is directly relevant to both an ETS scheme or a carbon tax approach, as these schemes can be designed with the appropriate range to achieve the goals of the policy makers.

The parameter space we optimise over is the carbon tax price over a 17 year period from 2018 to 2035. The carbon price is used to influence the objectives of average electricity price and relative carbon intensity in 2035. Grid and random search are approaches which trial parameters at evenly distributed spaces and random spaces respectively. These approaches are often inefficient, however, and require an increased number of simulations due to their static nature. They are inefficient due to the likelihood that they will continue to search a poor parameter space area, even with consistent negative feedback. Genetic Algorithms, in contrast, use an evolutionary computing approach to find global optimal solutions faster, however, this is not guaranteed, as there are some scenarios where grid search or random search may be quicker to find an optimal solution. This is of particular importance in cases with a large number of parameters or in simulations which require a long compute time, which is the case for ElecSim.

In order to optimise over two potentially competing objectives, i.e. average electricity price and relative carbon intensity, we use the Non-Dominated Sorting Genetic Algorithm II (NSGA-II) \cite{Valkanas2014}. The NSGA-II algorithm can approximate a Pareto frontier ~\cite{Pareto1927, Stadler1979}. A Pareto frontier is a curve in which there is no solution which is better than another along the curve for different sets of parameters. In this context, better means that a solution is closer to the optimal for a particular combination of objectives.

We find that the rewards of average electricity price and relative carbon intensity are not mutually exclusive. That is, it is possible to have both a lower average electricity price and a lower relative carbon price. This is due to the low short-run marginal cost of renewable technology, which has been shown to lower electricity prices \cite{OMahoney2011}.

The main contribution of this work is to explore carbon tax strategies using genetic algorithms for multi-objective optimisation.








\section{Literature review}
\label{carbonoptim:sec:litreview}

Multi-objective optimisation problems are commonplace. In this section, we review multiple applications that have used multi-objective optimisation, as well as explore the literature which focus on finding optimal carbon tax strategies.

\subsection{Examples of Optimization}

Similar to our work, Ascione  \textit{et al}. \cite{Ascione2016} use the NSGA-II algorithm to generate a Pareto front to optimise for two objectives: operating cost for space conditioning and thermal comfort. The aim of their work is to optimise the hourly set point temperatures with a day-ahead planning horizon. A Pareto front is generated, which allows a user to select a solution according to their comfort needs and economic constraints. This work showed a reduction in operating costs by up to 56\% as well as improved thermal comfort.

Gorza\l{}czany \textit{et al}. \cite{Gorzaczany2016a} also apply the NSGA-II algorithm. However, they apply it to the credit classification problem. The objectives optimised over were accuracy and interpretability when making financial decisions such as credit scoring and bankruptcy prediction. This technique was able to significantly outperform the alternative methods in terms of interpretability while remaining competitive or superior in terms of the accuracy and speed of decision making in comparison with the existing classification methods.

Ma \textit{et al}. \cite{Ma2015} use the multi-objective artificial immune optimisation algorithm for land use allocation (MOAIM-LUA model). They balance land use supply and demand based on the future dynamic demands from different land-use stakeholders in the region at the macro-level in Anlu County, China. The objectives to optimise were economic and ecological benefits. They found that for this application, they were able to obtain better solution sets than the NSGA-II algorithm.

In our previous work, not presented here, we use the NSGA-II algorithm to optimise average overhead and energy consumption of a condor system \cite{Kell2019}. We use the genetic algorithm to trial different parameters of a Q-learning reinforcement learning algorithm, which acted as a job scheduler. We found that we were able to generate a Pareto-front  which would allow stakeholders to select an optimum for their use case. 

Whilst these works use a similar approach to solving an optimisation problem using the NSGA-II algorithm, they optimise a different model and application. 

\subsection{Carbon Tax Strategies}

In this section, we explore different strategies employed in the literature to analyse the benefits and consequences of a carbon tax. To the best of our knowledge, we are the first to employ a multi-objective optimisation algorithm to minimise average electricity price and relative carbon density.

Levin \textit{et al}.~\cite{Levin2019} use an optimisation model to analyse market and investment impacts of several incentive mechanisms to support investment in renewable energy and carbon emission reductions. Carbon tax was found to be the most cost-efficient method of reducing emissions. 

Zhou \textit{et al.}~\cite{Zhou2019} construct a social welfare model based on a Stackelberg game. The differences and similarities between a flat carbon tax and an increasing block tariff carbon tax are analysed using a numerical simulation. This work shows that an increasing block tariff carbon tax policy can significantly reduce tax burdens for manufacturers and encourage low-carbon production. In contrast to Zhou \textit{et al}. we trial multiple different carbon tax strategies using a machine learning approach. 

Li \textit{et al}.\cite{Li2017} use a hierarchical carbon market scheduling model to reduce carbon emissions. Multi-objective optimisation was applied to discover optimal behaviours for policymakers, customers and generators to minimise the costs incurred by these actors. Our work, however, focuses on the different strategies of carbon tax as opposed to optimal actor behaviour.

These works do not use optimisation to find a carbon tax directly, instead seeing the impact of several different options. Our work has the ability to trial a vastly greater number of carbon tax strategies, to hopefully find an optimum.

\section{Optimization methods}
\label{carbonoptim:sec:optimisation}

Multi-objective optimisation allows practitioners to overcome the problems with optimising multiple objectives with classical optimisation techniques, such as the genetic algorithm used in Chapter \ref{chapter:elecsim}. Multi-objective optimisation algorithms are able to generate Pareto-optimal solutions as opposed to converting the multiple objectives into a single-objective problem. A single-objective problem assumes that there is only a single optimum, or combination of results, and that other combinations are inferior. This may not be the case, as different solutions are superior for a different set of circumstances. A Pareto frontier is made up of many Pareto-optimal solutions which can be displayed graphically. A user is then able to choose between various solutions and trade-offs according to their wishes. The NSGA-II algorithm, a multi-objective genetic optimisation algorithm, is able to generate a Pareto frontier in a single optimisation run. 

In the following sub-sections, we detail the NSGA-II algorithm. For an overview of the standard Genetic Algorithm, please refer to subsection \ref{elecsim:ssec:geneticalgorithm}.


\subsection{NSGA-II}

NSGA-II \cite{Valkanas2014} is efficient for multi-objective optimisation on a number of benchmark problems and finds a better spread of solutions than Pareto Archived Evolution Strategy (PAES)~\cite{Knowles1999} and Strength Pareto EA (SPEA)~\cite{Zitzler2006} when approximating the true Pareto-optimal front.

The majority of multi-objective optimisation algorithms use the concept of \emph{domination} during population selection \cite{Burke2014}. A non-dominated algorithm, however, seeks to achieve the Pareto-optimal solution. This is where no single solution should dominate another. An individual solution $\mathbf{x}^{1}$ is said to dominate another $\mathbf{x}^{2}$, if and only if there is no objective of $\mathbf{x}^{1}$ that is worse than the same objective of $\mathbf{x}^{2}$ and at least one objective of $\mathbf{x}^{1}$ is better than the same objective of $\mathbf{x}^{2}$ \cite{Bao2017}.

Non-domination sorting is the process of finding a set of solutions which do not dominate each other and make up the Pareto front. See Figure \ref{fig:pareto-layering}a for a visual representation, where $f_1$ and $f_2$ are two objectives to minimise and L1, L2 and L3 are dominated layers.

In this section, we define the processes used by the NSGA-II algorithm to determine which solutions to keep:
\subsubsection{Non-dominated sorting}
We assume that there are $M$ objective functions to minimise, and that ${\bf x^{1}} = \{x_j^{1}\}$ and $\bf x^{2}$ are two solutions. $x_j^{1}<x_j^{2}$ implies solution $\bf x^{1}$ is better than solution $\bf x^{2}$ on objective $j$. A solution $\mathbf{x}^{1}$ is said to dominate the solution $\mathbf{x}^{2}$ if the following conditions are true:
\begin{enumerate}
	\item The solution $\mathbf{x}^{1}$ is no worse than $\mathbf{x}^{2}$ in every objective. I.e. $x^{1}_j \leq x^{2}_j \;\;  \forall j \in\{1,2,\ldots,M\}$.
	\item The solution $\mathbf{x}^{1}$ is better than $\mathbf{x}^{2}$ in at least one objective. \\ i.e. $\exists\  {j}\in \{ 1,2,\ldots,M\} \;\; s.t. \;\;x^{1}_j < x^{2}_j$.
\end{enumerate}

\begin{figure}[t] 
	\vskip -10pt
	\centering
	\includegraphics[width=0.48\textwidth]{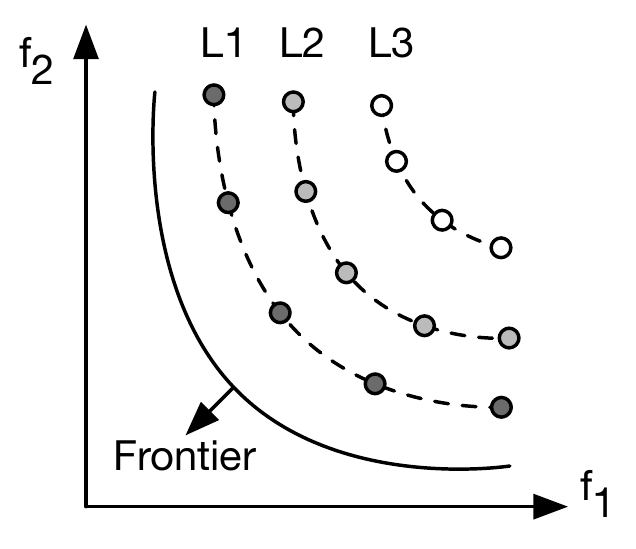}
	\includegraphics[width=0.480\textwidth]{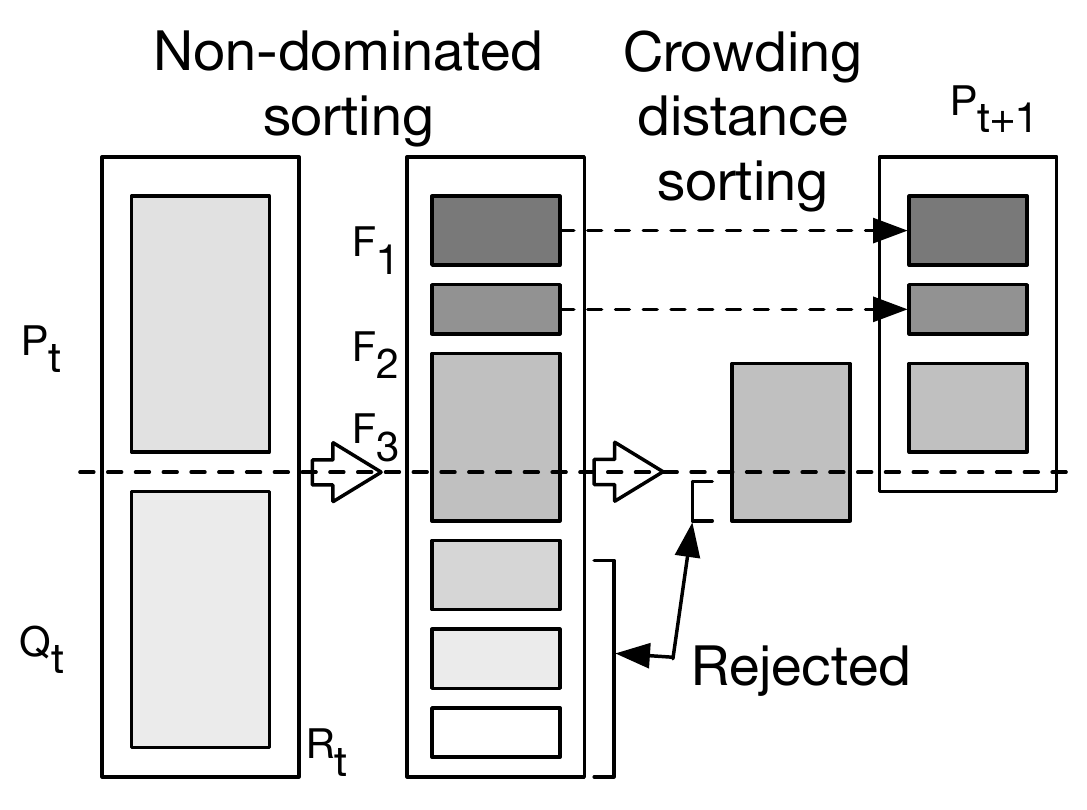}
	\vskip -8pt
	\caption{a) Schematic of non-dominated sorting with solution layering b) Schematic of the NSGA-II procedure}
	\label{fig:pareto-layering}
	\vskip -15pt
\end{figure}

Next, each of the solutions are ranked according to their level of non-domination. An example of this ranking is shown in Figure \ref{fig:pareto-layering}a. Here, $f_1$ and $f_2$ are the objectives to be minimised. The Pareto front is the first front. All of the solutions in the Pareto front are not dominated by any other solution. The solutions in layer 1, L1, are dominated only by those in the Pareto front, and are non-dominated by those in L2 and L3.

The solutions are then ranked according to their layer. For example, the solutions in the Pareto front are given a fitness rank ($i_{rank}$) of 1, solutions in L1 have an $i_{rank}$ of 2.

\subsubsection{Density Estimation}
($i_{distance}$) is calculated for each solution. This is the average distance between the two closest points to the solution in question.

\subsubsection{Crowded comparison operator}
($\prec_n$) is used to ensure that the final frontier is an evenly spread out Pareto-optimal front. This is achieved by using the two attributes: $(i_{rank})$ and$(i_{distance})$. 
The partial order is then defined as:\\    
$i\prec_nj$ if $(i_{rank}<j_{rank})$ or $((i_{rank}=j_{rank})$ and  $(i_{distance}>j_{distance}))$ \cite{Valkanas2014}.

This order prefers solutions with a lower rank $i_{rank}$. For solutions with the same rank, the solution in the less dense area is preferred.

\subsubsection{Overall algorithm}

Similarly to the standard GA presented in subsection \ref{elecsim:ssec:geneticalgorithm}, a population $P_{0}$ is created with random parameters. The solutions of $P_0$ are then sorted according to non-domination. The child population $C'_{1}$ of size $N$ is then created by binary tournament selection, recombination and mutation operators. In this case, tournament selection is the process of evaluating and comparing the fitness of various individuals within a population. In binary tournament selection, two individuals are chosen at random, their fitnesses are evaluated, and the individual with the better fitness is selected~\cite{AbdRahman2016}.

\begin{algorithm}
	\begin{algorithmic}[1]
		\State $R_t=P_t \cup C'_t$ combine parent and child population
		\State $\mathcal{F} = $ fast-non-dominated-sort $(R_t)$ 
		
		where $\mathcal{F}=(\mathcal{F}_1, \mathcal{F}_2,\ldots)$
		\State $P_{t+1}=\emptyset$
		\While $\left|P_{t+1}<N\right|$
		\State Calculate the crowding distance of $(\mathcal{F}_i)$)
		\State $P_{t+1}=P_{t+1}\cup \mathcal{F}_i$
		\EndWhile
		\State Sort($P_{t+1}, \prec_n$) sort in descending order using $\prec_n$
		\State $P_{t+1} = P_{t+1}[0:N]$ select the first $N$ elements of $P_{t+1}$
		\State $Q_{t+1} = $ make-new-population$(P_{t+1})$ using 
		
		selection, crossover and mutation to create 
		
		the new population $Q_{t+1}$
		\State $t=t+1$
		\caption{NSGA-II algorithm \cite{Valkanas2014}}
		\label{algo:nsga2}
	\end{algorithmic}
\end{algorithm}

Next, a new combined population is formed $R_{t}=P_{t} \cup C'_{t}$. $R_t$ has a size of $2N$. $R_t$ is then sorted according to non-domination. A new population is then formed $P_{t+1}$. Solutions are added from the sorted $R_t$ in order of non-domination. Solutions are added until the size of $P_{t+1}$ exceeds $N$. The solutions from the last layer are prioritised based on having the largest crowding distance~\cite{Valkanas2014}.

This process is shown in Figure \ref{fig:pareto-layering}b, which is repeated until the termination condition is met. A termination condition could be:  no significant improvement over $X$ iterations or a specified number of iterations have been performed. The full procedure is detailed formally by Algorithm \ref{algo:nsga2}.

\subsection{Carbon Optimization Application}

In this section, we describe how the genetic algorithm is applied in our carbon optimisation case. We use multi-objective optimisation to find a solution which has both a low carbon emission and low average electricity price. The parameter that we adjust is the carbon tax between the years 2018 and 2035.

The mating steps work by, initially, taking the sets of carbon prices over the 17 year period (2018 to 2035) which have the best rewards (lowest relative carbon emissions and average electricity price). These carbon prices are then mated with a probability of 90\%, creating child carbon prices. The children are mutated with a probability of 5\%. Therefore, 5\% of children have a carbon price which is not inherited from the parents. Over time, the mutations and inherited properties tend to a population with more desirable rewards.

\subsection{Simulation Environment}
In order to evaluate the different carbon strategies, we used ElecSim \cite{Kell,Kell2020}. For this work, we parametrised the model to data for the UK in 2018 to act as a digital twin of the UK electricity market. This includes the power plants in operation in 2018, and the funds available to their respective companies \cite{dukes_511, companies_house}. ElecSim is validated by being instantiated by data from 2013 and projected forward to 2018, with a mean absolute scaled error (MASE) below or equal to 0.701 for all generator types \cite{Kell2020}. 

The yearly income for each power plant is estimated for each generation company by running a merit-order dispatch electricity market ten years into the future. However, the expected cost of electricity ten years into the future is uncertain. We, therefore, use the reference scenario projected by BEIS and use the predicted costs of electricity calibrated by our previous work in Chapter \ref{chapter:elecsim} \cite{DBEIS2019, Kell2020}.

\subsection{Optimization}

In this section, we detail the optimisation approach taken. We modify the carbon tax each year, as we believe this is the most likely process taken by governments, giving generator companies and consumers the ability to understand market conditions during each year.

\label{ssec:optimization}
\subsubsection{Non-parametric carbon policy}
\label{sssec:non_parametric_strategy}
The optimization approach has two stages. First, we initialize the population of the NSGA-II algorithm $P_0$ with 18 attributes. These correspond to a separate carbon tax for each year, shown by Equation \ref{eq:eighteen_degrees_freedom}:
\begin{equation}
\label{eq:eighteen_degrees_freedom}
P_0=\{a_1,a_2,\ldots,a_{18}\}, 0\leq a_y\leq 250,
\end{equation} 

\noindent where $P_0$ is the first population, $a_y$ is the carbon price in year $y$. The constraints of the algorithm are that each of the carbon prices are bound between the values of \textsterling$0$ and \textsterling$250$. This provides the optimisation algorithm with the highest degree of freedom. The value \textsterling$250$ was chosen due to the relative costs of electricity, where \textsterling$250$ would be the upper bound for the cost of electricity. This high degree of freedom enables a high number of strategies to be trialled due to its non-parametric nature. This, however, comes with a large search space requiring a large number of iterations.

\subsubsection{Linear carbon policy}
\label{sssec:linear_carbon_strategy}
To reduce the search space for the carbon strategy, we also trial a linear carbon strategy, of the form:
\begin{equation}
p_c=a_1y_t+a_2, -14 \leq a_1\leq 14, 0 \leq a_2\leq 250,
\end{equation}
\noindent where $p_c$ is the carbon price, $y_t$ is the year, $a_1$ is the gradient or first attribute and $a_2$ is the intercept or second attribute. The constraints of the optimisation problem are that $a_1$ is bound by $-14$ and $14$, and $a_2$ by 0 and 250. These bounds are chosen to ensure that the carbon price does not exceed ${\sim}$\textsterling500 in the year 18 (2035) and is greater than about -\textsterling250, as well as ensuring that the carbon tax in the first year is greater than \textsterling0 but smaller than \textsterling250. The bounds for $a_1$ was chosen to make the mathematics simpler, whilst remaining in range.

\section{Results}
\label{carbonoptim:sec:results}

In this section, we explore the results of the optimisations, the optimum carbon strategies and the resultant electricity mixes.

\subsection{Non-parametric carbon policy}
\label{sssec:result_non_parametric_strategy}

\begin{figure}
	\centering
	\includegraphics[width=0.6\linewidth]{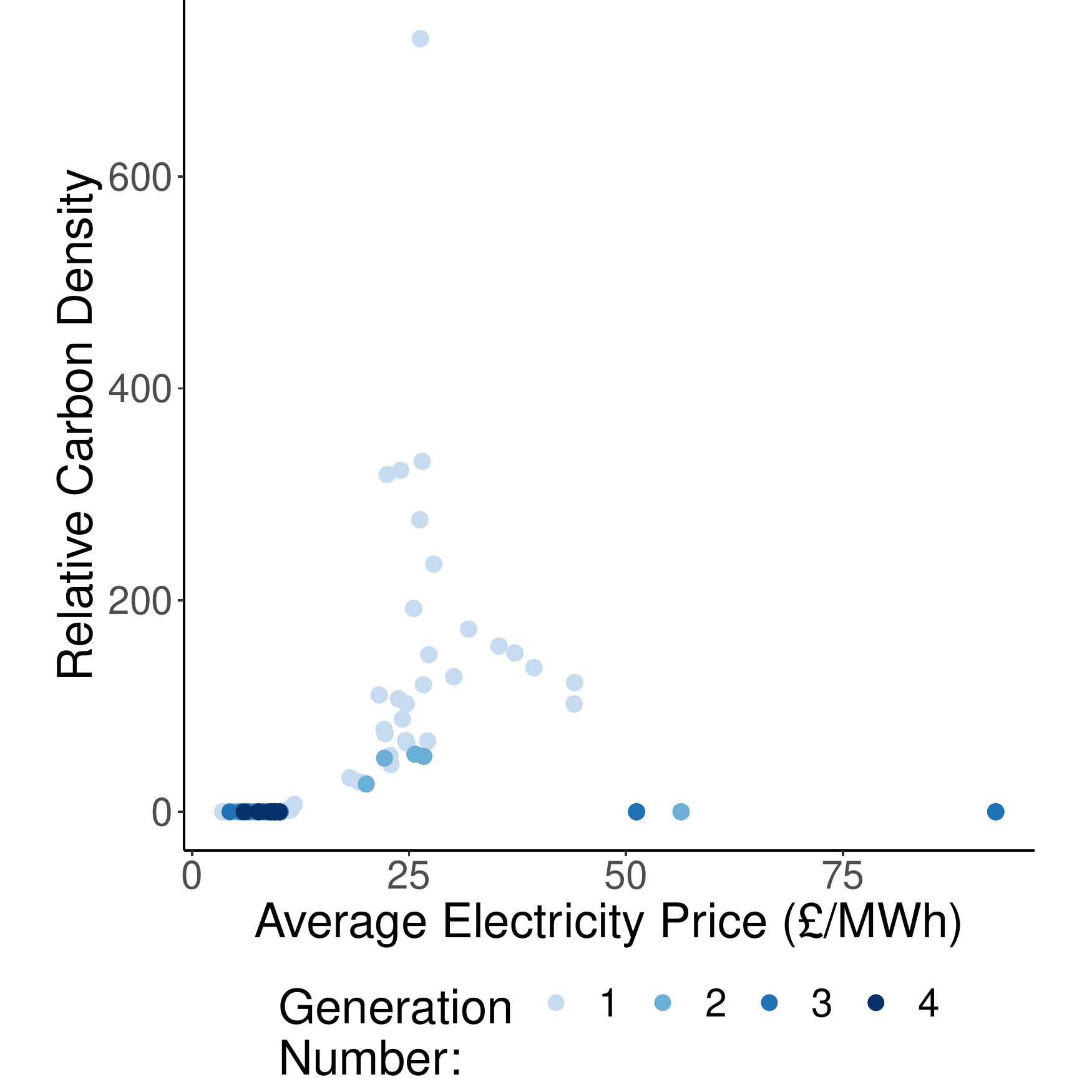}
	\caption{Development of genetic algorithm rewards for non-parametric carbon tax policy results in 2035.}
	\label{fig:free_points_ga_development}
\end{figure}

Figure \ref{fig:free_points_ga_development} displays the development of the genetic algorithm against the rewards, relative carbon density and average electricity price. Darker colours display higher generation numbers. The first generation shows a widespread in relative carbon density and average electricity price. However, over successive generations, the solutions converge to a relative carbon density of 0 and an average electricity price under \textsterling10MW/h. 

The rewards of relative carbon density and average electricity price are not mutually destructive. This could be seen as counter-intuitive, as historically low-carbon electricity generation was more expensive than fossil fuels. However, these results could be shown due to the low short-run marginal cost of renewable energy which reduces both electricity prices and carbon emissions~\cite{OMahoney2011}.

\begin{figure}
	\centering
	\includegraphics[width=0.6\linewidth]{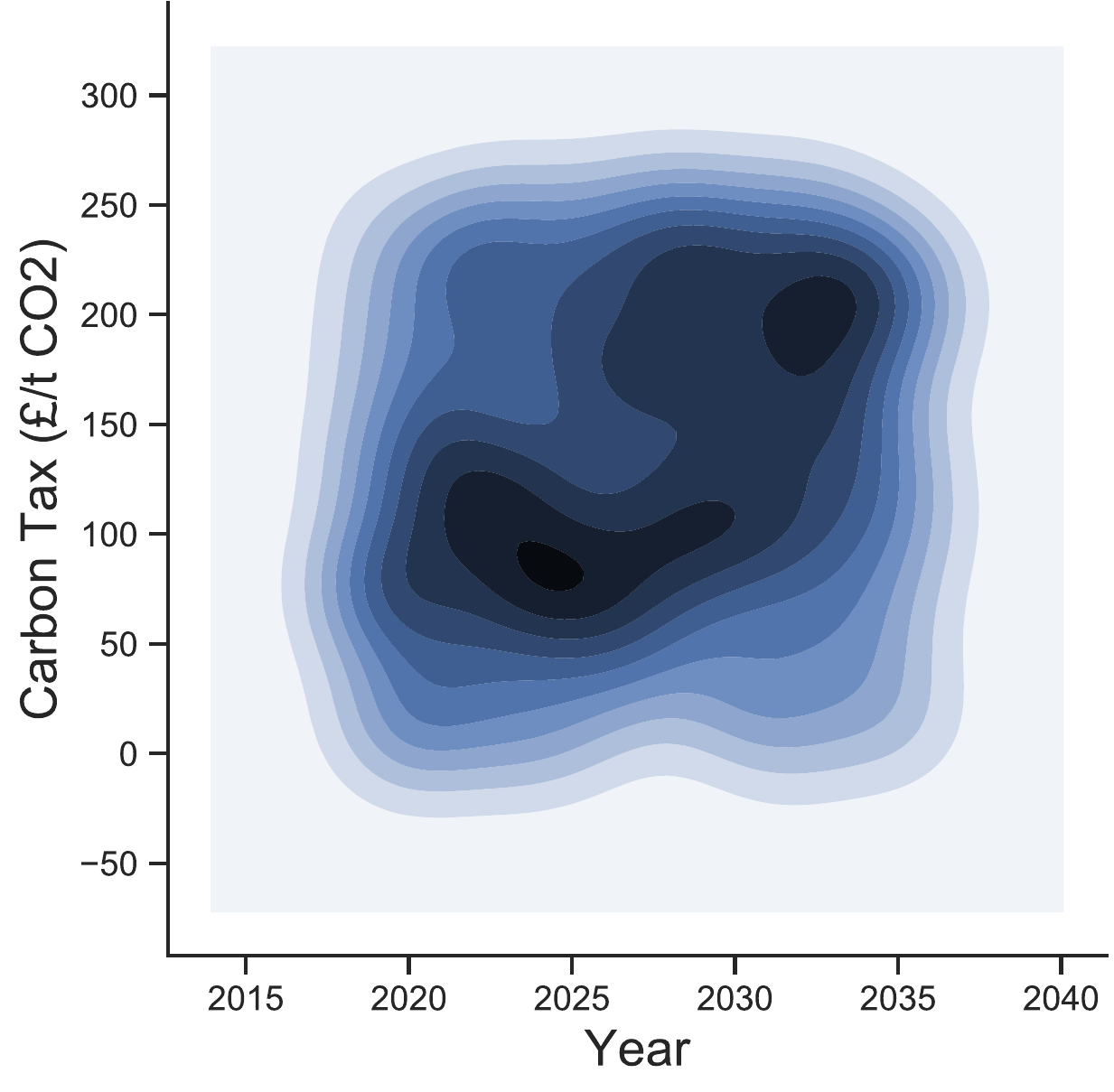}
	\caption{Density plot of points with an average electricity price <\textsterling5/MWh for non-parametric carbon tax policy results in 2035.}
	\label{fig:heatmap_of_free_points}
\end{figure}

To understand the optimum carbon strategies, we visualised the parameters that produced the lowest average electricity prices in Figure \ref{fig:heatmap_of_free_points}. Specifically, we filtered for electricity prices under \textsterling5/MWh and displayed the results using a heat map. The darker colours represent a higher density of points. 

Figure \ref{fig:heatmap_of_free_points} displays a general trend, where carbon tax starts at ${\sim}$\textsterling100 until the year 2030, where it increases to ${\sim}$\textsterling200 by 2035. This may be due to the fact that a lower initial carbon tax of ${\sim}$\textsterling100 encourages investment in low-carbon technologies before the higher rate of ${\sim}$\textsterling200 comes into force. This higher rate of carbon tax would allow GenCos to outcompete higher carbon-emitting generators over the lifetime of the plants.


\subsection{Linear carbon policy}
\label{sssec:result_linear_carbon_strategy}

\begin{figure}
	\centering
	\includegraphics[width=0.5\linewidth]{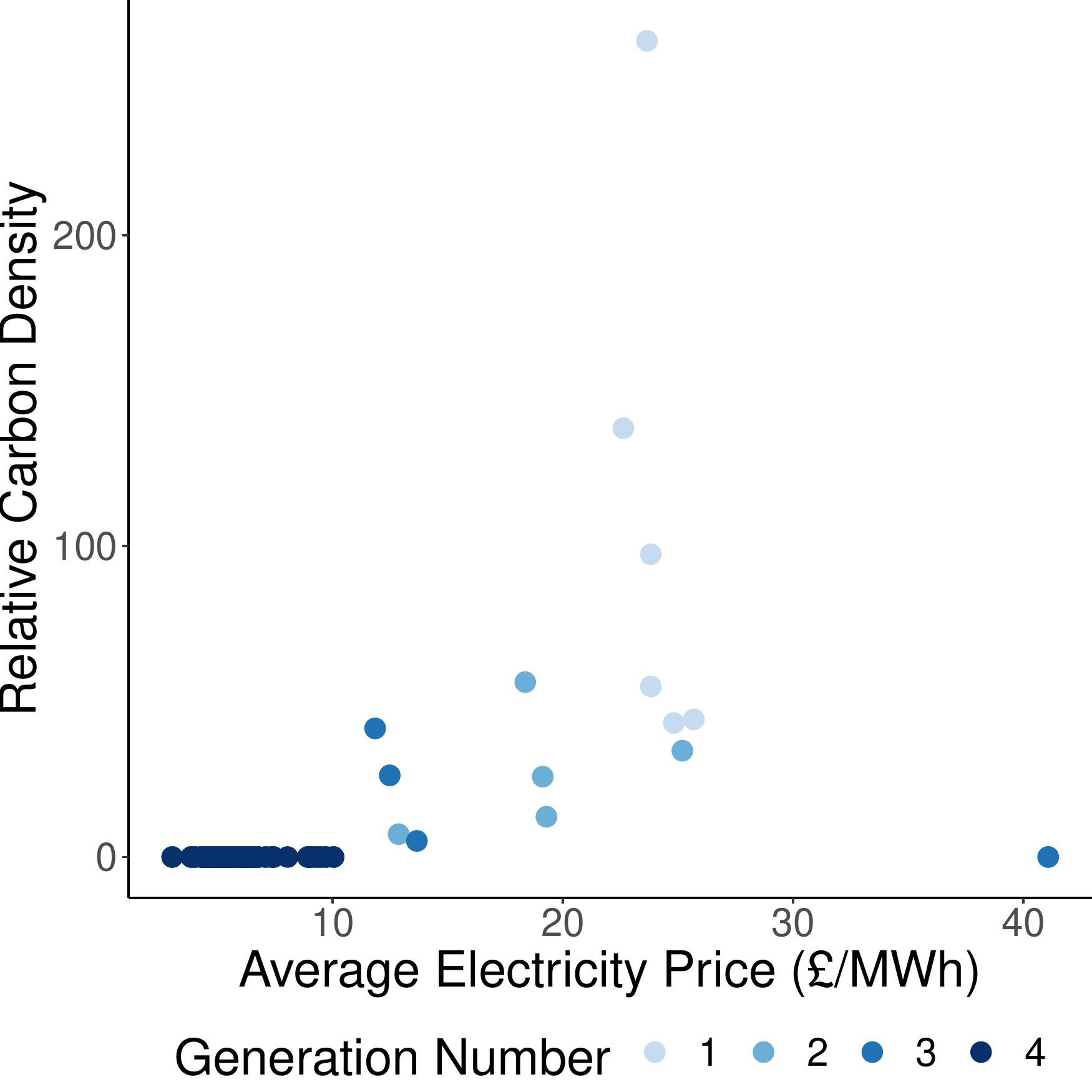}
	\caption{Development of genetic algorithm rewards in 2035 for linear carbon strategy.}
	\label{fig:linear_ga_development}
\end{figure}

Figure \ref{fig:linear_ga_development} displays the development of the genetic algorithm against the rewards: relative carbon density and average electricity price. Similarly to the non-parametric carbon policy shown in Figure \ref{fig:free_points_ga_development}, the first generation shows a wide spread of results. However, the spread is smaller than that of the linear carbon policy. This may be due to the fact that it is easier for the GenCos to predict the carbon policy, which increases confidence in the NPV calculations. The linear carbon policy also converges to a relative carbon density of 0, and an average electricity price smaller than \textsterling10MW/h.

\begin{figure}
	\centering
	\includegraphics[width=0.7\textwidth,]{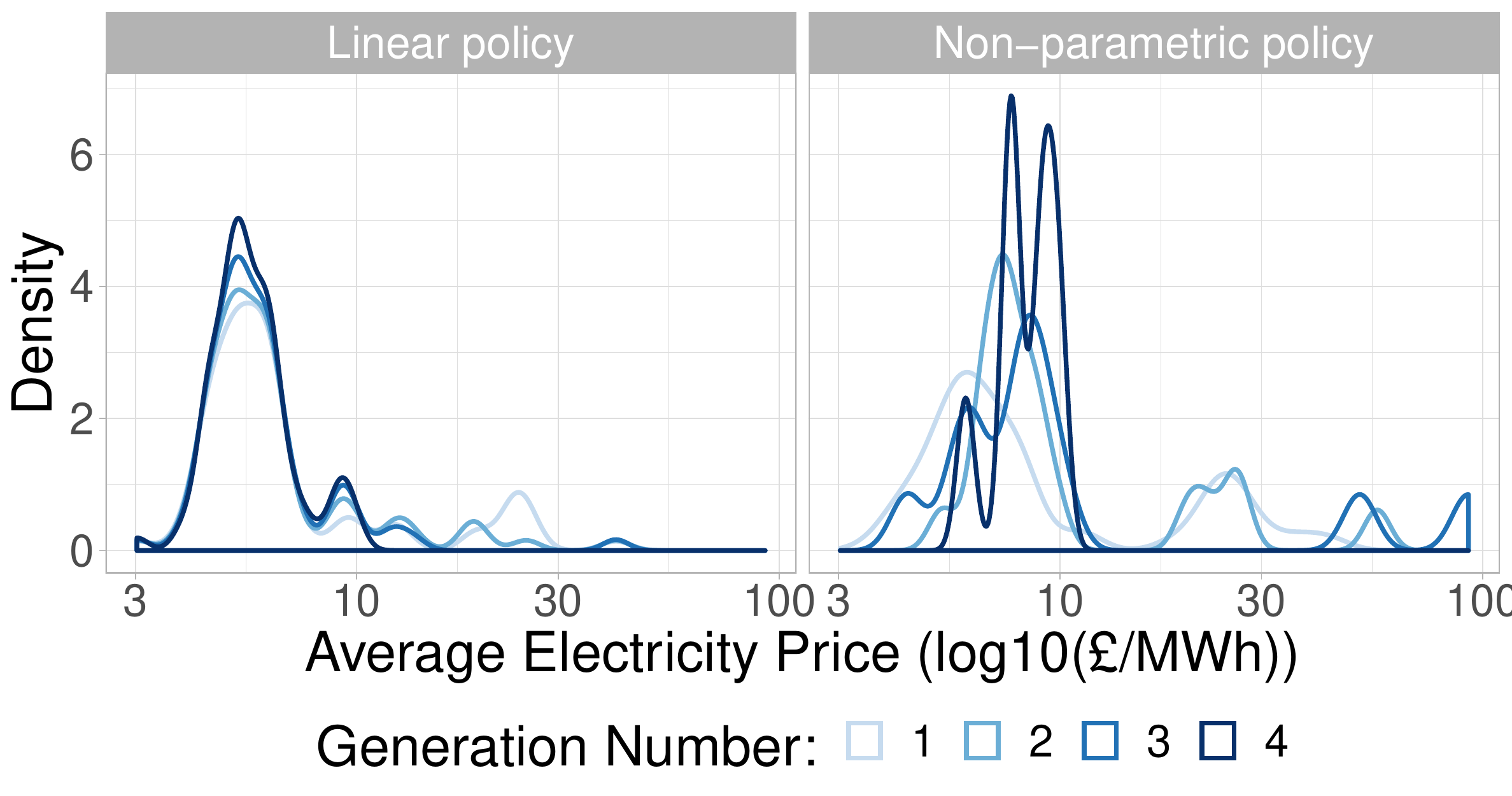}
	\caption{Density plot of average electricity price in 2035 over generation number of genetic algorithm for both linear and non-parametric policy.}
	\label{fig:comparison_of_distributions}
\end{figure}

Figure \ref{fig:comparison_of_distributions} compares the distributions of average electricity price for both techniques. Both methods show improvements as the number of generations of the genetic algorithm increase.  The linear policy, however, is able to more quickly converge to a low average electricity price, with a mode of ${\sim}$\textsterling5.4MW/h. The non-parametric policy has a number of poorer performing parameters, and Generation Number 4 has a bimodal distribution, with a mode of ${\sim}$\textsterling6.3MW/h.

\begin{figure}
	\begin{subfigure}[h]{0.6\linewidth}
		\includegraphics[width=\linewidth]{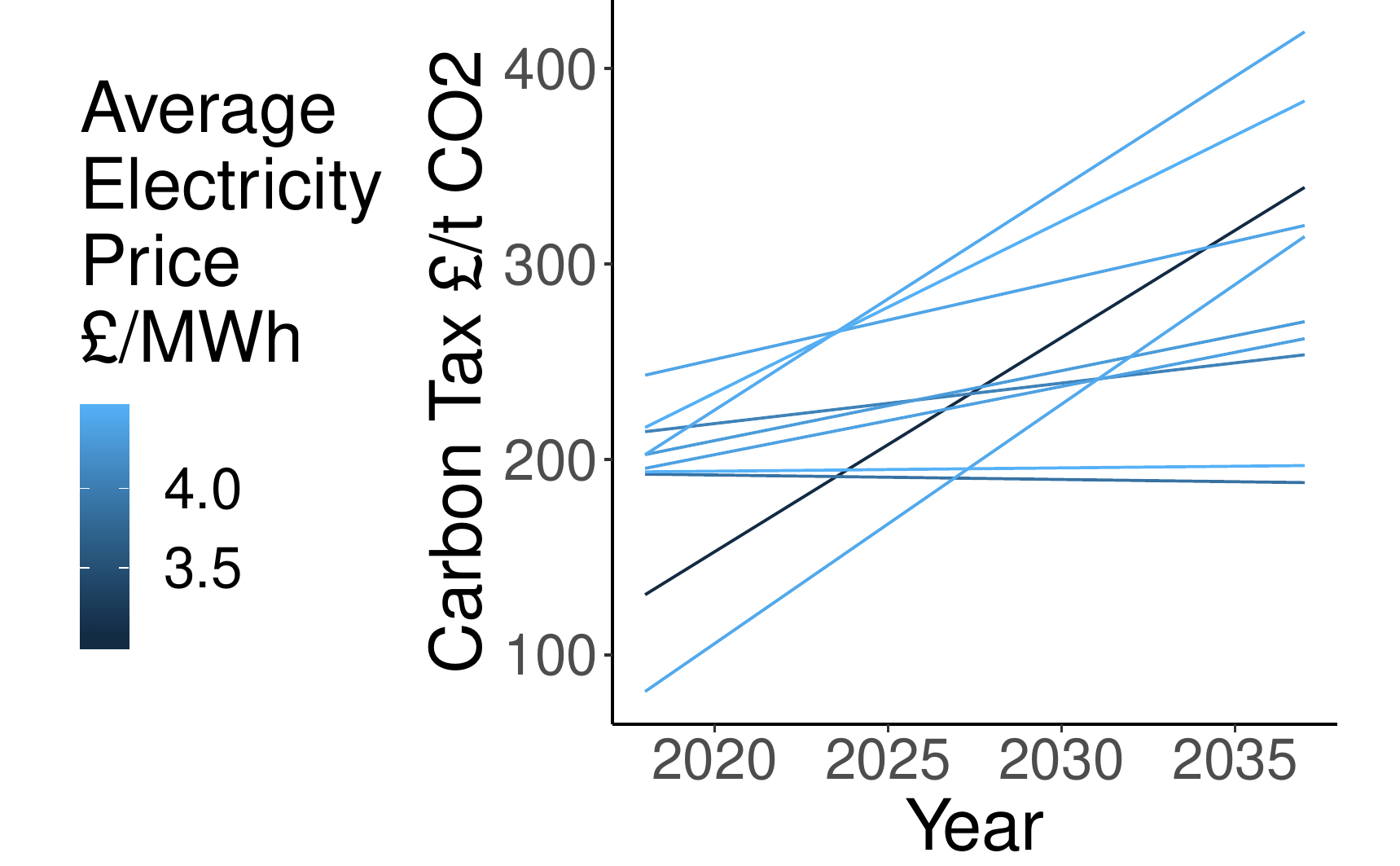}
		\caption{All carbon policies.}
		\label{fig:linear_actual_pdcs}
	\end{subfigure}
	\hfill
	\begin{subfigure}[h]{0.39\linewidth}
		\includegraphics[width=\linewidth]{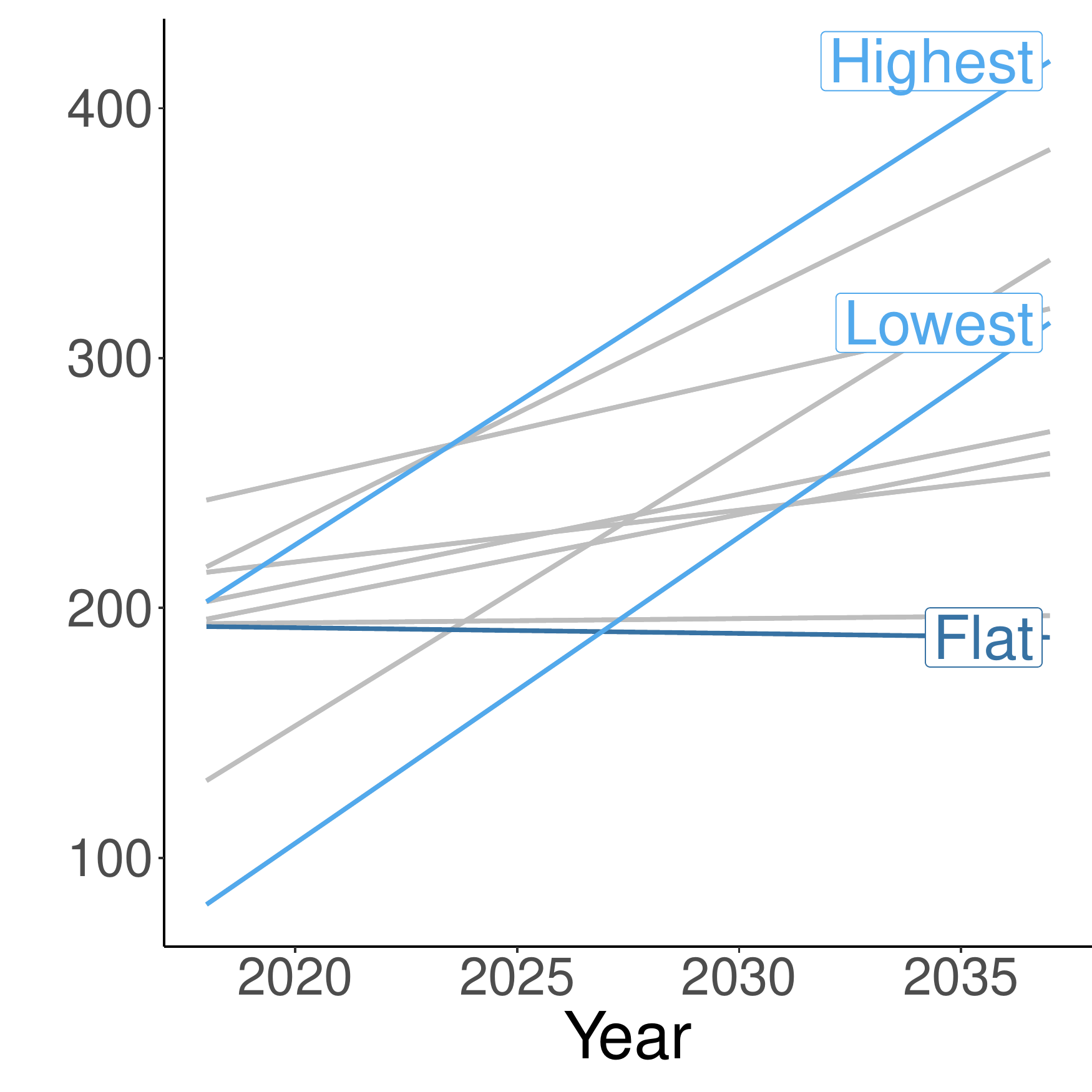}
		\caption{Highlighted policies.}
		\label{fig:highlighted_linear_actual_strategies}
	\end{subfigure}%
	\caption{Linear carbon policies under \textsterling4.5MW/h visualised.}
\end{figure}

Figure \ref{fig:linear_actual_pdcs} displays the linear carbon policies which had an average electricity price under \textsterling4.5MW/h. There is no single `optimum' carbon policy; a range of policies are able to achieve low carbon and a low average electricity price.

We explore the electricity mix generated of three different strategies shown in Figure \ref{fig:highlighted_linear_actual_strategies}. We selected the highest, lowest, and the lowest flat carbon strategy to show a range of possible strategies. This means that whilst a single carbon tax may be sought, in reality there are a number of different strategies which each perform equally well. 

It is therefore up to stakeholders and decision makers to select a strategy within the bounds presented here. For example, a government may decide to impose the lowest carbon tax policy, to avoid impacting different groups and stakeholders. Or, they may choose a flat carbon tax strategy to prevent an unwelcome increase.

%

\begin{figure}
	\centering
	\includegraphics[width=0.9\textwidth]{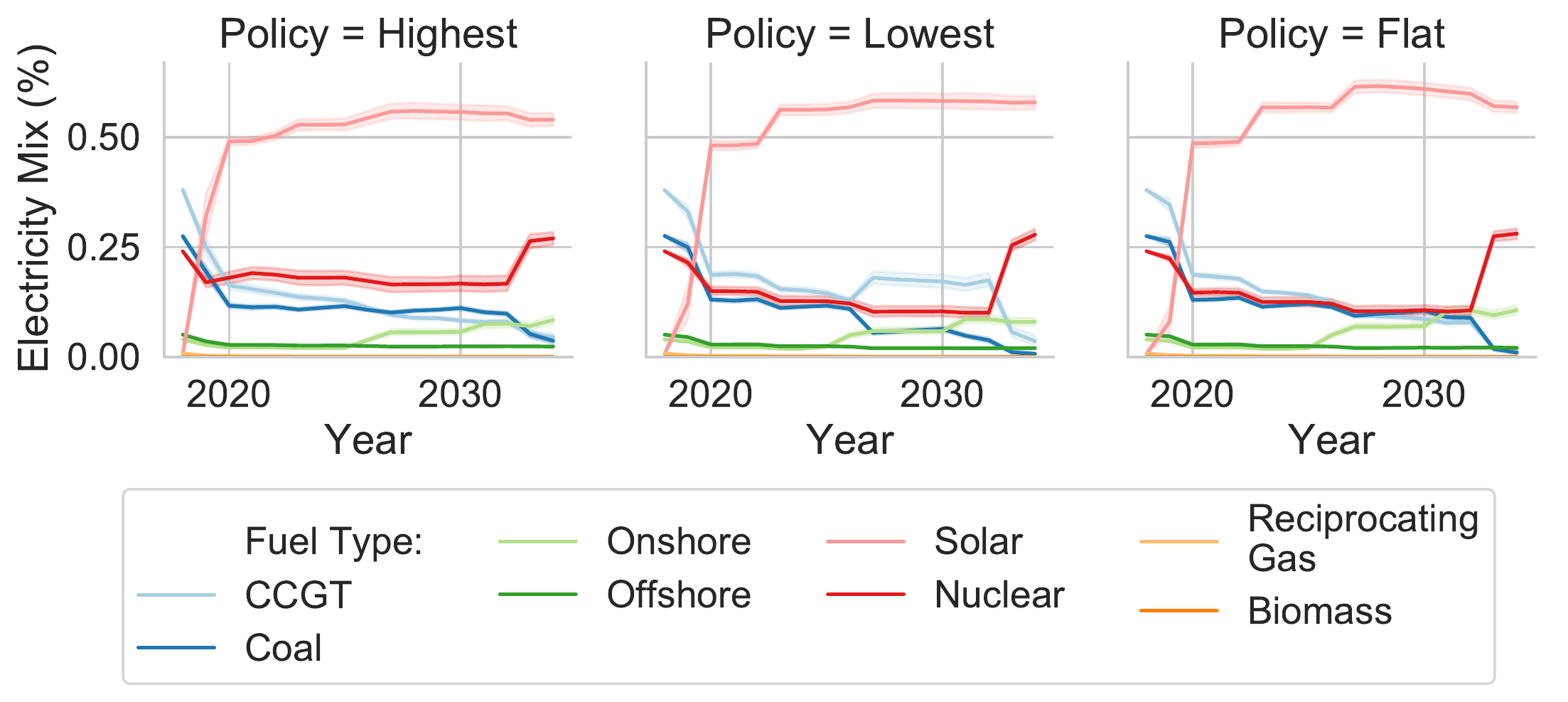}
	\caption{Electricity mixes under selected linear carbon policies.}
	\label{fig:best_electricity_mixes_facet}
\end{figure}

Figure \ref{fig:best_electricity_mixes_facet} displays the generated electricity mixes for each of the selected strategies. To generate these figures, we ran 80 scenarios to capture the variability between scenarios. 

Whilst there does not seem to be a significant difference between scenarios, with solar providing ${\sim}60\%$ of the electricity mix by 2035, there is an observable difference with the other generator types.

The `highest' carbon strategy exhibits a higher uptake in nuclear, possibly due to the fact that nuclear becomes more competitive when compared to coal or gas. The `lowest' carbon strategy shows a higher uptake in Combined Cycle Gas Turbines (CCGT) during the years of 2026 to 2031 as it outcompetes nuclear. The `flat' carbon policy shows a higher percentage of solar energy than any of the other scenarios, albeit with a lower percentage of nuclear. Onshore wind is shown to be consistent for these scenarios.

\subsection{UK Government Policy}

We compare our optimal carbon tax strategy to a scenario based on the UK Government's policy in this section from 2018 to 2034. Whilst it is not possible to know the future carbon tax strategy over such a long time horizon, we used a naive model approach to project a static carbon tax strategy. That is, the strategy is maintained at the current level of \textsterling18.08 throughout the projected horizon. We chose this level, due to the carbon price at the time of writing. We ran 40 simulations to capture the variance of the electricity mix. The work in this subsection is additional work to that which was published in \cite{Kell2020a}.

\begin{figure}
	\centering
	\includegraphics[width=0.6\textwidth]{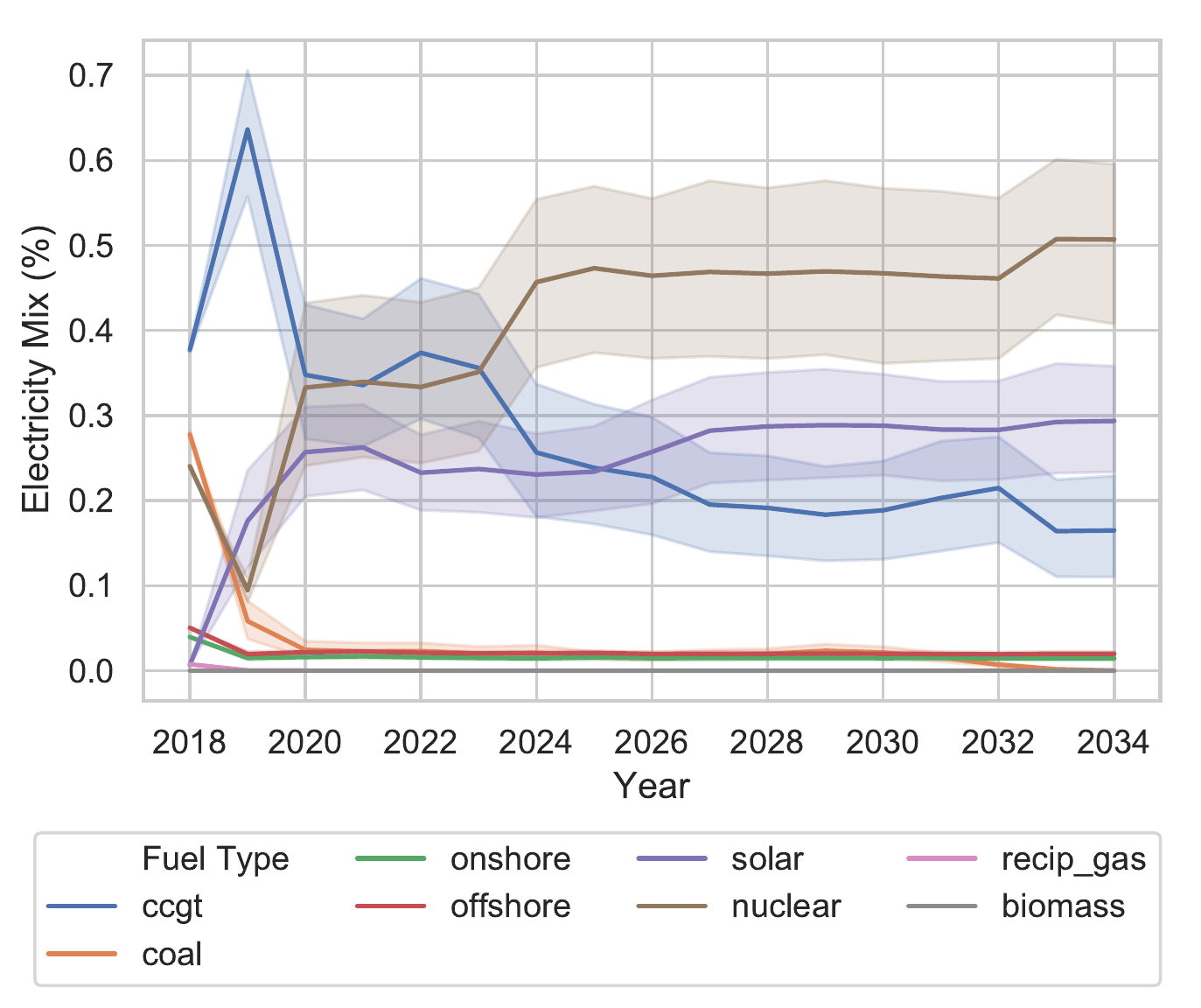}
	\caption{Electricity mixes under UK Government's carbon tax.}
	\label{fig:uk_government_carbon_tax_strategy_scenario}
\end{figure}

Figure \ref{fig:uk_government_carbon_tax_strategy_scenario} displays the resultant electricity mix of this carbon tax strategy. A significant difference can be seen when compared to the optimal carbon strategy. 50\% of the electricity mix is provided by nuclear energy by 2034. This is almost double that in the optimal carbon tax strategy. 

Solar, the second-largest source of electricity, provides 30\% of electricity supply, with CCGT providing ${\sim}$15\%. This is a marked difference to the electricity mix shown by Figure \ref{fig:best_electricity_mixes_facet}, where solar provides over 50\% of electricity supply, followed by nuclear, which provides ${\sim}$26\%. 

A larger variance can be seen in this scenario when compared to the optimal carbon tax. This may be because of the less defined differences between generator costs due to the lower carbon price, which does not distinguish so strongly between carbon-emitting and non-carbon emitting generators. 

Table \ref{carbonoptim:table:mix} displays the average electricity price and relative carbon density of this scenario. As is expected, both average electricity price and relative carbon density are above that of the optimal carbon price scenario. This is because of the high short-run marginal cost of nuclear, and the carbon intensity of CCGT. 

\begin{table}[]
	\centering	
	\begin{tabular}{@{}ll@{}}
		\toprule
		Metric & Value \\ \midrule
		Average electricity price (\textsterling/MWh) & 975.1 \\
		Relative carbon density                   & 46.6  \\ \bottomrule
	\end{tabular}
	\caption{Resulting average electricity price and relative carbon density under a scenario using the UK Government's carbon tax policy.}
	\label{carbonoptim:table:mix}
\end{table}

These results show the importance of carbon tax in deciding the electricity mix, where a difference is shown between high carbon taxes and low-carbon taxes. However, solar continues to do well without a carbon tax, as does nuclear, due to the high nuclear subsidy provided. 

These results can be used by policy makers who require a range of possible carbon tax levels to achieve certain objectives. These results do not prescribe a single solution which should be used, rather we present a range of different options and outcomes which could be chosen. The limitations of this work are the limited scenarios that were explored. For instance, different electricity demand trajectories could be explored. This would provide the policy makers with an increased level of confidence that the carbon tax policies would be robust in the face of uncertainty.

\section{Conclusion}
\label{carbonoptim:sec:conclusion}

In this work, we have demonstrated that it is possible to use the genetic algorithm technique NSGA-II to optimise carbon tax policy using an electricity market agent-based model. 

We trialled a non-parametric carbon policy by allowing the genetic algorithm to optimise a carbon price for each year. These results showed us that a linear carbon tax might be appropriate. We then used a linear model as a carbon tax policy to reduce the total number of parameters for the genetic algorithm to optimise.  

We were able to show that a range of linear carbon taxes were able to achieve both low average electricity price and a relative carbon intensity of zero in 2035. By exploring three different carbon tax policies, we saw that ${\sim}$60\% of electricity consumption in the UK would be provided by solar. The difference between these `optimal' carbon tax policies was largely shown by competition between CCGT, coal and nuclear.

This was largely due to the low short-run marginal cost of solar and nuclear energy, which means that they are often dispatched ahead of the fossil-fuel based generators. CCGT and coal, however, are useful for filling demand when there is low solar irradiance.

Additionally, we ran a carbon tax scenario based on that of the UK Government's. We found that an optimal carbon tax strategy had a much lower average electricity price due to the low short-run marginal cost of renewable energy. We also showed that we were able to achieve a low relative carbon density through a higher carbon tax. The majority of the electricity was provided by nuclear, which, in this scenario, had a high subsidy. Therefore, we believe that a lower carbon density, and lower average electricity price can be obtained without subsidies, but with a higher carbon tax. 

In future work, we would like to try additional scenarios with varying future generation costs and calculate a sensitivity analysis to carbon taxes. In addition to this, we would like to model the uncertain reactions by consumers and generation companies with regards to carbon taxes. The linear carbon tax approach is an introductory approach which can be expanded upon. In addition, we would like to trial this approach, as well as the future work to other countries, not just the United Kingdom.

Additionally, we believe that this work could allow for optimisation of other policies, and factors within the simulation. For example, it would be possible to optimise for a range of possibilities for subsidies, market designs and R\&D investment. Questions such as, should solar photovoltaics take investment priority for both R\&D and installation when compared to offshore wind in parts of the US.

We believe that we have created a framework in which different policy options can be explored, and the use of computational optimisation can trial many differing variants.

\chapter{Strategic bidding within Electricity Markets}
\label{chapter:reinforcement}
\ifpdf
\graphicspath{{Chapter3/Figs/Raster/}{Chapter3/Figs/PDF/}{Chapter3/Figs/}}
\else
\graphicspath{{Chapter3/Figs/Vector/}{Chapter3/Figs/}}
\fi

\section*{Summary}


Decentralised electricity markets are often dominated by a small set of generator companies who control the majority of the capacity. In this work, we explore the effect of collusion, or an oligopoly on electricity market prices. We demonstrate this through ElecSim. We develop an agent, representing a generator company, which uses a deep deterministic policy reinforcement learning algorithm to bid in a uniform pricing electricity market strategically. A uniform pricing market is one where all players are paid the highest accepted price. ElecSim is parametrised to the United Kingdom for the year 2018. However, this work could be applied to other countries. This work can help inform policy on how to best regulate a market to ensure that the price of electricity remains competitive.

We find that capacity has an impact on the average electricity price in a single year. If any single generator company, or a collaborating group of generator companies, control more than ${\sim}$11$\%$ of generation capacity, prices begin to increase by ${\sim}$25$\%$. The value of ${\sim}$25\% and ${\sim}$11\% may vary between market structures and countries. For instance, different load profiles may favour a particular type of generator or a different distribution of generation capacity. Once the capacity controlled by a generator company is higher than ${\sim}$35\% of the total capacity, prices increase rapidly. The use of a market cap of approximately double the average market price has the effect of significantly decreasing this effect and maintaining a competitive market.



In this chapter, we do not utilise the long-term features of the ElecSim model, focusing, instead on a single year. However, the objective of this thesis is to explore the wider impact of machine learning on the electricity market. The discoveries and recommendations found in this chapter can have a long-term impact on the wider market, through the design of an efficient market type for the long-term. Additionally, these types of recommendations are made more efficient through the methodology explored in previous chapters. Specifically Chapter \ref{chapter:elecsim}. For instance, agent-based models are required to model individual generator companies, and reinforcement learning is a useful methodology to make predictions and actions within an uncertain environment.

We introduce this work in Section \label{rl:sec:introduction}. In Section \ref{rl:sec:lit-review} we review the literature, and explore other uses of \acrfull{rl} in electricity markets. In Section \ref{rl:sec:material} we introduce the agent-based model used and the DDPG algorithm. Section \ref{rl:sec:methodology} explores the methodology taken for our case study. The results are presented in Section \ref{rl:sec:results}. We discuss and conclude our work in Sections \ref{rl:sec:discussion} and \ref{rl:sec:conclusion} respectively.

\section{Introduction}
\label{rl:sec:introduction}


Under perfectly competitive electricity markets, generator companies (GenCos) tend to bid their short-run marginal costs (SRMC) when bidding into the day-ahead electricity market. However, electricity markets are often oligopolistic, where a small subset of GenCos provide the majority of the capacity to the market. Under these conditions, it is possible that the assumption that GenCos are price-takers does not hold. That is, large GenCos artificially increase the price of electricity to gain an increased profit using their market power. If they were price-takers, they would have to accept the competitive price set by the market.

Reduced competition within electricity markets can lead to higher prices to consumers, with no societal benefit. It is, therefore, within the interests of the consumer and that of government to maintain a competitive market. Low energy costs enable innovation in other industries reliant on electricity, and in turn, make for a more productive economy. 

In this work, we explore the effect of total control over capacity on electricity prices. Specifically, we model different sizes of GenCos and groups of colluding GenCos, to bid strategically to maximise their profit using a reinforcement learning algorithm. This is in contrast to the strategy of bidding using the SRMC of their respective power plants, which would occur under perfect market conditions. To model this, we use deep reinforcement learning (RL) to generate a bidding strategy for GenCos in a day-ahead market. These GenCos are modelled as agents within the agent-based model, ElecSim \cite{Kell, Kell2020}. We use the UK electricity market instantiated on 2018 as a case study, similar to our work in \cite{Kell2019a}. That is, we model each GenCo with their respective power plants in the year 2018 to 2019. In total, we model 60 GenCos with 1,085 power plants between them. It is possible, however, to generalise this approach and model to any other decentralised electricity market. 

We use the \acrfull{ddpg} deep RL algorithm, which allows for a continuous action space \cite{Hunt2016a}. Conventional RL methods require discretisation of state or action spaces and therefore suffer from the curse of dimensionality \cite{Ye2020a}. As the number of discrete states and actions increases, the computational cost grows exponentially. However, too small a number of discrete states and actions will reduce the information available to the GenCos, leading to sub-optimal bidding strategies. Additionally, by using a continuous approach, we allow for GenCos to consider increasingly complex bidding strategies. 

Other works have considered a simplified model of an electricity market by modelling a small number of GenCos or plants \cite{EsmaeiliAliabadi2017,Tellidou2007}. We, however, model each GenCo as per the UK electricity market with their respective power plants in a day-ahead market. In addition, further work focuses on a bidding strategy to maximise profit for a GenCo. However, in our work, we focus on the impact that large GenCos, or colluding groups of GenCos, can have on electricity price.

Our approach does not require GenCos to formulate any knowledge of the data which informs the market-clearing algorithm or rival GenCo bidding strategies, unlike in game-theoretic approaches \cite{Wang2011}. This enables a more realistic simulation where the strategy of rival GenCos are unknown.

This work fits into the wider scope of this thesis through using machine learning to have an impact on the wider electricity market. In addition, it utilises the work carried out in Chapter \ref{chapter:elecsim} for an additional application. It is true that the long-term components of ElecSim are not required for this work, but utilising every aspect of ElecSim was not the central aim of this thesis.








\section{Literature Review}
\label{rl:sec:lit-review}

Intelligent bidding strategies for day-ahead electricity markets can be divided into two broad categories: simulation and game-theoretic models. Agent-based models (ABMs) allow for the simulation of heterogenous irrational actors with imperfect information. Additionally, ABMs allow for learning and adaption within a dynamic environment \cite{EsmaeiliAliabadi2017}. Game-theoretic approaches may struggle in complex electricity markets where Nash equilibriums do not exist \cite{Wang2011}.

\subsection{Game-theoretic approaches}

\begin{figure}
	\centering
	\includegraphics[width=0.6\columnwidth]{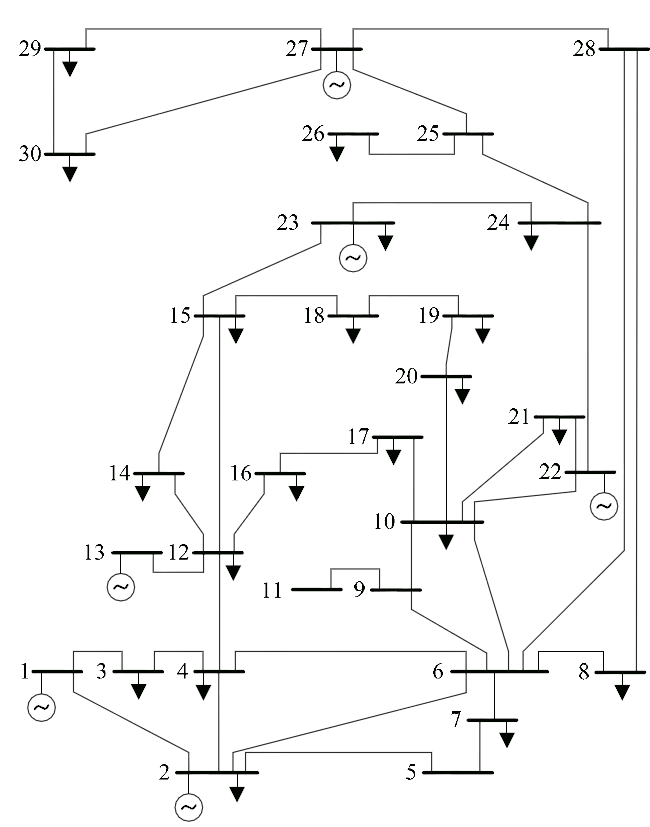}
	\caption{Example 30 bus system.}
	\label{fig:30-bus-system.}
\end{figure}

Here, we explore game-theoretic approaches. Kumar \textit{et al.} \cite{VijayaKumar2014} propose a Shuffled Frog Leaping Algorithm (SFLA) to find bidding strategies for GenCos in electricity markets. SFLA is a meta-heuristic that is based on the evolution of memes carried by active individuals, as well as a global exchange of information among the frog population. They test the effectiveness of the SFLA algorithm on an IEEE 30-bus system and a practical 75-bus Indian system. A bus in a power system is a vertical line which several components are connected in a power system. For example, generators, loads, and feeders can all be connected to a bus, an example is shown in Figure \ref{fig:30-bus-system.}. 

Kumar \textit{et al.} find superior results when compared to particle swarm optimisation and the genetic algorithm with respect to total profit and convergence with CPU time. They assume that each GenCo bids a linear supply function, and model the expectation of bids from rivals as a joint normal distribution. In contrast to their work, we do not require an estimation of the rivals bids.

Wang \textit{et al.} \cite{Wang2011} propose an evolutionary imperfect information game approach to analysing bidding strategies with price-elastic demand. Their evolutionary approach allows for GenCos to adapt and update their beliefs about an opponents' bidding strategy during the simulation. They model a 2-bus system with three GenCos. Our work, however, models a simulation with 60 GenCos across the entire UK, which would require a 28-bus system model \cite{Bell2010}. 

Our work differs from both of these approaches, in that their work investigates what each GenCo should do, but not what would happen to the entire market from a macro perspective.

\subsection{Simulation based approaches}

Next, we explore simulation based approaches, which utilise reinforcement learning to make intelligent bidding decisions in electricity markets. RL is a suitable method for analysing the dynamic behaviour of complex systems with uncertainties. RL can, therefore, be used to identify optimal bidding strategies in energy markets \cite{Yang2020}. Simulations are often used to provide an environment for the reinforcement learning algorithm, which is true for the following papers reviewed here.

Aliabadi \textit{et al.} \cite{EsmaeiliAliabadi2017} utilise an ABM and the Q-learning algorithm to study the impact of learning and risk aversion on GenCos in an oligopolistic electricity market with five GenCos. They find that some level of risk aversion is beneficial, however excessive risk degrades profits by causing an intense price competition. Our work focuses on the impact of the interaction of many GenCos within the UK electricity market. In addition, we extend the Q-learning algorithm to use the DDPG algorithm, which uses a continuous action space. This allows for a continuous action space for bids, providing a higher granularity to the possible actions taken by GenCos. This allows for a more nuanced, and precise understanding of the optimal behaviour.

Bertrand \textit{et al.} \cite{Bertrand2019} use RL in an intraday market. An intraday market, in this context, allows for bids to be made during the day, to aid in balancing supply and demand. Specifically, they use the REINFORCE algorithm to optimise the choice of price thresholds. The REINFORCE algorithm is a gradient-based method. They demonstrate an ability to outperform the traditionally used method, the rolling intrinsic method, by increasing profit per day by 4.2\%. The rolling intrinsic method accepts any trade, which gives a positive profit if the contracted quantity remains in the bounds of capacity. In our work, we model a day-ahead market and use a continuous action for price bids.

Ye \textit{et al.} \cite{Ye2020a} propose a novel deep RL based methodology which combines the DDPG algorithm \cite{Hunt2016a} with a prioritised experience replay (PER) strategy \cite{Schaul2016}. The PER samples from previous experience, but samples from the ``important'' ones more often \cite{Schaul2016}. The PER is a modification of the often used experience buffer, which is a buffer which stores previous transitions and samples uniformly. This helps to reduce the correlations between recent experiences. They use a day-ahead market with hourly resolution and show that they are able to achieve approximately 41\%, 20\% and 11\% higher profit for the GenCo than the MPEC, Q-learning and DQN methods, respectively. In our work, we instead look at how to prevent GenCos (or sets of colluding GenCos) from forcing higher prices above market rates.

Zhao \textit{et al.} \cite{Zhao2016} propose a modified RL method, known as the gradient descent continuous Actor-Critic (GDCAC) algorithm. This algorithm is used in a double-sided day-ahead electricity market simulation. Where in this case, a double-sided day-ahead market refers to GenCos selling their supply to distribution companies, retailers or large consumers. Their approach performs better in terms of participant's profit or social welfare compared with traditional table-based RL methods, such as Q-Learning. Our work also looks at improving on table-based methods by using function approximators, however our work looks at the entire electricity market as a whole, instead of from the point of view of GenCos.

\section{Methodology}
\label{rl:sec:material}

In this section, we describe the RL methodology used for the intelligent bidding process as well as the simulation model used as the environment.

\subsection{Reinforcement Learning background}


In reinforcement learning (RL) an agent interacts with an environment to maximise its cumulative reward. RL can be described as a \acrfull{mdp}. An MDP includes a state-space $\mathcal{S}$, action space $\mathcal{A}$, a transition dynamics distribution $p(s_{t+1}|s_t,a_t)$ and a reward function, where $r:S\times \mathcal{A} \rightarrow \mathbb{R}$. At each time step, an agent receives an observation of the current state which is used to modify the agent's behaviour.

An agent's behaviour is defined by a policy, $\pi$. $\pi$ maps states to a probability distribution over the actions $\pi:\mathcal{S}\rightarrow \mathcal{P}(\mathcal{A})$. The return from a state is defined as the sum of discounted future reward:

\begin{equation}
R_t=\sum_{i=t}^T\gamma^{(i-t)}r(s_i,a_i).
\end{equation}

Where $\gamma$ is a discounting factor $\gamma \in [0,1]$. The return is dependent on the action chosen, which is dependent on the policy $\pi$. Where $r(s_i, a_i)$ is the reward for state $s_i$ and action $a_i$. The goal in reinforcement learning is to learn a policy that maximizes the expected return from the start distribution $J=\mathbb{E}_{r_i,s_i \sim E,a_i \sim \pi}[R_1]$. 

The expected return after taking an action $a_t$ in state $s_t$ after following policy $\pi$ can be found by the action-value function. The action-value function is used in many reinforcement learning algorithms and is defined in Equation \ref{eq:action-value}.
\begin{equation}
\label{eq:action-value}
Q^{\pi}(s_t,a_t)=\mathbb{E}_{r_{i\geq t},s_{i>t}\sim \mathcal{E},a_{i>t}\sim\pi}[R_t|s_t,a_t].
\end{equation}
\noindent The action-value function defines the expected reward at time $t$, given a state $s_t$ and action $a_t$ when under the policy $\pi$.

\subsection{Q-Learning}

An optimal policy can be derived from the optimal $Q$-values $Q_*(s_t,a_t)=\max_\pi Q_\pi(s_t,a_t)$ by selecting the action corresponding to the highest Q-value in each state.

Many approaches in reinforcement learning use the recursive relationship known as the Bellman equation, as defined in Equation \ref{eq:bellman}:
\begin{dmath}
	\label{eq:bellman}
	Q_\pi(s_t,a_t)=\mathbb{E}_{{r_t},s_{t+1}\sim E} [r(s_t,a_t)+
	\gamma\mathbb{E}_{a_{t+1}\sim \pi}[Q_\pi(s_{t+1},\pi(s_{t+1}))]].
\end{dmath}
\noindent The Bellman equation is equal to the action which maximizes the reward plus the discount factor multiplied by the next state's value, by taking the action after following the policy in state $s_{t+1}$ or $\pi(s_{t+1})$.

The Q-value can therefore be improved by bootstrapping. This is where the current value of the estimate of $Q_\pi$ is used to improve its future estimate, using the known $r(s_t,a_t)$ value. This is the foundation of Q-learning \cite{Gay2007}, a form of \textit{temporal difference} (TD) learning \cite{Sutton2015}, where the update of the Q-value after taking action $a_t$ in state $s_t$ and observing reward $r_t$, which results in state $s_{t+1}$ is:
\begin{equation}
Q(s_t,a_t)\leftarrow Q(s_t,a_t)+\alpha\delta_t,
\end{equation}
\noindent where,
\begin{equation}
\delta_t=r_t+\gamma\max_{a_{t+1}}Q(s_{t+1},a_{t+1})-Q(s_{t},a_t),
\end{equation}
\noindent $\alpha\in [0,1]$ is the step size, $\delta_t$ represents the correction for the estimation of the Q-value function and $r_t+\gamma\max_{a_{t+1}}Q(s_{t+1},a_{t+1})$ represents the target Q-value at time step $t$.

It has been proven that if the Q-value for each state action pair is visited infinitely often, the learning rate $\alpha$ decreases over time step $t$, then as $t\rightarrow \infty$, $Q(s,a)$ converges to the optimal $Q_*(s,a)$ for every state-action pair \cite{Gay2007}.

However, it is often the case that Q-learning suffers from the curse of dimensionality. This is because Q-learning stores the Q-value function in a look-up table. This therefore requires the action and state spaces to be discretised. As the number of discretised states and actions increases, the computational cost increases exponentially, making the problem intractable. Many problems are naturally discretised which are well suited to a Q-learning approach, however this is not always the case. 

\subsection{Deep Deterministic Gradient Policy}

It is not straightforward to apply Q-learning to continuous action spaces. This is because in continuous spaces, finding the greedy policy requires an optimisation of $a_t$ at every time step. Optimising for $a_t$ at every time step would be too slow to be practical with large, unconstrained function approximators and nontrivial action spaces \cite{Hunt2016a}. To solve this, an actor-critic approach based on the deterministic policy gradient (DPG) algorithm is used \cite{Silver2014}.

The DPG algorithm maintains a parameterized actor function $\mu(s|\theta^\mu)$ which specifies the current policy by deterministically mapping states to a specific action. The critic $Q(s,a)$ is learned using the Bellman equation as in Q-learning. The actor is updated by applying the chain rule to the expected return from the start distribution $J$ with respect to the actor parameters:
\begin{align}
\begin{split}
\triangledown_{\theta^\mu}J\approx\mathbb{E}_{s_t\sim\rho^\beta}[\triangledown_{\theta^\mu}Q(s,a|\theta^Q)|_{s=s_t,a=\mu(s_t|\theta^\mu)}] \\
= \mathbb{E}_{s_t\sim\rho^\beta}[\triangledown_aQ(s,a|\theta^Q)|_{s=s_t,a=\mu(s_t)}\triangledown_{\theta_\mu}\mu(s|\theta^\mu)|_{s=s_t}]
\end{split}.
\end{align}

This is the policy gradient. The policy gradient is the gradient of the policy's performance. The policy gradient method optimises the policy directly by updating the weights, $\theta$, in such a way that an optimal policy is found within finite time. This is achieved by performing gradient ascent on the policy and its parameters $\pi^\theta$.

Introducing non-linear function approximators, however, means that convergence is no longer guaranteed. However, these function approximators are required in order to learn and generalise on large state spaces. The Deep Deterministic Gradient Policy (DDPG) modifies the DPG algorithm by using neural network function approximators to learn large state and action spaces online.

A replay buffer is utilised in the DDPG algorithm to address the issue of ensuring that samples are independently and identically distributed. The replay buffer is a finite-sized cache, $\mathcal{R}$. Transitions are sampled from the environment through the use of the exploration policy, and the tuple $(s_t,a_t,r_t,s_{t+1})$ is stored within this replay buffer. $\mathcal{R}$ discards older experiences as the replay buffer becomes full. The actor and critic are trained by sampling from $\mathcal{R}$ uniformly. 

A copy is made of the actor and critic networks, $Q'(s,a|\theta^{Q'})$ and $\mu'(s|\theta^{\mu'})$ respectively. These are used for calculating the target values. To ensure stability, the weights of these target networks are updated by slowly tracking the learned networks. Pseudo-code of the DDPG algorithm is presented in Algorithm \ref{alg:ddpg}.

\begin{algorithm}
	\caption{DDPG Algorithm \cite{Hunt2016a}}
	\begin{algorithmic}[1]
		\small
		\State Initialize critic network $Q(s,a|\theta^Q)$ and actor $\mu(s|\theta^\mu)$ with random weights $\theta^Q$ and $\theta^\mu$
		\State Initialize target network $Q'$ and $\mu'$ with weights $\theta^{Q'}\leftarrow\theta^Q,\theta^{\mu'}\leftarrow \theta^{\mu}$
		\State Initialize replay buffer $R$
		\For{\texttt{episode=1,M}}
		\State Initialize a random process $\mathcal{N}$ for action exploration
		\State Receive initial observation state $s_1$
		\For{\texttt{t=1,T}}
		\State Select action $a_t=\mu(s_t|\theta^{\mu})+\mathcal{N}_t$ according to the policy and exploration noise, $\mathcal{N}_t$
		\State Execute action $a_t$ and observe reward $r_t$ and new state $s_{t+1}$
		\State Store transition $(s_t, a_t, r_t, s_{t+1})$ in $R$
		\State Sample a random minibatch of $N$ transitions $(s_i, a_i, r_i, s_{i+1})$ from $R$
		\State Set $y_i=r_i+\gamma Q'(s_{i+1},\mu'(s_{i+1},\mu'(s_{i+1}|\theta^{\mu'})|\theta^{Q'})$
		\State Update critic by minimizing the loss: $$L=\frac{1}{N}\sum_i(y_i-Q(s_i,a_i|\theta^Q))^2$$
		\State Update the actor policy using the sampled policy gradient: $$\triangledown_{\theta^\mu}J\approx \frac{1}{N}\sum_i\triangledown_a Q(s,a|\theta^Q)|_{s=s_i,a=\mu(s_i)}\triangledown_{\theta^\mu}\mu(s|\theta^\mu)|_{s_i}$$
		\State Update the target networks:
		$$\theta^{Q'}\leftarrow\tau\theta^Q+(1-\tau)\theta^{Q'}$$
		$$\theta^{\mu'}\leftarrow\tau\theta^\mu+(1-\tau)\theta^{\mu'}$$
		\EndFor
		\EndFor
	\end{algorithmic}
	\label{alg:ddpg}
\end{algorithm}

\subsection{Simulation}

We utilized the long-term electricity market agent-based model, ElecSim \cite{Kell,Kell2020} discussed in Chapter \ref{chapter:elecsim}. The model was run using a short term approach by only iterating through a single year (2018), composed of eight representative days, each of 24-time steps.


In this work, we explore whether large GenCos, or group of GenCos, can manipulate the price of the electricity market through virtue of their size. We achieve this by allowing a subset of GenCos to bid away from their SRMC and allow them to learn an optimal bidding strategy for maximising their income. The GenCo agents adopt a DDPG RL algorithm to select their bids. This is to explore whether large GenCos, or a group of GenCos can manipulate the price of the electricity market through market power. The remaining GenCos, which fall outside of this group, maintain a bidding strategy based upon their SRMC.

For the purpose of this work, we do not consider flow constraints within the electricity mix. This is because we model the entire UK with more than one thousand generators, and many nodes and buses. This would make the optimisation problem intractable for the purpose of our simulation, especially when considering the many episodes required for training. It takes ${\sim}125$ seconds to run a single year in the simulation, or episode with our current setup. By increasing the simulation time further, we would make the compute time intractable due to the many episodes required for reinforcement learning to learn an effective policy. Additionally, we must train several reinforcement learning policies to account for each GenCo and market cap. Therefore a simulation that takes ${\sim}$2 minutes to run, which needs to be executed one thousand times takes ${\sim}$33 hours to complete, multiplied by 12 different GenCos takes ${\sim}$16 days, and with two market cap scenarios it would take ${\sim}$1 month.




\section{Experimental Setup}
\label{rl:sec:methodology}

To parametrise the simulation, we used data from the United Kingdom in 2018. This included 1,085 electricity generators and power plants with their respective GenCos. The data for this was taken from the BEIS DUKES dataset \cite{dukes_511}. The electricity load data was modelled using data from Elexon portal and Sheffield University \cite{gbnationalgridstatus2019}; offshore and onshore wind and solar irradiance data from renewables.ninja by Pfenninger \textit{et al.} \cite{Pfenninger2016}. It would be possible to adopt this approach to other decentralised markets in other countries.

By modelling bidding decisions as an RL algorithm, we hoped to observe the ability for RL to find the point at which market power artificially inflates electricity prices. To achieve this, we chose the six largest GenCos in the UK, as well as a smaller GenCo as a control. Groups of GenCos are modelled as a single GenCo with a single RL strategy for the purpose of this work. Table \ref{table:genco_table} displays the groups of GenCos, as well as individual GenCos, with their respective capacity and number of plants.

\begin{sidewaystable}
	\footnotesize
	\renewcommand{\arraystretch}{1.25}
	\centering
	\begin{tabular}{llllp{1.5cm}}
		\toprule
		GenCo Groups                                                                                                                  & Capacity (MW) & Num. of Plants \\ \midrule
		Orsted                                                                                                                       & 2738.7   & 11               \\
		Drax Power Ltd                                                                                                               & 4035.0   & 3                \\
		Scottish power                                                                                                               & 4471.5   & 49               \\
		Uniper UK Limited                                                                                                            & 6605.0   & 9                \\
		SSE                                                                                                                          & 8390.7   & 130              \\
		RWE Generation SE                                                                                                            & 8664.0   & 11               \\
		EDF Energy                                                                                                                   & 14763.0  & 14               \\
		\{EDF Energy, RWE Generation SE\}                                                                                              & 23427.0  & 25               \\
		\{EDF Energy, RWE Generation SE, SSE\}                                          & 31817.7  & 155              \\
		\{EDF Energy, RWE Generation SE, SSE, Uniper UK Ltd\}                              & 38422.7  & 164              \\
		\{EDF Energy, RWE Generation SE, SSE, Uniper UK Ltd, Scottish Power\}              & 42894.2  & 213              \\
		\{EDF Energy, RWE Generation SE, SSE, Uniper UK Ltd, Scottish Power, Drax Power Ltd\} & 46929.2  & 216              \\ 
		\bottomrule
	\end{tabular}
	\caption{Groups of GenCos that used reinforcement learning bidding strategies, number of plants and total electricity generating capacity.}
	\label{table:genco_table}
\end{sidewaystable}

For the reinforcement learning problem we have the following tuple: $(s_t,a_t,r_t,s_{t+1})$, where $(s_t, s_{t+1})$ is the state at time $t$ and $t+1$ respectively, $a_t$ is the action at time $t$ and $r_t$ is the reward at time $t$. For our problem the state space is given by the tuple shown in Equation \ref{eq:observation_tuple}:
\begin{equation}
\label{eq:observation_tuple}
s_t=(H_i,D_i,p_{gas},p_{coal},p_{C02},p_{c}),
\end{equation}
\noindent where $H_i$ is the segment hour to bid into at timestep $i$, $D_i$ is the demand of the segment hour at timestep $t$, $p_{gas}$ is the price of gas, $p_{coal}$ the price of coal, $p_{C02}$ is the carbon tax price, and $p_{c}$ is the clearing price. We set the reward, $r_t$ to be the average electricity price of that time step, $p_{avg}$.

For the action space, $a_t$, we modelled two scenarios. Where there was a price cap of \textsterling$150$/MWh and \textsterling$600$/MWh. Only these two values were chosen to reduce computational load. We chose \textsterling$150$/MWh as a reasonable price cap that may be introduced by a Government. This was roughly double the average accepted price in 2018, therefore allowed for higher prices in times of high demand or low supply. The \textsterling $600$/MWh was chosen to simulate a realistic unbounded price cap. This enabled us to see the price that an equilibrium is reached within a market with agents with market power. 

For this work, we assume that the action space $a_t$ only bids price, and not how much capacity to bid on the market. We assume this to reduce the dimensionality of $a_t$, and simplify the training process.

In this work, we assume that the GenCo groups have no information about the generation capacity, marginal cost, bid prices or profits of other GenCos \cite{EsmaeiliAliabadi2017}. They learn the maximum profit that can be made through experience within a particular market. We assume this because in real-life GenCo groups have little information on their competitors sensitive bidding data. If they were to have perfect information on all their competitors, they would be able to devise a perfect strategy which would always maximise their profit.




\section{Results}
\label{rl:sec:results}

In this section, we detail the results of the RL algorithm, and the effect that capacity has on average electricity price within the UK. Our approach could be generalised to any other decentralised electricity market in other countries. 

\begin{figure}
	\centering
	\includegraphics[width=0.6\textwidth]{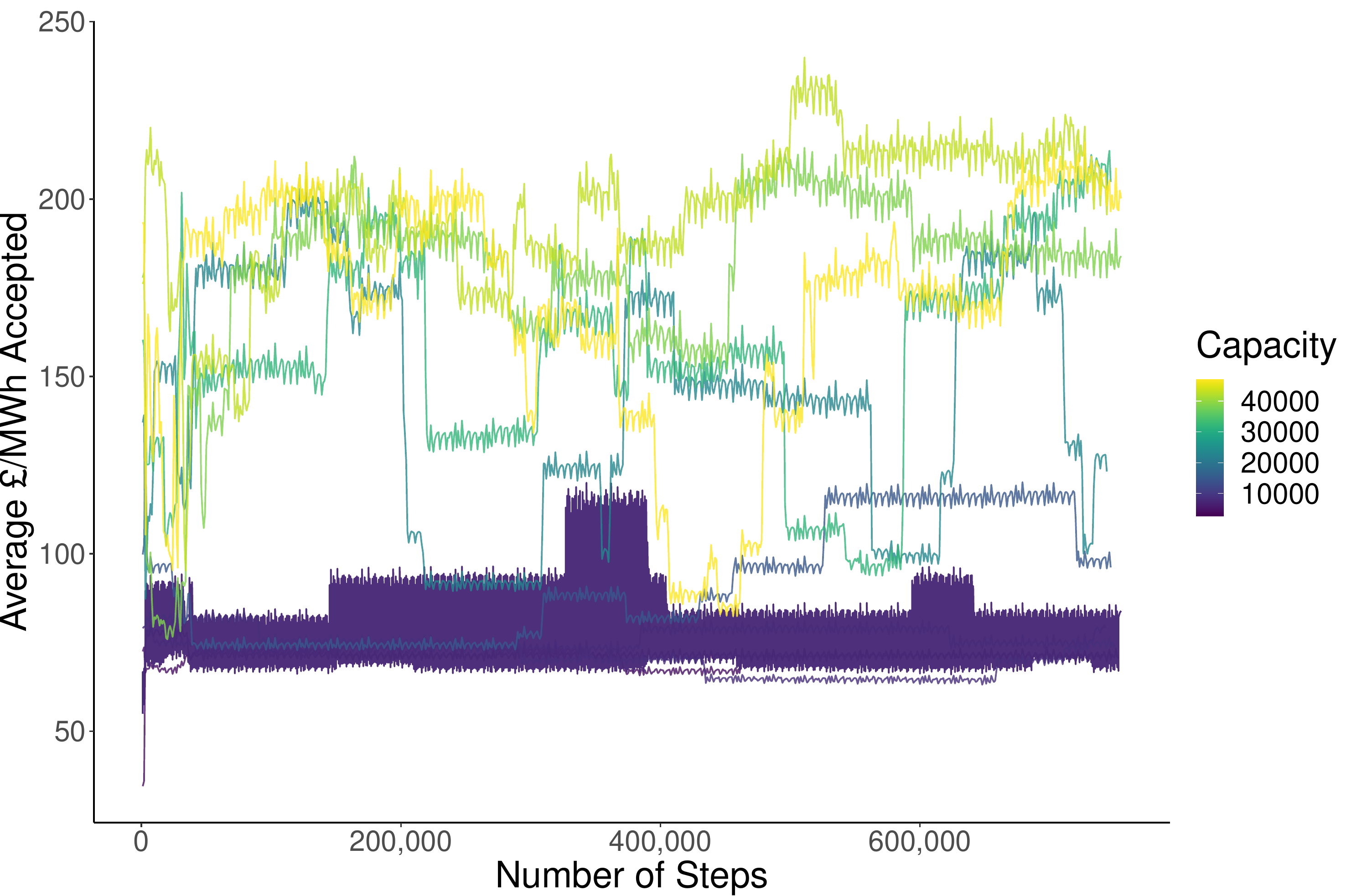}
	\caption{Reward over time for different groups of GenCos using Reinforcement Learning, max bid = \textsterling $600$/MWh.}
	\label{fig:unbounded_timesteps}
\end{figure}

\begin{figure}
	\centering
	\includegraphics[width=0.6\textwidth]{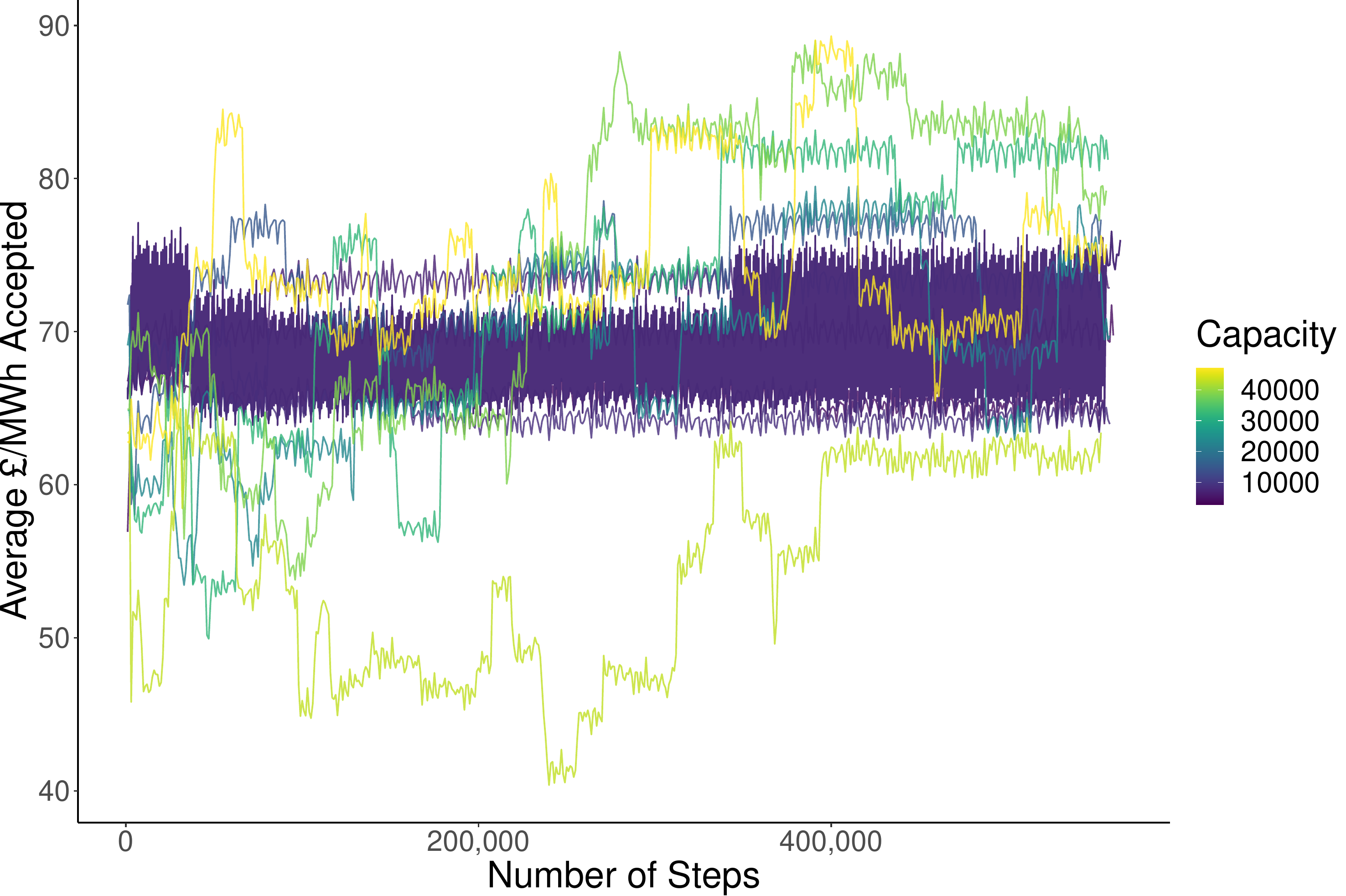}
	\caption{Reward over time for different groups of GenCos using Reinforcement Learning, max bid = \textsterling $150$/MWh.}
	\label{fig:bounded_timesteps}
\end{figure}

Figures \ref{fig:unbounded_timesteps} and \ref{fig:bounded_timesteps} show the rewards over a number of time steps for the unbounded and bounded cases respectively. Figure \ref{fig:unbounded_timesteps} shows a clear difference between agents which use the DDPG RL strategy and have a large capacity (green and yellow) compared to those which have a smaller capacity (dark purple). The axis in Figure \ref{fig:unbounded_timesteps} are much larger than those of \ref{fig:bounded_timesteps}, highlighting the effect of market power on an unbounded market.

The average electricity price for a capacity below 10,000MW, or ${\sim35\%}$ of total capacity, remains stable between \textsterling70/MWh and \textsterling100/MWh. This range may be due to the stochasticity in calculating the weights for the DDPG algorithm. The average electricity price does not change over the time steps or training. We, therefore, hypothesise that there is no market power as long as an individual GenCo owns below ${\sim}35\%$ of total electrical capacity. 

On the other hand, once the capacity of a GenCo or groups of GenCos is above 30,000MW, there is a significant increase in the average price for capacity. The average electricity price for capacity falls between the range of  \textsterling$170$/MWh and \textsterling$220$/MWh. 

\begin{figure}
	\centering
	\includegraphics[width=0.6\textwidth]{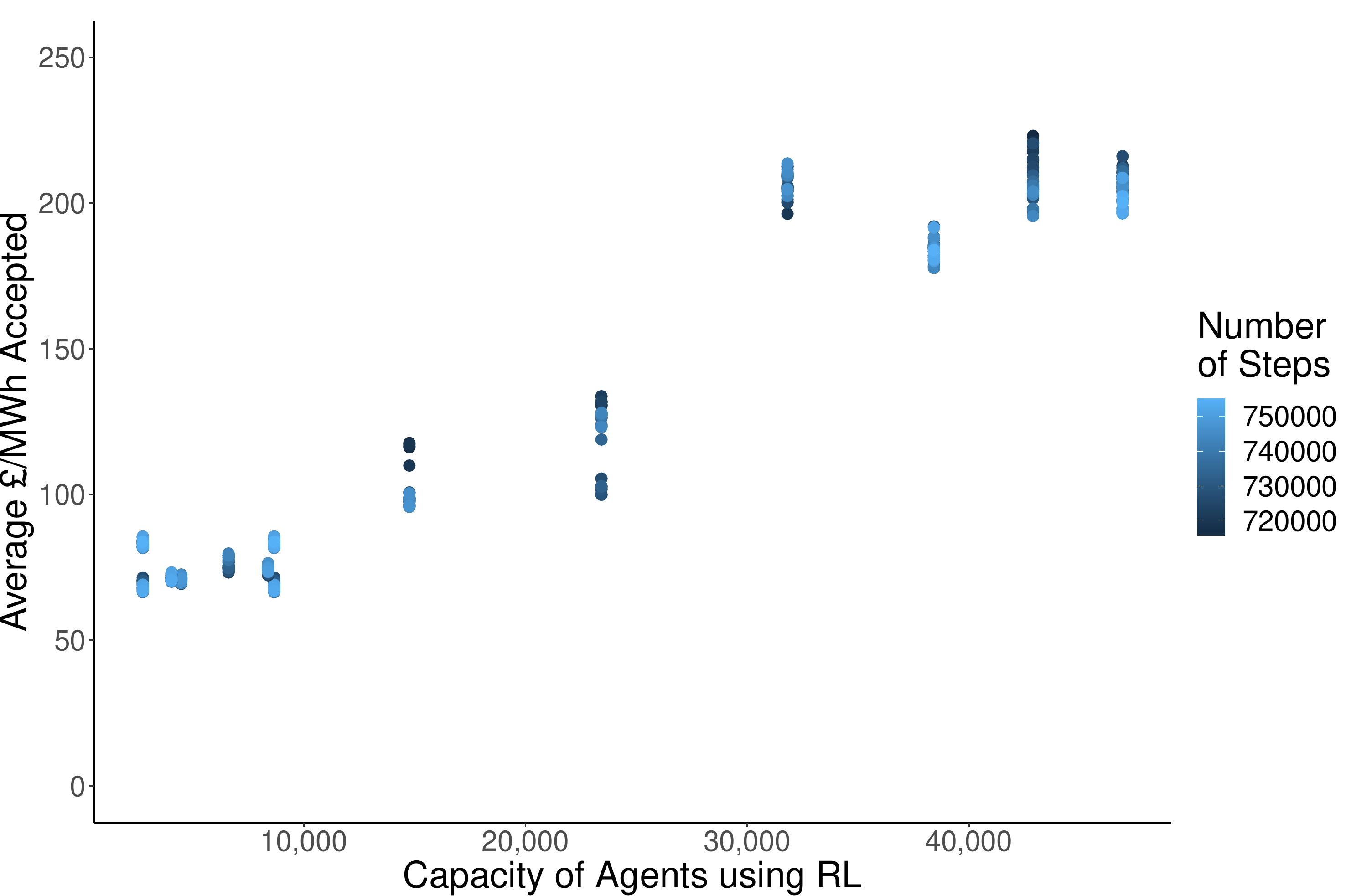}
	\caption{Capacity of agents using RL vs. average electricity price accepted, for unbounded agents.}
	\label{fig:unbounded_results_scatter}
\end{figure}

Figure \ref{fig:unbounded_results_scatter} displays the capacity controlled by the agents that use the RL strategy versus the average electricity price for the unbounded case. The colour displays the number of steps. The step-change, as shown in Figure \ref{fig:unbounded_timesteps} can be seen clearly here, with agents with a capacity larger than ${\sim}$25,000MW causing a step change in electricity price. Electricity prices seems to cluster below ${\sim}$10,000MW. However, after this point, the average electricity price begins to increase.

\begin{figure}
	\centering
	\includegraphics[width=0.6\textwidth]{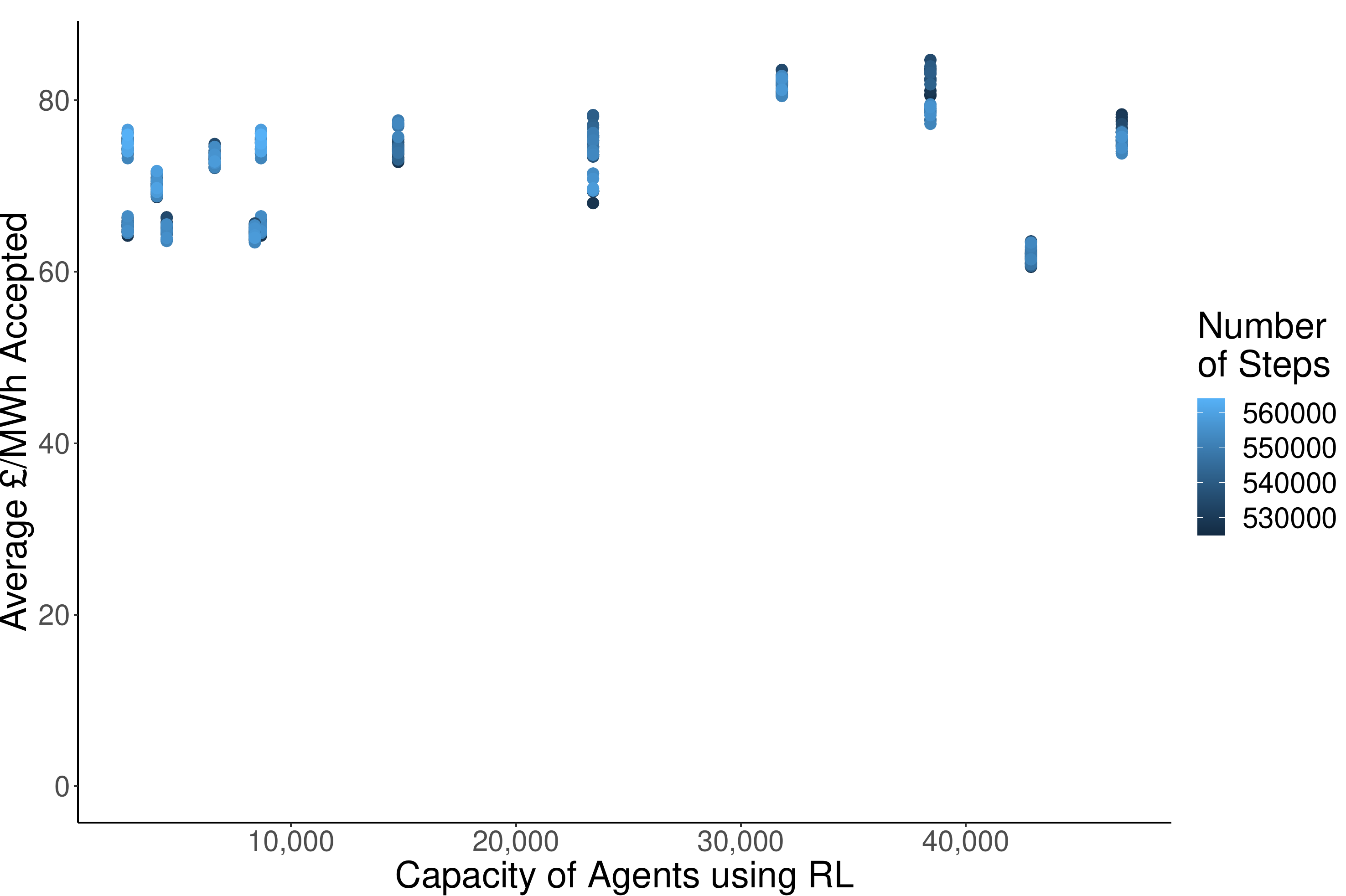}
	\caption{Capacity of agents using RL vs. average electricity price accepted, for bounded agents.}
	\label{fig:bounded_results_scatter}
\end{figure}

Figure \ref{fig:bounded_timesteps} shows a cluster between ${\sim}$\textsterling$60$/MWh and ${\sim}$\textsterling$80$/MWh irrespective of the capacity of the agents. This is verified by Figure \ref{fig:bounded_results_scatter}. This seems to suggest that setting a lower market cap reduces the ability for generators, irrespective of size, from influencing the electricity price.

\begin{figure}
	\centering
	\begin{subfigure}{0.6\textwidth}   
		\includegraphics[width=\columnwidth]{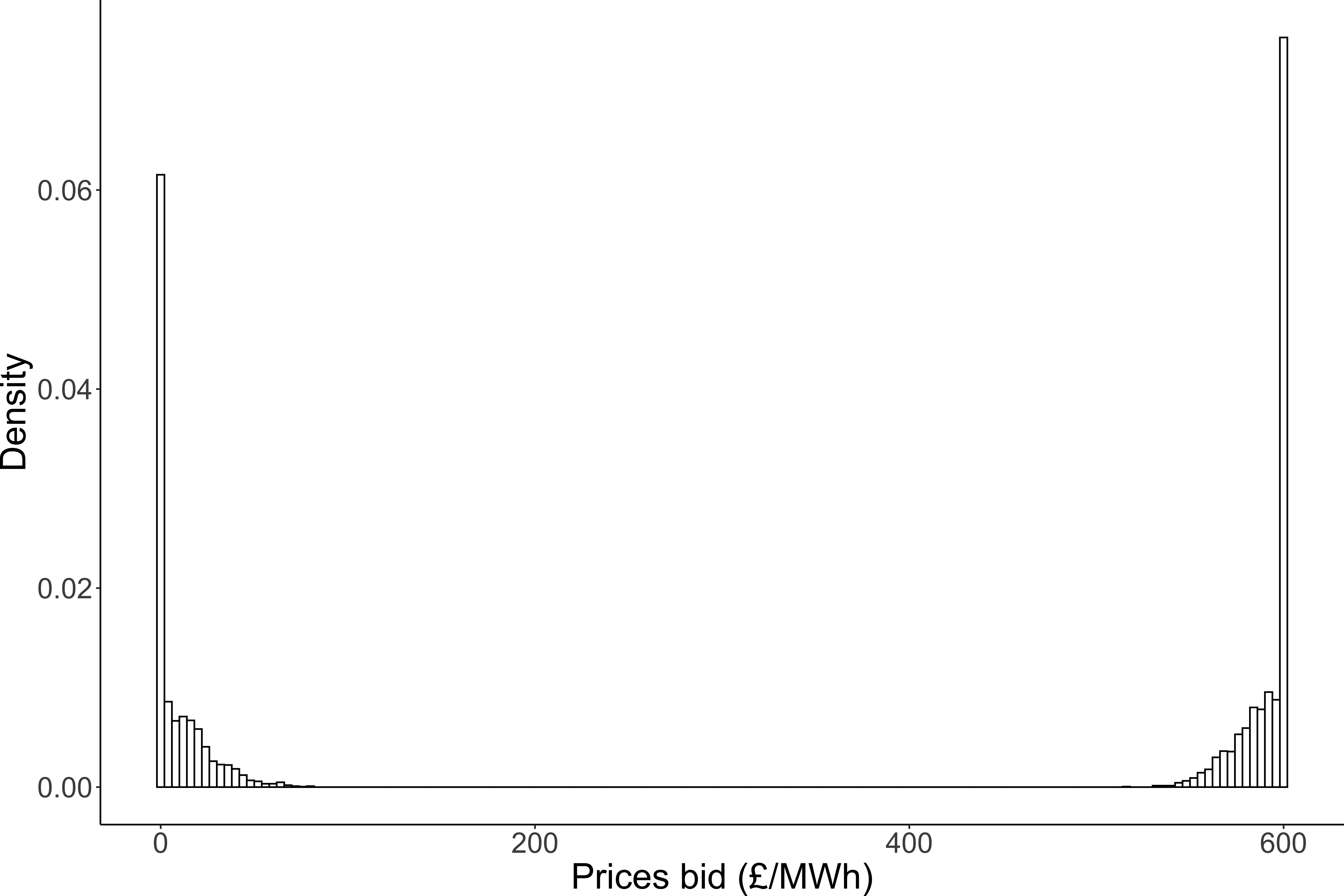}
		\caption{Largest group of GenCos with a total controlled capacity of 46929.2MW with a market cap of \textsterling600/MWh.}
		\label{fig:large_company_capped_600_bids}
	\end{subfigure}
	\hfil
	\begin{subfigure}{0.6\textwidth}   
		\includegraphics[width=\columnwidth]{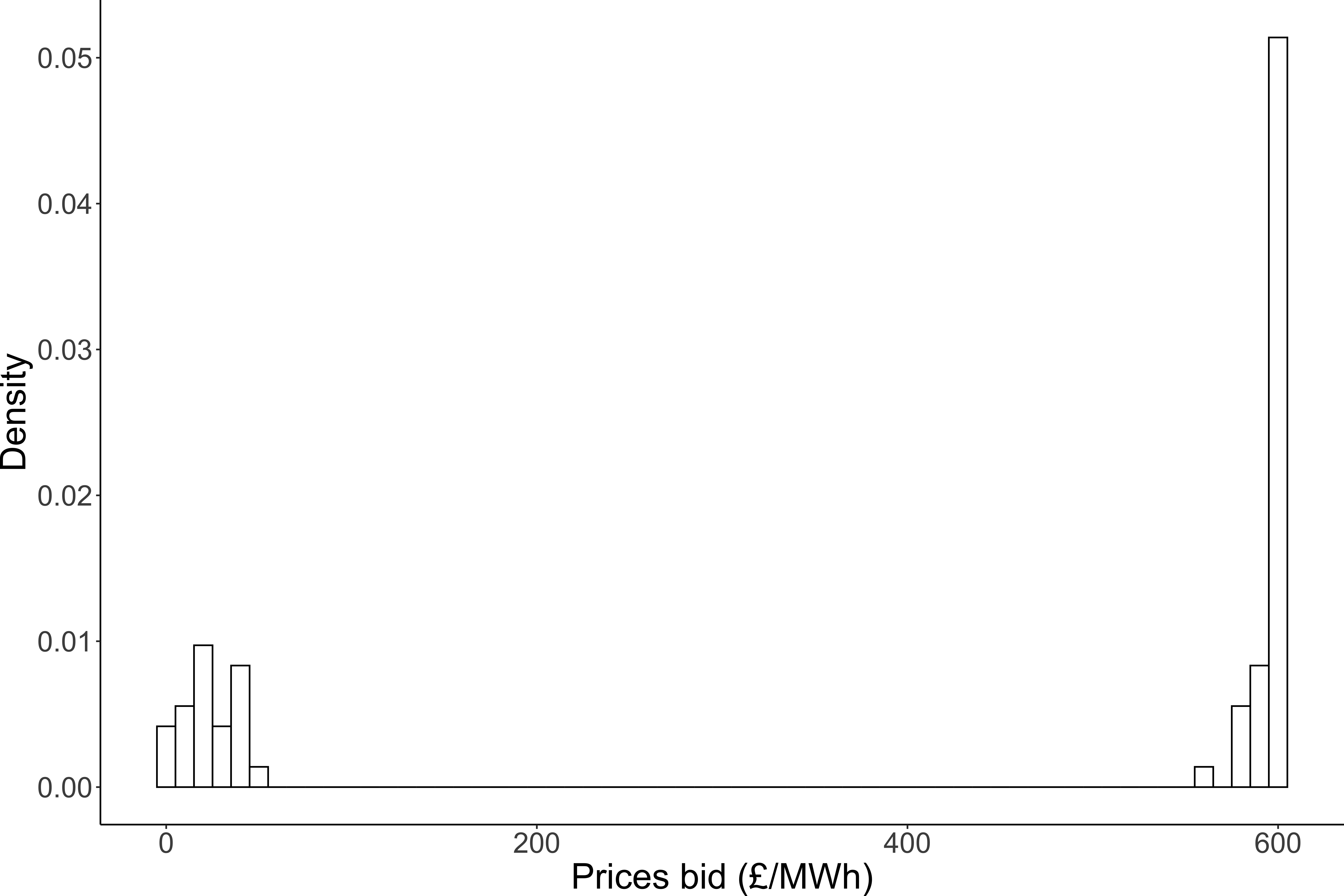}
		\caption{Smallest company with a total controlled capacity of 2738.7MW with a market cap of \textsterling600/MWh.}
		\label{fig:small_company_capped_600_bids}
	\end{subfigure}
	\caption{Bids made by generator companies using Reinforcement Learning with a market cap of \textsterling600/MWh.}
	\label{fig:capped_600_bids}
\end{figure}

\begin{figure}
	\centering
	\begin{subfigure}{0.9\textwidth}  
		\centering 
		\includegraphics[width=0.6\columnwidth]{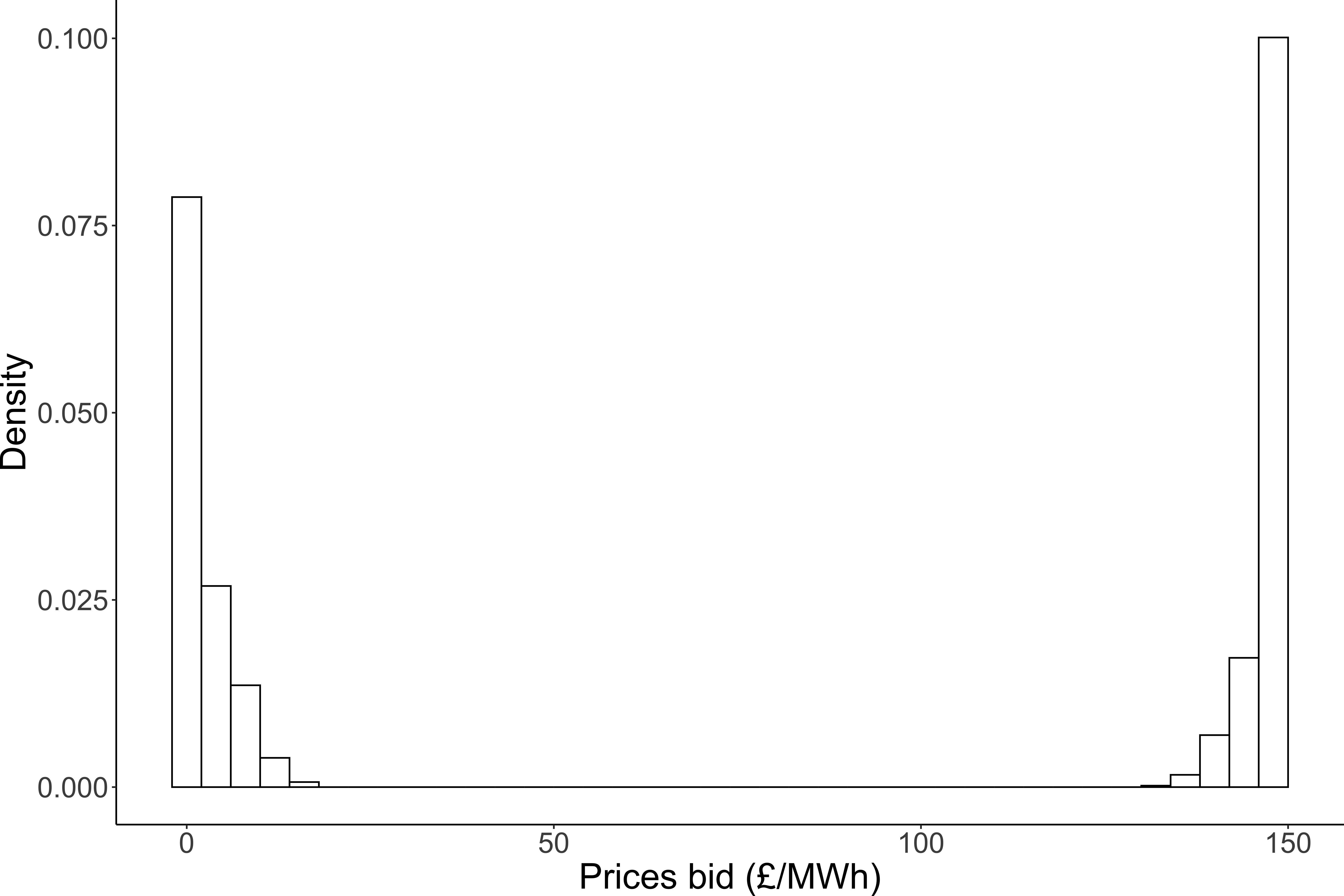}
		\caption{Largest group of GenCos with a total controlled capacity of 46929.2MW with a market cap of \textsterling150/MWh.}
		\label{fig:large_company_capped_150_bids}
	\end{subfigure}
	\hfil
	\begin{subfigure}{0.9\textwidth} 
		\centering  
		\includegraphics[width=0.6\columnwidth]{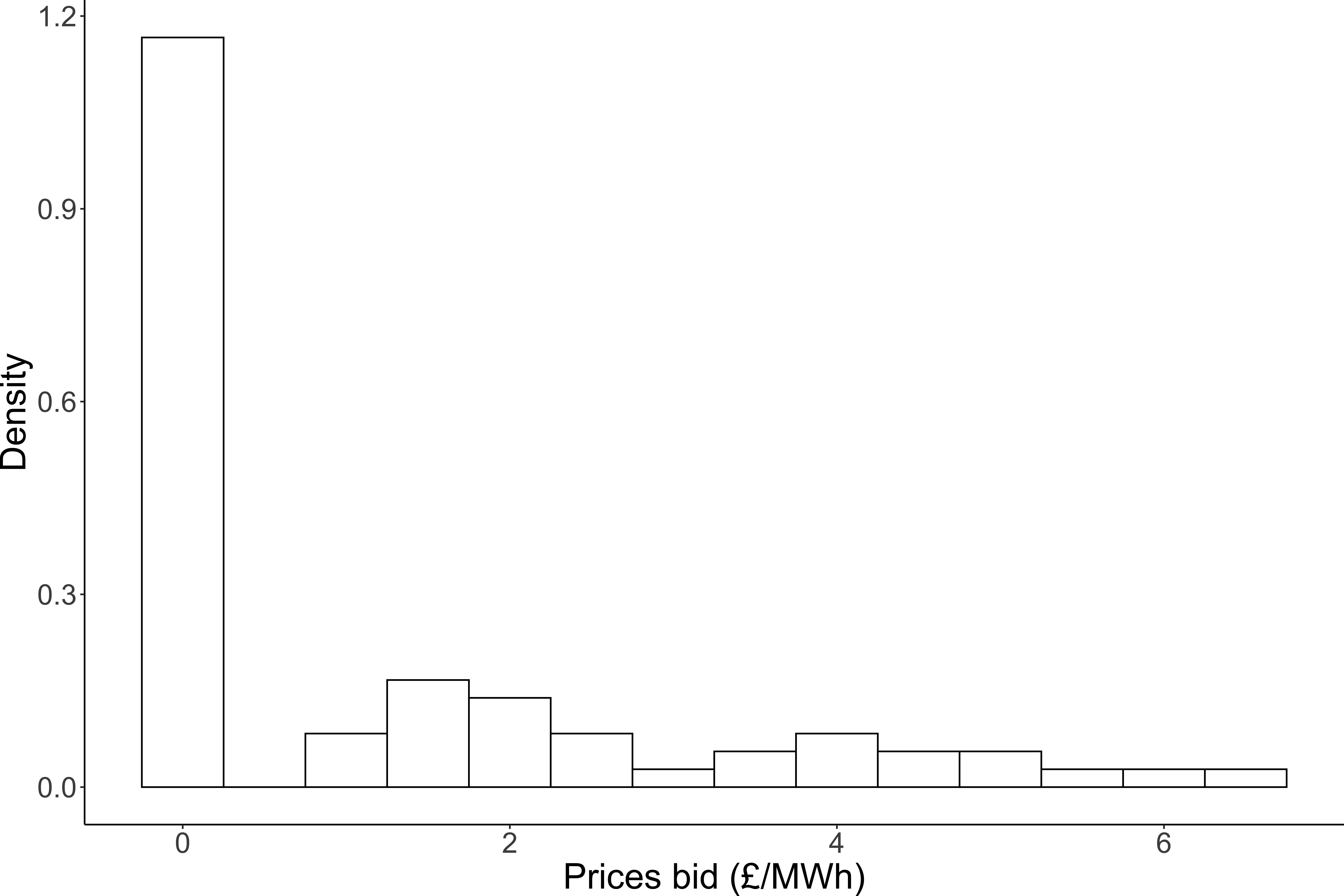}
		\caption{Smallest company with a total controlled capacity of 2738.7MW with a market cap of \textsterling150/MWh.}
		\label{fig:small_company_capped_150_bids}
	\end{subfigure}
	\caption{Bids made by generator companies using Reinforcement Learning with a market cap of \textsterling150/MWh.}
	\label{fig:capped_150_bids}
\end{figure}

Figures \ref{fig:capped_600_bids} and \ref{fig:capped_150_bids} display the actual bids made at the end of training within the electricity market for all of their power plants. The number of bids made by each GenCo changes dependent on the number of plants that they own, with the Orsted GenCo only making eleven bids per segment, and the largest group making 216 bids per segment. Figure \ref{fig:capped_600_bids} displays the uncapped scenario (\textsterling 600/MWh) and \ref{fig:capped_150_bids} displays the capped scenario (\textsterling 150/MWh).

Figure \ref{fig:large_company_capped_600_bids} shows the bids made by the largest group of GenCos as shown in Table \ref{table:genco_table}. A bimodal distribution can be seen, where the group of GenCos tend to bid either the maximum or the minimum bid. We hypothesise that they bid the maximum amount as this ensures that the market price is artificially raised, and that they are able to utilise their market power. The minimum price is bid the rest of the time to ensure that generators bids are always accepted, regardless of whether the market price has been artificially raised or not in each respective clearing segment.

Figure \ref{fig:small_company_capped_600_bids} displays the bids of the small company with a market cap of \textsterling 600/MWh. The small company also seems to have a bimodal distribution; however, bids the higher price more often. This may be due to the fact that it is able to influence the price at certain market segments, and the reward of the higher accepted reward outweighs the times in which it is not accepted on the market segments. Again, bidding low seems to be the strategy in which to take if the GenCo does not believe it will be able to influence the final price. As the market simulated is a uniform pricing market, having a \textsterling0 bid accepted does not mean that the GenCo will be paid \textsterling0. Rather, the GenCo will be paid the market clearing segment, which is set by the most expensive power plant accepted onto that market segment.

Figures \ref{fig:large_company_capped_150_bids} shows the bids made by the largest group of GenCos in each market segment. It seems to take a similar strategy to that of the large company in the uncapped scenario, as shown in Figure \ref{fig:large_company_capped_600_bids}. We believe, as similar in the uncapped scenario, that this is due to the market power that this group possesses. Being able to influence the market price enables the GenCo group to inflate the prices. It also takes a conservative strategy to bid the minimum price allowed, to ensure that the bid is accepted regardless of price.

Figure \ref{fig:small_company_capped_150_bids} displays the strategy of the smallest company in the capped scenario. Here, it seems that the GenCo is unable to influence the price at all, and therefore bids \textsterling0/MWh for the majority of the time. This is similar to the expected strategy of GenCos, who tend to bid their short-run marginal cost to ensure that they do not miss out on potential profit. The short-run marginal cost can often change based upon fuel, carbon and generator type. However, for renewable energy, it is near \textsterling0, and for fossil-fuel based plants it is near the cost of fuel and carbon at that point in time.

\begin{figure}
	\centering
	\includegraphics[width=0.6\textwidth]{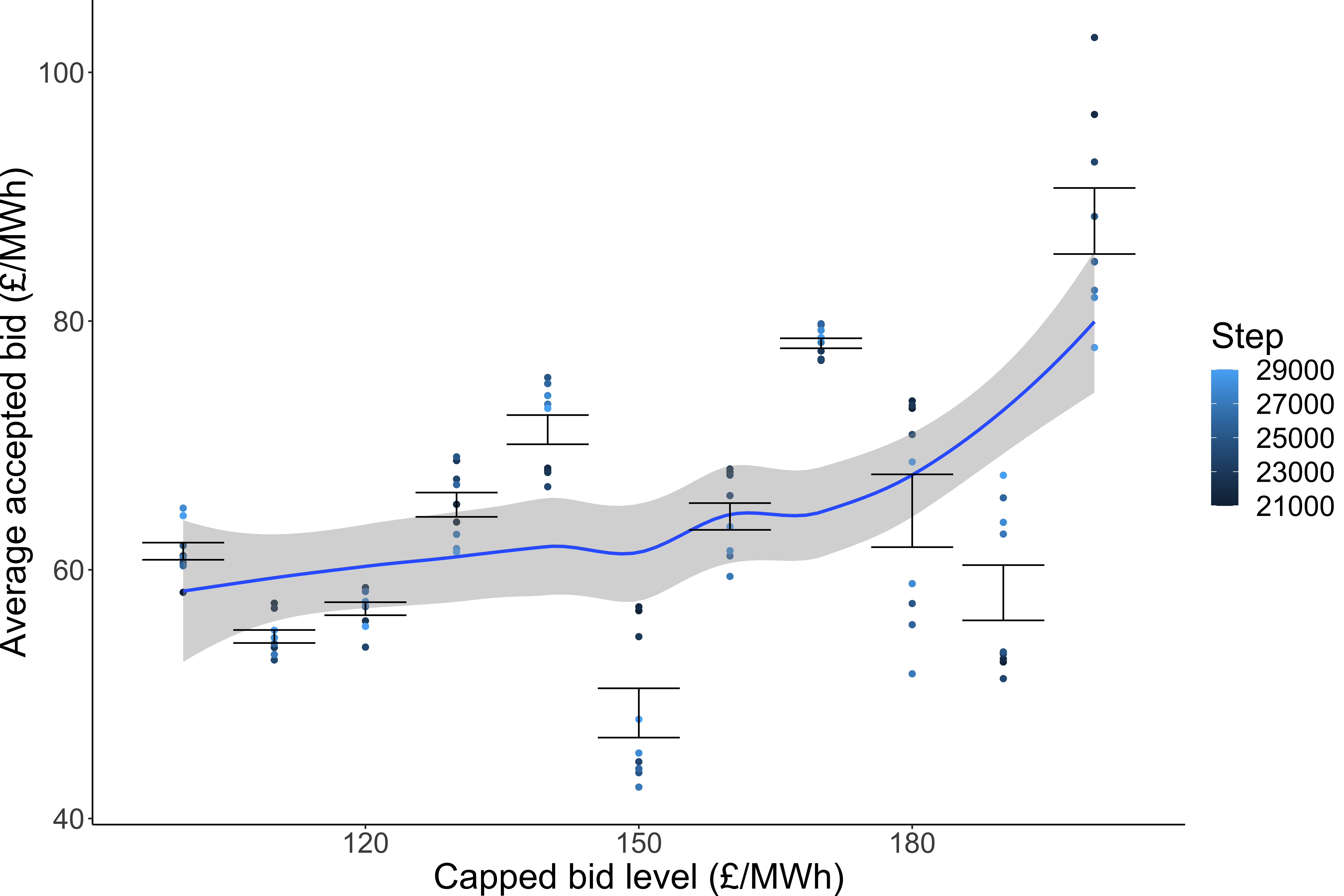}
	\caption{Capacity of agents using RL vs. average electricity price accepted, for bounded agents.}
	\label{fig:sensitivity_analysis}
\end{figure}

We ran a sensitivity analysis to observe the effects of the market cap on final average accepted bid price. The largest GenCo group used a strategy for this sensitivity analysis. Figure \ref{fig:sensitivity_analysis} displays the results. It seems that whilst the average accepted bid price increased with the capped bid level; there is a significant increase after a market cap of \textsterling190/MWh. This may be due to the case that the GenCos begin to outbid the SRMC bidding GenCos at this price point.

These results demonstrate a methodology for policy makers to investigate the ability for generator companies to artificially increase the electricity price. It gives quantitive advice on the level of capacity that a generator company can control before it becomes a problem from the perspective of consumers of the electricity market. These results also provide a method to control excessive electricity prices with the provision of a market cap.

The limitations from the perspective of the policy maker, is that we do not provide any information on how to prevent an oligopoly from occurring. We only provide advice on what to do once one has occurred.


\section{Discussion}
\label{rl:sec:discussion}

Our results demonstrate the ability for GenCos to artificially increase the electricity price through market power in an uncapped market. Our results have shown that in an uncapped market, any single agent or groups of agents who make bids using the same strategy and information, should have less than ${\sim}$10,000MW. This defines the optimal capacity of any single GenCo to have a fair level of competition. After this, the electricity price begins to rise with the same outcome and welfare. It is also worth noting that when the market is capped, the average accepted price does not simply become the capped price.

However, if there is an electricity market with a few large or colluding players, it is possible to remove their advantage through the introduction of a price cap. Our results show that whilst average accepted bids increase with market cap level, the value does not increase significantly until \textsterling$200$ is reached.


This information and approach can help to inform government policy to ensure fair competition within electricity markets, as well as run the model for their own scenario. It is hypothesised that the findings in this work are generalisable to other decentralised electricity markets in other geographies due to their similar market structures. Whilst the figures presented here may not be the same; we hypothesise that the region of interest will be similar.


\section{Conclusion}
\label{rl:sec:conclusion}

In this work, we investigated the ability for GenCos to make strategic bids within an electricity market. We did this using the deep deterministic policy gradient (DDPG) reinforcement learning method. We utilised the agent-based model ElecSim to model the UK electricity market. We utilised the DDPG algorithm only for a certain subset of agents, from small individual generation companies (GenCos) to large groups of GenCos. 

This enabled us to explore the ability for GenCos with a large capacity to artificially increase the price in the electricity market within the UK if they are in control of a sufficiently large generation capacity. Our results show that the optimum level of control of any single GenCo or groups of GenCo is below ${\sim}$10,000MW or ${\sim}$11\% of the total capacity. Above this, prices begin to increase with no real additional benefit to the consumer. After ${\sim}$25,000MW, or ${\sim}$35\% of the total capacity, the prices begin to increase substantially, to ${\sim}$\textsterling200, over triple the original cost without this market power. The introduction of a market cap of \textsterling$150$ reduces all market power and maintains electricity price at a reasonable level.

We found through a sensitivity analysis that the average electricity price in the market over a year remains low with a price cap smaller than \textsterling 190. However, after this level, the average electricity price begins to increase.


Our work has shown the ability for reinforcement learning to learn an optimal bidding strategy to maximise a GenCo's profit within an electricity market. The ability for GenCos to use their market power is also highlighted, and is dependent on electricity generation capacity of the respective GenCo.

In future work, we would like to enable GenCos to withhold the capacity on offer to the electricity market. This would enable further market power by reducing competition further.  Additionally, we would like to assess the market power in different countries with different market structures and total electricity supply. An other option would be to assess the effect of two large competing GenCos on the market.

\chapter{Conclusion}
\label{chapter:conclusion}
\ifpdf
    \graphicspath{{Chapter3/Figs/Raster/}{Chapter3/Figs/PDF/}{Chapter3/Figs/}}
\else
    \graphicspath{{Chapter3/Figs/Vector/}{Chapter3/Figs/}}
\fi

\section{Thesis summary}

In this thesis, we presented an open-sourced, long-term agent-based model for electricity markets called ElecSim. We were able to validate the model through two methods: cross-validation using historic data from 2013 and projecting forward, and comparing our scenario to that of the UK Government to 2035. We found positive results in both: we were able to accurately model the transition from coal to gas between 2013 and 2018, and closely match the scenario of the UK Government. 

For this validation, we used optimisation to find realistic parameters that would generate the desired scenarios. The primary parameter that was optimised was the predicted price duration curve. We found a predicted price duration curve which closely matched the price found in 2018 for the cross-validation using historic data. For the UK government forward scenario, we found a variety of predicted price duration curves, with a few boom and bust cycles in the price of electricity. 

In addition to designing and developing ElecSim, we used the forward scenario generated to optimise a carbon tax strategy. That is, to reduce both carbon emissions and electricity price from 2018 to 2035. To achieve this we used a multi-objective genetic algorithm. We found that we were able to reduce both electricity price and carbon emissions through a combined use of solar, nuclear and onshore energy.

Further, we predicted electricity demand consumption on two time scales: 30-minutes ahead and a day-ahead to use in ElecSim. We found that we were able to predict electricity demand at these time scales well. Through the use of an online learning method, we were able to reduce the required reserve capacity of the national grid in the UK. We also looked at the long-term impact of poor forecasting errors on the long-term market, and found that poor forecasts led to a high carbon density of electricity grid over the long-term.

Our final chapter focuses on the use of deep reinforcement learning to model a bidding strategy of a single generation company or group of generation company using ElecSim. The rest of the generation companies bid based upon the respective generator's short run marginal cost. We found that after a capacity controlled by a generation company of 10,000MW, the generation company with the reinforcement learning bidding strategy was able to influence electricity prices in their favour; effectively demonstrating market power within the electricity market. After a controlled capacity of 30,000MW the average electricity price triples from the equilibrium price. 

We utilised agent-based models in this work for several reasons:
\begin{enumerate}
	\item Ability to model scenarios in a way that emerges from the generator companies.
	\item The process which emerge a not a black box, unlike in equilibrium models.
	\item Optimisation based methods should be interpreted in a normative manner, for example, how policy choices should be carried out. Whereas agent-based models make no assumption on outcomes of scenarios.
	\item Ability to model decentralised agents with different desires.
	\item Ability to model outcomes which are not in equilibrium, for example boom and bust cycles.
\end{enumerate}

An additional reason was for the limited availability of an open-sourced agent-based model, and to ensure that this model is available to the community. Open-sourced models allow for transparency, and to garner greater acceptance and understanding within the wider community. 

Our contributions are the following:

\begin{enumerate}
	\item Developed a novel, open-source agent-based model for the electricity market called ElecSim. ElecSim and all related code can be accessed at: \url{https://github.com/alexanderkell/elecsim}.
	\item Validated this model by the means of cross-validation, and compared our model to that of the UK Government.
	\item Investigated the effect of online learning to improve electricity demand forecasting a day-ahead, and looked at the long-term impacts that this had on the electricity markets.
	\item Predicted electricity consumption 30-minutes ahead.
	\item Found a variety of optimal carbon tax strategies using genetic algorithm based optimisation and the ElecSim model.
	\item Investigated the impact of collusion within a oligopolistic electricity market.
\end{enumerate}

\section{Conclusions}

In this section we explore the conclusions that arose from this thesis. We explore each of the chapters where we present our primary technical contributions. We place our work in the context of our research question: what challenges can AI and ML tackle, and how do these methods relate back to the wider energy system? Each of the chapters explore a research subquestion, which are highlighted in Chapter \ref{chapter:intro}. We provide answers to these subquestions here.

\subsection{ElecSim Model}

\textbf{Can a simulation model an electricity market over the long-term?} Through the recreation of the salient features of the UK electricity market into an agent-based model, we show that we are able to derive similar outputs to an established, governmental model. Therefore, even though the underlying methodology between these two models are different, we have shown that it is possible to reach similar conclusions.

This has important ramifications for future work, as features that could not previously be modelled with traditional, optimisation based models, such as heterogenous generation companies, can be modelled. This finding adds a significantly different tool to the established models, and allows modellers and policy makers to make findings from an alternative kind of methodology. Whilst this does not provide a ground truth model to the energy modelling community, it does provide an additional tool to understand complexity within an electricity market.

\textbf{Is it possible to model the variability of an electricity system?} The variability within a modern day electricity system is significant. With the increase in solar photovoltaics and wind turbines a modern day electricity model must be able to model fluctuations in solar irradiance and wind speeds. This variability can change in minutes and therefore adds significant complexity for electricity system models. This is because a high temporal resolution is required to model this variability.

To take into account this variability, we used a k-means clustering approach to select representative days in Chapter \ref{chapter:elecsim}. Through the use of these representative days, we found that we were able to model significant peaks and troughs in demand, wind speed and solar irradiance. This was shown by comparing three different metrics: the normalised root mean squared error, correlation and relative energy error. This was demonstrated in Chapter \ref{chapter:elecsim}. We recommend that a similar approach is undertaken by other modellers for a realistic representation of the real-world, without compromising on runtime.

\textbf{Can we trust an electricity model's outputs?} Fundamentally, the ability to trust an electricity model's outputs is through the verification of its outputs with observed data. For long-term models this becomes a significant challenge because of all the uncertainties inherent in long-term investments. Further, a long-term electricity model produces scenarios and not predictions. This makes this problem even more difficult, as an output scenario could be both possible and realistic, but it does not mean that it is a certainty that it will occur.

To answer the subquestion of whether we can trust an electricity model's outputs, we investigated an observed shift in electricity mix from coal to gas from the previous five years. We calibrated our model to find parametric data that would generate the outputs required. We found that the parametric data required to generate these outputs were realistic, and we could model the shift in electricity mix from coal to gas over this time period. 

Whilst this approach has limitations, in that it raises an additional question: can short-term validation translate to long-term verification? However, this is the best solution we found without trying to predict many years in the future which is both infeasible and unrealistic.

From this work, we recommend that models undertake some form of validation using real-world, observed data. Without this, it is possible that some aspects of the real-world system is not accurately modelled. And so, an iterative process can be used to improve the functioning of the model, with this approach.

\subsection{Electricity Demand Prediction}

\textbf{Do poor short-term forecasts of electricity demand have a long-term impact?} Previous work on short-term forecasts of electricity demand has focused on improving the accuracy of such models. However, the long-term impact of these predictions has been studied to a lesser degree. To explore this question, we used a variety of different statistical and machine learning techniques to predict electricity demand. We then took the errors of these predictions and investigated the impact that these errors have on the long-term electricity market by perturbing the electricity demand perceived by the generator companies.

Our results show that with an increase in mean absolute error, there is an increase in CCGT, coal, nuclear, photovoltaics and reciprocal gas engines, with a decrease in onshore and offshore wind. Such a scenario changes the electricity mix landscape in the future, and is suboptimal if a low-carbon electricity mix is desired. 

We recommend that a high amount of research is placed into improving short-term forecasting for electricity demand. Whilst this can be expensive in the short-term, the long-term benefits are significant through the use of a more optimal and efficient system.

\subsection{Carbon Optimization}

\textbf{Is it possible to use an algorithm to set carbon policy?} A significant goal for policy makers is to achieve a low-cost and and low-carbon system. However, in practice, this can be a difficult system to achieve. Policy makers have limited levers in which to test, and those levers can have an uncertain impact on the market.

To simplify the process of finding a cost and carbon optimal carbon tax in the UK electricity market, we used a multi-objective genetic algorithm, NSGA-II, to vary carbon tax between 2018 and 2035. We found that it is possible to use a genetic algorithm to find a variety of carbon policies to minimise both carbon and cost of the electricity system. These carbon policies can be presented to policy makers to choose from, depending on their requirements and preferences. 

The recommendations that arise from this work is that a full quantitative analysis should be undertaken for long-term policy options, and that the search space can be explored using algorithms. This enables policy makers to find a variety of different options, from which they can choose from. Whilst the algorithm may not tell you the precise strategy, it can give a range of feasible options, in which policy makers can use their judgement with additional information.

\subsection{Strategic Bidding}

\textbf{Is it possible to limit the power of large generator companies?} Electricity markets can often be dominated by a small subset of generator companies. If given enough power, these generator companies can then use strategic bidding to artificially inflate the electricity price for their own benefit. 

With the use of reinforcement learning, we were able to simulate the strategic bidding behaviour that generator companies may exhibit. We found that if a generator company controls more than 11\% of total capacity, it is likely that price inflation will occur. However, we found that such price inflation can be limited through the means of a cap on bid levels. We found that a relatively high bid price cap of \textsterling 190/MWh prevents price inflation for the UK market.

We therefore recommend that policy makers place a price cap if they believe that an oligopoly or monopoly is beginning to form. This can limit the impact that these companies can have on consumers and the overall market.

\subsection{AI and ML in a Wider Electricity Market Context}

\textbf{What challenges can AI and ML tackle, and how do these methods relate back to the wider energy system?} In this thesis, we explored the different methods that AI and ML can tackle within the British electricity system. Whilst much work has focused on incremental improvements to targeted problems, such as short-term load forecasting or the possibility of using reinforcement learning for short-term bidding, the wider impact on the market has been explored to a lesser extent.

We found that AI and ML can be used for a wide range of different application domains, which are not limited to solely improving the accuracy of these domains. The subquestions answered previously in this section attest to this, and demonstrate that large-scale problems and systems can be optimised, improved and modelled with the use of state-of-the-art methods.

\section{Future research direction}

There remains significant research that can be undertaken as part of this work that we would like to carry out in the future. A major aspect of intermittent renewable energy sources such as wind and solar is their distributed nature. On a geographic scale as large as the UK, weather conditions change. Therefore, at any one time, capacity factors may change across the country. It would, therefore, be beneficial to model this, by integrating a higher temporal resolution into the model. This would increase the complexity of the model, and therefore compute time, however, the optimal placement of intermittent renewable energy sources could be modelled to ensure the maximum supply at times when it was required from such sources. Work could be done to find simplifications that could be made to other aspects of the model to maintain tractability. 

Additionally, the integration of a larger amount of countries could be carried out. For instance, coupling the UK market with that of Ireland, or with the EU. Whilst this would increase the compute time, the results would be more inclusive of other markets and enable for a larger temporal resolution still. Another area of interest is predicting the price of electricity in the future. Scenarios could be run to predict this price and integrate it into the model. Whilst increasing compute time, this would allow us to explore a wider range of scenarios. 

Another aspect is the comparison to further models. The traditional approach of optimisation models offer a variety of ``gold-standard'' scenarios that we could explore. These comparisons could be undertaken to find weaknesses in both our model as well as other models. 

An interesting area of research is the area of investment using optimisation or control algorithms such as reinforcement learning. Reinforcement learning suffers from a requirement to have multiple training epochs, and therefore we found this a difficult task to achieve in this work. However, through either speed improvements or use of another algorithm, this could be integrated into our work. 

Finally, the integration of fuel supply-demand curves could help us in better modelling electricity markets on both a local and global scale. Firstly, land is a limited resource, especially in the United Kingdom, and therefore limits to the amount of solar or wind that can be produced a real. Secondly, on a global scale, as demand for gas or coal reduces there are large effects on price. Firstly this may reduce due to the supply and demand relationship. However, if demand reduces sufficiently, the price may increase, due to the expense of extraction. This may turn out to be a large inflection point, where gas and coal are used much less. 

%
%

\backmatter


\begin{appendices} 

\end{appendices}

\begin{spacing}{0.9}


\bibliographystyle{apalike}
\cleardoublepage
\bibliography{library,custombibtex-carbon-optimiser,custombibtex-elecsim-1,custombibtex-elecsim-poster,custombibtex-validation-optimiser,forecasting,forecasting-full,forecasting-note,custombibtexfile,custom_bib,bib_custom,custombibtex,ftt-power-custom,systematic-review}



\end{spacing}


\end{document}